\documentclass[aps,showpacs,amsmath,amssymb]{revtex4}

\usepackage{latexsym}
\usepackage{graphicx}
\usepackage{dcolumn}
\usepackage{bm}
\usepackage{hyperref}

\newcommand{\beq}{\begin{equation}}
\newcommand{\eeq}{\end{equation}}
\newcommand{\beqa}{\begin{eqnarray}}
\newcommand{\eeqa}{\end{eqnarray}}
\newcommand{\beqar}{\begin{eqnarray*}}
\newcommand{\eeqar}{\end{eqnarray*}}
\newcommand{\bra}[1]{\mbox{$\left\langle{#1}\right|$}}
\newcommand{\ket}[1]{\mbox{$\left|{#1}\right\rangle$}}
\newcommand{\diracsp}[2]{\mbox{$\langle{#1}|{#2}\rangle$}}

\def\I{{\rm i}}
\def\d{{\rm d}}
\def\e{{\rm e}}

\newcounter{saveeqn}

\begin{document}

\title{Exact solution and perturbation theory in a general quantum system}
\author{An Min Wang}
\affiliation{Quantum Theory Group, Department of Modern Physics,
University of Science and Technology of China, Hefei, 230026,
P.R.China}\email{anmwang@ustc.edu.cn}

\begin{abstract}
By splitting a Hamiltonian into two parts, using the solvability of
eigenvalue problem of one part of the Hamiltonian, proving a useful
identity and deducing an expansion formula of power of operator
binomials, we obtain an explicit and general form of time evolution
operator in the representation of solvable part of the Hamiltonian.
Further we find out an exact solution of Schr\"{o}dinger equation in
a general time-independent quantum system, and write down its
concrete form when the solvable part of this Hamiltonian is taken as
the kinetic energy term. Comparing our exact solution with the usual
perturbation theory makes some features and significance of our
solution clear. Moreover, through deriving out the improved forms of
the zeroth, first, second and third order perturbed solutions
including the partial contributions from the higher order even all
order approximations, we obtain the improved transition probability.
In special, we propose the revised Fermi's golden rule. Then we
apply our scheme to obtain the improved forms of perturbed energy
and perturbed state. In addition, we study an easy understanding
example to illustrate our scheme and show its advantage. All of this
implies the physical reasons and evidences why our exact solution
and perturbative scheme are formally explicit, actually calculable,
operationally efficient, conclusively more accurate. Therefore our
exact solution and perturbative scheme can be thought of theoretical
developments of quantum dynamics. Further applications of our
results in quantum theory can be expected.
\end{abstract}


\maketitle

\section{Introduction}\label{sec1}

One of the most important tasks of physics is to obtain the time
evolution law and its form of a physical system, that is, so-called
dynamical equation and its solution. In a quantum system, the
dynamical equation is the Schr\"{o}dinger equation
\cite{seq,diracpqm} for a pure state or the von Neumann equation for
a mixed state \cite{vonneumann}. Suppose that, at time $t=0$, an
arbitrary initial state is $\ket{\Psi(0)}$, then at a given time
$t$, the state of quantum system will become \beq\label{eq.1}
\ket{\Psi(t)}=\e^{-\I H t}\ket{\Psi(0)}.\eeq where $H$ is the
Hamiltonian of a quantum system and it is assumed independent of
time, while $\e^{-\I H t}$ ($\hbar=1$) is called as the time
evolution operator. Above equation describes an arbitrary initial
state how to evolve to a final state, that is, it is a formal
solution of the Schr\"{o}dinger equation for a pure state.
Unfortunately, only finite several quantum systems are exactly
solvable at present. This implies that in a general quantum system,
the time evolution of expanding coefficients of final state in some
given representation can not be clearly and explicitly expressed by
the $c$-number function. From our point of view, this shortcoming
leads to that we can not fully determine and exactly calculate those
physical quantities dependent on the state (that is expanding
coefficients in a given representation) at any time. In other words,
above formal solution (\ref{eq.1}) of final state in the operator
form is not really useful for the many practical purposes. For
example, in the cases with environment-system interaction, how does
the final state decohere, and in the cases with an entangled initial
state, how does the entanglement of final state vary, and so on?

In order to overcome above shortcoming appearing in a general
quantum system, quantum perturbation theory is well developed and it
is successfully applied the calculations of perturbed energy,
perturbed state and transition probability etc. In the usual
perturbation theory, the key idea to research the time evolution of
system is to split the Hamiltonian of system into two parts, that is
\beq H=H_0+H_1,\eeq where the eigenvalue problem of $H_0$ is
solvable, and $H_1$ is the rest part of the Hamiltonian except for
$H_0$. In other words, this splitting is chosen in such a manner
that the solutions of $H_0$ are known as \beq\label{h0eeq}
H_0\ket{\Phi^\gamma}=E_\gamma \ket{\Phi^\gamma}, \eeq where
$\ket{\Phi^\gamma}$ is the eigenvector of $H_0$ and $E_\gamma$ is
the corresponding eigenvalue. However, when $H_1$ is not so small
compared with $H_0$ that we have to consider the higher order
approximations or when the partial contributions from higher order
even all order approximations become relatively important to the
studied problems, the usual perturbation theory might be difficult
to calculate to an enough precision in an effective way, even not
feasible practically since the lower approximation might break the
physical symmetries and/or constraints. In our point of view, the
physical reasons resulting in above difficulties and obstacles are
that the usual perturbation theory does not give a general term form
of expanding coefficient evolution with time for any order
approximation and does not explicitly express it as a $c$-number
function. Thus, it is necessary to find the perturbed solution from
the low to the high order approximation step by step up to some
order approximation for a needed precision. These problems motivate
us to focus on finding an explicit expression of general term for
any order approximation, that is an exact solution that is not in
the operator form, and then reasonably and physically deduce out a
scheme of perturbation theory. In fact, as soon as this purpose is
arrived at, we will obtain an exact solution with all order
contributions coming from $H_1$. This consequentially implies, in
principle, that this scheme can more effectively and more accurately
solve the physical problems than using the existing method. Then, we
concretely realize and demonstrate this physical idea by providing a
scheme of perturbation theory. In this process, just like the
accustomed way to study a more complicated quantum system, if $H_1$
is indeed small enough compared with $H_0$, its influence may be as
a perturbation and one can take up to some order approximation
conveniently at a needed precision. However, the distinct difference
between ours and usual one is that our perturbed solution will be
able to include the partial contributions from the higher order even
all order approximations based on the reasonable physical
considerations so that the physical problems are more accurately and
effectively calculated, and kept their symmetries and/or constraints
as possible since we have used the feature of our exact solution
with the general term form of time evolution of arbitrary initial
state and added systematically the contributions from the high order
approximations.

Actually, one of interesting and important tasks in the study of
quantum dynamics is to find clearly the the expanding coefficients
(amplitudes) before the basis vectors in a given representation how
to evolve. The typical examples are the calculations decoherence and
disentanglement of quantum systems. Although we split the
Hamiltonian into two parts in terms of the known method, the problem
has not yet been finally answered because that $H_0$ does not
commute with $H_1$ in general, and the matrix elements of time
evolution operator $\e^{-\I Ht}$ do not have the closed factor forms
about time $t$ in the given representations including $H_0$
representation spanned by $\{\ket{\Phi^\gamma}, \gamma=1,2,\cdots
\}$ in terms of the known perturbation theory. To solve this kind of
problems is also one of our main motivations.

Just well-known, quantum dynamics and its perturbation theory are
correct and have been sufficiently studied. Many famous physicists
created their nice formulism and obtained some marvelous results. An
attempt to improve its part content or increase some new content as
well as some new methods must be very difficult in their
realizations. This results in our this possibly lengthy manuscript.
However, our endeavor is only at the beginning. It must be pointed
out we still use the extensively accepted principles and laws of
quantum mechanics, and we try to use elementary mathematics as
possible. All of mathematics knowledge used here does not exceed the
university level.

Undoubtedly, the formalization of physical theory including quantum
dynamics often has its highly mathematical focus, but this can not
cover its real motivations, potential applications and related
conclusions in physics. We study the reexpression in form should not
be only a mathematical game and skill, we expect to find its
physical content and applications. This is leaded actually by our
more explicit and general form of the Schr\"{o}dinger equation and
our perturbative scheme. A series of conclusions done here make us
assure that we have partially arrived at our purpose.

In this paper, we start from finding an explicitly closed form about
time $t$ of matrix elements of time evolution operator $\e^{-\I Ht}$
in the $H_0$ representation spanned by $\{\ket{\Phi^\gamma},
\gamma=1,2,\cdots \}$. Our method is first to derive out an
expansion formula of power of operator binomials and then apply it
to the Taylor's expansion of time evolution operator $\e^{-\I Ht}$.
Moreover, by proving and using our identity, we derive out the
explicit and general form of representation matrix of $\e^{-\I Ht}$
in the representation of $H_0$. Consequently, we obtain an exact
solution of the Schr\"odinger equation in a general time-independent
quantum system, in particular, its concrete form when the solvable
part of Hamiltonian is taken as the kinetic energy term.
Furthermore, by comparing our solution with the usual perturbation
theory, we reveal their relation and show what is more as well as
what is different in our exact solution. The conclusions clearly
indicate that our exact solution is consistent with the usual
perturbation theory at any order approximation, but also in our
exact solution we explicitly calculate out the expanding
coefficients of unperturbed state in Lippmann-Schwinger equation
\cite{LSE} for the time-independent (stationary) perturbation theory
and/or fully solves the recurrence equation of the expansion
coefficients of final state in $H_0$ representation from a view of
time-dependent (dynamical) perturbation theory. After introducing
the perturbative method, in order to provide a new scheme of
perturbation theory based on our exact solution, we first propose
two useful skills. Then, we expressly demonstrate our solution is
not only formally explicit, but also actually more accurate and
effective via generally deriving out the improved forms of the
zeroth, first, second and third order approximation of perturbed
solution including partial contributions from the higher order even
all order approximations, finding the improved transition
probability, specially, the revised Fermi's golden rule, and
providing a operational scheme to calculate the perturbed energy and
perturbed state. Furthermore, by studying a concrete example of two
state system, we illustrate clearly that our solution is more
efficient and more accurate than the usual perturbative method. In
short, our exact solution and perturbative scheme are formally
explicit, actually calculable, operationally efficient, conclusively
more accurate (to the needed precision).

It is worth emphasizing that we obtain an explicit and general
expression of time evolution operator that is a summation over all
order contributions from the rest part of Hamiltonian except for the
solvable part, or all order approximations of the perturbative part
of the Hamiltonian. This makes us be able to build our perturbative
scheme via so-called ``dynamical rearrangement and summation"
technology, which is seen in Sec. \ref{sec7}. In a sentence, our
perturbative scheme comes from the our exact solution, and then show
our exact solution indeed contain useful and interesting physical
content besides its explicit mathematics form.

Based on the above all of reasons, our exact solution and
perturbative scheme can be thought of theoretical developments of
quantum dynamics. In our point of view, it is helpful for
understanding the dynamical behavior and related subjects of quantum
systems in theory. Specially, we think that the features and
advantages of our solution can not be fully revealed only by the
perturbative method, because it is an exact solution. We would like
to study the possible applications to the formulation of quantum
dynamics in the near future.

This paper consists of 11 sections and one appendix. Besides Sec.
\ref{sec1} is an introduction, Sec. \ref{sec11} is the conclusion
and discussion, the other 9 sections can be divided into three
parts. The first part made of sections from two to four is to deduce
and prove our exact solution in the general time-independent quantum
system; the second part made of section five is to compare our exact
solution including all of order approximations with the usual
perturbation theory and demonstrate the features and advantages of
our exact solution; the third part made of sections from six to ten
is to propose our perturbative scheme, obtain its application and
illustrate its results. More concretely, they are organized as the
following: in Sec. \ref{sec2} we first propose the expansion formula
of power of operator binomials; in Sec. \ref{sec3} we derive out an
explicit and general expression of time evolution operator by
proving and using our identity; in Sec. \ref{sec4} we obtain the
exact solution of the Schr\"odinger equation and present a concrete
example when the solvable part of Hamiltonian is taken as the
kinetic energy term; in Sec. \ref{sec5} we compare our solution with
the usual perturbation theory and prove their consistency, their
relations and explain what is more as well as what is different in
our solution; in Sec. \ref{sec6} we introduce two skills to relate
our exact solution and the perturbative scheme in order to include
the contributions from the high order approximations; in Sec.
\ref{sec7} we deduce out the improved perturbed solution of dynamics
including partial contributions from the higher even all order
approximations; in Sec. \ref{sec8} we obtain the improved transition
probability, specially, the revised Fermi's golden rule. In Sec.
\ref{sec9} we provide a scheme to calculate the perturbed energy and
the perturbed state; in Sec. \ref{sec10} we study an example of two
state system in order to concretely illustrate our solution to be
more effective and more accurate than the usual method; in Sec.
\ref{sec11} we summarize our conclusions and give some discussions.
Finally, we write an appendix where the proof of our identity is
presented and some expressions are calculated in order to derive out
our improved forms of perturbed solutions.

\section{Expansion formula of power of operator
binomials}\label{sec2}

In order to obtain the explicitly exact solution of the
Schr\"odinger equation in a general time-independent quantum system,
let us start with the derivation of expansion formula of power of
operator binomials. Without loss of generality, we are always able
to write the power of operator binomials as two parts
\beq\label{fnd} (A+B)^n=A^n+f^n(A,B). \eeq where $A$ and $B$ are two
operators and do not commute with each other in general. While the
second part $f^n(A,B)$ is a polynomials including at least first
power of $B$ and at most $n$th power of $B$ in its every term. Thus,
a general term with $l$th power of $B$ has the form $
\left(\prod_{i=1}^lA^{k_i}B\right)A^{n-l-\sum_{i=1}^l k_i}$. From
the symmetry of power of binomials we conclude that every $k_i$ take
the values from 0 to $(n-l)$, but it must keep $n-l-\sum_{i=1}^l
k_i\geq 0$. So we have \beqa\label{fne}
f^n(A,B)\!\!&=&\!\!\sum_{l=1}^n\!\!\sum_{\stackrel{\scriptstyle
k_1,\cdots,k_l=0}{\sum_{i=1}^l k_i+l\leq
n}}^{n-l}\!\!\!\left(\prod_{i=1}^lA^{k_i}B\right)\!A^{n-l-\sum_{i=1}^l
k_i}\\
&=&\label{fne1}\sum_{l=1}^n\sum_{k_1,\cdots,k_l=0}^{n-l}
\left(\prod_{i=1}^lA^{k_i}B\right)A^{n-l-\sum_{i=1}^l
k_i}\theta\left(n-l-\sum_{i=1}^l k_i\right). \eeqa where $\theta(x)$
is a step function, that is, $\theta(x)=1$ if $x\geq 0$, and
$\theta(x)=0$ if $x<0$. Obviously based on above definition, we
easily verify \beq f^1(A,B)=B,\quad f^2(A,B)=AB+B(A+B). \eeq They
imply that the expression (\ref{fne}) is correct for $n=1,2$.

Now we use the mathematical induction to proof the expression
(\ref{fne}) of $f^n(A,B)$, that is, let us assume that it is valid
for a given $n$, and then prove that it is valid too for $n+1$.
Denoting ${\mathcal{F}}^{n+1}(A,B)$ with the form of expression
(\ref{fne1}) where $n$ is replaced by $n+1$, and we extract the part
of $l=1$ in its finite summation for $l$ \beqa
\mathcal{F}^{n+1}(A,B)&=&\sum_{k_1=0}^{n}A^{k_1}BA^{n-k_1}+\sum_{l=2}^{n+1}
\sum_{k_1,\cdots,k_l=0}^{(n+1)-l}\left(\prod_{i=1}^lA^{k_i}B\right)A^{(n+1)-l-\sum_{i=1}^l
k_i}\nonumber\\
& &\times\theta\left((n+1)-l-\sum_{i=1}^l k_i\right).\eeqa We
extract the terms $k_1=0$ in the first and second summations, and
again replace $k_1$ by $k_1-1$ in the summation (the summations for
$k_i$ from 1 to $(n+1)-l$ change as one from 0 to $n-l$), the result
is \beqa \mathcal{F}^{n+1}(A,B)&=&B
A^n+\sum_{k_1=1}^{n}A^{k_1}BA^{n-k_1}+ B\sum_{l=2}^{n+1}
\sum_{k_2,\cdots,k_l=0}^{(n+1)-l}\left(\prod_{i=2}^lA^{k_i}B\right)\nonumber\\
& &\times A^{(n+1)-l-\sum_{i=2}^l
k_i}\theta\left((n+1)-l-\sum_{i=2}^l k_i\right) \nonumber\\
& & +\sum_{l=2}^{n+1}\sum_{k_1=1}^{(n+1)-l}
\sum_{k_2,\cdots,k_l=0}^{(n+1)-l}\left(\prod_{i=1}^lA^{k_i}B\right)\nonumber
\\ & &\times A^{(n+1)-l-\sum_{i=1}^l k_i}\theta\left((n+1)-l-\sum_{i=1}^l
k_i\right), \eeqa furthermore \beqa \mathcal{F}^{n+1}(A,B)&=&B
A^n+A\sum_{k_1=0}^{n-1}A^{k_1}BA^{n-1-k_1}+ B\sum_{l=2}^{n+1}
\sum_{k_2,\cdots,k_l=0}^{(n+1)-l}\left(\prod_{i=2}^lA^{k_i}B\right)\nonumber\\
& &\times A^{(n+1)-l-\sum_{i=2}^l
k_i}\theta\left((n+1)-l-\sum_{i=2}^l k_i\right) \nonumber\\
& & +A\sum_{l=2}^{n+1}\sum_{k_1=0}^{n-l}
\sum_{k_2,\cdots,k_l=0}^{(n+1)-l}\left(\prod_{i=1}^lA^{k_i}B\right)A^{n-l-\sum_{i=1}^l
k_i}\theta\left(n-l-\sum_{i=1}^l k_i\right).\label{fneproof}\eeqa
Considering the third term in above expression, we change the dummy
index $\{k_2,k_3,\cdots,k_l\}$ into $\{k_1,k_2,\cdots,k_{l-1}\}$,
rewrite $(n+1)-l$ as $n-(l-1)$, and finally replace $l$ by $l-1$ in
the summation (the summation for $l$ from 2 to $n+1$ changes as one
from 1 to $n$), we obtain \beqa & &B\sum_{l=2}^{n+1}
\sum_{k_2,\cdots,k_l=0}^{(n+1)-l}\left(\prod_{i=2}^lA^{k_i}B\right)A^{(n+1)-l-\sum_{i=2}^l
k_i}\theta\left((n+1)-l-\sum_{i=2}^l k_i\right)\nonumber\\
& &\quad = B\sum_{l=2}^{n+1}
\sum_{k_1,\cdots,k_{l-1}=0}^{n-(l-1)}\left(\prod_{i=1}^{l-1}A^{k_i}B\right)A^{n-(l-1)-\sum_{i=1}^{l-1}
k_i}\theta\left(n-(l-1)-\sum_{i=1}^{l-1} k_i\right)\nonumber\\
& &\quad = B\sum_{l=1}^{n}
\sum_{k_1,\cdots,k_{l}=0}^{n-l}\left(\prod_{i=1}^{l}A^{k_i}B\right)A^{n-l-\sum_{i=1}^{l}
k_i}\theta\left(n-l-\sum_{i=1}^{l} k_i\right)\nonumber\\ & &\quad=
Bf^n(A,B)\label{fnet3},\eeqa where we have used the expression
(\ref{fne1}) of $f^n(A,B)$. Because of the step function
$\theta\left(n-l-\sum_{i=1}^l k_i\right)$ in the fourth term of the
expression (\ref{fneproof}), the upper bound of summation for $l$ is
abated to $n$, and the upper bound of summation for
$k_2,k_3,\cdots,k_l$ is abated to $n-l$. Then, merging the fourth
term and the second term in eq.(\ref{fneproof}) gives \beqa &
&A\sum_{k_1=0}^{n-1}A^{k_1}BA^{n-1-k_1}+A\sum_{l=2}^{n+1}\sum_{k_1=0}^{n-l}
\sum_{k_2,\cdots,k_l=0}^{(n+1)-l}\left(\prod_{i=1}^lA^{k_i}B\right)\nonumber\\
& &\qquad \times A^{n-l-\sum_{i=1}^l
k_i}\theta\left(n-l-\sum_{i=1}^l k_i\right)\nonumber\\
& &\quad =A\sum_{k_1=0}^{n-1}A^{k_1}BA^{n-1-k_1}+A\sum_{l=2}^{n}
\sum_{k_1,k_2,\cdots,k_l=0}^{n-l}\left(\prod_{i=1}^lA^{k_i}B\right)\nonumber\\
& &\qquad \times A^{n-l-\sum_{i=1}^l
k_i}\theta\left(n-l-\sum_{i=1}^l k_i\right)\nonumber\\
& &\quad =A\sum_{l=1}^{n}
\sum_{k_1,k_2,\cdots,k_l=0}^{n-l}\left(\prod_{i=1}^lA^{k_i}B\right)A^{n-l-\sum_{i=1}^l
k_i}\theta\left(n-l-\sum_{i=1}^l k_i\right)\nonumber\\
& &\qquad =A f^n(A,B)\label{fnet24},\eeqa where we have used again
the expression (\ref{fne1}) of $f^n(A,B)$.

Substituting (\ref{fnet3}) and (\ref{fnet24}) into (\ref{fneproof}),
immediately leads to the following result \beq
\label{fnadd1org}\mathcal{F}^{n+1}=B A^n +(A+B)f^n(A,B).\eeq

Note that
\beqa (A+B)^{n+1}&=&(A+B)(A+B)^n\nonumber\\
&=&A^{n+1}+B A^{n}+(A+B)f^{n}(A,B),\eeqa we have the relation \beq
\label{fnadd1}f^{n+1}(A,B)=B A^{n}+(A+B)f^{n}(A,B).\eeq Therefore,
in terms of eqs.(\ref{fnadd1org}) and (\ref{fnadd1}) we have
finished our proof that  $f^n(A,B)$ has the expression (\ref{fne})
or (\ref{fne1}) for any $n$.

\section{Expression of the time evolution operator}\label{sec3}

Now we investigate the expression of the time evolution operator by
means of above expansion formula of power of operator binomials,
that is, we write \beq e^{-\I H t}=\sum_{n=0}^\infty\frac{(-\I
t)^n}{n!}(H_0+H_1)^n=\e^{-\I H_0 t}+\sum_{n=0}^\infty\frac{(-\I
t)^n}{n!}f^n(H_0,H_1). \eeq In above equation, inserting the
complete relation $
\sum_{\gamma}\ket{\Phi^{\gamma}}\bra{\Phi^{\gamma}}=1 $ before every
$H_0^{k_i}$, and using the eigen equation of $H_0$ (\ref{h0eeq}), it
is easy to see that \beqa \label{fk}
f^n(H_0,H_1)&=&\sum_{l=1}^n\sum_{\gamma_1,\cdots,\gamma_{l+1}}
\sum_{\stackrel{\scriptstyle k_1,\cdots,k_l=0}{\sum_{i=1}^l
k_i+l\leq n}}^{n-l}\left[\prod_{i=1}^l
E_{\gamma_i}^{k_i}\right]\nonumber\\ & & \times
E_{\gamma_{l+1}}^{n-\sum_{i=1}^lk_i-l}
\left[\prod_{i=1}^lH_1^{\gamma_i\gamma_{i+1}}\right]
\ket{\Phi^{\gamma_1}}\bra{\Phi^{\gamma_{l+1}}}\nonumber\\
&=&\sum_{l=1}^n\sum_{\gamma_1,\cdots,\gamma_{l+1}}C_l^n(E[\gamma,l])
\left[\prod_{i=1}^lH_1^{\gamma_i\gamma_{i+1}}\right]
\ket{\Phi^{\gamma_1}}\bra{\Phi^{\gamma_{l+1}}}, \label{fkc}\eeqa
where $H_1^{\gamma_i\gamma_{i+1}}
=\bra{\Phi^{\gamma_{i}}}H_1\ket{\Phi^{\gamma_{i+1}}}$, $E[\gamma,l]$
is a vector with $l+1$ components denoted by \beq E[\gamma,l]
=\{E_{\gamma_1},E_{\gamma_2},
\cdots,E_{\gamma_l},E_{\gamma_{l+1}}\}\eeq and we introduce the
definition of $C^n_l(E[\gamma,l])$ ($l\geq 1$) as the following
\beq\label{cd1} C_l^n(E[\gamma,l])=\sum_{\stackrel{\scriptstyle
k_1,\cdots,k_l=0}{\sum_{i=1}^l k_i+l\leq
n}}^{n-l}\!\!\!\left[\prod_{i=1}^l E_{\gamma_i}^{k_i}\right]
E_{\gamma_{l+1}}^{n-\sum_{i=1}^lk_i-l}.\eeq

In order to derive out an explicit and useful expression of
$C^n_l(E[\gamma,l])$, we first change the dummy index
$k_l\rightarrow n-l-\sum_{i=1}^lk_i$ in the eq.(\ref{fne1}). Note
that at the same time in spite of $\theta(n-l-\sum_{i=1}^lk_i)$
changes as $\theta(k_l)$, but a hiding factor $\theta(k_l)$ becomes
$\theta(n-l-\sum_{i=1}^lk_i)$. Thus, we can rewrite
\beqa\label{fennew}
f^n(A,B)&=&\sum_{l=1}^n\sum_{k_1,\cdots,k_l=0}^{n-l}
\left(\prod_{i=1}^{l-1}A^{k_i}B\right)A^{n-l-\sum_{i=1}^l
k_i}B\nonumber\\
& &\times A^{k_l}\theta\left(n-l-\sum_{i=1}^l k_i\right). \eeqa In
fact, this new expression of $f^n(A,B)$ is a result of the symmetry
of power of binomials for its every factor. Similarly, in terms of
the the above method to obtain the definition of
$C^n_l(E[\gamma,l])$ (\ref{cd1}), we have its new definition (where
$k_l$ is replaced by $k$) \beqa C_l^n(E[\gamma,l])
&=&\sum_{k=0}^{n-l}\left\{\sum_{k_1,\cdots,k_{l-1}=0}^{n-l}\left[\prod_{i=1}^{l-1}
E_{\gamma_i}^{k_i}\right]E_{\gamma_l}^{(n-k)-\sum_{i=1}^{l-1}k_i-l}\right.\nonumber\\
& &\left. \times
\theta\left((n-k)-l-\sum_{i=1}^{l-1}k_i\right)\right\}E^k_{\gamma_{l+1}}.
\eeqa Because that $\theta\left((n-k)-l-\sum_{i=1}^{l-1}k_i\right)$
abates the upper bound of summation for $k_i (i=1,2,\cdots,l-1)$
from $(n-l)$ to $(n-k-1)-(l-1)$, we obtain the recurrence equation
\beqa C_l^n(E[\gamma,l])
&=&\sum_{k=0}^{n-l}\left\{\sum_{k_1,\cdots,k_{l-1}=0}^{(n-k-1)-(l-1)}\left[\prod_{i=1}^{l-1}
E_{\gamma_i}^{k_i}\right]E_{\gamma_l}^{(n-k-1)-\sum_{i=1}^{l-1}k_i-(l-1)}\right.\nonumber\\
& &\left. \times
\theta\left((n-k-1)-(l-1)-\sum_{i=1}^{l-1}k_i\right)\right\}E^k_{\gamma_{l+1}}\nonumber\\
&=&\sum_{k=0}^{n-l}C_{l-1}^{n-k-1}(E[\gamma,l-1])
E_{\gamma_{l+1}}^k. \label{creq} \eeqa

In particular, when $l=1$, from the definition of
$C^n_l(E(\gamma,1))$ it follows that \beq
C^n_1(E(\gamma,1))=\sum_{k_1=0}^{n-1}E_{\gamma_1}^{k_1}E_{\gamma_2}^{n-1-k_1}
=E_{\gamma_2}^{n-1}\sum_{k_1=0}^{n-1}\left(\frac{E_{\gamma_1}}{E_{\gamma_2}}\right)^{k_1}.\eeq
By means of the summation formula of a geometric series, we find
that \beq
C^n_1(E(\gamma,1))=\frac{E_{\gamma_1}^n}{E_{\gamma_1}-E_{\gamma_2}}
-\frac{E_{\gamma_2}^n}{E_{\gamma_1}-E_{\gamma_2}}.\eeq Based on the
recurrence equation (\ref{creq}), we have \beqa
C^n_2(E(\gamma,2))&=&\sum_{k=0}^{n-2}\left(\frac{E_{\gamma_1}^{n-k-1}}{E_{\gamma_1}-E_{\gamma_2}}
-\frac{E_{\gamma_2}^{n-k-1}}{E_{\gamma_1}-E_{\gamma_2}}\right)E_{\gamma_3}^k\nonumber\\
&=&
\sum_{i=1}^2\frac{(-1)^{i-1}E_{\gamma_i}^{n-1}}{E_{\gamma_1}-E_{\gamma_2}}
\sum_{k=0}^{n-2}\left(\frac{E_{\gamma_3}}{E_{\gamma_i}}\right)^{k}\nonumber\\
&=&\sum_{i=1}^2\frac{(-1)^{i-1}(E_{\gamma_i}^n-E_{\gamma_i}E_{\gamma_3}^{n-1})}
{(E_{\gamma_1}-E_{\gamma_2})(E_{\gamma_i}-E_{\gamma_3})}\nonumber\\
&=&\frac{E_{\gamma_1}^n}{(E_{\gamma_1}-E_{\gamma_2})(E_{\gamma_1}-E_{\gamma_3})}
-\frac{E_{\gamma_2}^n}{(E_{\gamma_1}-E_{\gamma_2})(E_{\gamma_2}-E_{\gamma_3})}\nonumber\\
&
&+\frac{E_{\gamma_3}^n}{(E_{\gamma_1}-E_{\gamma_3})(E_{\gamma_2}-E_{\gamma_3})}.\eeqa
In the above calculations, the last step is important. In fact, in
order to obtain the concrete expression of $C_l^n$ ($l$ and $n$ are
both positive integers), we need our identity \beq
\label{myi}\sum_{i=1}^{l+1} (-1)^{i-1}
\frac{E_{\gamma_i}^K}{d_i(E[\gamma,l])}
=\left\{\begin{array}{c l}0 &\quad (\mbox{If $0\leq K<l$})\\[8pt] 1 &\quad (\mbox{If
$K=l$})\end{array}\right.. \eeq It is proved in Appendix A in
detail. The denominators $d_i(E[\gamma,l])$ in above identity are
defined by \vskip -0.2cm\beqa
d_1(E[\gamma,l])&=&\prod_{i=1}^{l}\left(E_{\gamma_{1}}
-E_{\gamma_{i+1}}\right),\\
 d_i(E[\gamma,l])&=&
\prod_{j=1}^{i-1}\left(E_{\gamma_{j}}
-E_{\gamma_{i}}\right)\!\!\!\prod_{k=i+1}^{l+1}\left(E_{\gamma_{i}}
-E_{\gamma_{k}}\right),\\[-3pt] d_{l+1}(E[\gamma,l])
&=&\prod_{i=1}^{l}\left(E_{\gamma_{i}}-E_{\gamma_{l+1}}\right),\eeqa
where $2\leq i \leq l$. Then, using the recurrence equation
(\ref{creq}) and our identity (\ref{myi}), we obtain \beq\label{cne}
C^n_l(E[\gamma,l])=\sum_{i=1}^{l+1} (-1)^{i-1}
\frac{E_{\gamma_i}^n}{d_i(E[\gamma,l])}. \eeq Here, the mathematical
induction shows its power again. If the expression (\ref{cne}) is
correct for a given $n$, for example $n=1,2$, then for $n+1$ from
the recurrence equation (\ref{creq}) it follows that \beqa
C^{n+1}_l(E[\gamma,l])\!\!&=&\!\!\sum_{k=0}^{n+1-l}\left(\sum_{i=1}^{l}
(-1)^{i-1}
\frac{E_{\gamma_i}^{n-k}}{d_i(E[\gamma,l-1])}\right)E_{\gamma_{l+1}}^k\nonumber\\
&=&
\sum_{i=1}^{l}\frac{(-1)^{i-1}E_{\gamma_i}^{n}}{d_i(E[\gamma,l-1])}\sum_{k=0}^{n+1-l}
\left(\frac{E_{\gamma_{l+1}}}{E_{\gamma_{i}}}\right)^k\nonumber\\
&=&\sum_{i=1}^{l}\frac{(-1)^{i-1}(E_{\gamma_i}^{n+1}-E_{\gamma_i}^{l-1}E_{\gamma_{l+1}}^{n+2-l})}
{d_i(E[\gamma,l-1])(E_{\gamma_i}-E_{\gamma_{l+1}})}.\label{cnproof}\eeqa
Because that
$d_i(E[\gamma,l-1])(E_{\gamma_i}-E_{\gamma_{l+1}})=d_i(E[\gamma,l])$
($i\leq l+1$) and our identity (\ref{myi}), it is easy to see\beq
\sum_{i=1}^{l}\frac{(-1)^{i-1}E_{\gamma_i}^{l-1}}
{d_i(E[\gamma,l])}= -(-1)^l \frac{E_{\gamma_{l+1}}^{l-1}}
{d_{l+1}(E[\gamma,l])}. \eeq Substitute it into eq.(\ref{cnproof})
yields \beq C^{n+1}_l(E[\gamma,l])=\sum_{i=1}^{l+1} (-1)^{i-1}
\frac{E_{\gamma_i}^{n+1}}{d_i(E[\gamma,l])}. \eeq That is, we have
proved that the expression (\ref{cne}) of $C^n_l(E[\gamma,l])$ is
valid for any $n$.

It must be emphasized that for the diagonal elements or in the
degeneration cases, we need to understand above expressions in the
sense of limitations. For instance, for $
C^n_1(E[\gamma,1])=\left(E_{\gamma_{1}}^{n}
-E_{\gamma_{2}}^{n}\right)/\left({E_{\gamma_{1}}
-E_{\gamma_{2}}}\right)$, we have that the expression $
\lim_{E_{\gamma_{2}} \rightarrow E_{\gamma_{1}}}C_1^n(E[\gamma,1])
=\left(\delta_{\gamma_1\gamma_2}
+\Theta(\left|\gamma_2-\gamma_1\right|)\delta_{E_{\gamma_{1}}
E_{\gamma_{2}}}\right)n E_{\gamma_{1}}^{n-1}$, where the step
function $\Theta(x)=1$, if $x>0$, and $\Theta(x)=0$, if $x\leq 0$.
Because of our identity (\ref{myi}), the summation to $l$ in
eq.(\ref{fkc}) can be extended to $\infty$. Thus, the expression of
the time evolution operator is changed to a summation according to
the order (or power) of the $H_1$ as the following \beqa \label{mk}
& &\bra{\Phi^\gamma}\e^{-\I H t}\ket{\Phi^{\gamma^\prime}}= \e^{-\I
E_{\gamma} t}\delta_{\gamma\gamma^\prime} + \sum_{l=1}^\infty
\sum_{\gamma_1,\cdots,\gamma_{l+1}}\left[
\sum_{i=1}^{l+1}(-1)^{i-1}\frac{\e^{-\I E_{\gamma_i}
t}}{d_i(E[\gamma,l])}\right]
\prod_{j=1}^{l}H_1^{\gamma_j\gamma_{j+1}}
\delta_{\gamma_1\gamma}\delta_{\gamma_{l+1}\gamma^\prime}.\eeqa It
is clear that it has the closed time evolution factors.

\section{Solution of the Schr\"odinger equation}\label{sec4}

Substituting the our expression of the time evolution operator
(\ref{mk}) into eq.(\ref{eq.1}), we obtain immediately our explicit
form of time evolution of an arbitrary initial state in a general
quantum system \beqa \label{ouress}
\ket{\Psi(t)}&=&\sum_{\gamma,\gamma^\prime}\left[\e^{-\I E_{\gamma}
t}\delta_{\gamma\gamma^\prime} + \sum_{l=1}^\infty
A_l^{\gamma\gamma^\prime}(t) \right]
\left[\diracsp{\Phi^{\gamma^\prime}}{\Psi(0)}\right]\ket{\Phi^\gamma},\\
\label{Aldefinition}
A_l^{\gamma\gamma^\prime}(t)&=&\sum_{\gamma_1,\cdots,\gamma_{l+1}}\left[
\sum_{i=1}^{l+1}(-1)^{i-1}\frac{\e^{-\I E_{\gamma_i}
t}}{d_i(E[\gamma,l])}\right]\left[
\prod_{j=1}^{l}H_1^{\gamma_j\gamma_{j+1}}\right]
\delta_{\gamma_1\gamma}\delta_{\gamma_{l+1}\gamma^\prime}.\eeqa Of
course, since the linearity of the evolution operator and the
completeness of eigenvector set of $H_0$, we can simply set the
initial state as an eigenvector of $H_0$. Up to now, everything is
exact and no any approximation enters. Therefore, it is an exact
solution in spite of its form is an infinity series. In other words,
it exactly includes all order approximations of $H_1$. Only when
$H_1$ is taken as a perturbation, it can be cut-off based on the
needed precision. From a view of formalized theory, it is explicit
and general, but is not compact. Moreover, if the convergence is
guaranteed, it is strict since its general term is known. Usually,
to a practical purpose, if only including the finite (often low)
order approximation of $H_1$, above expression is cut-off to the
finite terms and becomes a perturbed solution.

Actually, since our solution includes all order contributions from
the rest part of Hamiltonian except for the solvable part $H_0$, or
all order approximations of the perturbative part $H_1$ of
Hamiltonian, it is not very important whether $H_1$ is (relatively)
large or small when compared with $H_0$. In principle, for a general
quantum system with the normal form Hamiltonian, we always can write
down \beq \label{Hnf} H=\frac{\hat{\bm{k}}^2}{2m}+V .\eeq If taking
the solvable kinetic energy term as
$H_0=\displaystyle\frac{\hat{\bm{k}}^2}{2m}$ and the potential
energy part as $H_1=V$, our method is applicable to such quantum
system. It is clear that \beq \frac{\hat{\bm{k}}^2}{2m}\ket{\bm{k}}
=\frac{\bm{k}^2}{2m}\ket{\bm{k}}=E_{\bm{k}}\ket{\bm{k}},\quad
E_{\bm{k}}=\frac{\bm{k}^2}{2m}. \eeq And denoting \beqa
\diracsp{\bm{x}}{\bm{k}}&=&\frac{1}{L^{3/2}}\;\e^{\I\bm{k}\cdot\bm{x}},\\
\psi_{\bm{k}}(0)&=& \diracsp{\bm{k}}{\Psi(0)},\\
\Psi(\bm{x},t)&=&\diracsp{\bm{x}}{\Psi(t)},\eeqa we obtain the final
state as \beqa \label{ffs1} \Psi(\bm{x},t)&=&\sum_{\bm{k}}
\psi_{\bm{k}}(0)\frac{\e^{\I\left(\bm{k}\cdot\bm{x}-E_{\bm{k}}t\right)}}{L^{3/2}}\nonumber\\
& &+ \sum_{l=1}^\infty
\sum_{\bm{k},\bm{k}^\prime}\sum_{\bm{k}_1,\cdots,\bm{k}_{l+1}}
\psi_{\bm{k}^\prime}(0)
\left[\prod_{j=1}^{l}V^{\bm{k}_j\bm{k}_{j+1}}\right]
\delta_{\bm{k}_1\bm{k}}\delta_{\bm{k}_{l+1}\bm{k}^\prime}\nonumber\\
& &\times\left[ \sum_{i=1}^{l+1}\frac{(-1)^{i-1}}{d_i(E[k,l])}\;
\frac{1}{L^{3/2}}\;\e^{\I\left(\bm{k}\cdot\bm{x}-E_{\bm{k}_i}t\right)}
\right]\\ \label{ffs2} &=&\sum_{\bm{k}}
\psi_{\bm{k}}(0)\frac{\e^{\I\left(\bm{k}\cdot\bm{x}-E_{\bm{k}}t\right)}}{L^{3/2}}\nonumber\\
& &+ \sum_{l=1}^\infty
\sum_{\bm{k},\bm{k}^\prime}\sum_{\bm{k}_1,\cdots,\bm{k}_{l+1}}
\psi_{\bm{k}^\prime}(0)
\left[\prod_{j=1}^{l}\mathcal{V}(\bm{k}_j-\bm{k}_{j+1})\right]
\delta_{\bm{k}_1\bm{k}}\delta_{\bm{k}_{l+1}\bm{k}^\prime}\nonumber\\
& &\times\left[ \sum_{i=1}^{l+1}\frac{(-1)^{i-1}}{d_i(E[k,l])}\;
\frac{1}{L^{3/2}}\;\e^{\I\left(\bm{k}\cdot\bm{x}-E_{\bm{k}_i}t\right)}
\right].\eeqa In the second equal mark, we have used the fact that
$V=V(\bm{x})$ usually, thus \beqa
V^{\bm{k}_j\bm{k}_{j+1}}&=& \bra{\bm{k_j}}V(\bm{x})\ket{\bm{k}_{j+1}}\nonumber\\
&=&\int^\infty_{-\infty}\d^3x_j\d^3x_{j+1}
\diracsp{\bm{k_j}}{\bm{x}_j}\bra{\bm{x}_j}V(\bm{x})\ket{\bm{x}_{j+1}}
\diracsp{\bm{x}_{j+1}}{\bm{k_{j+1}}}\nonumber\\
&=&\frac{1}{L^3}\int^\infty_{-\infty}\d^3x_j\d^3x_{j+1}
\e^{-\I\bm{k}_j\cdot\bm{x}_j}V(\bm{x}_{j+1})\delta^3(\bm{x}_j-\bm{x}_{j+1})
\e^{\I\bm{k}_{j+1}\cdot\bm{x}_{+1}j}\nonumber \\
&=&\frac{1}{L^3}\int^\infty_{-\infty}\d^3x
V(\bm{x})\e^{-\I(\bm{k_{j}}-\bm{k_{j+1}})\cdot\bm{x}}\nonumber\\
 &=&\mathcal{V}(\bm{k}_{j}-\bm{k}_{j+1}). \eeqa It is clear
 that $\mathcal{V}(\bm{k})$ is the Fourier transformation of
 $V(\bm{x})$. The
eq.(\ref{ffs1}) or (\ref{ffs2}) is a concrete form of our solution
(\ref{ouress}) of Schr\"odinger equation when the solvable part of
Hamiltonian is taken as the kinetic energy term. Therefore, in
principle, if the Fourier transformation of $V(\bm{x})$ can be
found, the evolution of the arbitrarily initial state with time can
be obtained. It must be emphasized that it is often that the
solvable part of Hamiltonian will be not only $T=\displaystyle
\frac{\hat{\bm{k}}^2}{2m}$ for the practical purposes. Moreover,
since the general term is obtained, we can selectively include the
partial contributions from some high order even all order
approximations.

Furthermore, we can derive out the propagator \beqa
\bra{\bm{x}}\e^{-\I
Ht}\ket{\bm{x}^\prime}&=&\sum_{\bm{k},\bm{k}^\prime}\diracsp{\bm{x}}{\bm{k}}\bra{\bm{k}}\e^{-\I
Ht}\ket{\bm{k}^\prime}\diracsp{\bm{k}^\prime}{\bm{x}^\prime}\nonumber\\
&=& \sum_{\bm{k}}\frac{1}{L^3}\e^{\I\bm{k}(\bm{x}-\bm{x}^\prime)-\I
E_{\bm{k}}t}+\sum_{l=1}^\infty
\sum_{\bm{k},\bm{k}^\prime}\sum_{\bm{k}_1,\cdots,\bm{k}_{l+1}}
\left[\prod_{j=1}^{l}V^{\bm{k}_j\bm{k}_{j+1}}\right]
\delta_{\bm{k}_1\bm{k}}\delta_{\bm{k}_{l+1}\bm{k}^\prime}\nonumber\\
& &\times\left[ \sum_{i=1}^{l+1}\frac{(-1)^{i-1}}{d_i(E[k,l])}\;
\frac{1}{L^{3/2}}\;\e^{-\I E_{\bm{k}_i}t}
\right]\frac{1}{L^3}\e^{\I\bm{k}\cdot\bm{x}-\I\bm{k}^\prime\cdot\bm{x}^\prime}.
\eeqa It is different from the known propagator in the expressive
form, but they should be equivalent to the physical results. One of
its distinguished features is this matrix element to be fully
$c$-number function made of the matrix elements
$V^{\gamma_i\gamma_j}$ and unperturbed energy levels $E_{\gamma_i}$.

\section{Comparing with the usual perturbation theory}\label{sec5}

In this section, we will compare the usual perturbation theory with
our solution, reveal their consistency and relation, and point out
what is more in our solution and what is different among them.
Furthermore, we expect to reveal the features, significance and
possible applications of our solution in theory. We will
respectively investigate and discuss the cases comparing with the
time-independent, time-dependent perturbation theories as well as
the non-perturbed solution.

\subsection{Comparing with the time-independent perturbation theory}

The usual time-independent (stationary) perturbation theory is
mainly to study the stationary wave equation in order to obtain the
perturbative energies and perturbative states. However, our solution
focus on the development of quantum states, which is a solution of
the Schr\"{o}dinger dynamical equation. Although their main purposes
are different at their start point, but they are consistent.
Moreover, our exact solution is also able to apply to the
calculation of perturbed energy and perturbed state, which will be
seen clearly in Sec. \ref{sec9}.

In order to verify that the time-independent perturbation theory is
consistent with our solution, we suppose that in the initial state,
the system is in the eigen state of the total Hamiltonian, that is
\beq H\ket{\Psi_{E_T}(0)}={E_T}\ket{\Psi_{E_T}(0)}.\eeq It is clear
that \beqa \ket{\Psi_{E_T}(t)}&=&\e^{-\I {E_T}
t}\ket{\Psi_{E_T}(0)}=\sum_{\gamma,\gamma^\prime}\left\{\e^{-\I
E_{\gamma} t}\delta_{\gamma\gamma^\prime} + \sum_{l=1}^\infty
A_l^{\gamma\gamma^\prime}(t) \right\}
\left[\diracsp{\Phi^{\gamma^\prime}}{\Psi_{E_T}(0)}\right]\ket{\Phi^\gamma}.
\eeqa  To calculate the $K$th time derivative of this equation and
then set $t=0$, we obtain \beqa \label{ktdess}{E_T}^K
\ket{\Psi_{E_T}(0)}&=&\sum_{\gamma,\gamma^\prime}\left\{E^K_{\gamma}
\delta_{\gamma\gamma^\prime} + (\I)^K\sum_{l=1}^\infty
\left.\frac{\d^K {A}_l^{\gamma\gamma^\prime}(t)}{\d t^K}
\right|_{t=0}\right\} a_{\gamma^\prime}\ket{\Phi^\gamma}\\
&=& \sum_{\gamma,\gamma^\prime}\left\{E^K_{\gamma}
\delta_{\gamma\gamma^\prime} + \sum_{l=1}^\infty
B_l^{\gamma\gamma^\prime}(K)\right\}
a_{\gamma^\prime}\ket{\Phi^\gamma},\eeqa where we have used the fact
that \beq
\ket{\Psi_{E_T}(0)}=\sum_{\gamma}a_{\gamma}\ket{\Phi^\gamma}, \quad
a_\gamma= \diracsp{\Phi^{\gamma}}{\Psi_{E_T}(0)},\eeq \beq
\left.\frac{\d^K {A}_l^{\gamma\gamma^\prime}(t)}{\d t^K}
\right|_{t=0}=(-\I)^K B_l^{\gamma\gamma^\prime}(K).\eeq It is easy
to obtain that \beqa \left.\frac{\d^K
{A}_l^{\gamma\gamma^\prime}(t)}{\d t^K}
\right|_{t=0}&=&(-\I)^K\sum_{\gamma_1,\cdots,\gamma_{l+1}}\left[
\sum_{i=1}^{l+1}(-1)^{i-1}\frac{E^K_{\gamma_i}}{d_i(E[\gamma,l])}\right]\left[
\prod_{j=1}^{l}H_1^{\gamma_j\gamma_{j+1}}\right]
\delta_{\gamma_1\gamma}\delta_{\gamma_{l+1}\gamma^\prime},\\
B_l^{\gamma\gamma^\prime}(K)&=&\sum_{\gamma_1,\cdots,\gamma_{l+1}}
C_l^K(E[\gamma,l])\theta(K-l)\left[
\prod_{j=1}^{l}H_1^{\gamma_j\gamma_{j+1}}\right]
\delta_{\gamma_1\gamma}\delta_{\gamma_{l+1}\gamma^\prime}.\label{ktdbeta}\eeqa
where we have used our identity (\ref{myi}).  This means that if
$l>K$ \beq \left.\frac{\d^K {A}_{l>K}^{\gamma\gamma^\prime}(t)}{\d
t^K} \right|_{t=0}=(-\I)^K B_l^{\gamma\gamma^\prime}(K>l)=0.\eeq In
special \beq \left.\frac{\d {A}_{l}^{\gamma\gamma^\prime}(t)}{\d t}
\right|_{t=0}=-\I B_1^{\gamma\gamma^\prime}(1)=-\I
H_1^{\gamma\gamma^\prime}\delta_{l1}.\eeq \beq \left.\frac{\d^2
{A}_{l}^{\gamma\gamma^\prime}(t)}{\d t^2}
\right|_{t=0}=-B_l^{\gamma\gamma^\prime}(2)=-
\left(E_\gamma+E_{\gamma^\prime}\right)H_1^{\gamma\gamma^\prime}\delta_{l1}
-\sum_{\gamma_1}H_1^{\gamma\gamma_1}H_1^{\gamma_1\gamma^\prime}\delta_{l2}.\eeq
In other words, the summation to $l$ in the right side of
eq.(\ref{ktdess}) is cut-off to $K$. Therefore, by left multiplying
$\bra{\Phi^\gamma}$ to eq.(\ref{ktdess}) we obtain \beq
\label{pee}{E_T}^K a_\gamma = E^K_{\gamma}a_\gamma
 + \sum_{\gamma^\prime}\sum_{l=1}^K B_l^{\gamma\gamma^\prime}(K)
a_{\gamma^\prime}. \eeq When $K=1$, above equation backs to the
start point of time-independent perturbation theory \beq
\label{peek1} {E_T}a_\gamma =E_\gamma a_\gamma+\sum_{\gamma^\prime}
H_1^{\gamma\gamma^\prime}a_{\gamma}^\prime . \eeq It implies that it
is consistent with our solution. In order to certify that our
solution is compatible with the time-independent perturbation
theory, we have to consider the cases when $K\geq 2$. Note that
$E_T^K a_\gamma=E_T\left(E_T^{K-1} a_\gamma\right)$, we substitute
eq.(\ref{pee}) twice, then move the second term to the right side,
we have \beqa \label{cppee} E_\gamma^{K-1}E_T a_\gamma &=&E_\gamma^K
a_\gamma+\sum_{\gamma^\prime}
\sum_{l=1}^{K}B_l^{\gamma\gamma^\prime}(K)
a_{\gamma^\prime}-\sum_{\gamma^\prime}
\sum_{l=1}^{K-1}B_l^{\gamma\gamma^\prime}(K-1)
\sum_{\gamma^{\prime\prime}}\left(E_{\gamma^\prime}
\delta_{\gamma^\prime\gamma^{\prime\prime}}+H_1^{\gamma^\prime\gamma^{\prime\prime}}\right)
a_{\gamma^{\prime\prime}}\nonumber\\
&=&E_\gamma^K a_\gamma+\sum_{\gamma^\prime}
\sum_{l=1}^{K-1}\left[B_l^{\gamma\gamma^\prime}(K)-E_{\gamma^\prime}
B_l^{\gamma\gamma^\prime}(K-1)\right]
a_{\gamma^\prime}\nonumber\\
&
&+\sum_{\gamma^\prime}\left(\prod_{j=1}^KH_1^{\gamma_j\gamma_{j+1}}\right)
\delta_{\gamma\gamma_1}\delta_{\gamma_{l+1}\gamma^\prime}a_{\gamma^\prime}
-\sum_{\gamma^\prime} \sum_{l=1}^{K-1}B_l^{\gamma\gamma^\prime}(K-1)
\sum_{\gamma^{\prime\prime}}H_1^{\gamma^\prime\gamma^{\prime\prime}}
a_{\gamma^{\prime\prime}}.\eeqa Since when $K\geq 2$\beq
\label{c1minus}
C_1^K(E[\gamma,1])-E_{\gamma_2}C_1^{K-1}(E[\gamma,1])=E_{\gamma_1}^{K-1},\eeq
and again $l\geq 2$ \beq \label{clminus}
C_l^K(E[\gamma,l])-E_{\gamma_{l+1}}C_l^{K-1}(E[\gamma,l])=C_{l-1}^{K-1}(E[\gamma,l-1]),\eeq
which has been proved in appendix A. Obviously, when $K=2$,
eq.(\ref{cppee}) becomes \beq E_\gamma E_T a_\gamma =E_\gamma^2
a_\gamma+ E_\gamma\sum_{\gamma^\prime}
H_1^{\gamma\gamma^\prime}a_{\gamma^\prime} .\eeq It can back to
eq.(\ref{peek1}). Similarly, for $K\geq 3$, so do they. In fact,
from eqs.(\ref{c1minus},\ref{clminus}), eq.(\ref{cppee}) becomes
\beqa E_\gamma^{K-1}E_T a_\gamma&=&E_\gamma^K a_\gamma
+E_\gamma^{K-1}\sum_{\gamma^\prime}H_1^{\gamma\gamma^\prime}
a_{\gamma^\prime},\eeqa where we have used the fact \beqa
\label{cppee2}& &\sum_{\gamma^\prime}
\sum_{l=2}^{K-1}\left[B_l^{\gamma\gamma^\prime}(K)-E_{\gamma^\prime}
B_l^{\gamma\gamma^\prime}(K-1)\right]
a_{\gamma^\prime}+\sum_{\gamma^\prime}\left(\prod_{j=1}^KH_1^{\gamma_j\gamma_{j+1}}\right)
\delta_{\gamma\gamma_1}\delta_{\gamma_{l+1}\gamma^\prime}a_{\gamma^\prime}
\nonumber\\
& &-\sum_{\gamma^\prime}
\sum_{l=1}^{K-1}B_l^{\gamma\gamma^\prime}(K-1)
\sum_{\gamma^{\prime\prime}}H_1^{\gamma^\prime\gamma^{\prime\prime}}
a_{\gamma^{\prime\prime}}=0.\eeqa Its proof is not difficult because
we can derive out \beqa & &
\sum_{\gamma^\prime}\sum_{l=2}^{K-1}\left[B_l^{\gamma\gamma^\prime}(K)-E_{\gamma^\prime}
B_l^{\gamma\gamma^\prime}(K-1)\right] a_{\gamma^\prime}\nonumber\\
& & \quad =\sum_{\gamma^\prime}
\sum_{l=2}^{K-1}\sum_{\gamma_1,\cdots,\gamma_{l+1}}\left[C_l^K(E[\gamma,l])-E_{\gamma^\prime}
C_l^{K-1}(E[\gamma,l])\right]\left(\prod_{j=1}^l
H_1^{\gamma_j\gamma_{j+1}}\right)\delta_{\gamma\gamma_1}\delta_{\gamma_{l+1}\gamma^\prime}
a_{\gamma^\prime}\nonumber\\
& &\quad
=\sum_{\gamma^\prime}\sum_{l=2}^{K-1}\sum_{\gamma_1,\cdots,\gamma_{l+1}}
C_{l-1}^{K-1}(E[\gamma,l-1])\left(\prod_{j=1}^l
H_1^{\gamma_j\gamma_{j+1}}\right)\delta_{\gamma\gamma_1}\delta_{\gamma_{l+1}\gamma^\prime}
a_{\gamma^\prime}\nonumber\\
& &\quad
=\sum_{\gamma^\prime}\sum_{l=1}^{K-2}\sum_{\gamma_1,\cdots,\gamma_{l+1}}
C_{l}^{K-1}(E[\gamma,l])\left(\prod_{j=1}^{l}
H_1^{\gamma_j\gamma_{j+1}}\right)H_1^{\gamma_{l+1}\gamma^\prime}
\delta_{\gamma\gamma_1} a_{\gamma^\prime}\nonumber\\
& &\quad =\sum_{\gamma^\prime,\gamma^\prime}\sum_{l=1}^{K-2}
B_l^{\gamma\gamma^\prime}(K-1)H_1^{\gamma^\prime\gamma^{\prime\prime}}
a_{\gamma^{\prime\prime}}.\eeqa The first equality has used the
definitions of $B_l^{\gamma\gamma^\prime}(K)$, the second equality
has used eq.(\ref{clminus}), the third equality sets $l-1\rightarrow
l$ and sums over $\gamma_{l+2}$, the forth equality sets
$\gamma^\prime\rightarrow \gamma^{\prime\prime}$, inserts
$\sum_{\gamma^\prime}\delta_{\gamma_{l+1}\gamma^\prime}$ and uses
the definition of $B_l^{\gamma\gamma^\prime}(K)$ again. Substituting
it into the left side of eq.(\ref{cppee2}) and using
$B_{K-1}^{\gamma\gamma^\prime}(K-1)$ expression, we can finish the
proof of eq.(\ref{cppee2}). Actually, above proof further verify the
correctness of our solution from the usual perturbation theory.

In our point of view, only if the expression of the high order
approximation has been obtained, we can consider its contribution in
the time-independent perturbation theory since its method is to find
exactly a given order approximation. However, this task needs to
solve a simultaneous equation system, it will be heavy for the
enough high order approximation. In Sec. \ref{sec7}, we will give a
method to find the improved forms of perturbed energy and perturbed
state including partially high order even all order approximations.

Now, let us see what is more in our solution than the
nonperturbative method (sometime be called time-independent theory).
Actually, since $H$ is not explicitly time-dependent, its
eigenvectors can be given formally by so-called Lippmann-Schwinger
equations as follows \cite{LSE}: \beqa \ket{\Psi^\gamma_{\rm
S}(\pm)}&=& \ket{\Phi^\gamma}+\frac{1}{E^\gamma-H_0\pm
\I\eta}H_1\ket{\Psi^\gamma_{\rm S}(\pm)}\\
&=& \ket{\Phi^\gamma}+G_0^\gamma(\pm)H_1\ket{\Psi^\gamma_{\rm
S}(\pm)} \\
&=&\ket{\Phi^\gamma}+\frac{1}{E_\gamma-H\pm
\I\eta}H_1\ket{\Phi^\gamma}\\
&=&\ket{\Phi^\gamma}+G^\gamma(\pm)H_1\ket{\Phi^\gamma},\eeqa where
the complete Green's function $G^\gamma(\pm)=1/(E_\gamma-H\pm
\I\eta)$ and the unperturbed Green's function
$G^\gamma_0(\pm)=1/(E^\gamma-H_0\pm \I\eta)$ satisfy the Dyson's
equation \beqa
G^\gamma(\pm)&=&G_0^\gamma(\pm)+G_0^\gamma(\pm)H_1G^\gamma(\pm)\\
&=&\sum_{l=0}^\infty
\left(G_0^\gamma(\pm)H_1\right)^lG_0^\gamma(\pm). \eeqa Here, the
subscript``S" means the stationary solution: \beq H
\ket{\Psi^\gamma_{\rm S}(\pm)}=E_\gamma \ket{\Psi^\gamma_{\rm
S}(\pm)}.\eeq More strictly, we should use the ``in" and ``out"
states to express it \cite{inout}. In historical literature, this
solution is known as so-called non-perturbative one. It has played
an important role in the formal scatter theory.

Back to our attempt, for a given initial state $\ket{\Psi(0)}$, we
have \beqa
\ket{\Psi(0)}&=&\sum_{\gamma^\prime}\diracsp{\Psi^{\gamma^\prime}_{\rm
S}(\pm)}{\Psi(0)}\ket{\Psi^{\gamma^\prime}_{\rm S}(\pm)}.\eeqa
Acting the time evolution operator on it, we immediately obtain
\beqa
\ket{\Psi(t)}&=&\sum_{\gamma^\prime}\diracsp{\Psi^{\gamma^\prime}_{\rm
S}(\pm)}{\Psi(0)}\e^{-\I
E_{\gamma^\prime}t}\ket{\Psi^{\gamma^\prime}_{\rm
S}(\pm)}\nonumber\\&=&\sum_{\gamma^\prime}
\bra{\Phi^{\gamma^\prime}}\sum_{l=0}^\infty\left(
H_1G_0^{\gamma^\prime}(\mp) \right)^l\ket{\Psi(0)}\e^{-\I
E_{\gamma^\prime}t} \sum_{l^\prime=0}^\infty
\left(G^{\gamma^\prime}(\pm)
H_1\right)^{l^\prime}\ket{\Phi^{\gamma^\prime}}\nonumber\\
&=&\sum_{\gamma,\gamma^\prime}\left[\e^{-\I E_\gamma
t}\delta_{\gamma\gamma^\prime}\diracsp{\Phi^{\gamma^\prime}}{\Psi(0)}\right.\nonumber\\
& &\left. +\sum_{\stackrel{\scriptstyle l,l^\prime=0}{l+l^\prime\neq
0}}^\infty
\bra{\Phi^{\gamma^\prime}}\left(H_1G_0^{\gamma^\prime}(\mp)
\right)^l\ket{\Psi(0)}
\bra{\Phi^{\gamma}}\left(G_0^{\gamma^\prime}(\pm)
H_1\right)^{l^\prime}\ket{\Phi^{\gamma^\prime}}\e^{-\I
E_{\gamma^\prime} t}\right]\ket{\Phi^{\gamma}},\eeqa comparing this
result with our solution (\ref{ouress}), we can say that our
solution has finished the calculations of explicit form of the
expanding coefficients (matrix elements) in the second term of the
above equation using our method and rearrange the resulting terms
according with the power of elements of representation matrix
$H_1^{\gamma_j\gamma_{j+1}}$. It implies that our solution can have
more and more explicit physical content and significance. Therefore,
we think that our solution is a new development of the stationary
perturbation theory. In spite of its physical results which
consistent with the Feynman's diagram expansion of Dyson's equation,
its form is new, explicit, and convenient to calculate the time
evolution of states with time. Specially, it will be seen that our
perturbative scheme based on our exact solution has higher
efficiency and higher precision in the calculation from Secs.
\ref{sec8}, \ref{sec9} and \ref{sec10}.

\subsection{Comparing with the time-dependent perturbation theory}

The aim of our solution is similar to the time-dependent
perturbation theory. But their methods are different. The usual
time-dependent perturbation theory \cite{diracpqm} considers a
quantum state initially in the eigenvector of the Hamiltonian $H_0$,
and a subsequent evolution of system caused by the application of an
explicitly time-dependent potential $V=\lambda v(t)$, where
$\lambda$ is enough small to indicate $V$ as a perturbation, that
is,  \beq\label{seqintd} \I\frac{\partial}{\partial
t}\ket{\Psi(t)}=(H_0+\lambda v(t))\ket{\Psi(t)}\eeq with the initial
state \beq\label{inicintd} \ket{\Psi(0)}=\ket{\Phi^{\alpha}}. \eeq
It is often used to study the cases that the system returns to an
eigenvector $\ket{\Phi_f}$ of the Hamiltonian $H_0$ when the action
of the perturbing potential becomes negligible.

In order to compare it with our solution, let us first recall the
time-dependent perturbation method. Note that the state
$\ket{\Psi(t)}$ at time $t$ can be expanded by the complete
orthogonal system of the eigenvectors $\ket{\Phi^\gamma}$ of the
Hamiltonian $H_0$, that is \beq \label{tdpt0}
\ket{\Psi(t)}=\sum_\gamma c_\gamma(t)\ket{\Phi^\gamma}\eeq with \beq
c_\gamma(t)=\diracsp{\Phi^\gamma}{\Psi(t)},\eeq whereas the
evolution equation (\ref{seqintd}) will become \beq\label{Ceqintd}
\I\frac{\partial}{\partial t}c_\gamma(t)=E_\gamma
c_\gamma(t)+\lambda\sum_{\gamma^\prime}v^{\gamma\gamma^\prime}(t)c_{\gamma^\prime}(t),\eeq
where
$v^{\gamma\gamma^\prime}(t)=\bra{\Phi^\gamma}v(t)\ket{\Phi^{\gamma^\prime}}$.
Now setting \beq \label{cbr} c_\gamma(t)=b_\gamma(t)\e^{-\I E_\gamma
t}, \eeq and inserting this into (\ref{Ceqintd}) give
\beq\label{Beqintd} \I\frac{\partial}{\partial
t}b_\gamma(t)=\lambda\sum_{\gamma^\prime}\e^{\I(E_\gamma-E_{\gamma^\prime})t}
v^{\gamma\gamma^\prime}(t)b_{\gamma^\prime}(t).\eeq Then, making a
series expansion of $b_\gamma(t)$ according to the power of
$\lambda$: \beq b_\gamma(t)=\sum_{l=0}\lambda^l
b^{(l)}_\gamma(t),\eeq  and setting equal the coefficients of
$\lambda^l$ on the both sides of the equation (\ref{Beqintd}), ones
find: \beqa \I\frac{\partial}{\partial t}b^{(0)}_\gamma(t)&=&0,\\
\label{bleq} \I\frac{\partial}{\partial
t}b^{(l)}_\gamma(t)&=&\sum_{\gamma^\prime}\e^{\I(E_\gamma-E_{\gamma^\prime})t}
v^{\gamma\gamma^\prime}(t)b_{\gamma^\prime}^{(l-1)}(t), \quad (l\neq
0).\eeqa From the initial condition (\ref{inicintd}), it follows
that
 $c_\gamma(0)=b_\gamma(0)=\delta_{\gamma \alpha}=b^{(0)}_\gamma(0)$
and $b_\gamma^{(l)}(0)=0$ if $l\geq 1$. By integration over the
variable $t$ this will yield \beqa
b^{(0)}_\gamma(t)&=&\delta_{\gamma \alpha},\\
b^{(1)}_\gamma(t)&=&{-\I}\int_0^t \d
t_1\e^{\I(E_\gamma-E_{\alpha})t_1}
v^{\gamma \alpha}(t_1),\\
b^{(2)}_\gamma(t)&=&{-\I}\int_0^t \d
t_2\sum_{\gamma_2}\e^{\I(E_\gamma-E_{\gamma_2})t_2}
v^{\gamma \gamma_2}(t_2)b^{(1)}_{\gamma_2}(t_2)\nonumber\\
&=& - \sum_{\gamma_2}\int_0^t\d t_2\int_0^{t_2}\d t_1
\e^{\I(E_\gamma-E_{\gamma_2})t_2}\e^{\I(E_{\gamma_2}-E_{\alpha})t_1}
v^{\gamma\gamma_2}(t_2)v^{\gamma_2 \alpha}(t_1).\eeqa If we take $v$
independent of time, the above method in the time-dependent
perturbation theory is still valid. Thus, \beq b_\gamma^{(1)}(t)=
-\frac{1}{E_\gamma-E_\alpha}\left(\e^{-\I E_\alpha t}-\e^{-\I
E_\gamma t}\right)\e^{\I E_\gamma t}v^{\gamma \alpha},\eeq \beqa
b_\gamma^{(2)}(t)&=&
\sum_{\gamma_2}\frac{1}{(E_\gamma-E_\alpha)(E_{\gamma_2}-E_\alpha)}
\left(\e^{-\I E_\alpha t}-\e^{-\I E_\gamma t}\right) \e^{\I E_\gamma
t}v^{\gamma \gamma_2}v^{\gamma_2
\alpha}\nonumber\\
& &-\sum_{\gamma_2}
\frac{1}{(E_{\gamma_2}-E_\alpha)(E_{\gamma}-E_{\gamma_2})}
\left(\e^{-\I E_{\gamma_2} t}-\e^{-\I E_\gamma t}\right) \e^{\I
E_\gamma
t}v^{\gamma \gamma_2}v^{\gamma_2 \alpha}\nonumber\\
&=& \sum_{\gamma_2}\left[
\frac{1}{(E_{\gamma}-E_{\gamma_2})(E_{\gamma}-E_\alpha)}-
\frac{1}{(E_{\gamma}-E_{\gamma_2})(E_{\gamma_2}-E_\alpha)}\e^{-\I
E_{\gamma_2} t} \e^{\I E_\gamma
t}\right.\nonumber\\
&
&\left.+\frac{1}{(E_\gamma-E_\alpha)(E_{\gamma_2}-E_\alpha)}\e^{-\I
E_{\alpha} t} \e^{\I E_\gamma t}\right]v^{\gamma
\gamma_2}v^{\gamma_2 \alpha}.\eeqa This means that \beq
c^{(1)}_\gamma(t)=\sum_{\gamma_1\gamma_2}\sum_{i=1}^2(-1)^{i-1}\frac{\e^{-\I
E_{\gamma_i}}}{d_i(E[\gamma,1])}v^{\gamma_1\gamma_2}\delta_{\gamma\gamma_1}\delta_{\gamma_2\alpha},\eeq
\beq
c^{(2)}_\gamma(t)=\sum_{\gamma_1,\gamma_2,\gamma_3}\sum_{i=1}^3(-1)^{i-1}\frac{1}{d_i(E[\gamma,2])}\e^{-\I
E_{\gamma_i}t}\delta_{\gamma\gamma_1}\delta_{\gamma_3
\alpha}v^{\gamma_1 \gamma_2}v^{\gamma_2 \gamma_3}.\eeq Now, we use
the mathematical induction to prove \beq \label{cee}
c^{(l)}_\gamma(t)=\sum_{\gamma_1,\cdots,\gamma_{l+1}}\sum_{i=1}^{l+1}(-1)^{i-1}\frac{\e^{-\I
E_{\gamma_i}t}}{d_i(E[\gamma,l])}\left(\prod_{j=1}^l
v^{\gamma_{j}\gamma_{j+1}}\right)\delta_{\gamma\gamma_1}\delta_{\gamma_{l+1}\alpha}.\eeq
Obviously, we have seen that it is valid for $l=1,2$. Suppose it is
also valid for a given $n$, thus form eqs.(\ref{bleq}) and
$b_\gamma^{(l)}(0)=0,\;l\geq 1$ it follows that \beq
b^{(n+1)}_\beta(t)={-\I}\int_0^t \d \tau
\sum_{\gamma}\e^{\I(E_\beta-E_{\gamma})\tau} v^{\beta
\gamma}b_\gamma^{(n)}(\tau) .\eeq Substitute eqs.(\ref{cbr}) and
(\ref{cee}) we obtain \beqa \label{ceep2}
\!\!\!\!\!\!b^{(n+1)}_\beta(t)&=&{-\I}\int_0^t \d \tau
\sum_{\gamma}\e^{\I E_\beta\tau} v^{\beta
\gamma}c_\gamma^{(n)}(\tau)\nonumber\\
&=&{-\I}\int_0^t \d \tau \sum_{\gamma}\e^{\I E_\beta\tau} v^{\beta
\gamma}
\sum_{\gamma_1,\cdots,\gamma_{n+1}}\sum_{i=1}^{n+1}(-1)^{i-1}\frac{\e^{-\I
E_{\gamma_i}\tau}}{d_i(E[\gamma,n])}\left(\prod_{j=1}^n
v^{\gamma_{j}\gamma_{j+1}}\right)\delta_{\gamma\gamma_1}\delta_{\gamma_{n+1}\alpha}\nonumber\\
&=&\!\!\!\sum_{\gamma_1,\cdots,\gamma_{n+2}}\sum_{i=1}^{n+1}(-1)^{i-1}\frac{(\e^{-\I(
E_{\gamma_i}-E_{\gamma_{n+2}})t}-1)}{d_i(E[\gamma,n])
(E_{\gamma_i}-E_{\gamma_{n+2}})}\left(\prod_{j=1}^n
v^{\gamma_{j}\gamma_{j+1}}\right)v^{\gamma_{n+2}\gamma_1}
\delta_{\gamma_{n+2}\beta}\delta_{\gamma_{n+1}\alpha}.\eeqa In the
last equality, we have inserted
$\sum_{\gamma_{n+2}}\delta_{\gamma_{n+2}\beta}$, summed over
$\gamma$ and integral over $\tau$. In terms of the definition of
$d_i(E[\gamma,n]$, we know that
$d_i(E[\gamma,n])(E_{\gamma_i}-E_{\gamma_{n+2}})=d_i(E[\gamma,n+1])$.
Then based on our identity (\ref{myi}), we have \beq
\sum_{i=1}^{n+1}(-1)^{i-1}\frac{1}{d_i(E[\gamma,n+1])}
=-(-1)^{(n+2)-1}\frac{1}{d_{n+2}(E[\gamma,n+1])}.\eeq Thus
eq.(\ref{ceep2}) becomes \beqa \label{ceep3}
b^{(n+1)}_\beta(t)&=&\sum_{\gamma_1,\cdots,\gamma_{n+2}}\sum_{i=1}^{n+2}(-1)^{i-1}\frac{\e^{-\I
E_{\gamma_i}t}}{d_i(E[\gamma,n+1])}\left(\prod_{j=1}^n
v^{\gamma_{j}\gamma_{j+1}}\right)v^{\gamma_{n+2}\gamma_1}
\delta_{\gamma_{n+2}\beta}\delta_{\gamma_{n+1}\alpha}\e^{\I
E_{\beta}t}.\eeqa Set the index taking turns, that is,
$\gamma_1\rightarrow\gamma_{n+2}$ and
$\gamma_i+1\rightarrow\gamma_i$,$(n+1)\geq i\geq 1)$. Again from the
definition of $d_i(E[\gamma,n]$, we can verify easily under above
index taking turns, \beq \sum_{i=1}^{n+2}(-1)^{i-1}\frac{\e^{-\I
E_{\gamma_i}t}}{d_i(E[\gamma,n+1])}\longrightarrow
\sum_{i=1}^{n+2}(-1)^{i-1}\frac{\e^{-\I
E_{\gamma_i}t}}{d_i(E[\gamma,n+1])}.\eeq Therefore we obtain the
conclusion \beq c_\beta^{(n+1)}(t)=b^{(n+1)}_\beta(t)\e^{-\I E_\beta
t}=\sum_{\gamma_1,\cdots,\gamma_{n+2}}\sum_{i=1}^{n+2}(-1)^{i-1}\frac{\e^{-\I
E_{\gamma_i}t}}{d_i(E[\gamma,n+1])}\left(\prod_{j=1}^{n+1}
v^{\gamma_{j}\gamma_{j+1}}\right)
\delta_{\beta\gamma_1}\delta_{\gamma_{n+2}\alpha}.\eeq This implies
that we finish the proof of expression (\ref{cee}). Substitute it
into eq.(\ref{tdpt0}), we immediately obtain the same expression as
the our solution (\ref{ouress}).

From the statement above, it can be seen that our solution actually
finish the task to solve the recurrence equation of
$b^{l}_\gamma(t)$ by using our new method. Although the recurrence
equation of $b^{l}_\gamma(t)$ can be solved by integral in
principle, it is indeed not easy if one does not to know our
identity (\ref{myi}) and relevant relations. At our knowledge, the
general term form has not ever been obtained up to now. This is an
obvious difficult to attempt including high order approximations.
However, our solution has done it, and so it will make some
calculations of perturbation theory more convenient and more
accurate.

However, we pay for the price to require that $H$ is not explicitly
dependent on time. Obviously our solution is written as a symmetric
and whole form. Moreover, our method gives up, at least in the sense
of formalization, the requirement that $V$ is a small enough
compared wit $H_0$, and our proof is more general and more strict.
In addition, we do not need to consider the cases that the
perturbing potential is switched at the initial and final time, but
the perturbing potential is not explicitly dependent of time. Since
we obtain the general term, our solution must have more
applications, it is more efficient and more accurate for the
practical applications, because we can selectively include partial
contributions from high order even all order approximation. This can
be called as the improved forms of perturbed solutions, which will
be given in Sec. \ref{sec7}.

\section{Two skills in the improved scheme of perturbation
theory}\label{sec6}

For the practical applications, we indeed can take the finite order
approximations, that is, a finite $l$ is given in our solution
(\ref{ouress}). Moreover, in order to show the advantages of our
exact solution and provide the improved scheme of perturbation
theory, we will derive out the improved forms of perturbative
solution including the partial contributions from the higher order
approximations. Because all of steps are well-regulated and only
technology is to find the limitation of primary functions. In other
words, the calculation of our exact solution and perturbation theory
is operational in the practical applications, and further it can
advance the efficiency and precision in the calculations. Frankly
speaking, before we know our exact solution, we are puzzled by too
many irregular terms and very trouble dependence on previous
calculation steps. Moreover, we are often anxious about the result
precision in such some calculations because those terms proportional
to $t^a \e^{-\I E_{\gamma_i} t}$ $(a=1,2,\cdots)$ in the high order
approximation might not be ignorable with time increasing.
Considering the contributions these terms can obviously improve the
precision. However, the known perturbation theory does not give the
general term, considering this task to add reasonably the high order
approximations is actually impossible.

Since our exact solution has given the explicit form of any order
approximation, that is a general term of an arbitrary order
perturbed solution, and their forms are simply the summations of a
series. Just enlightened by this general term of arbitrary order
perturbed solution, we use two skills to develop the perturbation
theory, which are respectively expressed in the following two
subsections.

\subsection{Redivision of Hamiltonian}

The first skill is to decompose the matrix elements of $H_1$ in the
representation of $H_0$ into diagonal part and nondiagonal part:
\beq\label{H1d2}
H_1^{\gamma_j\gamma_{j+1}}=h_1^{\gamma_j}\delta_{\gamma_j\gamma_{j+1}}
+g_1^{\gamma_j\gamma_{j+1}},\eeq so that the concrete expression of
a given order approximation can be easily calculated. Note that
$h_1^{\gamma_j}$ has  been chosen as its diagonal elements and then
$g_1^{\gamma_j\gamma_{j+1}}$ has been set as its nondiagonal
elements: \beq g_1^{\gamma_j\gamma_{j+1}}=g_1^{\gamma_j\gamma_{j+1}}
(1-\delta_{\gamma_j\gamma_{j+1}}).\eeq

As examples, for the first order approximation, it is easy to
calculate out that \beq
\label{A1h}A_1^{\gamma\gamma^\prime}(h)=\sum_{\gamma_1,\gamma_{2}}\left[
\sum_{i=1}^{2}(-1)^{i-1}\frac{\e^{-\I E_{\gamma_i}
t}}{d_i(E[\gamma,l])}\right]
\left(h_1^{\gamma_1}\delta_{\gamma_1\gamma_{2}}\right)
\delta_{\gamma\gamma_1}\delta_{\gamma^\prime\gamma_{2}}=\frac{(-\I
h_1^{\gamma} t)}{1!}\e^{-\I E_{\gamma}
t}\delta_{\gamma\gamma^\prime},\eeq \beqa\label{A1g}
A_1^{\gamma\gamma^\prime}(g)&=&\sum_{\gamma_1,\gamma_{2}}\left[
\sum_{i=1}^{2}(-1)^{i-1}\frac{\e^{-\I E_{\gamma_i}
t}}{d_i(E[\gamma,l])}\right]
g_1^{\gamma_1}(1-\delta_{\gamma_1\gamma_2})
\delta_{\gamma\gamma_1}\delta_{\gamma^\prime\gamma_{2}}\nonumber\\
&=&\left[\frac{\e^{-\I E_{\gamma}
t}}{E_{\gamma}-E_{\gamma^\prime}}-\frac{\e^{-\I E_{\gamma^\prime}
t}}{E_{\gamma}-E_{\gamma^\prime}}\right](1-\delta_{\gamma\gamma^\prime}).\eeqa
Note that here and after we use the symbol
$A_i^{\gamma\gamma^\prime}$ denoting the contribution from $i$th
order approximation, which is defined by (\ref{Aldefinition}), while
its argument indicates the product form of matrix elements of $H_1$
in $H_0$ representation. However, for the second order
approximation, since \beqa
\label{H12deq}\prod_{j=1}^{2}H_1^{\gamma_j\gamma_{j+1}}&=&
\left(h_1^{\gamma_1}\right)^2\delta_{\gamma_1\gamma_{2}}\delta_{\gamma_2\gamma_{3}}
+h_1^{\gamma_1}g_1^{\gamma_2\gamma_{3}}\delta_{\gamma_1\gamma_2} +
g_1^{\gamma_1\gamma_{2}}h_1^{\gamma_2}\delta_{\gamma_2\gamma_3}
+g_1^{\gamma_1\gamma_{2}}g_1^{\gamma_2\gamma_{3}}.\eeqa we need to
calculate the mixed product of diagonal and nondiagonal elements of
$H_1$. Obviously, we have \beq \label{A2hh}
A_2^{\gamma\gamma^\prime}(hh)=\sum_{\gamma_1,\gamma_{2},\gamma_3}\left[
\sum_{i=1}^{3}(-1)^{i-1}\frac{\e^{-\I E_{\gamma_i}
t}}{d_i(E[\gamma,2])}\right]
\left(h_1^{\gamma_1}\delta_{\gamma_1\gamma_{2}}h_1^{\gamma_2}\delta_{\gamma_2\gamma_{3}}\right)
\delta_{\gamma\gamma_1}\delta_{\gamma^\prime\gamma_{3}}=\frac{(-\I
h_1^{\gamma} t)^2}{2!}\e^{-\I E_{\gamma}
t}\delta_{\gamma\gamma^\prime},\eeq \beqa \label{A2hg}
A_2^{\gamma\gamma^\prime}(hg)&=&\sum_{\gamma_1,\gamma_2,\gamma_{3}}\left[
\sum_{i=1}^{3}(-1)^{i-1}\frac{\e^{-\I E_{\gamma_i}
t}}{d_i(E[\gamma,2])}\right] h_1^{\gamma_1}g_1^{\gamma_2\gamma_{3}}
\delta_{\gamma_1\gamma_2}\delta_{\gamma\gamma_1}\delta_{\gamma^\prime\gamma_3}\nonumber\\
& =&\left[-\frac{\e^{-\I E_{\gamma}
t}}{(E_{\gamma}-E_{\gamma^\prime})^2}+\frac{\e^{-\I
E_{\gamma^\prime} t}}{(E_{\gamma}-E_{\gamma^\prime})^2}+(-\I
t)\frac{\e^{-\I E_{\gamma}
t}}{E_{\gamma}-E_{\gamma^\prime}}\right]h_1^{\gamma}g_1^{\gamma\gamma^\prime},\eeqa
\beqa
\label{A2gh}A_2^{\gamma\gamma^\prime}(gh)&=&\sum_{\gamma_1,\gamma_2,\gamma_{3}}\left[
\sum_{i=1}^{3}(-1)^{i-1}\frac{\e^{-\I E_{\gamma_i}
t}}{d_i(E[\gamma,2])}\right]
h_1^{\gamma_2}g_1^{\gamma_1\gamma_{2}}\delta_{\gamma_2\gamma_3}
\delta_{\gamma\gamma_1}\delta_{\gamma^\prime\gamma_3}\nonumber\\
& =&\left[\frac{\e^{-\I E_{\gamma}
t}}{(E_{\gamma}-E_{\gamma^\prime})^2}-\frac{\e^{-\I
E_{\gamma^\prime} t}}{(E_{\gamma}-E_{\gamma^\prime})^2}-(-\I
t)\frac{\e^{-\I E_{\gamma^\prime}
t}}{E_{\gamma}-E_{\gamma^\prime}}\right]g_1^{\gamma\gamma^\prime}h_1^{\gamma^\prime},\eeqa
\beqa
\label{A2gg}A_2^{\gamma\gamma^\prime}(gg)&=&\sum_{\gamma_1,\gamma_2,\gamma_{3}}\left[
\sum_{i=1}^{3}(-1)^{i-1}\frac{\e^{-\I E_{\gamma_i}
t}}{d_i(E[\gamma,2])}\right]
g_1^{\gamma_1\gamma_2}g_1^{\gamma_2\gamma_{3}}
\delta_{\gamma\gamma_1}\delta_{\gamma^\prime\gamma_3}\nonumber\\
&=&\sum_{\gamma_1}\left[\frac{\e^{-\I
E_{\gamma}t}}{(E_{\gamma}-E_{\gamma_1})(E_{\gamma}-E_{\gamma^\prime})}-\frac{\e^{-\I
E_{\gamma_1}t}}{(E_{\gamma}-E_{\gamma_1})(E_{\gamma_1}-E_{\gamma^\prime})}\right.\nonumber\\
& &\left.+\frac{\e^{-\I
E_{\gamma^\prime}t}}{(E_{\gamma}-E_{\gamma^\prime})(E_{\gamma_1}-E_{\gamma^\prime})}\right]
g_1^{\gamma\gamma_1}g_1^{\gamma_1\gamma^\prime}. \eeqa In the usual
time-dependent perturbation theory, the zeroth order approximation
of time evolution of quantum state keeps its original form \beq
\ket{\Psi^{(0)}(t)}=\e^{-\I E_\gamma t}\ket{\Phi^\gamma},\eeq where
we have set the initial state as $\ket{\Phi^\gamma}$ for simplicity.
By using our solution, we easily calculate out the contributions of
all of order approximations from the product of completely diagonal
elements $h$ to this zeroth order approximation \beqa &
&\sum_{\gamma_1,\cdots,\gamma_{l+1}}\left[
\sum_{i=1}^{l+1}(-1)^{i-1}\frac{\e^{-\I E_{\gamma_i}
t}}{d_i(E[\gamma,l])}\right]
\left(\prod_{j=1}^{l}h_1^{\gamma_j}\delta_{\gamma_j\gamma_{j+1}}\right)
\delta_{\gamma\gamma_1}\delta_{\gamma^\prime\gamma_{l+1}}\nonumber\\
& &=\frac{(-\I h_1^{\gamma} t)^l}{l!}\e^{-\I E_{\gamma}
t}\delta_{\gamma\gamma^\prime}.\eeqa Therefore, we can add the
contributions of all of order approximation parts from the product
of completely diagonal elements $h$ to this zeroth order
approximation to obtain \beq \label{0thawithde}
\ket{{\Psi^\prime}^{(0)}(t)}=\e^{-\I
\left(E_\gamma+h_1^\gamma\right) t}\ket{\Phi^\gamma}.\eeq Similarly,
by calculation, we can deduce out that up to the second
approximation, the perturbed solution has the following form \beqa
\label{2thawithde}
\ket{\Psi^\prime(t)}&=&\sum_{\gamma,\gamma^\prime}\left\{\e^{-\I
\left(E_{\gamma}+h_1^\gamma\right)t}\delta_{\gamma\gamma^\prime} +
\left[\frac{\e^{-\I \left(E_{\gamma}+h_1^\gamma\right) t}-\e^{-\I
\left(E_{\gamma^\prime}+h_1^{\gamma^\prime}\right)
t}}{\left(E_{\gamma}+h_1^{\gamma}\right)
-\left(E_{\gamma^\prime}+h_1^{\gamma^\prime}\right)}\right]g_1^{\gamma\gamma^\prime}\right.\nonumber\\
& & +\sum_{\gamma_1}\left[\frac{\e^{-\I
\left(E_{\gamma}+h_1^\gamma\right)t}}{\left[\left(E_{\gamma}+h_1^\gamma\right)
-\left(E_{\gamma_1}+h_1^{\gamma_1}\right)\right]\left[\left(E_{\gamma}+h_1^\gamma\right)
-\left(E_{\gamma^\prime}+h_1^{\gamma^\prime}\right)\right]}\right.\nonumber\\
& & -\frac{\e^{-\I
\left(E_{\gamma_1}+h_1^{\gamma_1}\right)t}}{\left[\left(E_{\gamma}+h_1^\gamma\right)
-\left(E_{\gamma_1}+h_1^{\gamma_1}\right)\right]\left[\left(E_{\gamma_1}+h_1{^\gamma_1}\right)
-\left(E_{\gamma^\prime}+h_1^{\gamma^\prime}\right)\right]}\nonumber\\
& &\left.\left.+\frac{\e^{-\I\left(
E_{\gamma^\prime}+h_1^{\gamma^\prime}\right)t}}{\left[\left(E_{\gamma}+h_1^\gamma\right)
-\left(E_{\gamma^\prime}+h_1^{\gamma^\prime}\right)\right]\left[\left(E_{\gamma_1}+h_1^{\gamma_1}\right)
-\left(E_{\gamma^\prime}+h_1^{\gamma^\prime}\right)\right]}\right]
g_1^{\gamma\gamma_1}g_1^{\gamma_1\gamma^\prime}\right\}\nonumber\\
& &
\left[\diracsp{\Phi^{\gamma^\prime}}{\Psi(0)}\right]\ket{\Phi^\gamma}+\mathcal{O}(H_1^3).
\eeqa However, for the higher order approximation, the corresponding
calculation is heavy. In fact, it is unnecessary to calculate the
contributions from those terms with the diagonal elements of $H_1$
since introducing the following skill. This is a reason why we omit
the relevant calculation details. Here we mention it only for
verifying the correctness of our exact solution in this way.

The results (\ref{0thawithde}) and (\ref{2thawithde}) are not
surprised because of the fact that the Hamiltonian is re-divisible.
Actually, we can furthermore use a trick of redivision of the
Hamiltonian so that the new $H_0$ contains the diagonal part of
$H_1$, that is \beqa
H_0^\prime&=&H_0+\sum_{\gamma}h_1^\gamma\ket{\Phi^\gamma}\bra{\Phi^\gamma},\\
H_1^\prime&=&H_1-\sum_{\gamma}h_1^\gamma\ket{\Phi^\gamma}\bra{\Phi^\gamma}
=\sum_{\gamma,\gamma^\prime}g_1^{\gamma\gamma^\prime}\ket{\Phi^\gamma}\bra{\Phi^{\gamma\prime}}.\eeqa
In other words, without loss of generality, we always can choose
that $H_1^\prime$ has only the nondiagonal elements in the
$H_0^\prime$ (or $H_0$) representation and \beq
H_0^\prime\ket{\Phi^\gamma}=\left(E_\gamma+h_1^\gamma\right)\ket{\Phi^\gamma}
=E_{\gamma}^\prime\ket{\Phi^\gamma}.\eeq In spite of our skill is so
simple, it seems not be sufficiently transpired and understood from
the fact that the recent some textbooks of quantum mechanics still
remain the diagonal contribution from the unperturbed part in the
expression of the second order perturbed state.

It must be emphasized that the skill to redivide Hamiltonian leads
to the fact that the new perturbed solution can be obtained by the
replacement \beq E_{\gamma_i}\rightarrow
E_{\gamma_i}+h_1^{\gamma_i}\eeq in the usual perturbed solution and
its conclusions. This implies that there are two equivalent ways to
obtain the same perturbed solution and its conclusions. One of them
is to redefine the energy level $E_{\gamma_i}$ as
$E^\prime_{\gamma_i}$, think $E^\prime_{\gamma_i}$ to be explicitly
independent on the perturbed parameter from a redefined view, and
then use the method in the usual perturbation theory to obtain the
result from the redivided $H_1^\prime$. The other way is directly
deduce out the perturbed solution from the original Hamiltonian by
using the standard procedure, but a rearrangement and summation are
carried out just like above done by us. From mathematical view, this
is because the perturbed parameter is only formal multiplier and it
is introduced after redefining $E^\prime_{\gamma_i}$. The first way
or technology will be again applied to our scheme to obtain improved
forms of perturbed energy and perturbed state in Sec. \ref{sec9}.

For simplicity, in the following, we omit the ${}^\prime$ in $H_0$,
$H_1$ as well as $E_\gamma$, and always let $H_1$ have only its
nondiagonal part unless particular claiming.

\subsection{Contraction and anti-contraction of nondiagonal element product}

In this subsection, we present the second important skill
enlightened by our exact solution, this is, a method to calculate
the contributions from the contractions and anti-contractions of
nondiagonal element product in a given order approximation. This
method is also the most important technology in our scheme of
perturbation theory.

Let us start with the second order approximation. Since we have
taken $H_1^{\gamma_j\gamma_{j+1}}$ only with the nondiagonal part
$g_1^{\gamma_j\gamma_{j+1}}$, the contribution from the second order
approximation is only $A_2^{\gamma\gamma^\prime}(gg)$ in
eq.(\ref{A2gg}). However, we find that the limitation in the
expression of $A_2^{\gamma\gamma^\prime}(gg)$ has not been
completely found out because we have not excluded the case
$E_\gamma=E_{\gamma^\prime}$ (or $\gamma=\gamma^\prime$). This
problem can be fixed by introducing a decomposition \beq
g_1^{\gamma_1\gamma_2}g_1^{\gamma_2\gamma_3}
=g_1^{\gamma_1\gamma_2}g_1^{\gamma_2\gamma_3}\delta_{\gamma_1\gamma_3}
+g_1^{\gamma_1\gamma_2}g_1^{\gamma_2\gamma_3}\eta_{\gamma_1\gamma_3},
\eeq where $\eta_{\gamma_1\gamma_3}=1-{\delta}_{\gamma_1\gamma_3}$.
Thus, the contribution from the second order approximation is made
of two terms, one so-called contraction term with the $\delta$
function and another so-called anti-contraction term with the $\eta$
function. It must be emphasized that we only consider the
non-degenerate case here and after for simplification. While when
the degeneration happens, two indexes with the same main energy
level number will not have the anti-contraction.

In terms of above skill, we find that the contribution from the
second order approximation is made of the corresponding contraction-
and anti-contraction- terms \beq
{A}_2^{\gamma\gamma^\prime}({gg})={A}_2^{\gamma\gamma^\prime}(gg;c)
+{A}_2^{\gamma\gamma^\prime}(gg;n),\eeq where \beqa \label{A2ggc}
{A}_2^{\gamma\gamma^\prime}(gg;c)&=&\sum_{\gamma_1,\gamma_2,\gamma_{3}}\left[
\sum_{i=1}^{3}(-1)^{i-1}\frac{\e^{-\I E_{\gamma_i}
t}}{d_i(E[\gamma,2])}\right]
g_1^{\gamma_1\gamma_2}g_1^{\gamma_2\gamma_{3}}\delta_{\gamma_1\gamma_3}
\delta_{\gamma\gamma_1}\delta_{\gamma^\prime\gamma_3}\nonumber\\
&=& \sum_{\gamma_1}\left[-\frac{\e^{-\I
E_{\gamma}t}}{\left(E_\gamma-E_{\gamma_1}\right)^2}+\frac{\e^{-\I
E_{\gamma_1}t}}{\left(E_\gamma-E_{\gamma_1}\right)^2}+(-\I
t)\frac{\e^{-\I E_{\gamma}t}}{E_\gamma-E_{\gamma_1}}\right]
\left|g_1^{\gamma\gamma_1}\right|^2\delta_{\gamma\gamma^\prime},
\eeqa \beqa \label{A2ggn}
{A}_2^{\gamma\gamma^\prime}(gg;n)&=&\sum_{\gamma_1,\gamma_2,\gamma_{3}}\left[
\sum_{i=1}^{3}(-1)^{i-1}\frac{\e^{-\I E_{\gamma_i}
t}}{d_i(E[\gamma,2])}\right]
g_1^{\gamma_1\gamma_2}g_1^{\gamma_2\gamma_{3}}\eta_{\gamma_1\gamma_3}
\delta_{\gamma\gamma_1}\delta_{\gamma^\prime\gamma_3}\nonumber\\
&=&\sum_{\gamma_1}\left[\frac{\e^{-\I
E_{\gamma}t}}{(E_{\gamma}-E_{\gamma_1})(E_{\gamma}-E_{\gamma^\prime})}-\frac{\e^{-\I
E_{\gamma_1}t}}{(E_{\gamma}-E_{\gamma_1})(E_{\gamma_1}-E_{\gamma^\prime})}\right.\nonumber\\
& &\left.+\frac{\e^{-\I
E_{\gamma^\prime}t}}{(E_{\gamma}-E_{\gamma^\prime})(E_{\gamma_1}-E_{\gamma^\prime})}\right]
g_1^{\gamma\gamma_1}g_1^{\gamma_1\gamma^\prime}\eta_{\gamma\gamma^\prime}.
\eeqa This method can be extended to the higher order approximation
by introducing a concept of $g$-product decomposition. For a
sequential product of nondiagonal elements $g$ with the form
$\prod_{k=1}^m g_1^{\gamma_k\gamma_{k+1}}$ ($m\geq 2$), we define
its $(m-1)$th decomposition by \beq \label{gpd} \prod_{k=1}^m
g_1^{\gamma_k\gamma_{k+1}}=\left(\prod_{k=1}^m
g_1^{\gamma_k\gamma_{k+1}}\right)\delta_{\gamma_1\gamma_{m+1}}
+\left(\prod_{k=1}^m
g_1^{\gamma_k\gamma_{k+1}}\right)\eta_{\gamma_1\gamma_{m+1}}.\eeq
When we calculate the contributions from the $n$th order
approximation, we first will carry out $n-1$ the first
decompositions, that is \beq \label{rdofgp}\prod_{k=1}^n
g_1^{\gamma_k\gamma_{k+1}}=\left(\prod_{k=1}^n
g_1^{\gamma_k\gamma_{k+1}}\right)\left[\prod_{k=1}^{n-1}
\left(\delta_{\gamma_k\gamma_{k+2}}+\eta_{\gamma_k\gamma_{k+2}}\right)\right].
\eeq Obviously, from the fact that $H_1$ is usually taken as Hermit
one, it follows that \beq
g_1^{\gamma_{j}\gamma_{j+1}}g_1^{\gamma_{j+1}\gamma_{j+2}}\delta_{\gamma_j\gamma_{j+2}}
=\left|g_1^{\gamma_{j}\gamma_{j+1}}\right|^2\delta_{\gamma_j\gamma_{j+2}}.\eeq
When considering the contributions a given order approximation, the
summation over one of two subscripts will lead in the contraction of
$g$-production. More generally, for the contraction of even number
$g$-production \beq \left(\prod_{j=1}^m g_1^{\gamma_j\gamma_{j+1}}
\prod_{k=1}^{m-1}\delta_{\gamma_k\gamma_{k+2}}\right)
\delta_{\gamma_1\gamma}\delta_{\gamma_{m+1}\gamma^\prime}
=\left|g_1^{\gamma\gamma_2}\right|^m\left(\prod_{k=1}^{m-1}\delta_{\gamma_k\gamma_{k+2}}\right)
\delta_{\gamma_1\gamma}\delta_{\gamma_{m+1}\gamma^\prime}\delta_{\gamma\gamma^\prime},\eeq
and for the contraction of odd number $g$-production, \beq
\left(\prod_{j=1}^m g_1^{\gamma_j\gamma_{j+1}}
\prod_{k=1}^{m-1}\delta_{\gamma_k\gamma_{k+2}}\right)
\delta_{\gamma_1\gamma}\delta_{\gamma_{m+1}\gamma^\prime}
=\left|g_1^{\gamma\gamma^\prime}\right|^{m-1}\left(\prod_{k=1}^{m-1}\delta_{\gamma_k\gamma_{k+2}}\right)
\delta_{\gamma_1\gamma}\delta_{\gamma_{m+1}\gamma^\prime}g_1^{\gamma\gamma^\prime},\eeq
where $\delta_{\gamma_1\gamma}\delta_{\gamma_{m+1}}$ is a factor
appearing in the expression of our solution.

Then, we consider, in turn, all possible the second decompositions,
the third decompositions, and up to the $(n-1)$th decompositions. It
must be emphasized that after calculating the contributions from
terms of lower decompositions, some of terms in the higher
decompositions may be trivial because there are some symmetric and
complementary symmetric indexes in the corresponding result, that
is, the product of this result and the given
$\delta_{\gamma_k\gamma_{k^\prime}}$ or
$\eta_{\gamma_k\gamma_{k^\prime}}$ is zero. In other words, such
some higher decompositions do not need to be considered. As an
example, let us analysis the contribution from the third order
approximation. It is clear that the first decomposition of a
sequential production of three nondiagonal elements becomes \beqa
g_1^{\gamma_1\gamma_2}g_1^{\gamma_2\gamma_3}g_1^{\gamma_3\gamma_4}
&=&g_1^{\gamma_1\gamma_2}g_1^{\gamma_2\gamma_3}g_1^{\gamma_3\gamma_4}
\delta_{\gamma_1\gamma_3}\delta_{\gamma_2\gamma_4}
+g_1^{\gamma_1\gamma_2}g_1^{\gamma_2\gamma_3}g_1^{\gamma_3\gamma_4}
\delta_{\gamma_1\gamma_3}\eta_{\gamma_2\gamma_4}
\nonumber\\
&
&+g_1^{\gamma_1\gamma_2}g_1^{\gamma_2\gamma_3}g_1^{\gamma_3\gamma_4}
\eta_{\gamma_1\gamma_3}\delta_{\gamma_2\gamma_4}
+g_1^{\gamma_1\gamma_2}g_1^{\gamma_2\gamma_3}g_1^{\gamma_3\gamma_4}
\eta_{\gamma_1\gamma_3}\eta_{\gamma_2\gamma_4}.\eeqa Thus, the
related contribution is just divided as $4$ terms \beq
A_3^{\gamma\gamma^\prime}(ggg)=A_3^{\gamma\gamma^\prime}(ggg;cc)
+A_3^{\gamma\gamma^\prime}(ggg;cn)+A_3^{\gamma\gamma^\prime}(ggg;nc)
+{A}_3^{\gamma\gamma^\prime}(ggg,nn).\eeq In fact, by calculating we
know that the second decomposition of the former three terms do not
need to be considered, only the second decomposition of the last
term is nontrivial. This means that \beq
{A}_3^{\gamma\gamma^\prime}(ggg;nn)={A}_3^{\gamma\gamma^\prime}(ggg;nn,c)
+{A}_3^{\gamma\gamma^\prime}(ggg;nn,n),\eeq where we have added
$\delta_{\gamma_1\gamma_3}$ in the definition of
${A}_3^{\gamma\gamma^\prime}(ggg;nn,c)$, and
$\eta_{\gamma_1\gamma_3}$ in the definition of
${A}_3^{\gamma\gamma^\prime}(ggg;nn,n)$. Obviously, in the practical
process, this feature largely simplifies the calculations. It is
easy to see that the number of all of terms with contractions and
anti-contractions is $5$. For convenience and clearness, we call the
contributions from the different terms in the decomposition of
$g$-product as the contractions and anti-contractions of
$g$-product. Of course, the contraction and anti-contraction refer
to the meaning after summation(s) over the subscript(s) in general.
Moreover, here and after, we first drop the argument $gg\cdots g$ in
the $i$th order approximation $A_i$ since its meaning has been
indicated by $i$ after the Hamiltonian is redivided. For example,
the explicit expressions of all contraction- and anti-contraction
terms in the third order approximation $A_3$ can be calculated as
the following: \beqa\label{A3cc} {A}_3^{\gamma\gamma^\prime}(cc)&=
&\sum_{\gamma_1,\cdots,\gamma_{4}}\left[
\sum_{i=1}^{4}(-1)^{i-1}\frac{\e^{-\I E_{\gamma_i}
t}}{d_i(E[\gamma,3])}\right] \left[\prod_{j=1}^3
g_1^{\gamma_j\gamma_{j+1}}\right]\left(
\prod_{k=1}^{2}\delta_{\gamma_k\gamma_{k+2}}\right)
\delta_{\gamma_1\gamma}\delta_{\gamma_{4}\gamma^\prime}\nonumber\\
& =& \left[-\frac{2\e^{-\I
E_{\gamma}t}}{\left(E_\gamma-E_{\gamma^\prime}\right)^3}+\frac{2\e^{-\I
E_{\gamma^\prime}t}}{\left(E_\gamma-E_{\gamma^\prime}\right)^3}+(-\I
t)\frac{\e^{-\I E_{\gamma}t}}
{\left(E_\gamma-E_{\gamma^\prime}\right)^2}\right.\nonumber\\ &
&\left.+(-\I t)\frac{\e^{-\I
E_{\gamma^\prime}t}}{\left(E_\gamma-E_{\gamma^\prime}\right)^2}\right]
\left|g_1^{\gamma\gamma^\prime}\right|^2g_1^{\gamma\gamma^\prime},\eeqa
\beqa\label{A3cn}
{A}_3^{\gamma\gamma^\prime}(cn)&=&\sum_{\gamma_1,\cdots,\gamma_{l+1}}\left[
\sum_{i=1}^{4}(-1)^{i-1}\frac{\e^{-\I E_{\gamma_i}
t}}{d_i(E[\gamma,3])}\right] \left[\prod_{j=1}^3
g_1^{\gamma_j\gamma_{j+1}}\right]\delta_{\gamma_1\gamma_3}\eta_{\gamma_2\gamma_4}
\delta_{\gamma_1\gamma}\delta_{\gamma_{l+1}\gamma^\prime}\nonumber\\
& =& \sum_{\gamma_1}\left[-\frac{\e^{-\I
E_{\gamma}t}}{\left(E_\gamma-E_{\gamma_1}\right)
\left(E_\gamma-E_{\gamma^\prime}\right)^2}-\frac{\e^{-\I
E_{\gamma}t}}{\left(E_\gamma-E_{\gamma_1}\right)^2
\left(E_\gamma-E_{\gamma^\prime}\right)}\right.\nonumber\\
& &+\frac{\e^{-\I
E_{\gamma_1}t}}{\left(E_\gamma-E_{\gamma_1}\right)^2
\left(E_{\gamma_1}-E_{\gamma^\prime}\right)}-\frac{\e^{-\I
E_{\gamma^\prime}t}}{\left(E_\gamma-E_{\gamma^\prime}\right)^2
\left(E_{\gamma_1}-E_{\gamma^\prime}\right)}\nonumber\\
& &\left.+(-\I t)\frac{\e^{-\I E_{\gamma}t}}
{\left(E_\gamma-E_{\gamma_1}\right)\left(E_\gamma-E_{\gamma^\prime}\right)}\right]
\left|g_1^{\gamma\gamma_1}\right|^2g_1^{\gamma\gamma^\prime}
\eta_{\gamma_1\gamma^\prime},\eeqa
\beqa \label{A3nc} {A}_3^{\gamma\gamma^\prime}(nc)&=
&\sum_{\gamma_1,\cdots,\gamma_{4}}\left[
\sum_{i=1}^{4}(-1)^{i-1}\frac{\e^{-\I E_{\gamma_i}
t}}{d_i(E[\gamma,3])}\right] \left[\prod_{j=1}^3
g_1^{\gamma_j\gamma_{j+1}}\right]\eta_{\gamma_1\gamma_3}\delta_{\gamma_2\gamma_4}
\delta_{\gamma_1\gamma}\delta_{\gamma_{l+1}\gamma^\prime}\nonumber\\
& =& \sum_{\gamma_1}\left[\frac{\e^{-\I
E_{\gamma}t}}{\left(E_\gamma-E_{\gamma_1}\right)\left(E_\gamma-E_{\gamma^\prime}\right)^2}-\frac{\e^{-\I
E_{\gamma_1}t}}{\left(E_\gamma-E_{\gamma_1}\right)
\left(E_{\gamma_1}-E_{\gamma^\prime}\right)^2}\right.\nonumber\\
& &-\frac{\e^{-\I
E_{\gamma^\prime}t}}{\left(E_\gamma-E_{\gamma_1}\right)\left(E_\gamma-E_{\gamma^\prime}\right)^2}+\frac{\e^{-\I
E_{\gamma^\prime}t}}{\left(E_\gamma-E_{\gamma_1}\right)\left(E_{\gamma_1}-E_{\gamma^\prime}\right)^2}\nonumber\\
& &\left.+(-\I t)\frac{\e^{-\I E_{\gamma^\prime}t}}
{\left(E_\gamma-E_{\gamma^\prime}\right)\left(E_{\gamma_1}-E_{\gamma^\prime}\right)}\right]
g_1^{\gamma\gamma^\prime}
\left|g_1^{\gamma_1\gamma^\prime}\right|^2\eta_{\gamma\gamma_1},\eeqa
\beqa \label{A3nn-c} {A}_3^{\gamma\gamma^\prime}(nn,c)
&=&\sum_{\gamma_1,\cdots,\gamma_{4}}\left[
\sum_{i=1}^{4}(-1)^{i-1}\frac{\e^{-\I E_{\gamma_i}
t}}{d_i(E[\gamma,3])}\right] \left[\prod_{j=1}^3
g_1^{\gamma_j\gamma_{j+1}}\right]
\delta_{\gamma_1\gamma}\delta_{\gamma_{4}\gamma^\prime}
\eta_{\gamma_1\gamma_3}
\eta_{\gamma_2\gamma_4}{\delta}_{\gamma\gamma^\prime}
\nonumber\\
&=& \sum_{\gamma_1\gamma_2}\left[-\frac{\e^{-\I
E_{\gamma}t}}{\left(E_{\gamma}-E_{\gamma_1}\right)^2
\left(E_{\gamma_1}-E_{\gamma_2}\right)}+\frac{\e^{-\I
E_{\gamma}t}}{\left(E_{\gamma}-E_{\gamma_2}\right)^2
\left(E_{\gamma_1}-E_{\gamma_2}\right)}\right.\nonumber\\
& & +\frac{\e^{-\I
E_{\gamma_1}t}}{\left(E_{\gamma}-E_{\gamma_1}\right)^2
\left(E_{\gamma_1}-E_{\gamma_2}\right)}-\frac{\e^{-\I
E_{\gamma_2}t}}{\left(E_{\gamma}-E_{\gamma_2}\right)^2
\left(E_{\gamma_1}-E_{\gamma_2}\right)}\nonumber\\
& &\left. +(-\I t)\frac{\e^{-\I
E_{\gamma}t}}{\left(E_{\gamma}-E_{\gamma_1}\right)
\left(E_{\gamma_1}-E_{\gamma_2}\right)}
\right]g_1^{\gamma\gamma_1}g_1^{\gamma_1\gamma_2}g_1^{\gamma_2\gamma^\prime}
\eta_{\gamma\gamma_2}{\delta}_{\gamma\gamma^\prime}, \eeqa \beqa
\label{A3nn-n} {A}_3^{\gamma\gamma^\prime}(nn,n)
&=&\sum_{\gamma_1,\cdots,\gamma_{4}}\left[
\sum_{i=1}^{4}(-1)^{i-1}\frac{\e^{-\I E_{\gamma_i}
t}}{d_i(E[\gamma,3])}\right] \left[\prod_{j=1}^3
g_1^{\gamma_j\gamma_{j+1}}\right]
\delta_{\gamma_1\gamma}\delta_{\gamma_{4}\gamma^\prime}
\eta_{\gamma_1\gamma_3}
\eta_{\gamma_2\gamma_4}\eta_{\gamma\gamma^\prime}
\nonumber\\
& &+ \sum_{\gamma_1\gamma_2}\left[\frac{\e^{-\I
E_{\gamma}t}}{\left(E_{\gamma}-E_{\gamma_1}\right)
\left(E_{\gamma}-E_{\gamma_2}\right)\left(E_{\gamma}-E_{\gamma^\prime}\right)}\right.\nonumber\\
& & -\frac{\e^{-\I
E_{\gamma_1}t}}{\left(E_{\gamma}-E_{\gamma_1}\right)
\left(E_{\gamma_1}-E_{\gamma_2}\right)\left(E_{\gamma_1}-E_{\gamma^\prime}\right)}\nonumber\\
& & +\frac{\e^{-\I
E_{\gamma_2}t}}{\left(E_{\gamma}-E_{\gamma_2}\right)
\left(E_{\gamma_1}-E_{\gamma_2}\right)\left(E_{\gamma_2}-E_{\gamma^\prime}\right)}\nonumber\\
& &\left. -\frac{\e^{-\I
E_{\gamma^\prime}t}}{\left(E_{\gamma}-E_{\gamma^\prime}\right)
\left(E_{\gamma_1}-E_{\gamma^\prime}\right)\left(E_{\gamma_2}-E_{\gamma^\prime}\right)}
\right]g_1^{\gamma\gamma_1}g_1^{\gamma_1\gamma_2}g_1^{\gamma_2\gamma^\prime}
\eta_{\gamma\gamma_2}\eta_{\gamma_1\gamma^\prime}
\eta_{\gamma\gamma^\prime}.\eeqa In above calculations, the used
technologies are mainly to find the limitation, dummy index changing
and summation, as well as the replacement
$g_1^{\gamma_i\gamma_j}\eta_{\gamma_i\gamma_j}=g_1^{\gamma_i\gamma_j}$
since $g_1^{\gamma_i\gamma_j}$ has been nondiagonal.

It must be emphasized that, in our notation,
$A_i^{\gamma\gamma^\prime}$ represents the contributions from the
$i$th order approximation. The other independent variables are
divided into $i-1$ groups and are arranged sequentially
corresponding to the order of $g$-product decomposition. That is,
the first variable group represents the first decompositions, the
second variable group represents the second decompositions, and so
on. Every variable group is a bit-string made of three possible
element $c,n,k$ and its length is equal to the number of the related
order of $g$-product decomposition, that is, for the $j$th
decompositions in the $i$th order approximation its length is $i-j$.
In each variable group, $c$ corresponds to a $\delta$ function, $n$
corresponds to a $\eta$ function and $k$ corresponds to $1$
(non-decomposition). Their sequence in the bit-string corresponds to
the sequence of contraction and/or anti-contraction index string.
From the above analysis and statement, the index string of the $j$th
decompositions in the $i$ order approximation is: \beq
\prod_{k=1}^{i-j}\left(\gamma_k,\gamma_{k+1+j}\right). \eeq For
example, the first variable group is $cccn$, which refers to the
first decomposition in five order approximation and the contribution
term to include the factor
$\delta_{\gamma_1\gamma_3}\delta_{\gamma_2\gamma_4}
\delta_{\gamma_3\gamma_5}\eta_{\gamma_4\gamma_6}$ in the definition
of $A_5(cccn)$. Similarly, $cncc$ means to insert the factor
$\delta_{\gamma_1\gamma_3}\eta_{\gamma_2\gamma_4}
\delta_{\gamma_3\gamma_5}\delta_{\gamma_4\gamma_6}$ in the
definition of $A_5(cncc)$. When there are non trivial second
contractions, for instance, two variable group $(ccnn,kkc)$
represents that the definition of $A_5(ccnn,kkc)$ has the factor
$\left(\delta_{\gamma_1\gamma_3}\delta_{\gamma_2\gamma_4}
\eta_{\gamma_3\gamma_5}\eta_{\gamma_4\gamma_6}\right)\delta_{\gamma_3\gamma_6}$.
Since there are fully trivial contraction (the bit-string is made of
only $k$), we omit their related variable group for simplicity.

Furthermore, we pack up all the contraction- and non-contraction
terms in the following way so that we can obtain conveniently the
improved forms of perturbed solution of dynamics including the
partial contributions from the high order approximation. We first
decompose $A_3^{\gamma\gamma^\prime}$, which is a summation of all
above terms, into the three parts according to $\e^{-\I
E_{\gamma_i}t}, (-\I t) \e^{-\I E_{\gamma_i}t} $ and $(-\I
t)^2\e^{-\I E_{\gamma_i}t}/2$: \beq
A_3^{\gamma\gamma^\prime}=A_3^{\gamma\gamma^\prime}(\e)+A_3^{\gamma\gamma^\prime}(t\e)
+A_3^{\gamma\gamma^\prime}(t^2\e).\eeq Secondly, we decompose its
every term into three parts according to $\e^{-\I E_{\gamma}t},
\e^{-\I E_{\gamma_1}t}$ ($\sum_{\gamma_1}\e^{-\I E_{\gamma_1}t}$)
and $\e^{-\I E_{\gamma^\prime}t}$: \beqa
A_3^{\gamma\gamma^\prime}(\e)&=&A_3^{\gamma\gamma^\prime}(\e^{-\I
E_{\gamma}t})+A_3^{\gamma\gamma^\prime}(\e^{-\I
E_{\gamma_1}t})+A_3^{\gamma\gamma^\prime}(\e^{-\I E_{\gamma^\prime}t}),\\
A_3^{\gamma\gamma^\prime}(t\e)&=&A_3^{\gamma\gamma^\prime}(t\e^{-\I
E_{\gamma}t})+A_3^{\gamma\gamma^\prime}(t\e^{-\I
E_{\gamma_1}t})+A_3^{\gamma\gamma^\prime}(t\e^{-\I E_{\gamma^\prime}t}),\\
A_4^{\gamma\gamma^\prime}(t^2\e)&=&A_3^{\gamma\gamma^\prime}(t^2\e^{-\I
E_{\gamma}t})+A_3^{\gamma\gamma^\prime}(t^2\e^{-\I
E_{\gamma_1}t})+A_3^{\gamma\gamma^\prime}(t^2\e^{-\I
E_{\gamma^\prime}t}). \eeqa Finally, we again decompose every term
in above equations into the diagonal and non-diagonal parts about
$\gamma$ and $\gamma^\prime$: \beqa
A_3^{\gamma\gamma^\prime}(\e^{-\I
E_{\gamma_i}t})&=&A_3^{\gamma\gamma^\prime}(\e^{-\I
E_{\gamma_i}t};{\rm D})+A_3^{\gamma\gamma^\prime}(\e^{-\I
E_{\gamma_i}t};{\rm N}),\\
A_3^{\gamma\gamma^\prime}(t\e^{-\I
E_{\gamma_i}t})&=&A_3^{\gamma\gamma^\prime}(t\e^{-\I
E_{\gamma_i}t};{\rm D})+A_3^{\gamma\gamma^\prime}(t\e^{-\I
E_{\gamma_i}t};{\rm N}) ,\\
A_3^{\gamma\gamma^\prime}(t^2\e^{-\I
E_{\gamma_i}t})&=&A_3^{\gamma\gamma^\prime}(t^2\e^{-\I
E_{\gamma_i}t};{\rm D})+A_3^{\gamma\gamma^\prime}(t^2\e^{-\I
E_{\gamma_i}t};{\rm N}),  \eeqa where $E_{\gamma_i}$ takes $
E_{\gamma}, E_{\gamma_1}$ and $E_{\gamma^\prime}$.

According above way, it is easy to obtain \beqa
A_3^{\gamma\gamma^\prime}(\e^{-\I E_{\gamma}t};{\rm D})&=&
-\sum_{\gamma_1,\gamma_2}\e^{-\I E_{\gamma}
t}\left[\frac{1}{\left(E_{\gamma}-E_{\gamma_1}\right)
\left(E_{\gamma}-E_{\gamma_1}\right)^2}\right.\nonumber\\
& &\left.+\frac{1}{\left(E_{\gamma}-E_{\gamma_1}\right)^2
\left(E_{\gamma}-E_{\gamma_1}\right)}\right]g_1^{\gamma\gamma_1}g_1^{\gamma_1\gamma_2}
g_1^{\gamma_2\gamma}\delta_{\gamma\gamma^\prime},\\
A_3^{\gamma\gamma^\prime}(\e^{-\I E_{\gamma}t};{\rm N})&=&
-\sum_{\gamma_1}\e^{-\I E_{\gamma}
t}\left[\frac{1}{\left(E_{\gamma}-E_{\gamma_1}\right)
\left(E_{\gamma}-E_{\gamma^\prime}\right)^2}\right.\nonumber\\
& &\left.+\frac{1}{\left(E_{\gamma}-E_{\gamma_1}\right)^2
\left(E_{\gamma}-E_{\gamma^\prime}\right)}\right]g_1^{\gamma\gamma_1}g_1^{\gamma_1\gamma}
g_1^{\gamma\gamma^\prime}\nonumber\\
& &+\sum_{\gamma_1,\gamma_2}\e^{-\I
E_{\gamma}t}\frac{g_1^{\gamma\gamma_1}g_1^{\gamma_1\gamma_2}
g_1^{\gamma_2\gamma^\prime}\eta_{\gamma\gamma_2}\eta_{\gamma\gamma^\prime}}
{\left(E_{\gamma}-E_{\gamma_1}\right)
\left(E_{\gamma}-E_{\gamma_2}\right)\left(E_{\gamma}-E_{\gamma^\prime}\right)},\eeqa
\beqa A_3^{\gamma\gamma^\prime}(\e^{-\I E_{\gamma_1}t};{\rm D})&=&
\sum_{\gamma_1,\gamma_2}\e^{-\I E_{\gamma_1}
t}\frac{g_1^{\gamma\gamma_1}g_1^{\gamma_1\gamma_2}
g_1^{\gamma_2\gamma}\delta_{\gamma\gamma^\prime}}{\left(E_{\gamma}-E_{\gamma_1}\right)^2
\left(E_{\gamma_1}-E_{\gamma_2}\right)},\\
A_3^{\gamma\gamma^\prime}(\e^{-\I E_{\gamma_1}t};{\rm N})&=&
-\sum_{\gamma_1,\gamma_2}\e^{-\I
E_{\gamma_1}t}\frac{g_1^{\gamma\gamma_1}g_1^{\gamma_1\gamma_2}
g_1^{\gamma_2\gamma^\prime}\eta_{\gamma_1\gamma^\prime}\eta_{\gamma\gamma^\prime}}
{\left(E_{\gamma}-E_{\gamma_1}\right)
\left(E_{\gamma_1}-E_{\gamma_2}\right)\left(E_{\gamma_1}-E_{\gamma^\prime}\right)},\eeqa
\beqa A_3^{\gamma\gamma^\prime}(\e^{-\I E_{\gamma_2}t};{\rm D})&=&
-\sum_{\gamma_1,\gamma_2}\e^{-\I E_{\gamma_2}
t}\frac{g_1^{\gamma\gamma_1}g_1^{\gamma_1\gamma_2}
g_1^{\gamma_2\gamma}\delta_{\gamma\gamma^\prime}}{\left(E_{\gamma}-E_{\gamma_2}\right)^2
\left(E_{\gamma_1}-E_{\gamma_2}\right)},\\
A_3^{\gamma\gamma^\prime}(\e^{-\I E_{\gamma_1}t};{\rm N})&=&
\sum_{\gamma_1,\gamma_2}\e^{-\I
E_{\gamma_2}t}\frac{g_1^{\gamma\gamma_1}g_1^{\gamma_1\gamma_2}
g_1^{\gamma_2\gamma^\prime}\eta_{\gamma\gamma_2}\eta_{\gamma\gamma^\prime}}
{\left(E_{\gamma}-E_{\gamma_2}\right)
\left(E_{\gamma_1}-E_{\gamma_2}\right)\left(E_{\gamma_2}-E_{\gamma^\prime}\right)},\eeqa
\beqa A_3^{\gamma\gamma^\prime}(\e^{-\I E_{\gamma^\prime}t};{\rm
D})&=&0,\\
A_3^{\gamma\gamma^\prime}(\e^{-\I E_{\gamma^\prime}t};{\rm N})&=&
\sum_{\gamma_1}\e^{-\I E_{\gamma^\prime}
t}\left[\frac{1}{\left(E_{\gamma}-E_{\gamma^\prime}\right)
\left(E_{\gamma_1}-E_{\gamma^\prime}\right)^2}\right.\nonumber\\
& &\left.+\frac{1}{\left(E_{\gamma}-E_{\gamma^\prime}\right)^2
\left(E_{\gamma_1}-E_{\gamma^\prime}\right)}\right]g_1^{\gamma^\prime\gamma_1}g_1^{\gamma_1\gamma^\prime}
g_1^{\gamma\gamma^\prime}\nonumber\\
& &-\sum_{\gamma_1,\gamma_2}\e^{-\I
E_{\gamma^\prime}t}\frac{g_1^{\gamma\gamma_1}g_1^{\gamma_1\gamma_2}
g_1^{\gamma_2\gamma^\prime}\eta_{\gamma_1\gamma^\prime}\eta_{\gamma\gamma^\prime}}
{\left(E_{\gamma}-E_{\gamma^\prime}\right)
\left(E_{\gamma_1}-E_{\gamma^\prime}\right)\left(E_{\gamma_2}-E_{\gamma^\prime}\right)}.\eeqa

In the end of this subsection, we would like to point out that one
of purposes introducing the decompositions of $g$-product and the
contractions and anti-contractions of $g$-product is to eliminate
the apparent (not real) singularities and find out all the
limitations from the contributions of $g$-product contractions. It
is important to express the results with the physical significance.

\section{Improved forms of perturbed solution of dynamics}\label{sec7}

In fact, the final aim using the decomposition and then obtaining
contraction and anti-contraction is to include partially
contributions from the high order approximations of nondiagonal part
of $H_1$ in our improved scheme of perturbation theory. In this
section, making use of the skill to calculate contractions and
anti-contractions of $g$-product, we can obtain the zeroth, first,
second and third order improved forms of perturbed solutions
including partial higher order approximations.

In mathematics, the process to obtain the improved forms of
perturbed solutions is a kind of skill to deal with an infinity
series, that is, according to some principles and the general term
form to rearrange those terms with the same features together
forming a group, then sum all of the terms in such a particular
group that they become a compact function at a given precision,
finally this infinity series is transformed into a new form that
directly relates with the studied problem. More concretely speaking,
since we concern the system developing with time $t$, we take those
terms with $(-\I y_i t) \e^{-\I x_i t}$, $(-\I y_i t)^2 \e^{-\I x_i
t}/2!$ and $(-\I y_i t)^3 \e^{-\I x_i t}/3!$, $\cdots$ with the same
factor $f$ together forming a group, then sum them to obtain an
exponential function $f\exp\left[-\I\left(x_i+y_i\right)t\right]$.
The physical reason to do this is such an exponential function
represents the system evolution in theory and has the obvious
physical significance in the calculation of transition probability
and perturbed energy. Through rearranging and summing, those terms
with factors $t^a\e^{-\I E_{\gamma_i}t}$,  $(a=1,2,\cdots)$ in the
higher order approximation are added to the improved lower
approximations, we thus can advance the precision, particular, when
the evolution time $t$ is longer. We can call it ``dynamical
rearrangement and summation" technology.

\subsection{Improved form of the zeroth order perturbed solution of dynamics}

Let us start with the zeroth order perturbed solution of dynamics.
In the usual perturbation theory, it is well-known \beq
\ket{\Psi^{(0)}(t)}=\sum_{\gamma}\e^{-\I E_\gamma
t}\diracsp{\Phi^\gamma}{\Psi(0)}\ket{\Phi^\gamma}=\sum_{\gamma\gamma^\prime}\e^{-\I
E_\gamma
t}\delta_{\gamma\gamma^\prime}a_{\gamma^\prime}\ket{\Phi^\gamma},\eeq
where $a_{\gamma^\prime}=\diracsp{\Phi^{\gamma^\prime}}{\Psi(0)}$.
Now, we would like to improve it so that it can include the partial
contributions from higher order approximations. Actually, we can
find that $A_2(c)$ and $A_3(nn,c)$ have the terms proportional to
$(-\I t)$ \beqa\label{0th1} & & (-\I t)\e^{-\I E_{\gamma}
t}\left[\sum_{\gamma_1}\frac{1}{E_{\gamma}-E_{\gamma_1}}
\left|g_1^{\gamma\gamma_1}\right|^2\right]
\delta_{\gamma\gamma^\prime},\\
\label{0th2} & &(-\I t)\e^{-\I E_{\gamma}
t}\left[\sum_{\gamma_1,\gamma_2}\frac{1}{(E_{\gamma}-E_{\gamma_1})(E_{\gamma}-E_{\gamma_2})}
g_1^{\gamma\gamma_1}g_1^{\gamma_1\gamma_2}g_1^{\gamma_2\gamma^\prime}\right]
\delta_{\gamma\gamma^\prime}.\eeqa Introduce the notation \beqa
G^{(2)}_{\gamma}&=&\sum_{\gamma_1}\frac{1}{E_{\gamma}-E_{\gamma_1}}\left|g_1^{\gamma\gamma_1}\right|^2,\\
G^{(3)}_{\gamma}&=&\sum_{\gamma_1,\gamma_2}\frac{1}{(E_{\gamma}-E_{\gamma_1})(E_{\gamma}-E_{\gamma_2})}
g_1^{\gamma\gamma_1}g_1^{\gamma_1\gamma_2}g_1^{\gamma_2\gamma}.\eeqa
It is clear that $G_\gamma^{(a)}$ has energy dimension, and we will
see that it can be called the $a$th revision energy. Let us add the
terms (\ref{0th1}), (\ref{0th2}) and the related terms in
$A_4(t\e^{-\I E_{\gamma} t},{\rm D}),A_4(t^2\e^{-\I
E_{\gamma}t},{\rm D}),A_5(t\e^{-\I E_{\gamma} t},{\rm D})$,
$A_5(t^2\e^{-\I E_{\gamma} t},{\rm D})$, $A_6(t^2\e^{-\I E_{\gamma}
t},{\rm D})$ and $A_6(t^3 \e^{-\I E_{\gamma} t})$ given in Appendix
B together, that is, \beqa A_{0;\rm
I}^{\gamma\gamma^\prime}&=&\e^{-\I E_{\gamma}t}\left[1+(-\I
t)\left(G^{(2)}_\gamma+G^{(3)}_\gamma+G^{(4)}_\gamma+G^{(5)}_\gamma\right)\right.\nonumber\\
& &\left.+ \frac{(-\I
t)^2}{2!}\left(G^{(2)}_\gamma+G^{(3)}_\gamma\right)^2+ \frac{(-\I
t)^2}{2!}2 G^{(2)}_\gamma G^{(4)}_\gamma+\cdots
\right]\delta_{\gamma\gamma^\prime} ,\eeqa Although we have not
finished the more calculations, from the mathematical symmetry and
physical concept, we can think \beqa A_{0;\rm
I}^{\gamma\gamma^\prime}&=&\e^{-\I E_{\gamma}t}\left[1+(-\I
t)\left(G^{(2)}_\gamma+G^{(3)}_\gamma+G^{(4)}_\gamma+G^{(5)}_\gamma\right)\right.\nonumber\\
& &\left.+ \frac{(-\I
t)^2}{2!}\left(G^{(2)}_\gamma+G^{(3)}_\gamma+G^{(4)}_\gamma+G^{(5)}_\gamma\right)^2+\cdots
\right]\delta_{\gamma\gamma^\prime} ,\eeqa New terms can appear at
$A_7$, $A_8$, $A_9$ and $A_{10}$. So we obtain the improved form of
zeroth order perturbed solution of dynamics \beq \label{ips0}
\ket{\Psi_{E_T}^{(0)}(t)}_{\rm
I}=\sum_{\gamma\gamma^\prime}\e^{-\I\left(E_\gamma+G^{(2)}_\gamma
+{G}^{(3)}_\gamma+G^{(4)}_\gamma+G^{(5)}_\gamma\right)
t}\delta_{\gamma\gamma^\prime}a_{\gamma^\prime}\ket{\Phi^\gamma}.\eeq

It is clear that $G^{(2)}_\gamma$ is real. In fact, $G^{(3)}_\gamma$
is also real. In order to prove it, we exchange the dummy indexes
$\gamma_1$ and $\gamma_2$ and take the complex conjugate of
$G^{(3)}_\gamma$, that is \beqa
{G^{(3)}_{\gamma}}^*&=&\sum_{\gamma_1,\gamma_2}\frac{1}{(E_{\gamma}-E_{\gamma_1})(E_{\gamma}-E_{\gamma_2})}
\left(g_1^{\gamma\gamma_2}\right)^*\left(g_1^{\gamma_2\gamma_1}\right)^*
\left(g_1^{\gamma_1\gamma}\right)^*\nonumber\\
&=&
\sum_{\gamma_1,\gamma_2}\frac{1}{(E_{\gamma}-E_{\gamma_1})(E_{\gamma}-E_{\gamma_2})}
g_1^{\gamma\gamma_1}g_1^{\gamma_1\gamma_2}g_1^{\gamma_2\gamma}\nonumber\\
&=& G^{(3)}_\gamma, \eeqa where we have used the relations
$\left(g_1^{\beta_1\beta_2}\right)^*=g_1^{\beta_2\beta_1}$ for any
$\beta_1$ and $\beta_2$ since $H_1$ is Hermit. Similar analyses can
be applied to $G^{(4)}_\gamma$ and $G^{(5)}_\gamma$. These mean that
$\e^{-\I
\left(G^{(2)}_\gamma+G^{(3)}_\gamma+G^{(4)}_\gamma+G^{(5)}_\gamma\right)
t}$ is still an oscillatory factor.

\subsection{Improved form of the first order perturbed solution of
dynamics}

Furthermore, in order to include the partial contributions from
higher than zeroth order approximation, we need to consider the
contributions from nondiagonal elements in the higher order
approximations.

Well-known usual first order perturbed part of dynamics is \beqa
\ket{\Psi^{(1)}(t)}&=&\sum_{\gamma,\gamma^\prime}\left[\frac{\e^{-\I
E_\gamma t}}{E_{\gamma}-E_{\gamma^\prime}}-\frac{\e^{-\I
E_{\gamma^\prime}
t}}{E_{\gamma}-E_{\gamma^\prime}}\right]H_1^{\gamma\gamma^\prime}\ket{\Phi^\gamma}
= \sum_{\gamma,\gamma^\prime}\left[\left(\frac{\e^{-\I E_\gamma
t}}{E_{\gamma}-E_{\gamma^\prime}}-\frac{\e^{-\I E_{\gamma^\prime}
t}}{E_{\gamma}-E_{\gamma^\prime}}\right)g_1^{\gamma\gamma^\prime}\right]\ket{\Phi^\gamma}.\eeqa
It must be emphasized that $H_1$ is taken as only with the
nondiagonal part for simplicity. That is, we have used the skill one
stated above.

Thus, from $A_3(t\e^{-\I E_{\gamma} t},{\rm N})$ and $A_4(t\e^{-\I
E_{\gamma} t},{\rm N})$, $A_4(t^2\e^{-\I E_{\gamma}t},{\rm D})$,
$A_5(t^2\e^{-\I E_{\gamma} t},{\rm N})$, $A_6(t^2\e^{-\I E_{\gamma}
t},{\rm N})$ in the Appendix B, it follows that \beqa A_{1;\rm
I}^{\gamma\gamma^\prime}&=&\frac{\e^{-\I
E_{\gamma}t}}{\left(E_{\gamma}-E_{\gamma^\prime}\right)}\left[1+(-\I
t)\left(G^{(2)}_\gamma+G^{(3)}_\gamma+G^{(4)}_\gamma\right)\right.\nonumber\\
& &\left.+ \frac{(-\I t)^2}{2!}\left(G^{(2)}_\gamma\right)^2+
\frac{(-\I t)^2}{2!}2 G^{(2)}_\gamma G^{(3)}_\gamma+\cdots
\right]g_1^{\gamma\gamma^\prime}\nonumber\\
& &-\frac{\e^{-\I
E_{\gamma^\prime}t}}{\left(E_{\gamma}-E_{\gamma^\prime}\right)}\left[1+(-\I
t)\left(G^{(2)}_{\gamma^\prime}+G^{(3)}_{\gamma^\prime}+G^{(4)}_{\gamma^\prime}\right)\right.\nonumber\\
& &\left.+ \frac{(-\I
t)^2}{2!}\left(G^{(2)}_{\gamma^\prime}\right)^2+ \frac{(-\I
t)^2}{2!}2 G^{(2)}_{\gamma^\prime} G^{(3)}_{\gamma^\prime}+\cdots
\right]g_1^{\gamma\gamma^\prime}.\eeqa Therefore, the improved
perturbed solution of dynamics is just \beqa \label{ips1}
\ket{\Psi^{(1)}(t)}_{\rm
I}&=&\sum_{\gamma,\gamma^\prime}\left[\left(\frac{\e^{-\I
\left(E_\gamma+G^{(2)}_\gamma+G^{(3)}_\gamma+G^{(4)}_\gamma\right)
t}}{E_{\gamma}-E_{\gamma^\prime}}-\frac{\e^{-\I
\left(E_{\gamma^\prime}+G^{(2)}_{\gamma^\prime}+G^{(3)}_{\gamma^\prime}+G^{(4)}_{\gamma^\prime}\right)
t}}{E_{\gamma}-E_{\gamma^\prime}}\right)g_1^{\gamma\gamma^\prime}\right]
a_{\gamma^\prime}\ket{\Phi^\gamma}.\hskip 0.5cm\eeqa

\subsection{Improved second order- and third order perturbed solution}

Likewise, it is not difficult to obtain \beqa \label{ips2}
\ket{\Psi^{(2)}(t)}_{\rm
I}&=&-\sum_{\gamma,\gamma_1\gamma^\prime}\left\{\left[\frac{\e^{-\I
\left(E_\gamma+G^{(2)}_\gamma+G^{(3)}_\gamma\right) t}-\e^{-\I
\left(E_{\gamma_1}+G^{(2)}_{\gamma_1}+G^{(3)}_{\gamma_1}\right)
t}}{\left(E_{\gamma}-E_{\gamma_1}\right)^2}
g_1^{\gamma\gamma_1}g_1^{\gamma_1\gamma}\delta_{\gamma\gamma^\prime}\right.\right.\nonumber\\
& &+\left[\frac{\e^{-\I
\left(E_{\gamma}+G^{(2)}_{\gamma}+G^{(3)}_{\gamma}\right)
t}}{\left(E_{\gamma}-E_{\gamma_1}\right)\left(E_{\gamma}-E_{\gamma^\prime}\right)}-\frac{\e^{-\I
\left(E_{\gamma_1}+G^{(2)}_{\gamma_1}+G^{(3)}_{\gamma_1}\right)
t}}{\left(E_{\gamma}-E_{\gamma_1}\right)\left(E_{\gamma_1}-E_{\gamma^\prime}\right)}\right.\nonumber\\
& &\left.\left. +\frac{\e^{-\I
\left(E_{\gamma^\prime}+G^{(2)}_{\gamma^\prime}+G^{(3)}_{\gamma^\prime}\right)
t}}{\left(E_{\gamma}-E_{\gamma^\prime}\right)\left(E_{\gamma_1}-E_{\gamma^\prime}\right)}\right]
g_1^{\gamma\gamma_1}g_1^{\gamma_1\gamma^\prime}\eta_{\gamma\gamma^\prime}\right\}
a_{\gamma^\prime}\ket{\Phi^\gamma}.\eeqa
\beqa \label{ips3} \ket{\Psi^{(3)}(t)}_{\rm
I}&=&\sum_{\gamma,\gamma_1,\gamma_2,\gamma^\prime}\left\{\left[-\frac{\e^{-\I
\left(E_\gamma+G^{(2)}_\gamma\right)
t}}{\left(E_{\gamma}-E_{\gamma_1}\right)\left(E_{\gamma}-E_{\gamma_2}\right)^2}-\frac{\e^{-\I
\left(E_\gamma+G^{(2)}_\gamma\right)
t}}{\left(E_{\gamma}-E_{\gamma_1}\right)^2\left(E_{\gamma}-E_{\gamma_2}\right)}\right.\right.\nonumber\\
& &\left. \left.+\frac{\e^{-\I
\left(E_{\gamma_1}+G^{(2)}_{\gamma_1}\right)
t}}{\left(E_{\gamma}-E_{\gamma_1}\right)^2\left(E_{\gamma_1}-E_{\gamma_2}\right)}-\frac{\e^{-\I
\left(E_{\gamma_2}+G^{(2)}_{\gamma_2}\right)
t}}{\left(E_{\gamma}-E_{\gamma_2}\right)^2\left(E_{\gamma_1}-E_{\gamma_2}\right)}\right]
g_1^{\gamma\gamma_1}g_1^{\gamma_1\gamma_2}g_1^{\gamma_2\gamma}\delta_{\gamma\gamma^\prime}\right\}
a_{\gamma^\prime}\ket{\Phi^\gamma}\nonumber\\
&
&+\sum_{\gamma,\gamma^\prime}\left\{-\sum_{\gamma_1}\left[\frac{\e^{-\I
\left(E_\gamma+G^{(2)}_\gamma\right)
t}}{\left(E_{\gamma}-E_{\gamma_1}\right)\left(E_{\gamma}-E_{\gamma^\prime}\right)^2}+\frac{\e^{-\I
\left(E_\gamma+G^{(2)}_\gamma\right)
t}}{\left(E_{\gamma}-E_{\gamma_1}\right)^2\left(E_{\gamma}-E_{\gamma^\prime}\right)}\right]
g_1^{\gamma\gamma_1}g_1^{\gamma_1\gamma}g_1^{\gamma\gamma^\prime}\right.\nonumber\\
& &+\sum_{\gamma_1,\gamma_2}\left[\frac{\e^{-\I
\left(E_{\gamma}+G^{(2)}_{\gamma}\right)
t}\eta_{\gamma\gamma_2}}{\left(E_{\gamma}-E_{\gamma_1}\right)\left(E_{\gamma}-E_{\gamma_2}\right)
\left(E_{\gamma}-E_{\gamma^\prime}\right)}-\frac{\e^{-\I
\left(E_{\gamma_1}+G^{(2)}_{\gamma_1}\right)
t}\eta_{\gamma_1\gamma^\prime}}{\left(E_{\gamma}-E_{\gamma_1}\right)\left(E_{\gamma_1}-E_{\gamma_2}\right)
\left(E_{\gamma_1}-E_{\gamma^\prime}\right)}\right.\nonumber\\
& &\left.\left. +\frac{\e^{-\I
\left(E_{\gamma_2}+G^{(2)}_{\gamma_2}\right)
t}\eta_{\gamma\gamma_2}}{\left(E_{\gamma}-E_{\gamma_2}\right)\left(E_{\gamma_1}-E_{\gamma_2}\right)
\left(E_{\gamma_2}-E_{\gamma^\prime}\right)}\right]
g_1^{\gamma\gamma_1}g_1^{\gamma_1\gamma_2}g_1^{\gamma_2\gamma^\prime}\eta_{\gamma\gamma^\prime}\right\}
a_{\gamma^\prime}\ket{\Phi^\gamma}.\eeqa

\subsection{Summary}

Obviously, our improved form of perturbed solution of dynamics in
the third order approximation is \beqa
\ket{\Psi(t)}&=&\sum_{i=0}^3\ket{\Psi^{(i)}(t)}_{\rm
I}+\mathcal{O}(H_1^4).\eeqa However, this solution including the
contributions from the whole $A_l^{\gamma\gamma^\prime}(t\e)$,
$A_l^{\gamma\gamma^\prime}(t^2\e)$ parts up to the fifth order
approximation and the whole $A_l^{\gamma\gamma^\prime}(t^2\e)$,
$A_l^{\gamma\gamma^\prime}(t^3\e)$ parts in the sixth order
approximation. After considering the contractions and
anti-contractions, we see the result corresponds to the replacement
\beq \e^{-\I E_{\gamma_i} t}\rightarrow \e^{-\I
\widetilde{E}_{\gamma_i} t}, \eeq in the
$A_l^{\gamma\gamma^\prime}(\e)$ part, where \beq
\widetilde{E}_{\gamma_i}={E}_{\gamma_i}+\sum_{a=2}G_{\gamma_i}^{(a)},
\eeq $i=0,1,2,\cdots$, and $\gamma_0=\gamma$. Although the upper
bound of summation index $a$ is different from the approximation
order in the finished calculations, we can conjecture it may be
taken to at least 5 based on the consideration from the physical
concept and mathematical symmetry. For $a\geq 5$, their forms should
be similar. From our point of view, such form is so delicate that
its form happens impossibly by accident. Perhaps, there is a
fundamental formula within it. Nevertheless, we have no idea of how
to prove it strictly and generally at present.

Actually, as soon as we carry out further calculations, we can
include the contributions from higher order approximations.
Moreover, these calculations are not difficult and are programmable
since we only need to calculate the limitation and summation.
Therefore, the advantages of our solution have been made clear in
our improved forms of perturbed solution of dynamics above. In other
words, they offer clear evidences to show it is better than the
existing method in the precision and efficiency. In the following
several sections, we will clearly demonstrate these problems.

\section{Improved transition probability and revised Fermi's golden rule}\label{sec8}

One of the interesting applications of our solution is the
calculation of transition probability in a general time-independent
quantum system. It improves the well-known conclusion because our
solution include the contributions from the high order
approximation. Moreover, in terms of our improved forms of perturbed
solution, it is easy to obtain the high order transition
probability. In addition, for the case of sudden perturbation, our
scheme is also suitable. Since considering time-independent system,
it is not necessary to using the interaction picture.

Let us start with the following perturbed expansion of state
evolution with time $t$,
 \beq
\ket{\Psi(t)}=\sum_{\gamma}c_\gamma(t)\ket{\Phi^\gamma}=\sum_{l=0}^\infty
\sum_{\gamma}c_\gamma^{(n)}(t)\ket{\Phi^\gamma}.\eeq When we take
the initial state as $\ket{\Phi^\beta}$, from our improved first
order perturbed solution, we immediately obtain \beq c_{\gamma,{\rm
I}}^{(1)}=\frac{\e^{-\I \widetilde{E}_\gamma t}-\e^{-\I
\widetilde{E}_{\beta}t}}{E_{\gamma}-E_{\beta}}
g_1^{\gamma\beta},\eeq where \beq
\widetilde{E}_{\gamma_i}=E_{\gamma_i}+G_{\gamma_i}^{(2)}
+G_{\gamma_i}^{(3)}+G_{\gamma_i}^{(4)}.\eeq Here, we use the
subscript ``I" for distinguishing it and the usual result. Omitting
a unimportant phase factor $\e^{-\I \widetilde{E}_{\gamma}t}$, we
can rewrite it as \beq c_{\gamma,{\rm I}}^{(1)}=\frac{
g_1^{\gamma\beta}}{E_{\gamma}-E_{\beta}}
\left(1-\e^{\I\widetilde{\omega}_{\gamma\beta}t}\right),\eeq where
\beq\widetilde{\omega}_{\gamma\beta}=\widetilde{E}_{\gamma}-\widetilde{E}_{\beta}.
\eeq Obviously it is different from the well known conclusion \beq
c_{\gamma}^{(1)}=\frac{ g_1^{\gamma\beta}}{E_{\gamma}-E_{\beta}}
\left(1-\e^{\I {\omega}_{\gamma\beta}t}\right),\eeq where \beq
\omega_{\gamma\beta}=E_{\gamma}-E_{\beta}.\eeq Therefore, our result
contains the partial contributions from the high order
approximation.

Considering the transition probability from $\ket{\Phi^\beta}$ to
$\ket{\Phi^\gamma}$ after time $T$, we have \beq P_{\rm
I}^{\gamma\beta}(t)=\frac{\left|g_1^{\gamma\beta}\right|^2}{\omega_{\gamma\beta}^2}
\left|1-\e^{\I\widetilde{\omega}_{\gamma\beta}T}\right|^2
=\left|g_1^{\gamma\beta}\right|^2\frac{\sin^2\left(\widetilde{\omega}_{\gamma\beta}T/2\right)}
{\left({\omega}_{\gamma\beta}/2\right)^2}.\eeq

In terms of the relation \beq
\sin^2x-\sin^2y=\frac{1}{2}\left[\cos(2y)-\cos(2x)\right],\eeq we
have the revision part of transition probability \beq
\vartriangle\!\!P_{\rm I}^{\gamma\beta}(t)
=2\left|g_1^{\gamma\beta}\right|^2\frac{\cos\left({\omega}_{\gamma\beta}T\right)
-\cos\left(\widetilde{\omega}_{\gamma\beta}T\right)}
{\left({\omega}_{\gamma\beta}\right)^2}.\eeq

If plotting \beq
\frac{\sin^2\left(\widetilde{\omega}_{\gamma\beta}T/2\right)}{\left({\omega}_{\gamma\beta}/2\right)^2}=
\left(\frac{\widetilde{\omega}_{\gamma\beta}}{{\omega}_{\gamma\beta}}\right)^2
\frac{\sin^2\left(\widetilde{\omega}_{\gamma\beta}T/2\right)}
{\left(\widetilde{\omega}_{\gamma\beta}/2\right)^2},\eeq we can see
that it has a well-defined peak centered at
$\widetilde{\omega}_{\gamma\beta}=0$. Just as what has been done in
the usual case, we can extend the integral range as
$-\infty\rightarrow\infty$. Thus, the revised Fermi's golden rule
\beq w=w_{\rm F}+\vartriangle\!\!w,\eeq where the usual Fermi's
golden rule is \cite{Fermi} \beq w_{\rm F}=2\pi\rho(E_\beta)
\left|g_1^{\gamma\beta}\right|^2,\eeq in which $w$ means the
transition velocity, $\rho(E_\gamma)$ is the density of final state
and we have used the integral formula \beq
\int_{-\infty}^\infty\frac{\sin^2x}{x^2}=\pi. \eeq while the
revision part is \beq \vartriangle\!\!w=2\int_{-\infty}^\infty \d
E_\gamma \rho\left(E_{\gamma}\right)\left|g_1^{\gamma\beta}\right|^2
\frac{\cos\left({\omega}_{\gamma\beta}T\right)
-\cos\left(\widetilde{\omega}_{\gamma\beta}T\right)}
{T\left({\omega}_{\gamma\beta}\right)^2}.\eeq It is clear that $
\widetilde{\omega}_{\gamma\beta}$ is a function of $E_\gamma$, and
then a function of $\omega_{\gamma\beta}$. For simplicity, we only
consider $\widetilde{\omega}_{\gamma\beta}$ to its second order
approximation, that is \beq
\widetilde{\omega}_{\gamma\beta}=\widetilde{\omega}\left(\omega_{\gamma\beta}\right)
={\omega}_{\gamma\beta}+\sum_{\gamma_1}\left[
\frac{\left|g_1^{\gamma\gamma_1}\right|^2}{{\omega}_{\gamma\beta}-{\omega}_{\gamma_1\beta}}
-\frac{\left|g_1^{\beta\gamma_1}\right|^2}{{\omega}_{\beta\gamma_1}}\right]+\mathcal{O}(H_1^3).\eeq
Again based on $\d E_\gamma=\d{\omega}_{\gamma\beta}$, we have \beq
\vartriangle\!\!w=2\int_{-\infty}^\infty \d \omega_{\gamma\beta}
\rho\left(\omega_{\gamma\beta}+E_{\beta}\right)\left|g_1^{\gamma\beta}\right|^2
\frac{\cos\left[{\omega}_{\gamma\beta}T\right]
-\cos\left[\widetilde{\omega}\left(\omega_{\gamma\beta}\right)T\right]}
{T\left({\omega}_{\gamma\beta}\right)^2}.\eeq It seems it is not
easy to deduce the general form of this integral. In order to
simplify it, we can use the fact that
$\widetilde{\omega}_{\gamma\beta}-{\omega}_{\gamma\beta}$ is a
smaller quantity since \beq
\vartriangle\!\!\omega_{\gamma\beta}=\widetilde{\omega}_{\gamma\beta}-{\omega}_{\gamma\beta}
=\sum_{i=2}^4 \left(G_{\gamma}^{(i)}-G_{\beta}^{(i)}\right).\eeq For
example, we can approximatively take \beq
\cos\left({\omega}_{\gamma\beta}T\right)
-\cos\left(\widetilde{\omega}_{\gamma\beta}T\right)\approx
T\left(\widetilde{\omega}_{\gamma\beta}-{\omega}_{\gamma\beta}\right)
\sin\left(\widetilde{\omega}_{\gamma\beta}T-{\omega}_{\gamma\beta}T\right),\eeq
then calculate the integral. We will study it in our other
manuscript (in preparing).

Obviously, the revision comes from the contributions of high order
approximations. The physical effect resulted from our solution,
whether is important or unimportant, should be investigated in some
concrete quantum systems. We will discuss them in our future
manuscripts (in preparing).

It is clear that the relevant results can be obtained from the usual
results via replacing $\omega_{\gamma\beta}$ in the exponential
power by using $\widetilde{\omega}_{\gamma\beta}$. To save the
space, we do not intend to mention them here.

In fact, there is not any difficult to obtain the second- and three
order transition probability in terms of our improved forms of
perturbed solution in the previous section. More higher order
transition probability can be given effectively and accurately by
our scheme.

\section{Improved forms of perturbed energy and perturbed state}\label{sec9}

Now we study how to calculate the improved forms of perturbed energy
and perturbed state. For simplicity, we only study them concerning
the improved second order approximation. Based on the experience
from the skill one in Sec. \ref{sec6}, we can, in fact, set a new
$\widetilde{E}$ and then use the technology in the usual
perturbative theory. That is, we denote \beq
\widetilde{E}_{\gamma_i}=E_{\gamma_i}+G_{\gamma_i}^{(2)}
+G_{\gamma_i}^{(3)}.\eeq
\beqa \label{ipsto2o}
\ket{\Psi(t)}&=&\sum_{\gamma,\gamma^\prime}\left\{\e^{-\I\widetilde{E}_\gamma
t}\delta_{\gamma\gamma^\prime}+\left[\frac{\e^{-\I
\widetilde{E}_\gamma t}-\e^{-\I
\widetilde{E}_{\gamma^\prime}t}}{E_{\gamma}-E_{\gamma^\prime}}\right]
g_1^{\gamma\gamma^\prime}-\sum_{\gamma_1}\frac{\e^{-\I
\widetilde{E}_\gamma t}-\e^{-\I
\widetilde{E}_{\gamma_1}t}}{\left(E_{\gamma}-E_{\gamma_1}\right)^2}
g_1^{\gamma\gamma_1}g_1^{\gamma_1\gamma}\delta_{\gamma\gamma^\prime}
\right.\nonumber\\
& & +\sum_{\gamma_1}\left[\frac{\e^{-\I
\widetilde{E}_{\gamma}t}}{\left(E_{\gamma}-E_{\gamma_1}\right)
\left(E_{\gamma}-E_{\gamma^\prime}\right)}-\frac{\e^{-\I
\widetilde{E}_{\gamma_1}t}}{\left(E_{\gamma}-E_{\gamma_1}\right)
\left(E_{\gamma_1}-E_{\gamma^\prime}\right)}\right.\nonumber\\
& &\left.\left. +\frac{\e^{-\I
\widetilde{E}_{\gamma^\prime}t}}{\left(E_{\gamma}-E_{\gamma^\prime}\right)
\left(E_{\gamma_1}-E_{\gamma^\prime}\right)}\right]
g_1^{\gamma\gamma_1}g_1^{\gamma_1\gamma}\eta_{\gamma\gamma^\prime}\right\}
a_{\gamma^\prime}\ket{\Phi^\gamma}+\mathcal{O}(H_1^3).\eeqa Because
that \beq \ket{\Psi(t)}_{\rm I}=\sum_{\gamma,\gamma^\prime}\e^{-\I
{E}_T
t}\delta_{\gamma\gamma^\prime}a_{\gamma^\prime}\ket{\Phi^\gamma},\eeq
we have \beqa\label{ipee} E_T
a_{\gamma}&=&\widetilde{E}_{\gamma}a_\gamma+\sum_{\gamma^\prime}\left\{\frac{
\widetilde{E}_\gamma-\widetilde{E}_{\gamma^\prime}}{E_{\gamma}-E_{\gamma^\prime}}
g_1^{\gamma\gamma^\prime}-\sum_{\gamma_1}\frac{\widetilde{E}_\gamma-\widetilde{E}_{\gamma_1}}
{\left(E_{\gamma}-E_{\gamma_1}\right)^2}
g_1^{\gamma\gamma_1}g_1^{\gamma_1\gamma}\delta_{\gamma\gamma^\prime}
\right.\nonumber\\
& &+\sum_{\gamma_1}\left[\frac{
\widetilde{E}_{\gamma}}{\left(E_{\gamma}-E_{\gamma_1}\right)
\left(E_{\gamma}-E_{\gamma^\prime}\right)}-\frac{
\widetilde{E}_{\gamma_1}}{\left(E_{\gamma}-E_{\gamma_1}\right)
\left(E_{\gamma_1}-E_{\gamma^\prime}\right)}\right.\nonumber\\
& &\left.\left.
+\frac{\widetilde{E}_{\gamma^\prime}}{\left(E_{\gamma}-E_{\gamma^\prime}\right)
\left(E_{\gamma_1}-E_{\gamma^\prime}\right)}\right]
g_1^{\gamma\gamma_1}g_1^{\gamma_1\gamma}\eta_{\gamma\gamma^\prime}\right\}
a_{\gamma^\prime}.\eeqa

In the usual perturbation theory, $H_1$ is taken as a perturbed part
with the form \beq H_1=\lambda \widetilde{H}_1,\eeq where $\lambda$
is a real number that is called the perturbation parameter. It must
be emphasized that $\widetilde{E}_{\gamma_i}$ can be taken as
explicitly independent perturbed parameter $\lambda$, because we
introduce $\lambda$ as a formal multiplier after redefinition. This
way has been seen in our skill one. Without loss of generality, we
further take $H_1$ only with the nondiagonal form, that is \beq
H_1^{\gamma_1\gamma_2}={g}_1^{\gamma_1\gamma_2}=\lambda
\widetilde{g}_1^{\gamma_1\gamma_2}.\eeq Then, we expand both the
desired expansion coefficients $a_\gamma$ and the energy eigenvalues
$E_T$ in a power series of perturbation parameter $\lambda$: \beqa
E_T&=&\sum_{l=0}^\infty \lambda^l E_{T,{\rm I}}^{(l)},\\
a_\gamma&=&\sum_{l=0}^\infty \lambda^l a_{\gamma;{\rm
I}}^{(l)}.\eeqa

\subsubsection{Improved 0th approximation}

If we set $\lambda=0$, eq.(\ref{ipee}) yields \beq E_{T,{\rm
I}}^{(0)} a_{\gamma;{\rm I}}^{(0)}=\widetilde{E}_\gamma
a_{\gamma;{\rm I}}^{(0)},\eeq where $\gamma$ runs over all levels.
Actually, let us focus on the level $\gamma=\beta$, then \beq
\label{0the} E_{T,{\rm I}}^{(0)}=\widetilde{E}_\beta .\eeq When the
initial state is taken as $\ket{\Phi^\beta}$, \beq \label{0ths}
a_{\gamma;{\rm I}}^{(0)}=\delta_{\gamma\beta}.\eeq Obviously, the
improved form of perturbed energy is different from the results in
the usual perturbative theory because it includes the contributions
from the higher order approximation. However, the so-call improved
form of perturbed state is the same as the usual result.

\subsubsection{Improved 1st approximation}

Again from eq.(\ref{ipee}) it follows that \beq E_{T,{\rm I}}^{(0)}
a_{\gamma;{\rm I}}^{(1)} +E_{T,{\rm I}}^{(1)}a_{\gamma;{\rm
I}}^{(0)} =\widetilde{E}_\gamma a_{\gamma;{\rm I}}^{(1)}
+\sum_{\gamma^\prime}\frac{\widetilde{E}_\gamma-\widetilde{E}_{\gamma^\prime}}{E_\gamma-E_{\gamma^\prime}}
\widetilde{g}_1^{\gamma\gamma^\prime}a_{\gamma^\prime;{\rm
I}}^{(0)}.\eeq When $\gamma=\beta$, it is easy to obtain \beq
E_{T,{\rm I}}^{(1)}=0. \eeq If $\gamma\neq\beta$, then \beq
a_{\gamma;{\rm I}}^{(1)}=-\frac{1}
{\left(E_\gamma-E_{\gamma^\beta}\right)}\widetilde{g}_1^{\gamma\beta}.\eeq
It is clear that the first order results are the same as the one in
the usual perturbative theory.

\subsubsection{Improved 2nd approximation}

Likewise, the following equation \beqa & & E_{T,{\rm I}}^{(2)}
a_{\gamma;{\rm I}}^{(0)}+ E_{T,{\rm
I}}^{(1)}a_{\gamma;I}^{(1)}+E_{T,{\rm I}}^{(0)}a_{\gamma;{\rm
I}}^{(2)} =\widetilde{E}_\gamma a_{\gamma;{\rm I}}^{(2)}
+\sum_{\gamma^\prime}\frac{\widetilde{E}_\gamma-\widetilde{E}_{\gamma^\prime}}{E_\gamma-E_{\gamma^\prime}}
\widetilde{g}_1^{\gamma\gamma^\prime}a_{\gamma^\prime;{\rm I}}^{(1)}\nonumber\\
& & \quad
-\sum_{\gamma_1,\gamma^\prime}\frac{\widetilde{E}_\gamma-\widetilde{E}_{\gamma_1}}
{\left(E_{\gamma}-E_{\gamma_1}\right)^2}
g_1^{\gamma\gamma_1}g_1^{\gamma_1\gamma}\delta_{\gamma\gamma^\prime}a_{\gamma^\prime;{\rm
I}}^{(0)} +\sum_{\gamma_1,\gamma^\prime}\left[\frac{
\widetilde{E}_{\gamma}}{\left(E_{\gamma}-E_{\gamma_1}\right)
\left(E_{\gamma}-E_{\gamma^\prime}\right)}\right.\nonumber\\
& &\quad \left. -\frac{
\widetilde{E}_{\gamma_1}}{\left(E_{\gamma}-E_{\gamma_1}\right)
\left(E_{\gamma_1}-E_{\gamma^\prime}\right)}
+\frac{\widetilde{E}_{\gamma^\prime}}{\left(E_{\gamma}-E_{\gamma^\prime}\right)
\left(E_{\gamma_1}-E_{\gamma^\prime}\right)}\right]
g_1^{\gamma\gamma_1}g_1^{\gamma_1\gamma}\eta_{\gamma\gamma^\prime}
a_{\gamma^\prime;{\rm I}}^{(0)}. \eeqa is obtained and it yields
\beq E_{T;{\rm I}}^{(2)}=0,\eeq if we take $\gamma=\beta$. When
$\gamma\neq\beta$, we have \beqa a_{\gamma;{\rm
I}}^{(2)}&=&\sum_{\gamma_1}\frac{1}
{\left(E_{\gamma}-E_{\beta}\right)
\left(E_{\gamma_1}-E_{\beta}\right)}
\widetilde{g}_1^{\gamma\gamma_1}\widetilde{g}_1^{\gamma_1\beta}\eta_{\gamma\beta}.\eeqa
It is consistent with the nondiagonal part of usual result. In fact,
since we have taken $H_1^{\gamma\gamma^\prime}$ to be nondiagonal,
it does not have a diagonal part. However, we think its form is more
appropriate. In addition, we do not consider the revision part
introduced by normalization. While $E_{T;{\rm I}}^{(2)}=0$ is a new
result.

\subsubsection{Summary}

Now we can see, up to the improved second order approximation: \beq
E_{T,\beta}\approx \widetilde{E}_\beta=E_{\beta}+G_{\beta}^{(2)}
+G_{\beta}^{(3)}.\eeq Comparing with the usual, they are consistent
at the former two orders. It is not strange since the physical law
is the same. However, our improved form of perturbed energy contains
a third order term. In other words, our solution might be effective
in order to obtain the contribution from high order approximation.
The possible physical reason is that a redefined form of the
solution is obtained.

In special, when we allow the $H_1^{\gamma\gamma^\prime}$ to have
the diagonal elements, the improved second order approximation
becomes \beq E_{T,\beta}\approx
E_{\beta}+h_1^{\beta}+G_{\beta}^{(2)} +G_{\beta}^{(3)}.\eeq

Likewise, if we redefine \beq
\widetilde{E}_{\gamma_i}=E_{\gamma_i}+G_{\gamma_i}^{(2)}
+G_{\gamma_i}^{(3)}+G_{\gamma_i}^{(4)}.\eeq Thus, only consider the
first order approximation, we can obtain \beq E_{T,\beta}\approx
E_{\beta}+h_1^{\beta}+G_{\beta}^{(2)}
+G_{\beta}^{(3)}+G_{\beta}^{(4)}. \eeq In fact, the reason is our
conjecture in the previous section. The correct form of redefined
$\widetilde{E}_{\gamma_i}$ should be \beq E_{T,\beta}\approx
E_{\beta}+h_1^{\beta}+G_{\beta}^{(2)}
+G_{\beta}^{(3)}+G_{\beta}^{(4)} +G_{\beta}^{(5)}+\cdots. \eeq This
implies that our improved scheme includes the partial even whole
significant contributions from the high order approximations. In
addition, based on the fact that the improved second approximation
is actually zero, it is possible that this implies our solution will
fade down more rapidly than the solution in the usual perturbative
theory.

Actually, the main advantage of our solution is in dynamical
development. The contributions from the high order approximation
play more important roles in the relevant physical problems such as
the entanglement dynamics and decoherence process. For the improved
perturbed energy, its high order part has obvious physical meaning.
But, for the improved form of perturbed state, we find them to be
the same.

\section{Example and application}\label{sec10}

In order to concretely illustrate that our exact solution and
perturbative scheme are indeed more effective and more accurate, let
us study an elementary example: two state system, which appears in
the most of quantum mechanics textbooks. Its Hamiltonian can be
written as\beq H=\left(\begin{array}{cc}
E_1&V_{12}\\
V_{21}& E_2\end{array}\right),\eeq where we have used the the basis
formed by the unperturbed energy eigenvectors, that is \beq
\ket{\Phi^1}=\left(\begin{array}{c}1\\0\end{array}\right),\quad
\ket{\Phi^2}=\left(\begin{array}{c}0\\1\end{array}\right). \eeq In
other words: \beq H_0\ket{\Phi^\gamma}=E_\gamma\ket{\Phi^\gamma},
\quad (\gamma=1,2)\eeq where \beq H_0=\left(\begin{array}{cc}
E_1&0\\
0& E_2\end{array}\right).\eeq Thus, this means the perturbed
Hamiltonian is taken as \beq H_1=\left(\begin{array}{cc}
0& V_{12}\\
V_{21}& 0\end{array}\right).\eeq This two state system has the
following eigen equation \beq
H\ket{\Psi^\gamma}=E^T_{\gamma}\ket{\Psi^\gamma}. \eeq It is easy to
obtain its solution: corresponding eigenvectors and eigenvalues
\beqa
\ket{\Psi^1}&=&\frac{1}{\sqrt{4|V|^2+(\omega_{21}+{\omega}_{21}^T)^2}}
\left(\begin{array}{c}\omega_{21}+{\omega}_{21}^T\\-2V_{21}\end{array}\right),\\
\ket{\Psi^2}&=&\frac{1}{\sqrt{4|V|^2+(\omega_{21}-{\omega}_{21}^T)^2}}
\left(\begin{array}{c}\omega_{21}-{\omega}_{21}^T\\-2V_{21}\end{array}\right);
\eeqa
\beqa E_1^T&=&\frac{1}{2}\left(E_1+E_2-{\omega}_{21}^T\right),\\
E_2^T&=&\frac{1}{2}\left(E_1+E_2+{\omega}_{21}^T\right); \eeqa where
$|V|=|V_{12}|=|V_{21}|$, $\omega_{21}=E_2-E_1$,
${\omega}_{21}^T=E_2^T-E_1^T=\sqrt{4|V|^2+\omega_{21}^2}$, and we
have set $E_2> E_1$ without loss of generality.

Obviously the transition probability from state 1 to state 2 is
\beqa P^T(1\rightarrow 2)&=&\left|\bra{\Phi^2}\e^{-\I H
t}\ket{\Phi^1}\right|^2=\left|\sum_{\gamma_1,\gamma_2=1}^2\diracsp{\Phi^2}{\Psi^{\gamma_1}}
\bra{\Psi^{\gamma_1}}\e^{-\I H
t}\ket{\Psi^{\gamma_2}}\diracsp{\Psi^{\gamma_2}}{\Phi^1}\right|^2\nonumber\\
&=&|V|^2\frac{\sin^2\left({\omega}_{21}^T
t/2\right)}{(\omega_{21}^T/2)^2}. \eeqa

In the usual perturbation theory, up to the second order
approximations, the well-known the perturbed energies are \beqa
E_1^P&=&E_1-\frac{|V|^2}{\omega_{21}},\\
E_2^P&=&E_1+\frac{|V|^2}{\omega_{21}}. \eeqa While, under the first
order approximation, the transition probability from state 1 to
state 2 is \beq P(1\rightarrow
2)=|V|^2\frac{\sin^2\left(\omega_{21}t/2\right)}{(\omega_{21}/2)^2}.
\eeq

Using our scheme, only to the first approximation, the corresponding
the perturbed energies are \beqa
\widetilde{E}_1&=&E_1-\frac{|V|^2}{\omega_{21}}+\frac{|V|^4}{\omega_{21}^3},\\
\widetilde{E}_2&=&E_1+\frac{|V|^2}{\omega_{21}}-\frac{|V|^4}{\omega_{21}^3},
\eeqa where we have used the facts that \beqa
G_1^{(2)}&=&-\frac{|V|^2}{\omega_{21}}=-G_2^{(2)},\\
G_1^{(3)}&=&G_2^{(3)}=0,\\
G_1^{(4)}&=&\frac{|V|^4}{\omega_{21}^3}=-G_2^{(4)}. \eeqa Obviously,
under the first order approximation, our scheme yields the
transition probability from state 1 to state 2 as \beq P_{\rm
I}(1\rightarrow
2)=|V|^2\frac{\sin^2\left(\widetilde{\omega}_{21}t/2\right)}{(\omega_{21}/2)^2}.
\eeq where
$\widetilde{\omega}_{21}=\widetilde{E}_2-\widetilde{E}_1$. Therefore
we can say our scheme is more effective. Moreover, we notice that
\beqa
{E}_1^T&=&E_1-\frac{|V|^2}{\omega_{21}}+\frac{|V|^4}{\omega_{21}^3}
+\mathcal{O}(|V|^6)\\
&=&\widetilde{E}_1+\mathcal{O}(|V|^6)\\
&=&{E}_1^P+\frac{|V|^4}{\omega_{21}^3}+\mathcal{O}(|V|^6),\\
\widetilde{E}_2&=&E_1+\frac{|V|^2}{\omega_{21}}
-\frac{|V|^4}{\omega_{21}^3}+\mathcal{O}(|V|^6)\\
&=&\widetilde{E}_2+\mathcal{O}(|V|^6)\\
&=& {E}_2^P-\frac{|V|^4}{\omega_{21}^3}+\mathcal{O}(|V|^6).
 \eeqa and \beqa
P^T(1\rightarrow
2)&=&|V|^2\frac{\sin^2\left(\omega_{21}t/2\right)}{(\omega_{21}/2)^2}
+|V|^2\left[\frac{\sin\left(
\omega_{21}t\right)}{2(\omega_{21}/2)^2}-\frac{\sin^2\left(
\omega_{21}t/2\right)}{(\omega_{21}/2)^3}\right]
\left(\widetilde{\omega}_{21}-{\omega}_{21}\right)
+\mathcal{O}[\left(\widetilde{\omega}_{21}-{\omega}_{21}\right)^2]\\
&=& P_{\rm I}(1\rightarrow 2)-|V|^2\frac{\sin^2\left(
\omega_{21}t/2\right)}{(\omega_{21}/2)^3}
\left(\widetilde{\omega}_{21}-{\omega}_{21}\right)
+\mathcal{O}[\left(\widetilde{\omega}_{21}-{\omega}_{21}\right)^2]\\
&=& P(1\rightarrow 2)+|V|^2\left[\frac{\sin\left(
\omega_{21}t\right)}{2(\omega_{21}/2)^2}-\frac{\sin^2\left(
\omega_{21}t/2\right)}{(\omega_{21}/2)^3}\right]
\left(\widetilde{\omega}_{21}-{\omega}_{21}\right)
+\mathcal{O}[\left(\widetilde{\omega}_{21}-{\omega}_{21}\right)^2].\eeqa
Therefore, we can say that our scheme is more accurate.

\section{Conclusion and discussion}\label{sec11}

In the end, we would like to point out that our general and explicit
solution (\ref{ouress}) or its particular forms
(\ref{ffs1},\ref{ffs2}) of the Schr\"{o}dinger equation in a general
time-independent quantum system is clear and exact in form in spite
of it is an infinity series. It must be emphasized that its
distinguished feature is to first express the exact solution in a
general time-independent quantum system by using the $c$-number
matrix elements rather than an operator form. We can say that our
exact solution is more explicit than the usual nonperturbed solution
of the Schr\"{o}dinger equation. Moreover, our deducing methods give
up some preconditions used in the usual scheme. So we can say our
solution is more general. Just as its explicit and general form, our
exact solution not only has the mathematical delicateness, but also
can contain more physical content, obtain the more efficiency and
higher precision and  result in new applications.

We obtain our exact solution based on our expansion formula of power
of operator binomials and the matrix representation of time
evolution operator, which is a new way to study the quantum dynamics
and the perturbation theory. Its idea is different from the usual
one. At our knowledge, this formula is first proposed and strictly
proved. Besides its theoretical value in mathematics, we are sure it
is interesting and important for expressing some useful operator
formula in quantum physics.

In the process of deducing our exact solution, we prove an identity
of fraction function. It should be interesting in mathematics.
Perhaps, it has other applications to be expected finding.

By virtu of the idea of quantum field theory, we build the concepts
of contraction and anti-contraction using in the limitation
computation. Moreover we develop two skills to deal with an infinity
series, in special, so-called dynamical rearrangement and summation.
These methods have played important roles in our scheme. We believe
they can be useful technologies in mathematical and physical
calculation.

As an example, we give out the concrete form of solution when the
solvable part of Hamiltonian is taken as the kinetic energy term in
a general quantum system. Its aim is to account for the fact that
our solution perhaps extends the applicable range of perturbation
theory and lose its preconditions.

We can see that in our solution, every expanding coefficient
(amplitude) before the basis vector $\ket{\Phi^\gamma}$ has some
closed time evolution factors with exponential form, and includes
all order approximations so that we can clearly understand the
dynamical behavior of quantum systems. Although our solution is
obtained in $H_0$ representation, its form in the other
representation can be given by the representation transformation and
the development factors with time $t$ in the expanding coefficients
(amplitude) do not change.

Because we obtain the general term form of time evolution of quantum
state, it provides the probability considering the partial
contributions from the high order even all of order approximations.
After developing the contraction and anti-contraction skills and
dynamical rearrangement and summation technology, we realize this
probability by present a perturbative scheme. From our exact
solution transferring to our perturbative scheme is physically
reasonable and mathematically clear. This provides the guarantee
achieving high efficiency and high precision. Through finding the
improved forms of perturbed solutions of dynamics, we generally
demonstrate this conclusion. Furthermore, we prove the correctness
of this conclusion via calculating the improved form of transition
probability, perturbed energy and perturbed state. Specially, we
obtain the revised Fermi's golden rule. Moreover, we illustrate the
advantages of our exact solution and perturbative scheme in an easy
understanding example of two state system. All of this implies the
physical reasons and evidences why our exact solution and
perturbative scheme are formally explicit, actually calculable,
operationally efficient, conclusively more accurate.

From the features of our solution, we believe that it will have
interesting applications in the calculation of entanglement dynamics
and decoherence process as well as the other physical quantities
dependent on the expanding coefficients. Of course, our solution is
an exact one, its advantages and features can not be fully revealed
only via the perturbative method.

Undoubtedly, the formalization of physical theory often has its
highly mathematical focus, but this can not cover its real
motivations, potential applications and related conclusions in
physics. Our exact solution and perturbative scheme are just so. In
order to account for what is more in our solution and reveal the
relations and differences between our solution and the existing
method, we compare our solution with the usual perturbation theory.
We find their consistency and relations. In fact, our solution has
finished the task to calculate the expanding coefficients of final
state in $H_0$ representation and obtain the general term up to any
order using our own method. But the usual perturbation theory only
carries out this task from some given order approximation to the
next order approximation step by step. In a sentence, more explicit
and general feature of our exact solution can lead to more physical
conclusions and applications.

It is worth pointing out that our solution, different from the
time-dependent perturbation theory, presents the explicit solution
of recurrence equation of expanding coefficients of final state in
$H_0$ representation, but we pay for the price that $H$ is not
explicitly dependent on the time. In short, there is gain and there
is lose.

In fact, a given lower order approximation of improved perturbation
solution including the partial contributions from the higher order
even all of order approximations is obtained by rearranging and
summing a kind of terms, just like it has been done in the quantum
field theory for the particular contributions over a given type of
the Feynman's diagrams. It is emphasized that these contributions
have to be significant in physics. Considering time development form
is our physical idea and adding the high order approximations with
the factors $t^a\e^{-\I E_{\gamma_i}t}$, $(a=1,2,\cdots)$ to the
improved lower order approximations definitely can advance the
precision. Therefore, our dynamical rearrangement and summation
technology is appropriate and reasonable from our view.

For a concrete example, except for some technological and
calculational works, it needs the extensive physical background
knowledge to account for the significance of related results. That
is, since the difference of the related conclusions between our
solution and the usual perturbation theory is in high order
approximation parts, we have to study the revisions (difference) to
find out whether they are important or unimportant to the studied
problem. In addition, our conjecture about the perturbed energy is
based on physical concept and mathematical consideration, it is
still open at the strict sense. As to the degenerate case, except
for the complicated expressions, there is no more new idea.

Based on the above statements, our results can be thought of as
theoretical developments of quantum dynamics. The extension of our
solution to mixed states is direct, and the extension of our
solution to open systems see our manuscript \cite{MyDD}.

It must be emphasized that the study on the time evolution operator
plays a central role in quantum dynamics. From our point of view,
one of the most main results in our method is to obtain a general
term form of any order of $H_1$ (perturbative part of Hamiltonian)
for the time evolution operation in the representation of $H_0$
(solvable part of Hamiltonian), however, the usual perturbation
theory has not really finished it and only has an expression in the
operator form. Because the universal significance of our new
expression of time evolution operator, we wish that it will have
more applications in quantum theory. Besides the above studies
through the perturbative method, it is more interesting to apply our
exact solution to the formalization study of quantum dynamics in
order to further and more powerfully show the advantages of our
exact solution.

In summary, we can say our present results are helpful to understand
the theory of quantum dynamics and provide some powerful tools in
the calculation of quantum dynamics. More investigations are on
progressing.

\section*{Acknowledgments}

We are grateful all the collaborators of our quantum theory group in
the Institute for Theoretical Physics of our university. This work
was funded by the National Fundamental Research Program of China
under No. 2001CB309310, and partially supported by the National
Natural Science Foundation of China under Grant No. 60573008.

\begin{appendix}
\renewcommand{\theequation}{\thesection\arabic{equation}}

\section{The proof of our identity}

In this appendix, we would like to prove our identity (\ref{myi}).
For simplicity in notation and universality, we replace the
variables $E_{\gamma_i}$ by $x_i$ as well as $E[\gamma,l]$ by
$x[l]$. It is clear that the common denominator $D(x[n])$ in the
above expression (\ref{myi}) (the index $l$ is replaced by $n$)
reads \beq D(x[n])=\prod_{i=1}^n\left[\prod_{j=i}^n
\left(x_i-x_{j+1}\right)\right], \eeq while the $i$-th numerator is
\beq n_i(x[n],K)=\frac{D(x[n])}{d_i(x[n])}x_i^K, \eeq and the total
numerator $N(x[n],K)$ is \beq
N(x[n],K]=\sum_{i=1}^{n+1}n_i(x[n],K).\eeq In order to simplify our
notation, we denote $n_i(x[n])=n_i(x[n],0)$. Again, introducing a
new vector \beqa x^D_1[n]&=&\{x_2,x_3,\cdots,x_n,x_{n+1}\},\\
x^D_i[n]&=&\{x_1,\cdots,x_{i-1},x_{i+1},\cdots,x_{n+1}\},\\
x^D_{n+1}[n]&=&\{x_1,x_2,\cdots,x_{n-1},x_{n}\}. \eeqa Obviously,
$x^D_i(x[n])$ with $n$ components is obtained by deleting the $i$-th
component from $x[n]$. From the definition of $n_i(x[n])$, it
follows that \beq\label{nDr} n_i(x[n])=D(x^D_i[n]). \eeq

Without loss of generality, for an arbitrary given $i$, we always
can rewrite $x^D_i[n]=y[n-1]=\{y_1,y_2,\cdots,y_{n-1},y_n\}$ and
consider the general expression of $D(y[n-1])$. It is easy to verify
that \beqa \!\!\!\!\!\!\!D(y[1])\!\!\!&=&\!\!\!-y_2n_1(y[1])+y_1n_2(y[1]), \\
 \!\!\!\!\!\!\!D(y[2])\!\!\!&=&\!\!\!y_2y_3n_1(y[2])-
y_1y_3n_2(y[2])+y_1y_2n_3(y[2]). \eeqa Thus, by mathematical
induction, we first assume that for $n\geq 2$, \beq \label{Dform}
D(y[n-1])=\sum_{i=1}^n (-1)^{(i-1)+(n-1)}p_i(y[n-1])n_i(y[n-1]),\eeq
where we have defined \beq p_i(y[n-1])=\prod_{\stackrel{\scriptstyle
j=1}{j\neq i}}^{n} y_j.\eeq As above, we have verified that the
expression (\ref{Dform}) is valid for $n=2,3$. Then, we need to
prove that the following expression \beq \label{Dformp}
D(y[n])=\sum_{i=1}^{n+1} (-1)^{(i-1)+n}p_i(y[n])n_i(y[n])\eeq is
correct.

To our purpose, we start from the proof of a conclusion of the
precondition (\ref{Dform}) as the following: \beq \label{nsum}
\sum_{i=1}^{n+1}(-1)^{i-1}n_i(y[n])=0.\eeq According to the relation
(\ref{nDr}) and substituting the precondition (\ref{Dform}), we have
\beqa
& &\sum_{i=1}^{n+1}(-1)^{i-1}n_i(y[n])=\sum_{i=1}^{n+1}(-1)^{i-1}D(y_i^D[n])\nonumber\\
&=&\sum_{i=1}^{n+1}(-1)^{i-1}\left[\sum_{j=1}^{n}(-1)^{(j-1)+(n-1)}
p_j(y^D_i[n])n_j(y^D_i[n])\right]\nonumber \\
&=& (-1)^{n-1} \sum_{j=1}^n \sum_{i=1}^{n+1}(-1)^{i+j}
p_j(y^D_i[n])n_j(y^D_i[n])\label{nsum1}.\eeqa Based on the
definitions of $p_i(x[n])$ and $n_i(x[n])$, we find that \beq
p_j(y^D_i[n])=\left\{
\begin{array}{ll} p_{i-1}(y^D_j[n]) &\quad \mbox{(If $i>j$)}\\[8pt]
p_i(y^D_{j+1}[n])&\quad \mbox{(If $i\leq j$)}\end{array}\right.,\eeq
\beq n_j(y^D_i[n])=\left\{
\begin{array}{ll} n_{i-1}(y^D_j[n]) &\quad \mbox{(If $i>j$)}\\[8pt]
n_i(y^D_{j+1}[n])&\quad \mbox{(If $i\leq j$)}\end{array}\right..
\eeq Thus, the right side of eq.(\ref{nsum1}) becomes \beqa &
&(-1)^{n-1} \sum_{j=1}^n \sum_{i=1}^{n+1}(-1)^{i+j}
p_j(y^D_i[n])n_j(y^D_i[n])\nonumber\\
&=&(-1)^{n-1}\sum_{j=1}^n\sum_{i=j+1}^{n+1}(-1)^{i+j}p_j(y^D_i[n])n_j(y^D_i[n])
\nonumber\\& & +
(-1)^{n-1}\sum_{j=1}^n\sum_{i=1}^{j}(-1)^{i+j}p_j(y^D_i[n])n_j(y^D_i[n])\nonumber\\
&=&(-1)^{n-1}\sum_{j=1}^n\sum_{i=j+1}^{n+1}(-1)^{i+j}p_{i-1}(y^D_j[n])n_{i-1}(y^D_j[n])
\nonumber\\ & &+
(-1)^{n-1}\sum_{j=1}^n\sum_{i=1}^{j}(-1)^{i+j}p_i(y^D_{j+1}[n])n_i(y^D_{j+1}[n]).
\eeqa Setting that $i-1\rightarrow i$ for the first term and
$j+1\rightarrow j$ for the second term in the right side of the
above equation, we obtain \beqa & &(-1)^{n-1} \sum_{j=1}^n
\sum_{i=1}^{n+1}(-1)^{i+j}
p_j(y^D_i[n])n_j(y^D_i[n])\nonumber \\
&=&(-1)^{n-1}\sum_{j=1}^n\sum_{i=j}^{n}(-1)^{i+j-1}p_{i}(y^D_j[n])n_{i}(y^D_j[n])
\nonumber\\ & &+
(-1)^{n-1}\sum_{j=2}^{n+1}\sum_{i=1}^{j-1}(-1)^{i+j-1}p_i(y^D_{j}[n])n_i(y^D_{j}[n])\nonumber \\
&=&(-1)^{n}\sum_{j=1}^n\sum_{i=1}^{n}(-1)^{i+j}p_{i}(y^D_j[n])n_{i}(y^D_j[n])
\nonumber\\ & &+
(-1)^{n}\sum_{i=1}^{n+1}(-1)^{i+(n+1)}p_i(y^D_{n+1}[n])n_i(y^D_{n+1}[n]).\eeqa
Again, setting that $i\leftrightarrow j$ in the right side of the
above equation gives \beqa & & (-1)^{n-1} \sum_{j=1}^n
\sum_{i=1}^{n+1}(-1)^{i+j} p_j(y^D_i[n])n_j(y^D_i[n])\nonumber \\ &
&= (-1)^{n} \sum_{j=1}^n \sum_{i=1}^{n+1}(-1)^{i+j}
p_j(y^D_i[n])n_j(y^D_i[n]). \eeqa Thus, it implies that \beq
(-1)^{n-1} \sum_{j=1}^n \sum_{i=1}^{n+1}(-1)^{i+j}
p_j(y^D_i[n])n_j(y^D_i[n])= 0. \eeq From the relation (\ref{nsum1})
it follows that the identity (\ref{nsum}) is valid.

Now, let us back to the proof of the expression (\ref{Dformp}).
Since the definition of $D(y[n])$, we can rewrite it as \beq
D(y[n])=D(y[n-1])d_{n+1}(y[n]). \eeq Substituting our precondition
(\ref{Dform}) yields \beqa
D(y[n])&=&\sum_{i=1}^n(-1)^{(i-1)+(n-1)}p_i(y[n-1])n_i(y[n-1])d_{n+1}(y[n]).\eeqa
Note that \beqa n_i(y[n-1])d_{n+1}(y[n])&=&n_i(y[n])(y_i-y_{n+1}),\\
p_i(y[n-1])y_{n+1}&=&p_i(y[n]),\\
p_i(y[n-1])y_{i}&=&p_{n+1}(y[n]),\eeqa we have that
\beqa\label{Dformm}
\!\!\!D(y[n])\!\!\!&=&\!\!\!\sum_{i=1}^n(-1)^{(i-1)+n}p_i(y[n])n_i(y[n])
+\sum_{i=1}^n(-1)^{(i-1)+(n-1)}p_{n+1}(y[n])n_i(y[n])\nonumber\\
&=& \sum_{i=1}^n (-1)^{(i-1)+n}p_i(y[n])n_i(y[n])
+(-1)^{n-1}p_{n+1}(y[n])\left[\sum_{i=1}^n(-1)^{i-1}n_i(y[n])\right]\nonumber\\
&=& \sum_{i=1}^n (-1)^{(i-1)+n}p_i(y[n])n_i(y[n])
+(-1)^{n-1}p_{n+1}(y[n])\left[-(-1)^{n}n_{n+1}(y[n])\right]. \eeqa
In the last equality we have used the conclusion (\ref{nsum}) of our
precondition, that is \beq \sum_{i=1}^n(-1)^{i-1}n_i(y[n])=-(-1)^n
n_{n+1}(y[n]).\eeq Thus, eq.(\ref{Dformm}) becomes \beq
D(y[n])=\sum_{i=1}^{n+1}(-1)^{(i-1)+n}p_i(y[n])n_i(y[n]).\eeq The
needed expression (\ref{Dformp}) is proved by mathematical
induction. That is, we have proved that for any $n\geq 1$, the
expression (\ref{Dformp}) is valid.

Since our proof of the conclusion (\ref{nsum}) of our precondition
is independent of $n$ ($n\geq 1$), we can, in the same way, prove
that the identity (\ref{nsum}) is correct for any $n\geq 1$.

Now, let us prove our identity (\ref{myi}). Obviously, when $K=0$ we
have \beq\label{cnzero}
\sum_{i=1}^{n+1}(-1)^{i-1}\frac{1}{d_i(x[n])}=\frac{1}{D(x[n])}
\left[\sum_{i=1}^{n+1}(-1)^{i-1}n_i(x[n])\right]=0,\eeq where we
have used the fact the identity (\ref{nsum}) is valid for any $n\geq
1$. Furthermore, we extend the definition domain of $C^K_n(x[n])$
from $K\geq n$ to $K\geq 0$, and still write its form as \beq
C^K_n(x[n])= \sum_{i=1}^{n+1}(-1)^{i-1}\frac{x_i^K}{d_i(x[n])}.\eeq
Obviously, eq.(\ref{cnzero}) means \beq C_n^0(x[n])=0,\quad (n\geq
1).\eeq In order to consider the cases when $K\neq 0$, by using of
$d_i(x[n])(x_i-x_{n+2})=d_i(x[n+1])$ ($i\leq n+1$), we obtain\beqa &
&C^K_n(x[n])= \sum_{i=1}^{n+1}(-1)^{i-1}\frac{x_i^K
}{d_i(x[n])}\nonumber\\&=&\sum_{i=1}^{n+1}(-1)^{i-1}\frac{x_i^K
}{d_i(x[n])}\left(\frac{x_i-x_{n+2}}{x_i-x_{n+2}}\right)\nonumber\\
&=&\sum_{i=1}^{n+1}(-1)^{i-1}\frac{x_i^{K+1}
}{d_i(x[n+1])}\nonumber\\
& &-x_{n+2}\sum_{i=1}^{n+1}(-1)^{i-1}\frac{x_i^K
}{d_i(x[n+1])}\nonumber\\ &=&
C^{K+1}_{n+1}(x[n+1])-(-1)^{(n+2)-1}\frac{x_{n+2}^{K+1}
}{d_{n+2}(x[n+1])}\nonumber \\&
&-x_{n+2}\sum_{i=1}^{n+1}(-1)^{i-1}\frac{x_i^K }{d_i(x[n+1])}. \eeqa
It follows the recurrence equation as the following
\beq\label{myirq}
C^K_n(x[n])=C^{K+1}_{n+1}(x[n+1])-x_{n+2}C^{K}_{n+1}(x[n+1]).\eeq It
implies that since $C_{n}^0(x[n])=0$ for any $n\geq 1$, then
$C_{n+1}^1(x[n+1])=0$ for any $n\geq 1$ or $C_{n}^1(x[n])=0$ for any
$n\geq 2$; since $C_{n}^1(x[n])=0$ for any $n\geq 2$, then
$C_{n+1}^2(x[n+1])=0$ for any $n\geq 2$ or $C_{n}^2(x[n])=0$ for any
$n\geq 3$; $\cdots$, since $C_{n}^k(x[n])=0$ for any $n\geq (k+1)$
($k\geq 0$), then $C_{n+1}^{k+1}(x[n+1])=0$ for any $n\geq (k+1)$ or
$C_{n}^{k+1}(x[n])=0$ for any $n\geq (k+2)$; $\cdots$. In fact, the
mathematical induction tells us this result. Obviously \beq
\label{myig1} C_n^K(x[n])=0,\quad (\mbox{If $0\leq K<n$}). \eeq
Taking $K=n$ in eq.(\ref{myirq}) and using eq.(\ref{myig1}), we have
 \beq C_n^n(x[n])=C_{n+1}^{n+1}(x[n+1])=1, \quad (n\geq 1),\eeq where we
have used the fact that $C_1^1(x[1])=1$. Therefore, the proof of our
identity (\ref{myi}) is finally finished.

\section{The calculations of the high order terms}

Since we have taken the $H_1$ only with the non diagonal part, it is
enough to calculate the contributions from them. In Sec. \ref{sec6}
the contributions from the first, second and third order
approximations. In this appendix, we would like to find the
contributions from the fourth to the sixth order approximations. The
used calculation technologies are mainly to find the limitation,
dummy index changing and summation, as well as the replacement
$g_1^{\gamma_i\gamma_j}\eta_{\gamma_i\gamma_j}=g_1^{\gamma_i\gamma_j}$
since $g_1^{\gamma_i\gamma_j}$ has been nondiagonal. They are not
difficult, but are a little lengthy.

\subsection{$l=4$ case}

For the fourth order approximation, its contributions from the first
decompositions consists of eight terms: \beqa \label{A4fc}
A_4^{\gamma\gamma^\prime}&=&A_4^{\gamma\gamma^\prime}(ccc)
+A_4^{\gamma\gamma^\prime}(ccn)+A_4^{\gamma\gamma^\prime}(cnc)
+A_4^{\gamma\gamma^\prime}(ncc)\nonumber\\
& & +A_4^{\gamma\gamma^\prime}(cnn)+A_4^{\gamma\gamma^\prime}(ncn)
+A_4^{\gamma\gamma^\prime}(nnc)+{A}_4^{\gamma\gamma^\prime}(nnn).
\eeqa We will see that the former four terms have no the nontrivial
second contractions, the fifth and seven terms have one nontrivial
second contraction, \beqa {A}_4^{\gamma\gamma^\prime}(cnn)
&=&{A}_4^{\gamma\gamma^\prime}(cnn,kc) +
{A}_4^{\gamma\gamma^\prime}(cnn;kn),\\
{A}_4^{\gamma\gamma^\prime}(ncn)
&=&{A}_4^{\gamma\gamma^\prime}(ncn,c)
+{A}_4^{\gamma\gamma^\prime}(ncn,n), \\
A_4^{\gamma\gamma^\prime}(nnc)
&=&{A}_4^{\gamma\gamma^\prime}(nnc,ck) +
{A}_4^{\gamma\gamma^\prime}(nnc,nk). \eeqa In addition, the last
term in eq.(\ref{A4fc}) has two nontrivial second contractions, thus
\beqa
{A}_4^{\gamma\gamma^\prime}(nnn)&=&{A}_4^{\gamma\gamma^\prime}(nnn,cc)
+{A}_4^{\gamma\gamma^\prime}(nnn,cn){A}_4^{\gamma\gamma^\prime}(nnn,nc)
+{A}_4^{\gamma\gamma^\prime}(nnn,nn),\eeqa its fourth term has also
the third contraction \beq
{A}_4^{\gamma\gamma^\prime}(nnn,nn)={A}_4^{\gamma\gamma^\prime}(nnn,nn,c)
+{A}_4^{\gamma\gamma^\prime}(nnn,nn,n).\eeq All together, we have
the fifteen expressions of the contributions from whole contractions
in the fourth order approximation.

First, let us calculate the former four terms only with the first
contractions and anti-contractions, that is, with more than two
$\delta$ functions (or less than two $\eta$ functions)
\beqa\label{A4ccc} {A}_4^{\gamma\gamma^\prime}(ccc)&=
&\sum_{\gamma_1,\cdots,\gamma_{5}}\left[
\sum_{i=1}^{5}(-1)^{i-1}\frac{\e^{-\I E_{\gamma_i}
t}}{d_i(E[\gamma,4])}\right]
\left[\prod_{j=1}^4g_1^{\gamma_j\gamma_{j+1}}\right]\left(
\prod_{k=1}^{3}\delta_{\gamma_k\gamma_{k+2}}\right)
\delta_{\gamma_1\gamma}\delta_{\gamma_{5}\gamma^\prime}\nonumber\\
&=&\sum_{\gamma_1}\left[\frac{3\e^{-\I
E_{\gamma}t}}{\left(E_\gamma-E_{\gamma_1}\right)^4}-\frac{3\e^{-\I
E_{\gamma_1}t}}{\left(E_\gamma-E_{\gamma_1}\right)^4}-(-\I
t)\frac{2\e^{-\I E_{\gamma}t}}{\left(E_\gamma-E_{\gamma_1}\right)^3}\right.\nonumber\\
& &\left.-(-\I t)\frac{\e^{-\I
E_{\gamma_1}t}}{\left(E_\gamma-E_{\gamma_1}\right)^3}+\frac{(-\I
t)^2}{2}\frac{\e^{-\I
E_{\gamma}t}}{\left(E_\gamma-E_{\gamma_1}\right)^2}\right]
\left|g_1^{\gamma\gamma_1}\right|^4\delta_{\gamma\gamma^\prime}.\eeqa
\beqa\label{A4ccn} {A}_4^{\gamma\gamma^\prime}(ccn)&=
&\sum_{\gamma_1,\cdots,\gamma_{5}}\left[
\sum_{i=1}^{5}(-1)^{i-1}\frac{\e^{-\I E_{\gamma_i}
t}}{d_i(E[\gamma,4])}\right]
\left[\prod_{j=1}^4g_1^{\gamma_j\gamma_{j+1}}\right]
\delta_{\gamma_1\gamma_3}{\delta}_{\gamma_2\gamma_4}\eta_{\gamma_3\gamma_5}
\delta_{\gamma_1\gamma}\delta_{\gamma_{5}\gamma^\prime}\nonumber\\
&=&\sum_{\gamma_1}\left[-\frac{\e^{-\I
E_{\gamma}t}}{\left(E_\gamma-E_{\gamma_1}\right)^2\left(E_{\gamma}-E_{\gamma^\prime}\right)^2}-\frac{2\e^{-\I
E_{\gamma}t}}{\left(E_\gamma-E_{\gamma_1}\right)^3\left(E_{\gamma}-E_{\gamma^\prime}\right)}\right.\nonumber\\
& &-\frac{\e^{-\I
E_{\gamma_1}t}}{\left(E_\gamma-E_{\gamma_1}\right)^2\left(E_{\gamma_1}-E_{\gamma^\prime}\right)^2}+\frac{2\e^{-\I
E_{\gamma_1}t}}{\left(E_\gamma-E_{\gamma_1}\right)^3\left(E_{\gamma_1}-E_{\gamma^\prime}\right)}\nonumber\\
& &+\frac{\e^{-\I
E_{\gamma^\prime}t}}{\left(E_\gamma-E_{\gamma^\prime}\right)^2\left(E_{\gamma_1}-E_{\gamma^\prime}\right)^2}+(-\I
t)\frac{\e^{-\I E_{\gamma}t}}{\left(E_\gamma-E_{\gamma_1}\right)^2
\left(E_{\gamma}-E_{\gamma^\prime}\right)}
\nonumber\\
& & \left. +(-\I t)\frac{\e^{-\I
E_{\gamma_1}t}}{\left(E_\gamma-E_{\gamma_1}\right)^2
\left(E_{\gamma_1}-E_{\gamma^\prime}\right)}
\right]\left|g_1^{\gamma\gamma_1}\right|^2
g_1^{\gamma\gamma_1}g_1^{\gamma_1\gamma^\prime}\eta_{\gamma\gamma^\prime}.
\eeqa
\beqa\label{A4cnc} {A}_4^{\gamma\gamma^\prime}(cnc)&=
&\sum_{\gamma_1,\cdots,\gamma_{5}}\left[
\sum_{i=1}^{5}(-1)^{i-1}\frac{\e^{-\I E_{\gamma_i}
t}}{d_i(E[\gamma,4])}\right]
\left[\prod_{j=1}^4g_1^{\gamma_j\gamma_{j+1}}\right]
\delta_{\gamma_1\gamma_3}\eta_{\gamma_2\gamma_4}\delta_{\gamma_3\gamma_5}
\delta_{\gamma_1\gamma}\delta_{\gamma_{5}\gamma^\prime}\nonumber\\
&=&\sum_{\gamma_1,\gamma_2}\left[\frac{\e^{-\I
E_{\gamma}t}}{\left(E_\gamma-E_{\gamma_1}\right)\left(E_{\gamma}-E_{\gamma_2}\right)^3}
+\frac{\e^{-\I
E_{\gamma}t}}{\left(E_\gamma-E_{\gamma_1}\right)^2\left(E_{\gamma}-E_{\gamma_2}\right)^2}\right.\nonumber\\
& &+\frac{\e^{-\I
E_{\gamma}t}}{\left(E_\gamma-E_{\gamma_1}\right)^3\left(E_{\gamma}-E_{\gamma_2}\right)}
-\frac{\e^{-\I
E_{\gamma_1}t}}{\left(E_\gamma-E_{\gamma_1}\right)^3\left(E_{\gamma_1}-E_{\gamma_2}\right)}
\nonumber\\
& &+\frac{\e^{-\I
E_{\gamma_2}t}}{\left(E_\gamma-E_{\gamma_2}\right)^3\left(E_{\gamma_1}-E_{\gamma_2}\right)}-(-\I
t)\frac{\e^{-\I
E_{\gamma}t}}{\left(E_\gamma-E_{\gamma_1}\right)\left(E_{\gamma}-E_{\gamma_2}\right)^2}\nonumber\\
& &\left. -(-\I t)\frac{\e^{-\I
E_{\gamma}t}}{\left(E_\gamma-E_{\gamma_1}\right)^2\left(E_{\gamma}-E_{\gamma_2}\right)}
+\frac{(-\I t)^2}{2}\frac{\e^{-\I
E_{\gamma}t}}{\left(E_\gamma-E_{\gamma_1}\right)
\left(E_{\gamma}-E_{\gamma_2}\right)}\right]\nonumber\\
& &\times\left|g_1^{\gamma\gamma_1}\right|^2
\left|g_1^{\gamma\gamma_2}\right|^2
\eta_{\gamma_1\gamma_2}\delta_{\gamma\gamma^\prime}. \eeqa
\beqa\label{A4ncc} {A}_4^{\gamma\gamma^\prime}(ncc)&=
&\sum_{\gamma_1,\cdots,\gamma_{5}}\left[
\sum_{i=1}^{5}(-1)^{i-1}\frac{\e^{-\I E_{\gamma_i}
t}}{d_i(E[\gamma,4])}\right]
\left[\prod_{j=1}^4g_1^{\gamma_j\gamma_{j+1}}\right]
\eta_{\gamma_1\gamma_3}{\delta}_{\gamma_2\gamma_4}{\delta}_{\gamma_3\gamma_5}
\delta_{\gamma_1\gamma}\delta_{\gamma_{5}\gamma^\prime}\nonumber\\
&=&\sum_{\gamma_1}\left[\frac{\e^{-\I
E_{\gamma}t}}{\left(E_\gamma-E_{\gamma_1}\right)^2
\left(E_{\gamma}-E_{\gamma^\prime}\right)^2}+\frac{2\e^{-\I
E_{\gamma_1}t}}{\left(E_\gamma-E_{\gamma_1}\right)
\left(E_{\gamma_1}-E_{\gamma^\prime}\right)^3}\right.\nonumber\\
& &-\frac{\e^{-\I
E_{\gamma_1}t}}{\left(E_\gamma-E_{\gamma_1}\right)^2
\left(E_{\gamma_1}-E_{\gamma^\prime}\right)^2}-\frac{2\e^{-\I
E_{\gamma^\prime}t}}{\left(E_\gamma-E_{\gamma^\prime}\right)
\left(E_{\gamma_1}-E_{\gamma^\prime}\right)^3}\nonumber\\
& &-\frac{\e^{-\I
E_{\gamma^\prime}t}}{\left(E_\gamma-E_{\gamma^\prime}\right)^2
\left(E_{\gamma_1}-E_{\gamma^\prime}\right)^2} -(-\I t)\frac{\e^{-\I
E_{\gamma_1}t}}{\left(E_\gamma-E_{\gamma_1}\right)
\left(E_{\gamma_1}-E_{\gamma^\prime}\right)^2}\nonumber\\
& & \left. -(-\I t)\frac{\e^{-\I
E_{\gamma^\prime}t}}{\left(E_\gamma-E_{\gamma^\prime}\right)
\left(E_{\gamma_1}-E_{\gamma^\prime}\right)^2}\right]\left|g_1^{\gamma_1\gamma^\prime}\right|^2
g_1^{\gamma\gamma_1}g_1^{\gamma_1\gamma^\prime}\eta_{\gamma\gamma^\prime}.
\eeqa

Then, we calculate the three terms with the single first
contraction, that is, with one $\delta$ function.  Because one
$\delta$ function can not eliminate the whole apparent singularity,
we also need to find out the nontrivial second contraction- and/or
anti-contraction terms. \beqa\label{A4cnn-kc}
{A}_4^{\gamma\gamma^\prime}(cnn,kc)&=
&\sum_{\gamma_1,\cdots,\gamma_{5}}\left[
\sum_{i=1}^{5}(-1)^{i-1}\frac{\e^{-\I E_{\gamma_i}
t}}{d_i(E[\gamma,4])}\right]
\left[\prod_{j=1}^4g_1^{\gamma_j\gamma_{j+1}}\right]
{\delta}_{\gamma_1\gamma_3}\eta_{\gamma_2\gamma_4}
\eta_{\gamma_3\gamma_5}{\delta}_{\gamma_2\gamma^\prime}
\delta_{\gamma_1\gamma}\delta_{\gamma_{5}\gamma^\prime}\nonumber\\
&=&\sum_{\gamma_1}\left[-\frac{2\e^{-\I
E_{\gamma}t}}{\left(E_\gamma-E_{\gamma_1}\right)
\left(E_{\gamma}-E_{\gamma^\prime}\right)^3}-\frac{\e^{-\I
E_{\gamma}t}}{\left(E_\gamma-E_{\gamma_1}\right)^2
\left(E_{\gamma}-E_{\gamma^\prime}\right)^2}\right.\nonumber\\
& &+\frac{\e^{-\I
E_{\gamma_1}t}}{\left(E_\gamma-E_{\gamma_1}\right)^2
\left(E_{\gamma_1}-E_{\gamma^\prime}\right)^2}-\frac{\e^{-\I
E_{\gamma^\prime}t}}{\left(E_\gamma-E_{\gamma^\prime}\right)^2
\left(E_{\gamma_1}-E_{\gamma^\prime}\right)^2}\nonumber\\
& &-\frac{2\e^{-\I
E_{\gamma^\prime}t}}{\left(E_\gamma-E_{\gamma^\prime}\right)^3
\left(E_{\gamma_1}-E_{\gamma^\prime}\right)}+(-\I t)\frac{\e^{-\I
E_{\gamma}t}}{\left(E_\gamma-E_{\gamma_1}\right)
\left(E_{\gamma}-E_{\gamma^\prime}\right)^2}\nonumber\\
& &\left. -(-\I t)\frac{\e^{-\I
E_{\gamma^\prime}t}}{\left(E_\gamma-E_{\gamma^\prime}\right)^2
\left(E_{\gamma_1}-E_{\gamma^\prime}\right)}\right]
\left|g_1^{\gamma\gamma^\prime}\right|^2
g_1^{\gamma\gamma_1}g_1^{\gamma_1\gamma^\prime}. \eeqa
\beqa\label{A4cnn-kn} {A}_4^{\gamma\gamma^\prime}(cnn,kn)&=
&\sum_{\gamma_1,\cdots,\gamma_{5}}\left[
\sum_{i=1}^{5}(-1)^{i-1}\frac{\e^{-\I E_{\gamma_i}
t}}{d_i(E[\gamma,4])}\right]
\left[\prod_{j=1}^4g_1^{\gamma_j\gamma_{j+1}}\right]
{\delta}_{\gamma_1\gamma_3}\eta_{\gamma_2\gamma_4}
\eta_{\gamma_3\gamma_5}\eta_{\gamma_2\gamma^\prime}
\delta_{\gamma_1\gamma}\delta_{\gamma_{5}\gamma^\prime}\nonumber\\
&=&\sum_{\gamma_1,\gamma_2}\left[-\frac{\e^{-\I
E_{\gamma}t}}{\left(E_\gamma-E_{\gamma_1}\right)
\left(E_\gamma-E_{\gamma_2}\right)
\left(E_{\gamma}-E_{\gamma^\prime}\right)^2}\right.\nonumber\\
& &-\frac{\e^{-\I E_{\gamma}t}}{\left(E_\gamma-E_{\gamma_1}\right)
\left(E_\gamma-E_{\gamma_2}\right)^2
\left(E_{\gamma}-E_{\gamma^\prime}\right)}\nonumber\\
& & -\frac{\e^{-\I
E_{\gamma}t}}{\left(E_\gamma-E_{\gamma_1}\right)^2
\left(E_\gamma-E_{\gamma_2}\right)
\left(E_{\gamma}-E_{\gamma^\prime}\right)}\nonumber\\
& & +\frac{\e^{-\I
E_{\gamma_1}t}}{\left(E_\gamma-E_{\gamma_1}\right)^2
\left(E_{\gamma_1}-E_{\gamma_2}\right)
\left(E_{\gamma_1}-E_{\gamma^\prime}\right)}\nonumber\\
& & -\frac{\e^{-\I
E_{\gamma_2}t}}{\left(E_\gamma-E_{\gamma_2}\right)^2
\left(E_{\gamma_1}-E_{\gamma_2}\right)
\left(E_{\gamma_2}-E_{\gamma^\prime}\right)}\nonumber\\
& &+\frac{\e^{-\I
E_{\gamma^\prime}t}}{\left(E_\gamma-E_{\gamma^\prime}\right)^2
\left(E_{\gamma_1}-E_{\gamma^\prime}\right)
\left(E_{\gamma_2}-E_{\gamma^\prime}\right)}\nonumber\\
& &\left. +(-\I t)\frac{\e^{-\I
E_{\gamma}t}}{\left(E_\gamma-E_{\gamma_1}\right)
\left(E_{\gamma}-E_{\gamma_2}\right)\left(E_{\gamma}-E_{\gamma^\prime}\right)}\right]\nonumber\\
& & \times\left|g_1^{\gamma\gamma_1}\right|^2
g_1^{\gamma\gamma_2}g_1^{\gamma_2\gamma^\prime}
\eta_{\gamma_1\gamma_2}\eta_{\gamma_1\gamma^\prime}
\eta_{\gamma\gamma^\prime}. \eeqa
\beqa\label{A4ncn-c} {A}_4^{\gamma\gamma^\prime}(ncn,c)&=
&\sum_{\gamma_1,\cdots,\gamma_{5}}\left[
\sum_{i=1}^{5}(-1)^{i-1}\frac{\e^{-\I E_{\gamma_i}
t}}{d_i(E[\gamma,4])}\right]
\left[\prod_{j=1}^4g_1^{\gamma_j\gamma_{j+1}}\right]
\eta_{\gamma_1\gamma_3}{\delta}_{\gamma_2\gamma_4}\eta_{\gamma_3\gamma_5}
\delta_{\gamma_1\gamma}\delta_{\gamma_{5}\gamma^\prime}{\delta}_{\gamma\gamma^\prime}\nonumber\\
&=&\sum_{\gamma_1,\gamma_2}\left[-\frac{\e^{-\I
E_{\gamma}t}}{\left(E_{\gamma}-E_{\gamma_1}\right)^2
\left(E_{\gamma}-E_{\gamma_2}\right)^2}-\frac{2\e^{-\I
E_{\gamma}t}}{\left(E_{\gamma}-E_{\gamma_1}\right)^3
\left(E_{\gamma}-E_{\gamma_2}\right)}\right.\nonumber\\
& &-\frac{\e^{-\I
E_{\gamma_1}t}}{\left(E_{\gamma}-E_{\gamma_1}\right)^2
\left(E_{\gamma_1}-E_{\gamma_2}\right)^2}+\frac{2\e^{-\I
E_{\gamma_1}t}}{\left(E_{\gamma}-E_{\gamma_1}\right)^3
\left(E_{\gamma_1}-E_{\gamma_2}\right)}\nonumber\\
& & +\frac{\e^{-\I
E_{\gamma_2}t}}{\left(E_{\gamma}-E_{\gamma_2}\right)^2
\left(E_{\gamma_1}-E_{\gamma_2}\right)^2}+(-\I t)\frac{\e^{-\I
E_{\gamma}t}}{\left(E_{\gamma}-E_{\gamma_1}\right)^2
\left(E_{\gamma}-E_{\gamma_2}\right)}\nonumber\\
& &\left. +(-\I t)\frac{\e^{-\I
E_{\gamma_1}t}}{\left(E_{\gamma}-E_{\gamma_1}\right)^2
\left(E_{\gamma_1}-E_{\gamma_2}\right)}\right]\left|g_1^{\gamma\gamma_1}\right|^2
\left|g_1^{\gamma_1\gamma_2}\right|^2\eta_{\gamma\gamma_2}\delta_{\gamma\gamma^\prime}.
\eeqa
\beqa\label{A4ncn-n} {A}_4^{\gamma\gamma^\prime}(ncn,n)&=
&\sum_{\gamma_1,\cdots,\gamma_{5}}\left[
\sum_{i=1}^{5}(-1)^{i-1}\frac{\e^{-\I E_{\gamma_i}
t}}{d_i(E[\gamma,4])}\right]
\left[\prod_{j=1}^4g_1^{\gamma_j\gamma_{j+1}}\right]
\eta_{\gamma_1\gamma_3}{\delta}_{\gamma_2\gamma_4}\eta_{\gamma_3\gamma_5}
\delta_{\gamma_1\gamma}\delta_{\gamma_{5}\gamma^\prime}\eta_{\gamma\gamma^\prime}\nonumber\\
&=&\sum_{\gamma_1,\gamma_2}\left[\frac{\e^{-\I
E_{\gamma}t}}{\left(E_\gamma-E_{\gamma_1}\right)^2
\left(E_\gamma-E_{\gamma_2}\right)
\left(E_{\gamma}-E_{\gamma^\prime}\right)}\right.\nonumber\\
& &+\frac{\e^{-\I E_{\gamma_1}t}}{\left(E_\gamma-E_{\gamma_1}\right)
\left(E_{\gamma_1}-E_{\gamma_2}\right)
\left(E_{\gamma_1}-E_{\gamma^\prime}\right)^2}\nonumber\\
& & +\frac{\e^{-\I
E_{\gamma_1}t}}{\left(E_\gamma-E_{\gamma_1}\right)
\left(E_{\gamma_1}-E_{\gamma_2}\right)^2
\left(E_{\gamma_1}-E_{\gamma^\prime}\right)}\nonumber\\
& &-\frac{\e^{-\I
E_{\gamma_1}t}}{\left(E_\gamma-E_{\gamma_1}\right)^2
\left(E_{\gamma_1}-E_{\gamma_2}\right)
\left(E_{\gamma_1}-E_{\gamma^\prime}\right)}\nonumber\\
& & -\frac{\e^{-\I
E_{\gamma_2}t}}{\left(E_\gamma-E_{\gamma_2}\right)
\left(E_{\gamma_1}-E_{\gamma_2}\right)^2
\left(E_{\gamma_2}-E_{\gamma^\prime}\right)}\nonumber\\
& &+\frac{\e^{-\I
E_{\gamma^\prime}t}}{\left(E_\gamma-E_{\gamma^\prime}\right)
\left(E_{\gamma_1}-E_{\gamma^\prime}\right)^2
\left(E_{\gamma_2}-E_{\gamma^\prime}\right)}\nonumber\\
& &\left. -(-\I t)\frac{\e^{-\I
E_{\gamma_1}t}}{\left(E_\gamma-E_{\gamma_1}\right)
\left(E_{\gamma_1}-E_{\gamma_2}\right)\left(E_{\gamma_1}-E_{\gamma^\prime}\right)}\right]\nonumber\\
& & \times\left|g_1^{\gamma_1\gamma_2}\right|^2
g_1^{\gamma\gamma_1}g_1^{\gamma_1\gamma^\prime}
\eta_{\gamma\gamma_2}\eta_{\gamma_2\gamma^\prime}
\eta_{\gamma\gamma^\prime}. \eeqa
\beqa\label{A4nnc-ck} {A}_4^{\gamma\gamma^\prime}(nnc,ck)&=
&\sum_{\gamma_1,\cdots,\gamma_{5}}\left[
\sum_{i=1}^{5}(-1)^{i-1}\frac{\e^{-\I E_{\gamma_i}
t}}{d_i(E[\gamma,4])}\right]
\left[\prod_{j=1}^4g_1^{\gamma_j\gamma_{j+1}}\right]
\eta_{\gamma_1\gamma_3}
\eta_{\gamma_2\gamma_4}{\delta}_{\gamma_3\gamma_5}
{\delta}_{\gamma_1\gamma_4}
\delta_{\gamma_1\gamma}\delta_{\gamma_{5}\gamma^\prime}\nonumber\\
&=&\sum_{\gamma_1}\left[-\frac{2\e^{-\I
E_{\gamma}t}}{\left(E_\gamma-E_{\gamma_1}\right)
\left(E_{\gamma}-E_{\gamma^\prime}\right)^3}-\frac{\e^{-\I
E_{\gamma}t}}{\left(E_\gamma-E_{\gamma_1}\right)^2
\left(E_{\gamma}-E_{\gamma^\prime}\right)^2}\right.\nonumber\\
& &+\frac{\e^{-\I
E_{\gamma_1}t}}{\left(E_\gamma-E_{\gamma_1}\right)^2
\left(E_{\gamma_1}-E_{\gamma^\prime}\right)^2}-\frac{\e^{-\I
E_{\gamma^\prime}t}}{\left(E_\gamma-E_{\gamma^\prime}\right)^2
\left(E_{\gamma_1}-E_{\gamma^\prime}\right)^2}\nonumber\\
& &-\frac{2\e^{-\I
E_{\gamma^\prime}t}}{\left(E_\gamma-E_{\gamma^\prime}\right)^3
\left(E_{\gamma_1}-E_{\gamma^\prime}\right)}+(-\I t)\frac{\e^{-\I
E_{\gamma}t}}{\left(E_\gamma-E_{\gamma_1}\right)
\left(E_{\gamma}-E_{\gamma^\prime}\right)^2}\nonumber\\
& &\left. -(-\I t)\frac{\e^{-\I
E_{\gamma^\prime}t}}{\left(E_\gamma-E_{\gamma^\prime}\right)^2
\left(E_{\gamma_1}-E_{\gamma^\prime}\right)}\right]
\left|g_1^{\gamma\gamma^\prime}\right|^2
g_1^{\gamma\gamma_1}g_1^{\gamma_1\gamma^\prime}. \eeqa
\beqa\label{A4nnc-nk} {A}_4^{\gamma\gamma^\prime}(nnc,nk)&=
&\sum_{\gamma_1,\cdots,\gamma_{5}}\left[
\sum_{i=1}^{5}(-1)^{i-1}\frac{\e^{-\I E_{\gamma_i}
t}}{d_i(E[\gamma,4])}\right]
\left[\prod_{j=1}^4g_1^{\gamma_j\gamma_{j+1}}\right]
\eta_{\gamma_1\gamma_3}
\eta_{\gamma_2\gamma_4}{\delta}_{\gamma_3\gamma_5}
\eta_{\gamma_1\gamma_4}
\delta_{\gamma_1\gamma}\delta_{\gamma_{5}\gamma^\prime}\nonumber\\
&=&\sum_{\gamma_1,\gamma_2}\left[\frac{\e^{-\I
E_{\gamma}t}}{\left(E_\gamma-E_{\gamma_1}\right)
\left(E_\gamma-E_{\gamma_2}\right)
\left(E_{\gamma}-E_{\gamma^\prime}\right)^2}\right.\nonumber\\
& &-\frac{\e^{-\I E_{\gamma_1}t}}{\left(E_\gamma-E_{\gamma_1}\right)
\left(E_{\gamma_1}-E_{\gamma_2}\right)
\left(E_{\gamma_1}-E_{\gamma^\prime}\right)^2}\nonumber\\
& & +\frac{\e^{-\I
E_{\gamma_2}t}}{\left(E_\gamma-E_{\gamma_2}\right)
\left(E_{\gamma_1}-E_{\gamma_2}\right)
\left(E_{\gamma_2}-E_{\gamma^\prime}\right)^2}\nonumber\\
& &-\frac{\e^{-\I
E_{\gamma^\prime}t}}{\left(E_\gamma-E_{\gamma^\prime}\right)
\left(E_{\gamma_1}-E_{\gamma^\prime}\right)
\left(E_{\gamma_2}-E_{\gamma^\prime}\right)^2}\nonumber\\
& & -\frac{\e^{-\I
E_{\gamma^\prime}t}}{\left(E_\gamma-E_{\gamma^\prime}\right)
\left(E_{\gamma_1}-E_{\gamma^\prime}\right)^2
\left(E_{\gamma_2}-E_{\gamma^\prime}\right)}\nonumber\\
& &-\frac{\e^{-\I
E_{\gamma^\prime}t}}{\left(E_\gamma-E_{\gamma^\prime}\right)^2
\left(E_{\gamma_1}-E_{\gamma^\prime}\right)
\left(E_{\gamma_2}-E_{\gamma^\prime}\right)}\nonumber\\
& &\left. -(-\I t)\frac{\e^{-\I
E_{\gamma^\prime}t}}{\left(E_\gamma-E_{\gamma^\prime}\right)
\left(E_{\gamma_1}-E_{\gamma^\prime}\right)\left(E_{\gamma_2}-E_{\gamma^\prime}\right)}\right]
\nonumber\\
& &\times\left|g_1^{\gamma_2\gamma^\prime}\right|^2
g_1^{\gamma\gamma_1}g_1^{\gamma_1\gamma^\prime}
\eta_{\gamma\gamma_2}\eta_{\gamma_1\gamma_2}
\eta_{\gamma\gamma^\prime}. \eeqa

Finally, we calculate the ${A}_4^{\gamma\gamma^\prime}(nnn)$ by
considering the two second decompositions, that is, its former three
terms \beqa\label{A4nnn-cc} {A}_4^{\gamma\gamma^\prime}(nnn,cc)&=
&\sum_{\gamma_1,\cdots,\gamma_{5}}\left[
\sum_{i=1}^{5}(-1)^{i-1}\frac{\e^{-\I E_{\gamma_i}
t}}{d_i(E[\gamma,4])}\right]
\left[\prod_{j=1}^4g_1^{\gamma_j\gamma_{j+1}}\right]\left(
\prod_{k=1}^{3}\eta_{\gamma_k\gamma_{k+2}}\right)
\delta_{\gamma_1\gamma_4}\delta_{\gamma_2\gamma_5}
\delta_{\gamma_1\gamma}\delta_{\gamma_{5}\gamma^\prime}
\nonumber\\
&=&\sum_{\gamma_1}\left[-\frac{2\e^{-\I
E_{\gamma}t}}{\left(E_{\gamma}-E_{\gamma_1}\right)
\left(E_{\gamma}-E_{\gamma^\prime}\right)^3}-\frac{\e^{-\I
E_{\gamma}t}}{\left(E_{\gamma}-E_{\gamma_1}\right)^2
\left(E_{\gamma}-E_{\gamma^\prime}\right)^2}\right.\nonumber\\
& &+\frac{\e^{-\I
E_{\gamma_1}t}}{\left(E_{\gamma}-E_{\gamma_1}\right)^2
\left(E_{\gamma_1}-E_{\gamma^\prime}\right)^2}-\frac{\e^{-\I
E_{\gamma^\prime}t}}{\left(E_{\gamma}-E_{\gamma^\prime}\right)^2
\left(E_{\gamma_1}-E_{\gamma^\prime}\right)^2}\nonumber\\
& &-\frac{2\e^{-\I
E_{\gamma^\prime}t}}{\left(E_{\gamma}-E_{\gamma^\prime}\right)^3
\left(E_{\gamma_1}-E_{\gamma^\prime}\right)}+(-\I t)\frac{\e^{-\I
E_{\gamma}t}}{\left(E_{\gamma}-E_{\gamma_1}\right)
\left(E_{\gamma}-E_{\gamma^\prime}\right)^2}\nonumber\\
& &\left.-(-\I t)\frac{\e^{-\I
E_{\gamma^\prime}t}}{\left(E_{\gamma}-E_{\gamma^\prime}\right)^2
\left(E_{\gamma_1}-E_{\gamma^\prime}\right)}\right]g_1^{\gamma\gamma^\prime}
g_1^{\gamma^\prime\gamma_1}g_1^{\gamma_1\gamma}g_1^{\gamma\gamma^\prime}.\eeqa
\beqa\label{A4nnn-cn} {A}_4^{\gamma\gamma^\prime}(nnn,cn)&=
&\sum_{\gamma_1,\cdots,\gamma_{5}}\left[
\sum_{i=1}^{5}(-1)^{i-1}\frac{\e^{-\I E_{\gamma_i}
t}}{d_i(E[\gamma,4])}\right]
\left[\prod_{j=1}^4g_1^{\gamma_j\gamma_{j+1}}\right]\left(
\prod_{k=1}^{3}\eta_{\gamma_k\gamma_{k+2}}\right)
{\delta}_{\gamma_1\gamma_4}\eta_{\gamma_2\gamma_5}
\delta_{\gamma_1\gamma}\delta_{\gamma_{5}\gamma^\prime}
\nonumber\\
&=&\sum_{\gamma_1,\gamma_2}\left[-\frac{\e^{-\I
E_{\gamma}t}}{\left(E_{\gamma}-E_{\gamma_1}\right)\left(E_{\gamma}-E_{\gamma_2}\right)
\left(E_{\gamma}-E_{\gamma^\prime}\right)^2}-\frac{\e^{-\I
E_{\gamma}t}}{\left(E_{\gamma}-E_{\gamma_1}\right)\left(E_{\gamma}-E_{\gamma_2}\right)^2
\left(E_{\gamma}-E_{\gamma^\prime}\right)}\right.\nonumber\\
& &-\frac{\e^{-\I
E_{\gamma}t}}{\left(E_{\gamma}-E_{\gamma_1}\right)^2\left(E_{\gamma}-E_{\gamma_2}\right)
\left(E_{\gamma}-E_{\gamma^\prime}\right)}+\frac{\e^{-\I
E_{\gamma_1}t}}{\left(E_{\gamma}-E_{\gamma_1}\right)^2\left(E_{\gamma_1}-E_{\gamma_2}\right)
\left(E_{\gamma_1}-E_{\gamma^\prime}\right)}\nonumber\\
& &-\frac{\e^{-\I
E_{\gamma_2}t}}{\left(E_{\gamma}-E_{\gamma_2}\right)^2\left(E_{\gamma_1}-E_{\gamma_2}\right)
\left(E_{\gamma_2}-E_{\gamma^\prime}\right)}+\frac{\e^{-\I
E_{\gamma^\prime}t}}{\left(E_{\gamma}-E_{\gamma^\prime}\right)^2\left(E_{\gamma_1}-E_{\gamma^\prime}\right)
\left(E_{\gamma_2}-E_{\gamma^\prime}\right)} \nonumber\\
& & \left. +(-\I t)\frac{\e^{-\I
E_{\gamma}t}}{\left(E_{\gamma}-E_{\gamma_1}\right)\left(E_{\gamma}-E_{\gamma_2}\right)
\left(E_{\gamma}-E_{\gamma^\prime}\right)}\right]
g_1^{\gamma\gamma_1}g_1^{\gamma_1\gamma_2}g_1^{\gamma_2\gamma}
g_1^{\gamma\gamma^\prime}\eta_{\gamma_1\gamma^\prime}\eta_{\gamma_2\gamma^\prime}.
\hskip 0.8cm\eeqa
\beqa\label{A4nnn-nc} {A}_4^{\gamma\gamma^\prime}(nnn,nc)&=
&\sum_{\gamma_1,\cdots,\gamma_{5}}\left[
\sum_{i=1}^{5}(-1)^{i-1}\frac{\e^{-\I E_{\gamma_i}
t}}{d_i(E[\gamma,4])}\right]
\left[\prod_{j=1}^4g_1^{\gamma_j\gamma_{j+1}}\right]\left(
\prod_{k=1}^{3}\eta_{\gamma_k\gamma_{k+2}}\right)
\eta_{\gamma_1\gamma_4}{\delta}_{\gamma_2\gamma_5}
\delta_{\gamma_1\gamma}\delta_{\gamma_{5}\gamma^\prime}
\nonumber\\
&=&\sum_{\gamma_1,\gamma_2}\left[\frac{\e^{-\I
E_{\gamma}t}}{\left(E_{\gamma}-E_{\gamma_1}\right)\left(E_{\gamma}-E_{\gamma_2}\right)
\left(E_{\gamma}-E_{\gamma^\prime}\right)^2}-\frac{\e^{-\I
E_{\gamma_1}t}}{\left(E_{\gamma}-E_{\gamma_1}\right)\left(E_{\gamma_1}-E_{\gamma_2}\right)
\left(E_{\gamma_1}-E_{\gamma^\prime}\right)^2}\right.\nonumber\\
& &+\frac{\e^{-\I
E_{\gamma_2}t}}{\left(E_{\gamma}-E_{\gamma_2}\right)\left(E_{\gamma_1}-E_{\gamma_2}\right)
\left(E_{\gamma_2}-E_{\gamma^\prime}\right)^2}-\frac{\e^{-\I
E_{\gamma^\prime}t}}{\left(E_{\gamma}-E_{\gamma^\prime}\right)
\left(E_{\gamma_1}-E_{\gamma^\prime}\right)
\left(E_{\gamma_2}-E_{\gamma^\prime}\right)^2}\nonumber\\
& & -\frac{\e^{-\I
E_{\gamma^\prime}t}}{\left(E_{\gamma}-E_{\gamma^\prime}\right)
\left(E_{\gamma_1}-E_{\gamma^\prime}\right)^2
\left(E_{\gamma_2}-E_{\gamma^\prime}\right)}-\frac{\e^{-\I
E_{\gamma^\prime}t}}{\left(E_{\gamma}-E_{\gamma^\prime}\right)^2
\left(E_{\gamma_1}-E_{\gamma^\prime}\right)
\left(E_{\gamma_2}-E_{\gamma^\prime}\right)}\nonumber\\
 & &\left. -(-\I t)\frac{\e^{-\I
E_{\gamma^\prime}t}}{\left(E_{\gamma}-E_{\gamma^\prime}\right)\left(E_{\gamma_1}-E_{\gamma^\prime}\right)
\left(E_{\gamma_2}-E_{\gamma^\prime}\right)}\right]
g_1^{\gamma\gamma^\prime}g_1^{\gamma^\prime\gamma_1}g_1^{\gamma_1\gamma_2}
g_1^{\gamma_2\gamma^\prime}\eta_{\gamma\gamma_1}\eta_{\gamma\gamma_2}.
\hskip 0.8cm\eeqa
while the fourth term has the third decomposition,
that is \beqa\label{A4nnn-nn-c}
{A}_4^{\gamma\gamma^\prime}(nnn,nn,c)&=
&\sum_{\gamma_1,\cdots,\gamma_{5}}\left[
\sum_{i=1}^{5}(-1)^{i-1}\frac{\e^{-\I E_{\gamma_i}
t}}{d_i(E[\gamma,4])}\right]
\left[\prod_{j=1}^4g_1^{\gamma_j\gamma_{j+1}}\right]\left(
\prod_{k=1}^{3}\eta_{\gamma_k\gamma_{k+2}}\right)
\delta_{\gamma_1\gamma}\delta_{\gamma_{5}\gamma^\prime}
\delta_{\gamma\gamma^\prime}\nonumber\\
&=&\sum_{\gamma_1,\gamma_2,\gamma_3}\left[-\frac{\e^{-\I
E_{\gamma}t}}{\left(E_{\gamma}-E_{\gamma_1}\right)\left(E_{\gamma}-E_{\gamma_2}\right)
\left(E_{\gamma}-E_{\gamma_3}\right)^2}\right.\nonumber\\
& &-\frac{\e^{-\I
E_{\gamma}t}}{\left(E_{\gamma}-E_{\gamma_1}\right)\left(E_{\gamma}-E_{\gamma_2}\right)^2
\left(E_{\gamma}-E_{\gamma_3}\right)}\nonumber\\
& &-\frac{\e^{-\I
E_{\gamma}t}}{\left(E_{\gamma}-E_{\gamma_1}\right)^2\left(E_{\gamma}-E_{\gamma_2}\right)
\left(E_{\gamma}-E_{\gamma_3}\right)}\nonumber\\
& &+\frac{\e^{-\I
E_{\gamma_1}t}}{\left(E_{\gamma}-E_{\gamma_1}\right)^2\left(E_{\gamma_1}-E_{\gamma_2}\right)
\left(E_{\gamma_1}-E_{\gamma_3}\right)}\nonumber\\
& &-\frac{\e^{-\I
E_{\gamma_2}t}}{\left(E_{\gamma}-E_{\gamma_2}\right)^2\left(E_{\gamma_1}-E_{\gamma_2}\right)
\left(E_{\gamma_2}-E_{\gamma_3}\right)}\nonumber\\
& &+\frac{\e^{-\I
E_{\gamma_3}t}}{\left(E_{\gamma}-E_{\gamma_3}\right)^2\left(E_{\gamma_1}-E_{\gamma_3}\right)
\left(E_{\gamma_2}-E_{\gamma_3}\right)}\nonumber\\
& &\left. +(-\I t)\frac{\e^{-\I
E_{\gamma}t}}{\left(E_{\gamma}-E_{\gamma_1}\right)\left(E_{\gamma}-E_{\gamma_2}\right)
\left(E_{\gamma}-E_{\gamma_3}\right)}\right]\nonumber\\
& & \times
g_1^{\gamma\gamma_1}g_1^{\gamma_1\gamma_2}g_1^{\gamma_2\gamma_3}
g_1^{\gamma_3\gamma}\eta_{\gamma\gamma_2} \eta_{\gamma_1\gamma_3}
{\delta}_{\gamma\gamma^\prime}. \eeqa
\beqa\label{A4nnn-nn-n} {A}_4^{\gamma\gamma^\prime}(nnn,nn,n)&=
&\sum_{\gamma_1,\cdots,\gamma_{5}}\left[
\sum_{i=1}^{5}(-1)^{i-1}\frac{\e^{-\I E_{\gamma_i}
t}}{d_i(E[\gamma,4])}\right]
\left[\prod_{j=1}^4g_1^{\gamma_j\gamma_{j+1}}\right]\left(
\prod_{k=1}^{3}\eta_{\gamma_k\gamma_{k+2}}\right)
\delta_{\gamma_1\gamma}\delta_{\gamma_{5}\gamma^\prime}
\eta_{\gamma\gamma^\prime}\nonumber\\
&=&\sum_{\gamma_1,\gamma_2,\gamma_3}\left[\frac{\e^{-\I
E_{\gamma}t}}{\left(E_{\gamma}-E_{\gamma_1}\right)\left(E_{\gamma}-E_{\gamma_2}\right)
\left(E_{\gamma}-E_{\gamma_3}\right)\left(E_{\gamma}-E_{\gamma^\prime}\right)}\right.\nonumber\\
& &-\frac{\e^{-\I
E_{\gamma_1}t}}{\left(E_{\gamma}-E_{\gamma_1}\right)\left(E_{\gamma_1}-E_{\gamma_2}\right)
\left(E_{\gamma_1}-E_{\gamma_3}\right)\left(E_{\gamma_1}-E_{\gamma^\prime}\right)}\nonumber\\
& &+\frac{\e^{-\I
E_{\gamma_2}t}}{\left(E_{\gamma}-E_{\gamma_2}\right)\left(E_{\gamma_1}-E_{\gamma_2}\right)
\left(E_{\gamma_2}-E_{\gamma_3}\right)\left(E_{\gamma_2}-E_{\gamma^\prime}\right)}\nonumber\\
& &-\frac{\e^{-\I
E_{\gamma_3}t}}{\left(E_{\gamma}-E_{\gamma_3}\right)\left(E_{\gamma_1}-E_{\gamma_3}\right)
\left(E_{\gamma_2}-E_{\gamma_3}\right)\left(E_{\gamma_3}-E_{\gamma^\prime}\right)}\nonumber\\
& &\left. +\frac{\e^{-\I
E_{\gamma^\prime}t}}{\left(E_{\gamma}-E_{\gamma^\prime}\right)\left(E_{\gamma_1}-E_{\gamma^\prime}\right)
\left(E_{\gamma_2}-E_{\gamma^\prime}\right)\left(E_{\gamma_3}-E_{\gamma^\prime}\right)}\right]\nonumber\\[7pt]
& & \times
g_1^{\gamma\gamma_1}g_1^{\gamma_1\gamma_2}g_1^{\gamma_2\gamma_3}
g_1^{\gamma_3\gamma^\prime}\eta_{\gamma\gamma_2}\eta_{\gamma\gamma_3}\eta_{\gamma\gamma^\prime}
\eta_{\gamma_1\gamma_3}\eta_{\gamma_1\gamma^\prime}
\eta_{\gamma_2\gamma^\prime}. \eeqa

Now, all 15 contractions and/or anti-contractions in the fourth
order approximation have been calculated out.

In order to add the contributions from the fourth order
approximation to the improved forms of lower order perturbed
solutions, we first decompose $A_4^{\gamma\gamma^\prime}$, which is
a summation of all above terms, into the three parts according to
$\e^{-\I E_{\gamma_i}t}, (-\I t) \e^{-\I E_{\gamma_i}t} $ and $(-\I
t)^2\e^{-\I E_{\gamma_i}t}/2$, that is \beq\label{A4dto}
A_4^{\gamma\gamma^\prime}=A_4^{\gamma\gamma^\prime}(\e)+A_4^{\gamma\gamma^\prime}(t\e)
+A_4^{\gamma\gamma^\prime}(t^2\e).\eeq Secondly, we decompose its
every term into three parts according to $\e^{-\I E_{\gamma}t},
\e^{-\I E_{\gamma_1}t}$ ($\sum_{\gamma_1}\e^{-\I E_{\gamma_1}t}$)
and $\e^{-\I E_{\gamma^\prime}t}$, that is \beqa
A_4^{\gamma\gamma^\prime}(\e)&=&A_4^{\gamma\gamma^\prime}(\e^{-\I
E_{\gamma}t})+A_4^{\gamma\gamma^\prime}(\e^{-\I
E_{\gamma_1}t})+A_4^{\gamma\gamma^\prime}(\e^{-\I E_{\gamma^\prime}t}),\\
A_4^{\gamma\gamma^\prime}(t\e)&=&A_4^{\gamma\gamma^\prime}(t\e^{-\I
E_{\gamma}t})+A_4^{\gamma\gamma^\prime}(t\e^{-\I
E_{\gamma_1}t})+A_4^{\gamma\gamma^\prime}(t\e^{-\I E_{\gamma^\prime}t}),\\
A_4^{\gamma\gamma^\prime}(t^2\e)&=&A_4^{\gamma\gamma^\prime}(t^2\e^{-\I
E_{\gamma}t})+A_4^{\gamma\gamma^\prime}(t^2\e^{-\I
E_{\gamma_1}t})+A_4^{\gamma\gamma^\prime}(t^2\e^{-\I
E_{\gamma^\prime}t}). \eeqa Finally, we again decompose every term
in above equations into the diagonal and non-diagonal parts about
$\gamma$ and $\gamma^\prime$, that is \beqa
A_4^{\gamma\gamma^\prime}(\e^{-\I
E_{\gamma_i}t})&=&A_4^{\gamma\gamma^\prime}(\e^{-\I
E_{\gamma_i}t};{\rm D})+A_4^{\gamma\gamma^\prime}(\e^{-\I
E_{\gamma_i}t};{\rm N}),\\
A_4^{\gamma\gamma^\prime}(t\e^{-\I
E_{\gamma_i}t})&=&A_4^{\gamma\gamma^\prime}(t\e^{-\I
E_{\gamma_i}t};{\rm D})+A_4^{\gamma\gamma^\prime}(t\e^{-\I
E_{\gamma_i}t};{\rm N}), \\
A_4^{\gamma\gamma^\prime}(t^2\e^{-\I
E_{\gamma_i}t})&=&A_4^{\gamma\gamma^\prime}(t^2\e^{-\I
E_{\gamma_i}t};{\rm D})+A_4^{\gamma\gamma^\prime}(t^2\e^{-\I
E_{\gamma_i}t};{\rm N}).  \eeqa where $E_{\gamma_i}$ takes $
E_{\gamma}, E_{\gamma_1}$ and $E_{\gamma^\prime}$.

If we do not concern the improved forms of perturbed solutions equal
to or more than the fourth order one, we only need to write down the
second and third terms in eq.(\ref{A4dto}) and calculate their
diagonal and nondiagonal parts respectively. Based on the calculated
results above, it is easy to obtain \beqa \label{A4gammaD}
A_4^{\gamma\gamma^\prime}\left(t\e^{-\I E_\gamma t};{\rm
D}\right)&=& \left(-\I t\right)\e^{-\I E_\gamma
t}\left[\sum_{\gamma_1}\frac{-2
\left|g_1^{\gamma\gamma_1}\right|^4}{\left(E_{\gamma}-E_{\gamma_1}\right)^3}
-\sum_{\gamma_1,\gamma_2}\frac{\left|g_1^{\gamma\gamma_1}\right|^2
\left|g_1^{\gamma\gamma_2}\right|^2\eta_{\gamma_1\gamma_2}}
{\left(E_{\gamma}-E_{\gamma_1}\right)\left(E_{\gamma}-E_{\gamma_2}\right)^2}\right.
\nonumber\\
& &
-\sum_{\gamma_1,\gamma_2}\frac{\left|g_1^{\gamma\gamma_1}\right|^2
\left|g_1^{\gamma\gamma_2}\right|^2\eta_{\gamma_1\gamma_2}}
{\left(E_{\gamma}-E_{\gamma_1}\right)^2\left(E_{\gamma}-E_{\gamma_2}\right)}
+\sum_{\gamma_1,\gamma_2}\frac{\left|g_1^{\gamma\gamma_1}\right|^2
\left|g_1^{\gamma_1\gamma_2}\right|^2\eta_{\gamma\gamma_2}}
{\left(E_{\gamma}-E_{\gamma_1}\right)^2\left(E_{\gamma}-E_{\gamma_2}\right)}\nonumber\\
&
&\left.+\sum_{\gamma_1,\gamma_2,\gamma_3}\frac{g_1^{\gamma\gamma_1}
g_1^{\gamma_1\gamma_2}g_1^{\gamma_2\gamma_3}g_1^{\gamma_3\gamma}\eta_{\gamma\gamma_2}\eta_{\gamma_1\gamma_3}}
{\left(E_{\gamma}-E_{\gamma_1}\right)\left(E_{\gamma}-E_{\gamma_2}\right)
\left(E_{\gamma}-E_{\gamma_3}\right)}\right]\delta_{\gamma\gamma^\prime}.\eeqa
Substituting the relation
$\eta_{\beta_1\beta_2}=1-\delta_{\beta_1\beta_2}$, using the
technology of index exchanging and introducing the definitions of
so-called revision energy $G_\gamma^{(a)}$: \beqa
{G}^{(2)}_\gamma&=&\sum_{\gamma_1}\frac{\left|g_1^{\gamma\gamma_1}\right|^2}
{E_{\gamma}-E_{\gamma_1}}\eeqa
\beqa G^{(4)}_\gamma &=
&\sum_{\gamma_1,\gamma_2,\gamma_3}\frac{g_1^{\gamma\gamma_1}
g_1^{\gamma_1\gamma_2}g_1^{\gamma_2\gamma_3}g_1^{\gamma_3\gamma}\eta_{\gamma\gamma_2}}
{\left(E_{\gamma}-E_{\gamma_1}\right)\left(E_{\gamma}-E_{\gamma_2}\right)
\left(E_{\gamma}-E_{\gamma_3}\right)}-\sum_{\gamma_1,\gamma_2}
\frac{g_1^{\gamma\gamma_1}g_1^{\gamma_1\gamma}g_1^{\gamma\gamma_2}g_1^{\gamma_2\gamma}}
{\left(E_{\gamma}-E_{\gamma_1}\right)^2\left(E_{\gamma}-E_{\gamma_2}\right)},
\eeqa we can simplify eq.(ref{A4gammaD}) to the following concise
form:
\beqa A_4^{\gamma\gamma^\prime}\left(t\e^{-\I E_\gamma t};{\rm
D}\right)&=&-\left(-\I t\right)\e^{-\I E_\gamma
t}\left[\sum_{\gamma_1}\frac{
G^{(2)}_\gamma}{\left(E_{\gamma}-E_{\gamma_1}\right)^2}
\left|g_1^{\gamma\gamma_1}\right|^2-G^{(4)}_\gamma\right]\delta_{\gamma\gamma^\prime}.
\eeqa

Similar calculation and simplification lead to \beqa
A_4^{\gamma\gamma^\prime}\left(t\e^{-\I E_{\gamma} t}; {\rm
N}\right)=(-\I t)\e^{-\I E_{\gamma} t}\left[
\frac{{G}^{(3)}_{\gamma}g_1^{\gamma\gamma^\prime}}{\left(E_{\gamma}-E_{\gamma^\prime}\right)}
+\sum_{\gamma_1} \frac{{G}^{(2)}_{\gamma}}
{\left(E_{\gamma}-E_{\gamma_1}\right)\left(E_{\gamma}-E_{\gamma^\prime}\right)}
\right], \eeqa where \beqa G^{(3)}_\gamma &=
&\sum_{\gamma_1,\gamma_2}\frac{g_1^{\gamma\gamma_1}
g_1^{\gamma_1\gamma_2}g_1^{\gamma_2\gamma}}
{\left(E_{\gamma}-E_{\gamma_1}\right)\left(E_{\gamma}-E_{\gamma_2}\right)}.
\eeqa For saving the space, the corresponding detail is omitted. In
fact, it is not difficult, but it is necessary to be careful enough,
specially in the cases of higher order approximations.

In the same way, we can obtain: \beqa
A_4^{\gamma\gamma^\prime}\left(t\e^{-\I E_{\gamma_1} t}; {\rm D}
\right)&=&\left(-\I
t\right)\sum_{\gamma_1}\frac{G^{(2)}_{\gamma_1}\e^{-\I E_{\gamma_1}
t}}{\left(E_{\gamma}-E_{\gamma_1}\right)^2}
\left|g_1^{\gamma\gamma_1}\right|^2\delta_{\gamma\gamma^\prime},\\
A_4^{\gamma\gamma^\prime}\left(t\e^{-\I E_{\gamma_1} t} ; {\rm
N}\right)&=&-(-\I t)\sum_{\gamma_1}
\frac{{G}^{(2)}_{\gamma_1}\e^{-\I E_{\gamma_1} t}}
{\left(E_{\gamma}-E_{\gamma_1}\right)\left(E_{\gamma_1}-E_{\gamma^\prime}\right)}
g_1^{\gamma\gamma_1}g_1^{\gamma_1\gamma^\prime}\eta_{\gamma\gamma^\prime}.
\eeqa
\beqa A_4^{\gamma\gamma^\prime}\left(t\e^{-\I E_{\gamma^\prime}t};
{\rm D}\right)&=&0,\\
A_4^{\gamma\gamma^\prime}\left(t\e^{-\I E_{\gamma^\prime}t}; {\rm
N}\right)&=&-(-\I t)\e^{-\I E_{\gamma^\prime}
t}\left[\sum_{\gamma_1}
\frac{{G}^{(3)}_{\gamma^\prime}}{\left(E_{\gamma}-E_{\gamma^\prime}\right)}
g_1^{\gamma\gamma^\prime}\right.\nonumber\\
& &\left. +\sum_{\gamma_1} \frac{{G}^{(2)}_{\gamma^\prime}}
{\left(E_{\gamma}-E_{\gamma^\prime}\right)\left(E_{\gamma_1}-E_{\gamma^\prime}\right)}
g_1^{\gamma\gamma_1}g_1^{\gamma_1\gamma^\prime}\eta_{\gamma\gamma^\prime}\right].
\eeqa

For $t^2\e$ terms, only one term is nonzero, that is \beqa
A_4^{\gamma\gamma^\prime}\left(t^2\e\right)
&=&A_4^{\gamma\gamma^\prime}\left(t^2\e^{-\I E_{\gamma} t}; D\right)
=\frac{(-\I G^{(2)}_\gamma t)^2}{2!}\e^{-\I E_{\gamma} t}. \eeqa
since \beqa A_4^{\gamma\gamma^\prime}\left(t^2\e^{-\I E_{\gamma_1}
t};{\rm D}\right)&=&A_4^{\gamma\gamma^\prime}\left(t^2\e^{-\I
E_{\gamma^\prime} t};{\rm D}\right)=0,\\
A_4^{\gamma\gamma^\prime}\left(t^2\e^{-\I E_{\gamma} t};{\rm
N}\right)&=&A_4^{\gamma\gamma^\prime}\left(t^2\e^{-\I E_{\gamma_1}
t};{\rm D}\right)=A_4^{\gamma\gamma^\prime}\left(t^2\e^{-\I
E_{\gamma^\prime} t};{\rm D}\right)=0. \eeqa

We can see that these terms can be merged with the lower order
approximation to obtain the improved forms of perturbed solutions.

\subsection{l=5 case}

Now let we consider the case of the fifth order approximation
($l=5$). From eq.(\ref{gpd}) it follows that the first
decompositions of $g$-product have $2^4=16$ terms. They can be
divided into 5 group \beq A_5^{\gamma\gamma^\prime}= \sum_{i=0}^4
\mathcal{A}_5^{\gamma\gamma^\prime}(i;\eta),\eeq where $i$ indicates
the number of $\eta$ functions. Obviously \beq
\mathcal{A}_5^{\gamma\gamma^\prime}(0;\eta)={A}_5^{\gamma\gamma^\prime}(cccc),\eeq
\beqa
\mathcal{A}_5^{\gamma\gamma^\prime}(1;\eta)={A}_5^{\gamma\gamma^\prime}(cccn)
+{A}_5^{\gamma\gamma^\prime}(ccnc)+{A}_5^{\gamma\gamma^\prime}(cncc)
+{A}_5^{\gamma\gamma^\prime}(nccc),\eeqa \beqa
\mathcal{A}_5^{\gamma\gamma^\prime}(2;\eta)&=&{A}_5^{\gamma\gamma^\prime}(ccnn)
+{A}_5^{\gamma\gamma^\prime}(cncn)+{A}_5^{\gamma\gamma^\prime}(cnnc)\nonumber\\
& & +{A}_5^{\gamma\gamma^\prime}(nccn)
+{A}_5^{\gamma\gamma^\prime}(ncnc)
+{A}_5^{\gamma\gamma^\prime}(nncc),\eeqa \beqa
\mathcal{A}_5^{\gamma\gamma^\prime}(3;\eta)&=&{A}_5^{\gamma\gamma^\prime}(cnnn)
+{A}_5^{\gamma\gamma^\prime}(ncnn)
+{A}_5^{\gamma\gamma^\prime}(nncn)
+{A}_5^{\gamma\gamma^\prime}(nnnc), \eeqa \beq
\mathcal{A}_5^{\gamma\gamma^\prime}(4;\eta)={A}_5^{\gamma\gamma^\prime}(nnnn).\eeq
Here, we have used the notations stated in Sec. \ref{sec6}.

By calculation, we obtain the
$\mathcal{A}^{\gamma\gamma^\prime}_5(0,\eta)$ and every term of
$\mathcal{A}^{\gamma\gamma^\prime}_5(1,\eta)$ have only nontrivial
first contractions and/or anti-contractions. But, we can find that
every term of $\mathcal{A}^{\gamma\gamma^\prime}_5(2,\eta)$ can have
one nontrivial second or third or fourth contraction or
anti-contraction, that is \beqa
A^{\gamma\gamma^\prime}_5(ccnn)&=&A^{\gamma\gamma^\prime}_5(ccnn,kkc)+A^{\gamma\gamma^\prime}_5(ccnn,kkn),\\
A^{\gamma\gamma^\prime}_5(cncn)&=&A^{\gamma\gamma^\prime}_5(cncn,kc)+A^{\gamma\gamma^\prime}_5(cncn,kc),\\
A^{\gamma\gamma^\prime}_5(cnnc)&=&A^{\gamma\gamma^\prime}_5(cnnc,kck)+A^{\gamma\gamma^\prime}_5(cnnc,kck),\\
A^{\gamma\gamma^\prime}_5(nccn)&=&A^{\gamma\gamma^\prime}_5(nccn,c)+A^{\gamma\gamma^\prime}_5(nccn,n),\\
A^{\gamma\gamma^\prime}_5(ncnc)&=&A^{\gamma\gamma^\prime}_5(ncnc,ck)+A^{\gamma\gamma^\prime}_5(ncnc,nk),\\
A^{\gamma\gamma^\prime}_5(nncc)&=&A^{\gamma\gamma^\prime}_5(nncc,ckk)+A^{\gamma\gamma^\prime}_5(nncc,nkk).
\eeqa Similarly, every term of
$\mathcal{A}^{\gamma\gamma^\prime}_5(3,\eta)$ can have two higher
order contractions and/or anti-contraction: \beqa
A^{\gamma\gamma^\prime}_5(cnnn)&=&A^{\gamma\gamma^\prime}_5(cnnn,kcc)
+A^{\gamma\gamma^\prime}_5(cnnn,kcn)\nonumber\\
& &+A^{\gamma\gamma^\prime}_5(cnnn,knc)
+A^{\gamma\gamma^\prime}_5(cnnn,knn),\\
A^{\gamma\gamma^\prime}_5(ncnn)&=&A^{\gamma\gamma^\prime}_5(ncnn,kkc,ck)
+A^{\gamma\gamma^\prime}_5(ncnn,kkn,ck)\nonumber\\
& & +A^{\gamma\gamma^\prime}_5(ncnn,kkc,nk)+A^{\gamma\gamma^\prime}_5(ncnn,kkn,nk),\\
A^{\gamma\gamma^\prime}_5(nncn)&=&A^{\gamma\gamma^\prime}_5(nncn,ckk,kc)
+A^{\gamma\gamma^\prime}_5(nncn,ckk,kn)\nonumber\\
& &+A^{\gamma\gamma^\prime}_5(nncn,nkk,kc)+A^{\gamma\gamma^\prime}_5(nncn,nkk,kn),\\
A^{\gamma\gamma^\prime}_5(nnnc)&=&A^{\gamma\gamma^\prime}_5(nnnc,cck)
+A^{\gamma\gamma^\prime}_5(nnnc,cnk)\nonumber\\
& &+A^{\gamma\gamma^\prime}_5(nnnc,nck)
+A^{\gamma\gamma^\prime}_5(nnnc,nnk).\eeqa Moreover, their last
terms, with two higher order anti-contractions, can have one
nontrivial more higher contraction or anti-contraction: \beqa
A^{\gamma\gamma^\prime}_5(cnnn,knn)&=&A^{\gamma\gamma^\prime}_5(cnnn,knn,kc)
+A^{\gamma\gamma^\prime}_5(cnnn,knn.kn),\\
A^{\gamma\gamma^\prime}_5(ncnn,kkn,nk)&=&A^{\gamma\gamma^\prime}_5(ncnn,kkn,nk,c)
+A^{\gamma\gamma^\prime}_5(ncnn,kkn,nk,n),\\
A^{\gamma\gamma^\prime}_5(nncn,nkk,kn)&=&A^{\gamma\gamma^\prime}_5(nncn,nkk,kn,c)
+A^{\gamma\gamma^\prime}_5(nncn,nkk,kn,n),\\
A^{\gamma\gamma^\prime}_5(nnnc,nnk)&=&A^{\gamma\gamma^\prime}_5(nnnc,nnk,ck)
+A^{\gamma\gamma^\prime}_5(nnnc,nnk,nk). \eeqa In the case of
$A^{\gamma\gamma^\prime}_5(nnnn)$, there are three second
decompositions that result in \beqa \hskip -0.3in
A^{\gamma\gamma^\prime}_5(nnnn)&=&A^{\gamma\gamma^\prime}_5(nnnn,ccc)
+A^{\gamma\gamma^\prime}_5(nnnn,ccn)+A^{\gamma\gamma^\prime}_5(nnnn,cnc)
+A^{\gamma\gamma^\prime}_5(nnnn,ncc)\nonumber\\
&
&+A^{\gamma\gamma^\prime}_5(nnnn,cnn)+A^{\gamma\gamma^\prime}_5(nnnn,ncn)
+A^{\gamma\gamma^\prime}_5(nnnn,nnc)+A^{\gamma\gamma^\prime}_5(nnnn,nnn).
\eeqa In the above expression, from the fifth term to the seventh
term have the third- or fourth- contraction and anti-contraction,
the eighth term has two third contractions and anti-contractions:
\beqa
A^{\gamma\gamma^\prime}_5(nnnn,cnn)&=&A^{\gamma\gamma^\prime}_5(nnnn,cnn,kc)
+A^{\gamma\gamma^\prime}_5(nnnn,cnn,kn),\\
A^{\gamma\gamma^\prime}_5(nnnn,ncn)&=&A^{\gamma\gamma^\prime}_5(nnnn,ncn,c)
+A^{\gamma\gamma^\prime}_5(nnnn,ncn,n),\\
A^{\gamma\gamma^\prime}_5(nnnn,nnc)&=&A^{\gamma\gamma^\prime}_5(nnnn,nnc,ck)
+A^{\gamma\gamma^\prime}_5(nnnn,nnc,nk),\\
A^{\gamma\gamma^\prime}_5(nnnn,nnn)&=&A^{\gamma\gamma^\prime}_5(nnnn,nnn,cc)
+A^{\gamma\gamma^\prime}_5(nnnn,nnn,cn)\nonumber\\
& &
+A^{\gamma\gamma^\prime}_5(nnnn,nnn,nc)+A^{\gamma\gamma^\prime}_5(nnnn,nnn,nn).
\eeqa In addition, $A^{\gamma\gamma^\prime}_5(nnnn,nnn,nn)$ consists
of the fourth contraction and anti-contraction \beq
A^{\gamma\gamma^\prime}_5(nnnn,nnn,nn)=A^{\gamma\gamma^\prime}_5(nnnn,nnn,nn,c)
+A^{\gamma\gamma^\prime}_5(nnnn,nnn,nn,n).\eeq According to above
analysis, we obtain the contribution from the five order
approximation made of $52$ terms after finding out all of
contractions and anti-contractions.

Just like we have done in the $l=4$ case, we decompose
\beq\label{A5dto}
A_5^{\gamma\gamma^\prime}=A_5^{\gamma\gamma^\prime}(\e)+A_5^{\gamma\gamma^\prime}(t\e)
+A_5^{\gamma\gamma^\prime}(t^2\e),\eeq where \beqa
A_5^{\gamma\gamma^\prime}(\e)&=&A_5^{\gamma\gamma^\prime}(\e^{-\I
E_{\gamma}t})+A_5^{\gamma\gamma^\prime}(\e^{-\I
E_{\gamma_1}t})+A_5^{\gamma\gamma^\prime}(\e^{-\I
E_{\gamma_2}t})+A_5^{\gamma\gamma^\prime}(\e^{-\I E_{\gamma^\prime}t}),\\
A_4^{\gamma\gamma^\prime}(t\e)&=&A_5^{\gamma\gamma^\prime}(t\e^{-\I
E_{\gamma}t})+A_5^{\gamma\gamma^\prime}(t\e^{-\I
E_{\gamma_1}t})+A_5^{\gamma\gamma^\prime}(t\e^{-\I
E_{\gamma_2}t})+A_5^{\gamma\gamma^\prime}(t\e^{-\I E_{\gamma^\prime}t}),\\
A_5^{\gamma\gamma^\prime}(t^2\e)&=&A_5^{\gamma\gamma^\prime}(t^2\e^{-\I
E_{\gamma}t})+A_5^{\gamma\gamma^\prime}(t^2\e^{-\I
E_{\gamma_1}t})+A_5^{\gamma\gamma^\prime}(t^2\e^{-\I
E_{\gamma_2}t})+A_5^{\gamma\gamma^\prime}(t^2\e^{-\I
E_{\gamma^\prime}t}). \eeqa While, every term in above equations has
its diagonal and non-diagonal parts about $\gamma$ and
$\gamma^\prime$, that is \beqa A_5^{\gamma\gamma^\prime}(\e^{-\I
E_{\gamma_i}t})&=&A_5^{\gamma\gamma^\prime}(\e^{-\I
E_{\gamma_i}t};{\rm D})+A_5^{\gamma\gamma^\prime}(\e^{-\I
E_{\gamma_i}t};{\rm N}),\\
A_5^{\gamma\gamma^\prime}(t\e^{-\I
E_{\gamma_i}t})&=&A_5^{\gamma\gamma^\prime}(t\e^{-\I
E_{\gamma_i}t};{\rm D})+A_5^{\gamma\gamma^\prime}(t\e^{-\I
E_{\gamma_i}t};{\rm N}), \\
A_5^{\gamma\gamma^\prime}(t^2\e^{-\I
E_{\gamma_i}t})&=&A_5^{\gamma\gamma^\prime}(t^2\e^{-\I
E_{\gamma_i}t};{\rm D})+A_5^{\gamma\gamma^\prime}(t^2\e^{-\I
E_{\gamma_i}t};{\rm N}).  \eeqa where $E_{\gamma_i}$ takes $
E_{\gamma}, E_{\gamma_1}, E_{\gamma_2}$ and $E_{\gamma^\prime}$.

If we do not concern the improved forms of perturbed solution more
than the fourth order one, we only need to write down the second and
third terms in eq.(\ref{A5dto}). In the following, we respectively
calculate them term by term, and put the second and third terms in
eq.(\ref{A5dto}) together as $A_5^{\gamma\gamma^\prime}(t\e,t^2\e)$.
\beqa {A}_5^{\gamma\gamma^\prime}(cccc;t\e,t^2\e)&=&\left[-(-\I
t)\frac{3\e^{-\I E_{\gamma}t}+3\e^{-\I
E_{\gamma^\prime}t}}{\left(E_\gamma-E_{\gamma^\prime}\right)^4}
+\frac{(-\I t)^2}{2}\frac{\e^{-\I
E_{\gamma}t}}{\left(E_\gamma-E_{\gamma^\prime}\right)^3}\right.\nonumber\\
& &\left.-\frac{(-\I t)^2}{2}\frac{\e^{-\I
E_{\gamma^\prime}t}}{\left(E_\gamma-E_{\gamma^\prime}\right)^3}\right]
\left|g_1^{\gamma\gamma^\prime}\right|^4g_1^{\gamma\gamma^\prime}.\eeqa
\beqa {A}_5^{\gamma\gamma^\prime}(cccn;t\e,t^2\e)&=&\sum_{\gamma_1}
\left[-(-\I t)\frac{2\e^{-\I
E_{\gamma}t}}{\left(E_\gamma-E_{\gamma_1}\right)^3
\left(E_\gamma-E_{\gamma^\prime}\right)}-(-\I t)\frac{\e^{-\I
E_{\gamma}t}}{\left(E_\gamma-E_{\gamma_1}\right)^2
\left(E_\gamma-E_{\gamma^\prime}\right)^2}\right.\nonumber\\
& &\left.-(-\I t)\frac{\e^{-\I
E_{\gamma_1}t}}{\left(E_\gamma-E_{\gamma_1}\right)^3
\left(E_{\gamma_1}-E_{\gamma^\prime}\right)}+\frac{(-\I
t)^2}{2!}\frac{\e^{-\I
E_{\gamma}t}}{\left(E_\gamma-E_{\gamma_1}\right)^2\left(E_\gamma-E_{\gamma^\prime}\right)}\right]\nonumber\\
& & \times
\left|g_1^{\gamma\gamma_1}\right|^4g_1^{\gamma\gamma^\prime}\eta_{\gamma_1\gamma^\prime}.\eeqa
\beqa {A}_5^{\gamma\gamma^\prime}(ccnc;t\e,t^2\e)&=&\sum_{\gamma_1}
\left[(-\I t)\frac{\e^{-\I
E_{\gamma}t}}{\left(E_\gamma-E_{\gamma_1}\right)
\left(E_\gamma-E_{\gamma^\prime}\right)^3}-(-\I t)\frac{\e^{-\I
E_{\gamma^\prime}t}}{\left(E_\gamma-E_{\gamma^\prime}\right)^2
\left(E_{\gamma_1}-E_{\gamma^\prime}\right)^2}\right.\nonumber\\
& &\left.-(-\I t)\frac{2\e^{-\I
E_{\gamma^\prime}t}}{\left(E_\gamma-E_{\gamma^\prime}\right)^3
\left(E_{\gamma_1}-E_{\gamma^\prime}\right)}-\frac{(-\I
t)^2}{2!}\frac{\e^{-\I
E_{\gamma^\prime}t}}{\left(E_\gamma-E_{\gamma^\prime}\right)^2
\left(E_{\gamma_1}-E_{\gamma^\prime}\right)}\right]\nonumber\\
& & \times \left|g_1^{\gamma_1\gamma^\prime}\right|^2
\left|g_1^{\gamma\gamma^\prime}\right|^2g_1^{\gamma\gamma^\prime}\eta_{\gamma\gamma_1}.\eeqa
\beqa {A}_5^{\gamma\gamma^\prime}(cncc;t\e,t^2\e)&=&\sum_{\gamma_1}
\left[-(-\I t)\frac{2\e^{-\I
E_{\gamma}t}}{\left(E_\gamma-E_{\gamma_1}\right)
\left(E_\gamma-E_{\gamma^\prime}\right)^3}-(-\I t)\frac{\e^{-\I
E_{\gamma}t}}{\left(E_\gamma-E_{\gamma_1}\right)^2
\left(E_{\gamma}-E_{\gamma^\prime}\right)^2}\right.\nonumber\\
& &\left.+(-\I t)\frac{\e^{-\I
E_{\gamma^\prime}t}}{\left(E_\gamma-E_{\gamma^\prime}\right)^3
\left(E_{\gamma_1}-E_{\gamma^\prime}\right)}+\frac{(-\I
t)^2}{2!}\frac{\e^{-\I
E_{\gamma}t}}{\left(E_\gamma-E_{\gamma_1}\right)
\left(E_{\gamma}-E_{\gamma^\prime}\right)^2}\right]\nonumber\\
& & \times \left|g_1^{\gamma\gamma_1}\right|^2
\left|g_1^{\gamma\gamma^\prime}\right|^2g_1^{\gamma\gamma^\prime}\eta_{\gamma_1\gamma^\prime}.\eeqa
\beqa {A}_5^{\gamma\gamma^\prime}(nccc;t\e,t^2\e)&=&\sum_{\gamma_1}
\left[-(-\I t)\frac{\e^{-\I
E_{\gamma_1}t}}{\left(E_\gamma-E_{\gamma_1}\right)
\left(E_{\gamma_1}-E_{\gamma^\prime}\right)^3}-(-\I t)\frac{2\e^{-\I
E_{\gamma^\prime}t}}{\left(E_\gamma-E_{\gamma^\prime}\right)
\left(E_{\gamma_1}-E_{\gamma^\prime}\right)^3}\right.\nonumber\\
& &\left.-(-\I t)\frac{\e^{-\I
E_{\gamma^\prime}t}}{\left(E_\gamma-E_{\gamma^\prime}\right)^2
\left(E_{\gamma_1}-E_{\gamma^\prime}\right)^2}-\frac{(-\I
t)^2}{2!}\frac{\e^{-\I
E_{\gamma^\prime}t}}{\left(E_\gamma-E_{\gamma^\prime}\right)
\left(E_{\gamma_1}-E_{\gamma^\prime}\right)^2}\right]\nonumber\\
& & \times
\left|g_1^{\gamma_1\gamma^\prime}\right|^4g_1^{\gamma\gamma^\prime}\eta_{\gamma\gamma_1}.\eeqa
\beqa & &
{A}_5^{\gamma\gamma^\prime}(ccnn,kkc;t\e,t^2\e)\nonumber\\
& & \quad =\sum_{\gamma_1,\gamma_2} \left[-(-\I t)\frac{\e^{-\I
E_{\gamma}t}}{\left(E_\gamma-E_{\gamma_1}\right)^2
\left(E_{\gamma}-E_{\gamma_2}\right)^2}-(-\I t)\frac{2\e^{-\I
E_{\gamma}t}}{\left(E_\gamma-E_{\gamma_1}\right)^3
\left(E_{\gamma}-E_{\gamma_2}\right)}\right.\nonumber\\
& &\qquad \left.-(-\I t)\frac{\e^{-\I
E_{\gamma_1}t}}{\left(E_\gamma-E_{\gamma_1}\right)^3
\left(E_{\gamma_1}-E_{\gamma_2}\right)}+\frac{(-\I
t)^2}{2!}\frac{\e^{-\I
E_{\gamma}t}}{\left(E_\gamma-E_{\gamma_1}\right)^2
\left(E_{\gamma}-E_{\gamma_2}\right)}\right]\nonumber\\
& &\qquad \times
\left|g_1^{\gamma\gamma_1}\right|^2g_1^{\gamma\gamma_1}g_1^{\gamma_1\gamma_2}
g_1^{\gamma_2\gamma}\delta_{\gamma\gamma^\prime}.\eeqa
\beqa {A}_5^{\gamma\gamma^\prime}(ccnn,kkn;t\e,t^2\e)
&=&\sum_{\gamma_1,\gamma_2} \left[(-\I t)\frac{\e^{-\I
E_{\gamma}t}}{\left(E_\gamma-E_{\gamma_1}\right)^2
\left(E_{\gamma}-E_{\gamma_2}\right)\left(E_{\gamma}-E_{\gamma^\prime}\right)}\right.\nonumber\\
& &\left.+(-\I t)\frac{\e^{-\I
E_{\gamma_1}t}}{\left(E_\gamma-E_{\gamma_1}\right)^2
\left(E_{\gamma_1}-E_{\gamma_2}\right)\left(E_{\gamma_1}-E_{\gamma^\prime}\right)}\right]\nonumber\\
& & \times
\left|g_1^{\gamma\gamma_1}\right|^2g_1^{\gamma\gamma_1}g_1^{\gamma_1\gamma_2}
g_1^{\gamma_2\gamma^\prime}\eta_{\gamma_1\gamma^\prime}\eta_{\gamma\gamma_2}\eta_{\gamma\gamma^\prime}.\eeqa
\beqa & &
{A}_5^{\gamma\gamma^\prime}(cncn,kc;t\e,t^2\e)\nonumber\\
& & \quad =\sum_{\gamma_1} \left[-(-\I t)\frac{2\e^{-\I
E_{\gamma}t}}{\left(E_\gamma-E_{\gamma_1}\right)
\left(E_{\gamma}-E_{\gamma^\prime}\right)^3}-(-\I t)\frac{\e^{-\I
E_{\gamma}t}}{\left(E_\gamma-E_{\gamma_1}\right)^2
\left(E_{\gamma}-E_{\gamma^\prime}\right)^2}\right.\nonumber\\
& &\qquad \left.+(-\I t)\frac{\e^{-\I
E_{\gamma^\prime}t}}{\left(E_\gamma-E_{\gamma^\prime}\right)^3
\left(E_{\gamma_1}-E_{\gamma^\prime}\right)}+\frac{(-\I
t)^2}{2!}\frac{\e^{-\I
E_{\gamma}t}}{\left(E_\gamma-E_{\gamma_1}\right)
\left(E_{\gamma}-E_{\gamma^\prime}\right)^2}\right]\nonumber\\
& &\qquad \times \left|g_1^{\gamma\gamma_1}\right|^2
\left|g_1^{\gamma\gamma^\prime}\right|^2
g_1^{\gamma\gamma^\prime}\eta_{\gamma_1\gamma^\prime}.\eeqa
\beqa {A}_5^{\gamma\gamma^\prime}(cncn,kn;t\e,t^2\e)
&=&\sum_{\gamma_1,\gamma_2} \left[-(-\I t)\frac{\e^{-\I
E_{\gamma}t}}{\left(E_\gamma-E_{\gamma_1}\right)\left(E_\gamma-E_{\gamma_2}\right)
\left(E_{\gamma}-E_{\gamma^\prime}\right)^2}\right.\nonumber\\
& & -(-\I t)\frac{\e^{-\I
E_{\gamma}t}}{\left(E_\gamma-E_{\gamma_1}\right)\left(E_\gamma-E_{\gamma_2}\right)^2
\left(E_{\gamma}-E_{\gamma^\prime}\right)}\nonumber\\
& &-(-\I t)\frac{\e^{-\I
E_{\gamma}t}}{\left(E_\gamma-E_{\gamma_1}\right)^2\left(E_\gamma-E_{\gamma_2}\right)
\left(E_{\gamma}-E_{\gamma^\prime}\right)}\nonumber\\
& &\left. +\frac{(-\I t)^2}{2!}\frac{\e^{-\I
E_{\gamma}t}}{\left(E_\gamma-E_{\gamma_1}\right)\left(E_\gamma-E_{\gamma_2}\right)
\left(E_{\gamma}-E_{\gamma^\prime}\right)}\right]\nonumber\\
& & \times \left|g_1^{\gamma\gamma_1}\right|^2
\left|g_1^{\gamma\gamma_2}\right|^2
g_1^{\gamma\gamma^\prime}\eta_{\gamma_1\gamma_2}
\eta_{\gamma_1\gamma^\prime}\eta_{\gamma_2\gamma^\prime}.\eeqa
\beqa & &
{A}_5^{\gamma\gamma^\prime}(cnnc,kck;t\e,t^2\e)\nonumber\\
& & \quad =\sum_{\gamma_1} \left[(-\I t)\frac{\e^{-\I
E_{\gamma}t}}{\left(E_\gamma-E_{\gamma_1}\right)^2
\left(E_{\gamma}-E_{\gamma^\prime}\right)^2}+(-\I t)\frac{\e^{-\I
E_{\gamma_1}t}}{\left(E_\gamma-E_{\gamma_1}\right)^2
\left(E_{\gamma_1}-E_{\gamma^\prime}\right)^2}\right.\nonumber\\
& &\qquad \left.+(-\I t)\frac{\e^{-\I
E_{\gamma^\prime}t}}{\left(E_\gamma-E_{\gamma^\prime}\right)^2
\left(E_{\gamma_1}-E_{\gamma^\prime}\right)^2}\right]\left|g_1^{\gamma\gamma_1}\right|^2
\left|g_1^{\gamma_1\gamma^\prime}\right|^2
g_1^{\gamma\gamma^\prime}.\eeqa \beqa
{A}_5^{\gamma\gamma^\prime}(cnnc,knk;t\e,t^2\e)
&=&\sum_{\gamma_1,\gamma_2} \left[(-\I t)\frac{\e^{-\I
E_{\gamma}t}}{\left(E_\gamma-E_{\gamma_1}\right)
\left(E_{\gamma}-E_{\gamma_2}\right)\left(E_{\gamma}-E_{\gamma^\prime}\right)^2}\right.\nonumber\\
& &\left.+(-\I t)\frac{\e^{-\I
E_{\gamma^\prime}t}}{\left(E_\gamma-E_{\gamma^\prime}\right)^2
\left(E_{\gamma_1}-E_{\gamma^\prime}\right)\left(E_{\gamma_2}-E_{\gamma^\prime}\right)}\right]\nonumber\\
& & \times \left|g_1^{\gamma\gamma_1}\right|^2
\left|g_1^{\gamma_2\gamma^\prime}\right|^2
g_1^{\gamma\gamma^\prime}\eta_{\gamma_1\gamma^\prime}\eta_{\gamma_1\gamma_2}\eta_{\gamma\gamma_2}.\eeqa
\beqa & &
{A}_5^{\gamma\gamma^\prime}(nccn,c;t\e,t^2\e)\nonumber\\
& & \quad =\sum_{\gamma_1,\gamma_2} \left[(-\I t)\frac{\e^{-\I
E_{\gamma}t}}{\left(E_\gamma-E_{\gamma_1}\right)^2
\left(E_{\gamma}-E_{\gamma_2}\right)^2}+(-\I t)\frac{\e^{-\I
E_{\gamma_1}t}}{\left(E_\gamma-E_{\gamma_1}\right)^2
\left(E_{\gamma_1}-E_{\gamma_2}\right)^2}\right.\nonumber\\
& &\qquad \left.+(-\I t)\frac{\e^{-\I
E_{\gamma_2}t}}{\left(E_\gamma-E_{\gamma_2}\right)^2
\left(E_{\gamma_1}-E_{\gamma_2}\right)^2}\right]\left|g_1^{\gamma_1\gamma_2}\right|^2
g_1^{\gamma\gamma_1}g_1^{\gamma_1\gamma_2}g_1^{\gamma_2\gamma}
\delta_{\gamma\gamma^\prime}.\eeqa \beqa
{A}_5^{\gamma\gamma^\prime}(nccn,n;t\e,t^2\e)
&=&\sum_{\gamma_1,\gamma_2} \left[-(-\I t)\frac{\e^{-\I
E_{\gamma_1}t}}{\left(E_\gamma-E_{\gamma_1}\right)
\left(E_{\gamma_1}-E_{\gamma_2}\right)^2\left(E_{\gamma_1}-E_{\gamma^\prime}\right)}\right.\nonumber\\
& &\left.-(-\I t)\frac{\e^{-\I
E_{\gamma_2}t}}{\left(E_\gamma-E_{\gamma_2}\right)
\left(E_{\gamma_1}-E_{\gamma_2}\right)^2\left(E_{\gamma_2}-E_{\gamma^\prime}\right)}\right]\nonumber\\
& & \times
\left|g_1^{\gamma_1\gamma_2}\right|^2g_1^{\gamma\gamma_1}g_1^{\gamma_1\gamma_2}
g_1^{\gamma_2\gamma^\prime}\eta_{\gamma_1\gamma^\prime}\eta_{\gamma\gamma_2}\eta_{\gamma\gamma^\prime}.\eeqa
\beqa & &
{A}_5^{\gamma\gamma^\prime}(ncnc,ck;t\e,t^2\e)\nonumber\\
& & \quad =\sum_{\gamma_1} \left[(-\I t)\frac{\e^{-\I
E_{\gamma}t}}{\left(E_\gamma-E_{\gamma_1}\right)
\left(E_{\gamma}-E_{\gamma^\prime}\right)^3}-(-\I t)\frac{\e^{-\I
E_{\gamma^\prime}t}}{\left(E_\gamma-E_{\gamma^\prime}\right)^2
\left(E_{\gamma_1}-E_{\gamma^\prime}\right)^2}\right.\nonumber\\
& &\qquad \left.-(-\I t)\frac{2\e^{-\I
E_{\gamma^\prime}t}}{\left(E_\gamma-E_{\gamma^\prime}\right)^3
\left(E_{\gamma_1}-E_{\gamma^\prime}\right)}-\frac{(-\I
t)^2}{2!}\frac{\e^{-\I
E_{\gamma^\prime}t}}{\left(E_\gamma-E_{\gamma^\prime}\right)^2
\left(E_{\gamma_1}-E_{\gamma^\prime}\right)}\right]\nonumber\\
& &\qquad\times \left|g_1^{\gamma_1\gamma^\prime}\right|^2
\left|g_1^{\gamma\gamma^\prime}\right|^2
g_1^{\gamma\gamma^\prime}\eta_{\gamma\gamma_1}.\eeqa \beqa
{A}_5^{\gamma\gamma^\prime}(ncnc,nk;t\e,t^2\e)
&=&\sum_{\gamma_1,\gamma_2} \left[-(-\I t)\frac{\e^{-\I
E_{\gamma^\prime}t}}{\left(E_\gamma-E_{\gamma^\prime}\right)
\left(E_{\gamma_1}-E_{\gamma^\prime}\right)
\left(E_{\gamma_2}-E_{\gamma^\prime}\right)^2}\right.\nonumber\\
& & -(-\I t)\frac{\e^{-\I
E_{\gamma^\prime}t}}{\left(E_\gamma-E_{\gamma^\prime}\right)
\left(E_{\gamma_1}-E_{\gamma^\prime}\right)^2
\left(E_{\gamma_2}-E_{\gamma^\prime}\right)}\nonumber\\
& &-(-\I t)\frac{\e^{-\I
E_{\gamma^\prime}t}}{\left(E_\gamma-E_{\gamma^\prime}\right)^2
\left(E_{\gamma_1}-E_{\gamma^\prime}\right)
\left(E_{\gamma_2}-E_{\gamma^\prime}\right)}\nonumber\\
& &\left. -\frac{(-\I t)^2}{2!}\frac{\e^{-\I
E_{\gamma^\prime}t}}{\left(E_\gamma-E_{\gamma^\prime}\right)
\left(E_{\gamma_1}-E_{\gamma^\prime}\right)
\left(E_{\gamma_2}-E_{\gamma^\prime}\right)}\right]\nonumber\\
& & \times \left|g_1^{\gamma_1\gamma^\prime}\right|^2
\left|g_1^{\gamma_2\gamma^\prime}\right|^2
g_1^{\gamma\gamma^\prime}\eta_{\gamma\gamma_1}
\eta_{\gamma\gamma_2}\eta_{\gamma_1\gamma_2}.\eeqa \beqa & &
{A}_5^{\gamma\gamma^\prime}(nncc,ckk;t\e,t^2\e)\nonumber\\
& & \quad =\sum_{\gamma_1,\gamma_2} \left[-(-\I t)\frac{2\e^{-\I
E_{\gamma}t}}{\left(E_\gamma-E_{\gamma_1}\right)
\left(E_{\gamma}-E_{\gamma_2}\right)^3}-(-\I t)\frac{\e^{-\I
E_{\gamma}t}}{\left(E_\gamma-E_{\gamma_1}\right)^2
\left(E_{\gamma}-E_{\gamma_2}\right)^2}\right.\nonumber\\
& &\qquad \left.+(-\I t)\frac{\e^{-\I
E_{\gamma_2}t}}{\left(E_\gamma-E_{\gamma_2}\right)^3
\left(E_{\gamma_1}-E_{\gamma_2}\right)}+\frac{(-\I
t)^2}{2!}\frac{\e^{-\I
E_{\gamma}t}}{\left(E_\gamma-E_{\gamma_1}\right)
\left(E_{\gamma}-E_{\gamma_2}\right)^2}\right]\nonumber\\
& &\qquad \times \left|g_1^{\gamma\gamma_2}\right|^2
g_1^{\gamma\gamma_1}g_1^{\gamma_1\gamma_2}g_1^{\gamma_2\gamma}\delta_{\gamma\gamma^\prime}.\eeqa
\beqa {A}_5^{\gamma\gamma^\prime}(nncc,nkk;t\e,t^2\e)
&=&\sum_{\gamma_1,\gamma_2} \left[(-\I t)\frac{\e^{-\I
E_{\gamma_2}t}}{\left(E_\gamma-E_{\gamma_2}\right)
\left(E_{\gamma_1}-E_{\gamma_2}\right)\left(E_{\gamma_2}-E_{\gamma^\prime}\right)^2}\right.\nonumber\\
& &\left.+(-\I t)\frac{\e^{-\I
E_{\gamma^\prime}t}}{\left(E_\gamma-E_{\gamma^\prime}\right)
\left(E_{\gamma_1}-E_{\gamma^\prime}\right)\left(E_{\gamma_2}-E_{\gamma^\prime}\right)^2}\right]\nonumber\\
& & \times
\left|g_1^{\gamma_2\gamma^\prime}\right|^2g_1^{\gamma\gamma_1}g_1^{\gamma_1\gamma_2}
g_1^{\gamma_2\gamma^\prime}\eta_{\gamma_1\gamma^\prime}\eta_{\gamma\gamma_2}\eta_{\gamma\gamma^\prime}.\eeqa
\beqa & &
{A}_5^{\gamma\gamma^\prime}(cnnn,kcc;t\e,t^2\e)\nonumber\\
& & \quad =\sum_{\gamma_1,\gamma_2} \left[-(-\I t)\frac{2\e^{-\I
E_{\gamma}t}}{\left(E_\gamma-E_{\gamma_1}\right)^3
\left(E_{\gamma}-E_{\gamma_2}\right)}-(-\I t)\frac{\e^{-\I
E_{\gamma}t}}{\left(E_\gamma-E_{\gamma_1}\right)^2
\left(E_{\gamma}-E_{\gamma_2}\right)^2}\right.\nonumber\\
& &\qquad \left.-(-\I t)\frac{\e^{-\I
E_{\gamma_1}t}}{\left(E_\gamma-E_{\gamma_1}\right)^3
\left(E_{\gamma_1}-E_{\gamma_2}\right)}+\frac{(-\I
t)^2}{2!}\frac{\e^{-\I
E_{\gamma}t}}{\left(E_\gamma-E_{\gamma_1}\right)^2
\left(E_{\gamma}-E_{\gamma_2}\right)}\right]\nonumber\\
& &\qquad \times \left|g_1^{\gamma\gamma_1}\right|^2
g_1^{\gamma\gamma_2}g_1^{\gamma_2\gamma_1}g_1^{\gamma_1\gamma}\delta_{\gamma\gamma^\prime}.\eeqa
\beqa {A}_5^{\gamma\gamma^\prime}(cnnn,kcn;t\e,t^2\e)
&=&\sum_{\gamma_1,\gamma_2} \left[(-\I t)\frac{\e^{-\I
E_{\gamma}t}}{\left(E_\gamma-E_{\gamma_1}\right)^2
\left(E_{\gamma}-E_{\gamma_2}\right)\left(E_{\gamma}-E_{\gamma^\prime}\right)}\right.\nonumber\\
& &\left.+(-\I t)\frac{\e^{-\I
E_{\gamma_1}t}}{\left(E_\gamma-E_{\gamma_1}\right)^2
\left(E_{\gamma_1}-E_{\gamma_2}\right)\left(E_{\gamma_1}-E_{\gamma^\prime}\right)}\right]\nonumber\\
& & \times
\left|g_1^{\gamma\gamma_1}\right|^2g_1^{\gamma\gamma_2}g_1^{\gamma_2\gamma_1}
g_1^{\gamma_1\gamma^\prime}\eta_{\gamma_2\gamma^\prime}
\eta_{\gamma\gamma^\prime}.\eeqa \beqa
{A}_5^{\gamma\gamma^\prime}(cnnn,knc;t\e,t^2\e)
&=&\sum_{\gamma_1,\gamma_2,\gamma_3} \left[-(-\I t)\frac{\e^{-\I
E_{\gamma}t}}{\left(E_\gamma-E_{\gamma_1}\right)^2\left(E_\gamma-E_{\gamma_2}\right)
\left(E_{\gamma}-E_{\gamma_3}\right)}\right.\nonumber\\
& & -(-\I t)\frac{\e^{-\I
E_{\gamma}t}}{\left(E_\gamma-E_{\gamma_1}\right)\left(E_\gamma-E_{\gamma_2}\right)^2
\left(E_{\gamma}-E_{\gamma_3}\right)}\nonumber\\
& &-(-\I t)\frac{\e^{-\I
E_{\gamma}t}}{\left(E_\gamma-E_{\gamma_1}\right)\left(E_\gamma-E_{\gamma_2}\right)
\left(E_{\gamma}-E_{\gamma_3}\right)^2}\nonumber\\
& &\left. +\frac{(-\I t)^2}{2!}\frac{\e^{-\I
E_{\gamma}t}}{\left(E_\gamma-E_{\gamma_1}\right)\left(E_\gamma-E_{\gamma_2}\right)
\left(E_{\gamma}-E_{\gamma_3}\right)}\right]\nonumber\\
& & \times \left|g_1^{\gamma\gamma_1}\right|^2
g_1^{\gamma\gamma_2}g_1^{\gamma_2\gamma_3}g_1^{\gamma_3\gamma}\eta_{\gamma_1\gamma_2}
\eta_{\gamma_1\gamma_3}\delta_{\gamma\gamma^\prime}.\eeqa
\beqa {A}_5^{\gamma\gamma^\prime}(cnnn,knn,kc;t\e,t^2\e)
&=&\sum_{\gamma_1,\gamma_2} \left[(-\I t)\frac{\e^{-\I
E_{\gamma}t}}{\left(E_\gamma-E_{\gamma_1}\right)
\left(E_{\gamma}-E_{\gamma_2}\right)\left(E_{\gamma}-E_{\gamma^\prime}\right)^2}\right.\nonumber\\
& &\left.+(-\I t)\frac{\e^{-\I
E_{\gamma^\prime}t}}{\left(E_\gamma-E_{\gamma^\prime}\right)^2
\left(E_{\gamma_1}-E_{\gamma^\prime}\right)\left(E_{\gamma_2}-E_{\gamma^\prime}\right)}\right]\nonumber\\
& &
\times\left|g_1^{\gamma\gamma^\prime}\right|^2g_1^{\gamma\gamma_1}g_1^{\gamma_1\gamma_2}
g_1^{\gamma_2\gamma^\prime}\eta_{\gamma\gamma_2}
\eta_{\gamma_1\gamma^\prime}.\eeqa \beqa &
&{A}_5^{\gamma\gamma^\prime}(cnnn,knn,kn;t\e,t^2\e) \nonumber\\ & &
\quad =\sum_{\gamma_1,\gamma_2,\gamma_3}(-\I t)\frac{\e^{-\I
E_{\gamma}t}\left|g_1^{\gamma\gamma_1}\right|^2
g_1^{\gamma\gamma_2}g_1^{\gamma_2\gamma_3}g_1^{\gamma_3\gamma^\prime}\eta_{\gamma\gamma_3}
\eta_{\gamma\gamma^\prime}\eta_{\gamma_1\gamma_2}
\eta_{\gamma_1\gamma_3}\eta_{\gamma_1\gamma^\prime}\eta_{\gamma_2\gamma^\prime}}{\left(E_\gamma-E_{\gamma_1}\right)
\left(E_{\gamma}-E_{\gamma_2}\right)\left(E_{\gamma}-E_{\gamma_3}\right)
\left(E_{\gamma}-E_{\gamma^\prime}\right)}. \eeqa
\beqa & &
{A}_5^{\gamma\gamma^\prime}(ncnn,kkc,ck;t\e,t^2\e)\nonumber\\
& & \quad =\sum_{\gamma_1} \left[(-\I t)\frac{\e^{-\I
E_{\gamma}t}}{\left(E_\gamma-E_{\gamma_1}\right)^2
\left(E_{\gamma}-E_{\gamma^\prime}\right)^2}+(-\I t)\frac{\e^{-\I
E_{\gamma_1}t}}{\left(E_\gamma-E_{\gamma_1}\right)^2
\left(E_{\gamma_1}-E_{\gamma^\prime}\right)^2}\right.\nonumber\\
& &\qquad \left.+(-\I t)\frac{\e^{-\I
E_{\gamma^\prime}t}}{\left(E_\gamma-E_{\gamma^\prime}\right)^2
\left(E_{\gamma_1}-E_{\gamma^\prime}\right)^2}\right]\left|g_1^{\gamma\gamma_1}\right|^2
\left|g_1^{\gamma_1\gamma^\prime}\right|^2
g_1^{\gamma\gamma^\prime}.\eeqa
\beqa {A}_5^{\gamma\gamma^\prime}(ncnn,kkc,nk;t\e,t^2\e)
&=&\sum_{\gamma_1,\gamma_2} \left[-(-\I t)\frac{\e^{-\I
E_{\gamma_1}t}}{\left(E_\gamma-E_{\gamma_1}\right)
\left(E_{\gamma_1}-E_{\gamma_2}\right)\left(E_{\gamma_1}-E_{\gamma^\prime}\right)^2}\right.\nonumber\\
& &\left.+(-\I t)\frac{\e^{-\I
E_{\gamma^\prime}t}}{\left(E_\gamma-E_{\gamma^\prime}\right)
\left(E_{\gamma_1}-E_{\gamma^\prime}\right)^2\left(E_{\gamma_2}-E_{\gamma^\prime}\right)}\right]\nonumber\\
& &
\times\left|g_1^{\gamma_1\gamma^\prime}\right|^2g_1^{\gamma\gamma_1}g_1^{\gamma_1\gamma_2}
g_1^{\gamma_2\gamma^\prime}\eta_{\gamma\gamma_2}\eta_{\gamma\gamma^\prime}.\eeqa
\beqa {A}_5^{\gamma\gamma^\prime}(ncnn,kkn,ck;t\e,t^2\e)
&=&\sum_{\gamma_1,\gamma_2} \left[(-\I t)\frac{\e^{-\I
E_{\gamma}t}}{\left(E_\gamma-E_{\gamma_1}\right)^2
\left(E_{\gamma}-E_{\gamma_2}\right)\left(E_{\gamma}-E_{\gamma^\prime}\right)}\right.\nonumber\\
& &\left.+(-\I t)\frac{\e^{-\I
E_{\gamma_1}t}}{\left(E_\gamma-E_{\gamma_1}\right)^2
\left(E_{\gamma_1}-E_{\gamma_2}\right)\left(E_{\gamma_1}-E_{\gamma^\prime}\right)}\right]\nonumber\\
& & \times\left|g_1^{\gamma\gamma_1}\right|^2
\left|g_1^{\gamma_1\gamma_2}\right|^2
g_1^{\gamma\gamma^\prime}\eta_{\gamma\gamma_2}\eta_{\gamma_1\gamma^\prime}\eta_{\gamma_2\gamma^\prime}.\eeqa
\beqa {A}_5^{\gamma\gamma^\prime}(ncnn,kkn,nk,c;t\e,t^2\e)
&=&\sum_{\gamma_1,\gamma_2,\gamma_3} \left[(-\I t)\frac{\e^{-\I
E_{\gamma}t}}{\left(E_\gamma-E_{\gamma_1}\right)^2
\left(E_{\gamma}-E_{\gamma_2}\right)\left(E_{\gamma}-E_{\gamma_3}\right)}\right.\nonumber\\
& &\left.+(-\I t)\frac{\e^{-\I
E_{\gamma_1}t}}{\left(E_\gamma-E_{\gamma_1}\right)^2
\left(E_{\gamma_1}-E_{\gamma_2}\right)\left(E_{\gamma_1}-E_{\gamma_3}\right)}\right]\nonumber\\
& & \times\left|g_1^{\gamma_1\gamma_2}\right|^2
g_1^{\gamma\gamma_1}g_1^{\gamma_1\gamma_3}
g_1^{\gamma_3\gamma}\eta_{\gamma\gamma_2}\eta_{\gamma_2\gamma_3}\delta_{\gamma\gamma^\prime}.\eeqa
\beqa & &{A}_5^{\gamma\gamma^\prime}(ncnn,kkn,nk,n;t\e,t^2\e)
\nonumber\\ & & \quad =-\sum_{\gamma_1,\gamma_2,\gamma_3}(-\I
t)\frac{\e^{-\I E_{\gamma_1}t}\left|g_1^{\gamma_1\gamma_2}\right|^2
g_1^{\gamma\gamma_1}g_1^{\gamma_1\gamma_3}g_1^{\gamma_3\gamma^\prime}\eta_{\gamma\gamma_2}\eta_{\gamma\gamma_3}
\eta_{\gamma\gamma^\prime}\eta_{\gamma_1\gamma^\prime}
\eta_{\gamma_2\gamma_3}\eta_{\gamma_2\gamma^\prime}}{\left(E_\gamma-E_{\gamma_1}\right)
\left(E_{\gamma_1}-E_{\gamma_2}\right)\left(E_{\gamma_1}-E_{\gamma_3}\right)
\left(E_{\gamma_1}-E_{\gamma^\prime}\right)}. \eeqa
\beqa & &
{A}_5^{\gamma\gamma^\prime}(nncn,ckk;kc,t\e,t^2\e)\nonumber\\
& & \quad =\sum_{\gamma_1} \left[(-\I t)\frac{\e^{-\I
E_{\gamma}t}}{\left(E_\gamma-E_{\gamma_1}\right)^2
\left(E_{\gamma}-E_{\gamma^\prime}\right)^2}+(-\I t)\frac{\e^{-\I
E_{\gamma_1}t}}{\left(E_\gamma-E_{\gamma_1}\right)^2
\left(E_{\gamma_1}-E_{\gamma^\prime}\right)^2}\right.\nonumber\\
& &\qquad \left.+(-\I t)\frac{\e^{-\I
E_{\gamma^\prime}t}}{\left(E_\gamma-E_{\gamma^\prime}\right)^2
\left(E_{\gamma_1}-E_{\gamma^\prime}\right)^2}\right]\left|g_1^{\gamma\gamma_1}\right|^2
\left|g_1^{\gamma_1\gamma^\prime}\right|^2
g_1^{\gamma\gamma^\prime}.\eeqa \beqa
{A}_5^{\gamma\gamma^\prime}(nncn,ckk,kn;t\e,t^2\e)
&=&\sum_{\gamma_1,\gamma_2} \left[(-\I t)\frac{\e^{-\I
E_{\gamma}t}}{\left(E_\gamma-E_{\gamma_1}\right)
\left(E_{\gamma}-E_{\gamma_2}\right)^2\left(E_{\gamma}-E_{\gamma^\prime}\right)}\right.\nonumber\\
& &\left.-(-\I t)\frac{\e^{-\I
E_{\gamma_2}t}}{\left(E_\gamma-E_{\gamma_2}\right)^2
\left(E_{\gamma_1}-E_{\gamma_2}\right)\left(E_{\gamma_2}-E_{\gamma^\prime}\right)}\right]\nonumber\\
& & \times\left|g_1^{\gamma\gamma_2}\right|^2
g_1^{\gamma\gamma_1}g_1^{\gamma_1\gamma_2}
g_1^{\gamma_2\gamma^\prime}\eta_{\gamma\gamma^\prime}\eta_{\gamma_1\gamma^\prime}.\eeqa
\beqa {A}_5^{\gamma\gamma^\prime}(nncn,nkk,kc;t\e,t^2\e)
&=&\sum_{\gamma_1,\gamma_2} \left[-(-\I t)\frac{\e^{-\I
E_{\gamma_1}t}}{\left(E_\gamma-E_{\gamma_1}\right)
\left(E_{\gamma_1}-E_{\gamma_2}\right)\left(E_{\gamma_1}-E_{\gamma^\prime}\right)^2}\right.\nonumber\\
& &\left.+(-\I t)\frac{\e^{-\I
E_{\gamma^\prime}t}}{\left(E_{\gamma}-E_{\gamma^\prime}\right)
\left(E_{\gamma_1}-E_{\gamma^\prime}\right)^2\left(E_{\gamma_2}-E_{\gamma^\prime}\right)}\right]\nonumber\\
& &
\times\left|g_1^{\gamma_1\gamma_2}\right|^2\left|g_1^{\gamma_1\gamma^\prime}\right|^2
g_1^{\gamma\gamma^\prime}\eta_{\gamma\gamma_1}\eta_{\gamma\gamma_2}\eta_{\gamma_2\gamma^\prime}.\eeqa
\beqa {A}_5^{\gamma\gamma^\prime}(nncn,nkk,kn,c;t\e,t^2\e)
&=&\sum_{\gamma_1,\gamma_2\gamma_3} \left[(-\I t)\frac{\e^{-\I
E_{\gamma}t}}{\left(E_\gamma-E_{\gamma_1}\right)
\left(E_{\gamma}-E_{\gamma_2}\right)^2
\left(E_{\gamma}-E_{\gamma_3}\right)}\right.\nonumber\\
& &\left.-(-\I t)\frac{\e^{-\I
E_{\gamma_2}t}}{\left(E_{\gamma}-E_{\gamma_2}\right)^2
\left(E_{\gamma_1}-E_{\gamma_2}\right)\left(E_{\gamma_2}-E_{\gamma_3}\right)}\right]\nonumber\\
& &
\times\left|g_1^{\gamma_2\gamma_3}\right|^2g_1^{\gamma\gamma_1}g_1^{\gamma_1\gamma_2}
g_1^{\gamma_2\gamma}\eta_{\gamma\gamma_3}\eta_{\gamma_1\gamma_3}\delta_{\gamma\gamma^\prime}.\eeqa
\beqa & &{A}_5^{\gamma\gamma^\prime}(nncn,nkk,kn,n;t\e,t^2\e)
\nonumber\\ & & \quad =\sum_{\gamma_1,\gamma_2,\gamma_3}(-\I
t)\frac{\e^{-\I E_{\gamma_2}t}\left|g_1^{\gamma_2\gamma_3}\right|^2
g_1^{\gamma\gamma_1}g_1^{\gamma_1\gamma_2}g_1^{\gamma_2\gamma^\prime}\eta_{\gamma\gamma_2}\eta_{\gamma\gamma_3}
\eta_{\gamma\gamma^\prime}
\eta_{\gamma_1\gamma_3}\eta_{\gamma_1\gamma^\prime}
\eta_{\gamma_3\gamma^\prime}}{\left(E_\gamma-E_{\gamma_2}\right)
\left(E_{\gamma_1}-E_{\gamma_2}\right)\left(E_{\gamma_2}-E_{\gamma_3}\right)
\left(E_{\gamma_2}-E_{\gamma^\prime}\right)}. \eeqa
\beqa & &
{A}_5^{\gamma\gamma^\prime}(nnnc,cck;t\e,t^2\e)\nonumber\\
& & \quad =\sum_{\gamma_1,\gamma_2} \left[-(-\I t)\frac{\e^{-\I
E_{\gamma}t}}{\left(E_\gamma-E_{\gamma_1}\right)^2
\left(E_{\gamma}-E_{\gamma_2}\right)^2}-(-\I t)\frac{2\e^{-\I
E_{\gamma}t}}{\left(E_\gamma-E_{\gamma_1}\right)^3
\left(E_{\gamma}-E_{\gamma_2}\right)}\right.\nonumber\\
& &\qquad \left.-(-\I t)\frac{\e^{-\I
E_{\gamma_1}t}}{\left(E_\gamma-E_{\gamma_1}\right)^3
\left(E_{\gamma_1}-E_{\gamma_2}\right)}+\frac{(-\I
t)^2}{2!}\frac{\e^{-\I
E_{\gamma}t}}{\left(E_\gamma-E_{\gamma_1}\right)^2
\left(E_{\gamma}-E_{\gamma_2}\right)}\right]\nonumber\\
& &\qquad \times \left|g_1^{\gamma\gamma_1}\right|^2
g_1^{\gamma\gamma_1}g_1^{\gamma_1\gamma_2}g_1^{\gamma_2\gamma}\delta_{\gamma\gamma^\prime}.\eeqa
\beqa {A}_5^{\gamma\gamma^\prime}(nnnc,cnk;t\e,t^2\e)
&=&\sum_{\gamma_1,\gamma_2,\gamma_3} \left[-(-\I t)\frac{\e^{-\I
E_{\gamma}t}}{\left(E_\gamma-E_{\gamma_1}\right)\left(E_\gamma-E_{\gamma_2}\right)
\left(E_{\gamma}-E_{\gamma_3}\right)^2}\right.\nonumber\\
& & -(-\I t)\frac{\e^{-\I
E_{\gamma}t}}{\left(E_\gamma-E_{\gamma_1}\right)\left(E_\gamma-E_{\gamma_2}\right)^2
\left(E_{\gamma}-E_{\gamma_3}\right)}\nonumber\\
& &-(-\I t)\frac{\e^{-\I
E_{\gamma}t}}{\left(E_\gamma-E_{\gamma_1}\right)^2\left(E_\gamma-E_{\gamma_2}\right)
\left(E_{\gamma}-E_{\gamma_3}\right)}\nonumber\\
& &\left. +\frac{(-\I t)^2}{2!}\frac{\e^{-\I
E_{\gamma}t}}{\left(E_\gamma-E_{\gamma_1}\right)\left(E_\gamma-E_{\gamma_2}\right)
\left(E_{\gamma}-E_{\gamma_3}\right)}\right]\nonumber\\
& & \times \left|g_1^{\gamma\gamma_3}\right|^2
g_1^{\gamma\gamma_1}g_1^{\gamma_1\gamma_2}g_1^{\gamma_2\gamma}\eta_{\gamma_1\gamma_3}
\eta_{\gamma_2\gamma_3}\delta_{\gamma\gamma^\prime}.\eeqa
\beqa {A}_5^{\gamma\gamma^\prime}(nnnc,nck;t\e,t^2\e)
&=&\sum_{\gamma_1,\gamma_2} \left[-(-\I t)\frac{\e^{-\I
E_{\gamma_1}t}}{\left(E_\gamma-E_{\gamma_1}\right)
\left(E_{\gamma_1}-E_{\gamma_2}\right)\left(E_{\gamma_1}-E_{\gamma^\prime}\right)^2}\right.\nonumber\\
& &\left.+(-\I t)\frac{\e^{-\I
E_{\gamma^\prime}t}}{\left(E_{\gamma}-E_{\gamma^\prime}\right)\left(E_{\gamma_1}-E_{\gamma^\prime}\right)^2
\left(E_{\gamma_2}-E_{\gamma^\prime}\right)}\right]\nonumber\\
& & \times
\left|g_1^{\gamma_1\gamma^\prime}\right|^2g_1^{\gamma\gamma_1}g_1^{\gamma_1\gamma_2}
g_1^{\gamma_2\gamma^\prime}\eta_{\gamma\gamma_2}
\eta_{\gamma\gamma^\prime}.\eeqa
\beqa {A}_5^{\gamma\gamma^\prime}(nnnc,nnk,ck;t\e,t^2\e)
&=&\sum_{\gamma_1,\gamma_2} \left[(-\I t)\frac{\e^{-\I
E_{\gamma}t}}{\left(E_\gamma-E_{\gamma_1}\right)
\left(E_{\gamma}-E_{\gamma_2}\right)\left(E_{\gamma}-E_{\gamma^\prime}\right)^2}\right.\nonumber\\
& &\left.+(-\I t)\frac{\e^{-\I
E_{\gamma^\prime}t}}{\left(E_{\gamma}-E_{\gamma^\prime}\right)^2\left(E_{\gamma_1}-E_{\gamma^\prime}\right)
\left(E_{\gamma_2}-E_{\gamma^\prime}\right)}\right]\nonumber\\
& & \times
\left|g_1^{\gamma\gamma^\prime}\right|^2g_1^{\gamma\gamma_1}g_1^{\gamma_1\gamma_2}
g_1^{\gamma_2\gamma^\prime}\eta_{\gamma\gamma_2}
\eta_{\gamma_1\gamma^\prime}.\eeqa
\beqa & &{A}_5^{\gamma\gamma^\prime}(nnnc,nnk,nk;t\e,t^2\e)
\nonumber\\ & & \quad =\sum_{\gamma_1,\gamma_2,\gamma_3}(-\I
t)\frac{\e^{-\I
E_{\gamma^\prime}t}\left|g_1^{\gamma_3\gamma^\prime}\right|^2
g_1^{\gamma\gamma_1}g_1^{\gamma_1\gamma_2}g_1^{\gamma_2\gamma^\prime}\eta_{\gamma\gamma_2}\eta_{\gamma\gamma_3}
\eta_{\gamma\gamma^\prime}
\eta_{\gamma_1\gamma_3}\eta_{\gamma_1\gamma^\prime}
\eta_{\gamma_2\gamma_3}}{\left(E_\gamma-E_{\gamma^\prime}\right)
\left(E_{\gamma_1}-E_{\gamma^\prime}\right)\left(E_{\gamma_2}-E_{\gamma^\prime}\right)
\left(E_{\gamma_3}-E_{\gamma^\prime}\right)}. \eeqa
\beqa & &
{A}_5^{\gamma\gamma^\prime}(nnnn,ccc;t\e,t^2\e)\nonumber\\
& & \quad =\sum_{\gamma_1} \left[(-\I t)\frac{\e^{-\I
E_{\gamma}t}}{\left(E_\gamma-E_{\gamma_1}\right)^2
\left(E_{\gamma}-E_{\gamma^\prime}\right)^2}+(-\I t)\frac{\e^{-\I
E_{\gamma_1}t}}{\left(E_\gamma-E_{\gamma_1}\right)^2
\left(E_{\gamma_1}-E_{\gamma^\prime}\right)^2}\right.\nonumber\\
& &\qquad \left.+(-\I t)\frac{\e^{-\I
E_{\gamma^\prime}t}}{\left(E_\gamma-E_{\gamma^\prime}\right)^2
\left(E_{\gamma_1}-E_{\gamma^\prime}\right)^2}\right]\left|g_1^{\gamma\gamma_1}\right|^2
\left(g_1^{\gamma\gamma_1}g_1^{\gamma_1\gamma^\prime}\right)^2
g_1^{\gamma\gamma^\prime}.\eeqa
\beqa {A}_5^{\gamma\gamma^\prime}(nnnn,ccn;t\e,t^2\e)
&=&\sum_{\gamma_1,\gamma_2} \left[(-\I t)\frac{\e^{-\I
E_{\gamma}t}}{\left(E_\gamma-E_{\gamma_1}\right)^2
\left(E_{\gamma}-E_{\gamma_2}\right)\left(E_{\gamma}-E_{\gamma^\prime}\right)}\right.\nonumber\\
& &\left.+(-\I t)\frac{\e^{-\I
E_{\gamma_1}t}}{\left(E_{\gamma}-E_{\gamma_1}\right)^2\left(E_{\gamma_1}-E_{\gamma_2}\right)
\left(E_{\gamma_1}-E_{\gamma^\prime}\right)}\right]\nonumber\\
& & \times \left(g_1^{\gamma\gamma_1}g_1^{\gamma_1\gamma_2}
g_1^{\gamma_2\gamma}\right)g_1^{\gamma\gamma_1}g_1^{\gamma_1\gamma^\prime}\eta_{\gamma\gamma^\prime}
\eta_{\gamma_2\gamma^\prime}.\eeqa
\beqa {A}_5^{\gamma\gamma^\prime}(nnnn,cnc;t\e,t^2\e)
&=&\sum_{\gamma_1,\gamma_2} \left[(-\I t)\frac{\e^{-\I
E_{\gamma}t}}{\left(E_\gamma-E_{\gamma_1}\right)
\left(E_{\gamma}-E_{\gamma_2}\right)\left(E_{\gamma}-E_{\gamma^\prime}\right)^2}\right.\nonumber\\
& &\left.+(-\I t)\frac{\e^{-\I
E_{\gamma^\prime}t}}{\left(E_{\gamma}-E_{\gamma^\prime}\right)^2\left(E_{\gamma_1}-E_{\gamma^\prime}\right)
\left(E_{\gamma_2}-E_{\gamma^\prime}\right)}\right]\nonumber\\
& & \times
\left(g_1^{\gamma\gamma_1}g_1^{\gamma_1\gamma^\prime}\right)\left(
g_1^{\gamma\gamma_2}g_1^{\gamma_2\gamma^\prime}\right)g_1^{\gamma\gamma^\prime}\eta_{\gamma_1\gamma_2}
.\eeqa
\beqa {A}_5^{\gamma\gamma^\prime}(nnnn,ncc;t\e,t^2\e)
&=&\sum_{\gamma_1,\gamma_2} \left[-(-\I t)\frac{\e^{-\I
E_{\gamma_1}t}}{\left(E_\gamma-E_{\gamma_1}\right)
\left(E_{\gamma_1}-E_{\gamma_2}\right)\left(E_{\gamma_1}-E_{\gamma^\prime}\right)^2}\right.\nonumber\\
& &\left.+(-\I t)\frac{\e^{-\I
E_{\gamma^\prime}t}}{\left(E_{\gamma}-E_{\gamma^\prime}\right)\left(E_{\gamma_1}-E_{\gamma^\prime}\right)^2
\left(E_{\gamma_2}-E_{\gamma^\prime}\right)}\right]\nonumber\\
& & \times \left(g_1^{\gamma^\prime\gamma_2}g_1^{\gamma_2\gamma_1}
g_1^{\gamma_1\gamma^\prime}\right)g_1^{\gamma\gamma_1}g_1^{\gamma_1\gamma^\prime}
\eta_{\gamma\gamma_2}\eta_{\gamma\gamma^\prime} .\eeqa
\beqa {A}_5^{\gamma\gamma^\prime}(nnnn,cnn,kc;t\e,t^2\e)
&=&\sum_{\gamma_1,\gamma_2} \left[(-\I t)\frac{\e^{-\I
E_{\gamma}t}}{\left(E_\gamma-E_{\gamma_1}\right)
\left(E_{\gamma}-E_{\gamma_2}\right)\left(E_{\gamma}-E_{\gamma^\prime}\right)^2}\right.\nonumber\\
& &\left.+(-\I t)\frac{\e^{-\I
E_{\gamma^\prime}t}}{\left(E_{\gamma}-E_{\gamma^\prime}\right)^2\left(E_{\gamma_1}-E_{\gamma^\prime}\right)
\left(E_{\gamma_2}-E_{\gamma^\prime}\right)}\right]\nonumber\\
& & \times
\left(g_1^{\gamma^\prime\gamma_1}g_1^{\gamma_1\gamma}\right)\left(
g_1^{\gamma\gamma_2}g_1^{\gamma_2\gamma^\prime}\right)g_1^{\gamma\gamma^\prime}
\eta_{\gamma_1\gamma_2}.\eeqa
\beqa & &{A}_5^{\gamma\gamma^\prime}(nnnn,cnn,kn;t\e,t^2\e)
\nonumber\\ & & \quad =\sum_{\gamma_1,\gamma_2,\gamma_3}(-\I
t)\frac{\e^{-\I E_{\gamma}t}
g_1^{\gamma\gamma_1}g_1^{\gamma_1\gamma_2}g_1^{\gamma_2\gamma}
g_1^{\gamma\gamma_3}g_1^{\gamma_3\gamma^\prime}\eta_{\gamma\gamma^\prime}\eta_{\gamma_1\gamma_3}
\eta_{\gamma_2\gamma_3}\eta_{\gamma_1\gamma^\prime}
\eta_{\gamma_2\gamma^\prime}}{\left(E_\gamma-E_{\gamma_1}\right)
\left(E_{\gamma}-E_{\gamma_2}\right)\left(E_{\gamma}-E_{\gamma_3}\right)
\left(E_{\gamma}-E_{\gamma^\prime}\right)}. \eeqa
\beqa {A}_5^{\gamma\gamma^\prime}(nnnn,ncn,c;t\e,t^2\e)
&=&\sum_{\gamma_1,\gamma_2\gamma_3} \left[(-\I t)\frac{\e^{-\I
E_{\gamma}t}}{\left(E_\gamma-E_{\gamma_1}\right)^2
\left(E_{\gamma}-E_{\gamma_2}\right)
\left(E_{\gamma}-E_{\gamma_3}\right)}\right.\nonumber\\
& &\left.+(-\I t)\frac{\e^{-\I
E_{\gamma_1}t}}{\left(E_{\gamma}-E_{\gamma_1}\right)^2
\left(E_{\gamma_1}-E_{\gamma_2}\right)\left(E_{\gamma_1}-E_{\gamma_3}\right)}\right]\nonumber\\
& &
\times\left|g_1^{\gamma\gamma_1}\right|^2g_1^{\gamma_1\gamma_2}g_1^{\gamma_2\gamma_3}
g_1^{\gamma_3\gamma_1}\eta_{\gamma\gamma_2}\eta_{\gamma\gamma_3}\delta_{\gamma\gamma^\prime}.\eeqa
\beqa & &{A}_5^{\gamma\gamma^\prime}(nnnn,ncn,n;t\e,t^2\e)
\nonumber\\ & & \quad =-\sum_{\gamma_1,\gamma_2,\gamma_3}(-\I
t)\frac{\e^{-\I E_{\gamma_1}t}
g_1^{\gamma\gamma_1}g_1^{\gamma_1\gamma_2}
g_1^{\gamma_2\gamma_3}g_1^{\gamma_3\gamma_1}g_1^{\gamma_1\gamma^\prime}
\eta_{\gamma\gamma^\prime}\eta_{\gamma\gamma_2}
\eta_{\gamma\gamma_3}\eta_{\gamma_2\gamma^\prime}
\eta_{\gamma_3\gamma^\prime}}{\left(E_\gamma-E_{\gamma_1}\right)
\left(E_{\gamma_1}-E_{\gamma_2}\right)\left(E_{\gamma_1}-E_{\gamma_3}\right)
\left(E_{\gamma_1}-E_{\gamma^\prime}\right)}. \eeqa
\beqa {A}_5^{\gamma\gamma^\prime}(nnnn,nnc,ck;t\e,t^2\e)
&=&\sum_{\gamma_1,\gamma_2} \left[(-\I t)\frac{\e^{-\I
E_{\gamma}t}}{\left(E_\gamma-E_{\gamma_1}\right)
\left(E_{\gamma}-E_{\gamma_2}\right)\left(E_{\gamma}-E_{\gamma^\prime}\right)^2}\right.\nonumber\\
& &\left.+(-\I t)\frac{\e^{-\I
E_{\gamma^\prime}t}}{\left(E_{\gamma}-E_{\gamma^\prime}\right)^2\left(E_{\gamma_1}-E_{\gamma^\prime}\right)
\left(E_{\gamma_2}-E_{\gamma^\prime}\right)}\right]\nonumber\\
& & \times
\left(g_1^{\gamma\gamma_1}g_1^{\gamma_1\gamma^\prime}\right)\left(
g_1^{\gamma^\prime\gamma_2}g_1^{\gamma_2\gamma}\right)g_1^{\gamma\gamma^\prime}\eta_{\gamma_1\gamma_2}
.\eeqa
\beqa & &{A}_5^{\gamma\gamma^\prime}(nnnn,nnc,nk;t\e,t^2\e)
\nonumber\\ & & \quad =\sum_{\gamma_1,\gamma_2,\gamma_3}(-\I
t)\frac{\e^{-\I E_{\gamma^\prime}t} g_1^{\gamma^\prime\gamma_2}
g_1^{\gamma_2\gamma_3}g_1^{\gamma_3\gamma^\prime}g_1^{\gamma\gamma_1}g_1^{\gamma_1\gamma^\prime}
\eta_{\gamma\gamma^\prime}\eta_{\gamma\gamma_2}
\eta_{\gamma\gamma_3}\eta_{\gamma_1\gamma_2}
\eta_{\gamma_1\gamma_3}}{\left(E_\gamma-E_{\gamma^\prime}\right)
\left(E_{\gamma_1}-E_{\gamma^\prime}\right)\left(E_{\gamma_2}-E_{\gamma^\prime}\right)
\left(E_{\gamma_3}-E_{\gamma^\prime}\right)}. \eeqa
\beqa {A}_5^{\gamma\gamma^\prime}(nnnn,nnn,cc;t\e,t^2\e)
&=&\sum_{\gamma_1,\gamma_2} \left[(-\I t)\frac{\e^{-\I
E_{\gamma}t}}{\left(E_\gamma-E_{\gamma_1}\right)
\left(E_{\gamma}-E_{\gamma_2}\right)\left(E_{\gamma}-E_{\gamma^\prime}\right)^2}\right.\nonumber\\
& &\left.+(-\I t)\frac{\e^{-\I
E_{\gamma^\prime}t}}{\left(E_{\gamma}-E_{\gamma^\prime}\right)^2\left(E_{\gamma_1}-E_{\gamma^\prime}\right)
\left(E_{\gamma_2}-E_{\gamma^\prime}\right)}\right]\nonumber\\
& & \times
g_1^{\gamma\gamma^\prime}g_1^{\gamma^\prime\gamma_1}g_1^{\gamma_1\gamma_2}
g_1^{\gamma_2\gamma}g_1^{\gamma\gamma^\prime}\eta_{\gamma\gamma_1}\eta_{\gamma_2\gamma^\prime}
.\eeqa
\beqa & &{A}_5^{\gamma\gamma^\prime}(nnnn,nnn,cn;t\e,t^2\e)
\nonumber\\ & & \quad =\sum_{\gamma_1,\gamma_2,\gamma_3}(-\I
t)\frac{\e^{-\I E_{\gamma}t}
g_1^{\gamma\gamma_1}g_1^{\gamma_1\gamma_2}
g_1^{\gamma_2\gamma_3}g_1^{\gamma_3\gamma}g_1^{\gamma\gamma^\prime}
\eta_{\gamma\gamma_2}
\eta_{\gamma_1\gamma_3}\eta_{\gamma_1\gamma^\prime}
\eta_{\gamma_2\gamma^\prime}\eta_{\gamma_3\gamma^\prime}}{\left(E_\gamma-E_{\gamma_1}\right)
\left(E_{\gamma}-E_{\gamma_2}\right)\left(E_{\gamma}-E_{\gamma_3}\right)
\left(E_{\gamma}-E_{\gamma^\prime}\right)}. \eeqa
\beqa & &{A}_5^{\gamma\gamma^\prime}(nnnn,nnn,nc;t\e,t^2\e)
\nonumber\\ & & \quad =\sum_{\gamma_1,\gamma_2,\gamma_3}(-\I
t)\frac{\e^{-\I E_{\gamma^\prime}t} g_1^{\gamma^\prime\gamma_1}
g_1^{\gamma_1\gamma_2}g_1^{\gamma_2\gamma_3}g_1^{\gamma_3\gamma^\prime}g_1^{\gamma\gamma^\prime}
\eta_{\gamma\gamma_1}\eta_{\gamma\gamma_2}
\eta_{\gamma\gamma_3}\eta_{\gamma_1\gamma_3}
\eta_{\gamma_2\gamma^\prime}}{\left(E_\gamma-E_{\gamma^\prime}\right)
\left(E_{\gamma_1}-E_{\gamma^\prime}\right)\left(E_{\gamma_2}-E_{\gamma^\prime}\right)
\left(E_{\gamma_3}-E_{\gamma^\prime}\right)}. \eeqa
\beqa & &{A}_5^{\gamma\gamma^\prime}(nnnn,nnn,nn,c;t\e,t^2\e)
\nonumber\\ & & \quad
=\sum_{\gamma_1,\gamma_2,\gamma_3,\gamma_4}(-\I t)\frac{\e^{-\I
E_{\gamma}t} g_1^{\gamma\gamma_1}g_1^{\gamma_1\gamma_2}
g_1^{\gamma_2\gamma_3}g_1^{\gamma_3\gamma_4}g_1^{\gamma_4\gamma}
\eta_{\gamma\gamma_2}
\eta_{\gamma\gamma_3}\eta_{\gamma_1\gamma_3}\eta_{\gamma_1\gamma_4}
\eta_{\gamma_2\gamma_4}\delta_{\gamma\gamma^\prime}}{\left(E_\gamma-E_{\gamma_1}\right)
\left(E_{\gamma}-E_{\gamma_2}\right)\left(E_{\gamma}-E_{\gamma_3}\right)
\left(E_{\gamma}-E_{\gamma_4}\right)}. \eeqa
\beqa {A}_5^{\gamma\gamma^\prime}(nnnn,nnn,nn,n;t\e,t^2\e)&=& 0
\eeqa

Based on above 52 contraction- and anti contraction- expressions, we
can, via the reorganization and summation, obtain \beqa A_5(t\e^{-\I
E_{\gamma} t},{\rm D})&=& -(-\I
G_\gamma^{(3)}t)\sum_{\gamma_1}\frac{\e^{-\I E_{\gamma}
t}}{\left(E_{\gamma}-E_{\gamma_1}\right)^2}g_1^{\gamma\gamma_1}g_1^{\gamma_1\gamma}
\delta_{\gamma\gamma^\prime}\nonumber\\
& &  -(-\I
G_\gamma^{(2)}t)\sum_{\gamma_1,\gamma_2}\left[\frac{\e^{-\I
E_{\gamma}
t}}{\left(E_{\gamma}-E_{\gamma_1}\right)^2\left(E_{\gamma}-E_{\gamma_2}\right)}\right.
\nonumber\\ & & \left. +\frac{\e^{-\I E_{\gamma}
t}}{\left(E_{\gamma}-E_{\gamma_1}\right)\left(E_{\gamma}-E_{\gamma_2}\right)^2}\right]
g_1^{\gamma\gamma_1}g_1^{\gamma_1\gamma_2} g_1^{\gamma_2\gamma}
\delta_{\gamma\gamma^\prime}+ (-\I
G_\gamma^{(5)}t)\delta_{\gamma\gamma^\prime}\eeqa where \beqa
G_\gamma^{(5)}&=&\sum_{\gamma_1,\gamma_2,\gamma_3,\gamma_4}
\frac{g_1^{\gamma\gamma_1}g_1^{\gamma_1\gamma_2}
g_1^{\gamma_2\gamma_3}g_1^{\gamma_3\gamma_4}g_1^{\gamma_4\gamma}
\eta_{\gamma\gamma_2}\eta_{\gamma\gamma_3}}
{\left(E_{\gamma}-E_{\gamma_1}\right)\left(E_{\gamma}-E_{\gamma_2}\right)
\left(E_{\gamma}-E_{\gamma_3}\right)\left(E_{\gamma}-E_{\gamma_4}\right)}\nonumber\\
& & -\sum_{\gamma_1,\gamma_2,\gamma_3}\left[
\frac{g_1^{\gamma\gamma_1}g_1^{\gamma\gamma_2}
g_1^{\gamma_1\gamma}g_1^{\gamma_2\gamma_3}g_1^{\gamma_3\gamma}}
{\left(E_{\gamma}-E_{\gamma_1}\right)^2\left(E_{\gamma}-E_{\gamma_2}\right)
\left(E_{\gamma}-E_{\gamma_3}\right)}
+\frac{g_1^{\gamma\gamma_1}g_1^{\gamma\gamma_2}
g_1^{\gamma_1\gamma}g_1^{\gamma_2\gamma_3}g_1^{\gamma_3\gamma}}
{\left(E_{\gamma}-E_{\gamma_1}\right)\left(E_{\gamma}-E_{\gamma_2}\right)^2
\left(E_{\gamma}-E_{\gamma_3}\right)}\right.\nonumber\\
& &\left.+\frac{g_1^{\gamma\gamma_1}g_1^{\gamma\gamma_2}
g_1^{\gamma_1\gamma}g_1^{\gamma_2\gamma_3}g_1^{\gamma_3\gamma}}
{\left(E_{\gamma}-E_{\gamma_1}\right)\left(E_{\gamma}-E_{\gamma_2}\right)
\left(E_{\gamma}-E_{\gamma_3}\right)^2}\right].\hskip 1.2cm\eeqa
\beqa A_5(t\e^{-\I E_{\gamma_1} t},{\rm D})&=& (-\I
G_\gamma^{(3)}t)\sum_{\gamma_1}\frac{\e^{-\I E_{\gamma_1}
t}}{\left(E_{\gamma}-E_{\gamma_1}\right)^2}g_1^{\gamma\gamma_1}g_1^{\gamma_1\gamma}
\delta_{\gamma\gamma^\prime}\nonumber\\
& &  +(-\I G_\gamma^{(2)}t)\sum_{\gamma_1,\gamma_2}\frac{\e^{-\I
E_{\gamma_1}
t}}{\left(E_{\gamma}-E_{\gamma_1}\right)^2\left(E_{\gamma_1}-E_{\gamma_2}\right)}
g_1^{\gamma\gamma_1}g_1^{\gamma_1\gamma_2} g_1^{\gamma_2\gamma}
\delta_{\gamma\gamma^\prime}\eeqa
\beqa A_5(t\e^{-\I E_{\gamma_2} t},{\rm D})&=& -(-\I
G_\gamma^{(2)}t)\sum_{\gamma_1,\gamma_2}\frac{\e^{-\I E_{\gamma_2}
t}}{\left(E_{\gamma}-E_{\gamma_2}\right)^2\left(E_{\gamma_1}-E_{\gamma_2}\right)}
g_1^{\gamma\gamma_1}g_1^{\gamma_1\gamma_2} g_1^{\gamma_2\gamma}
\delta_{\gamma\gamma^\prime}\eeqa
\beqa A_5(t\e^{-\I E_{\gamma^\prime} t},{\rm D})&=& 0\eeqa
\beqa A_5(t\e^{-\I E_{\gamma} t},{\rm N})&=&(-\I G_\gamma^{(4)}
t)\frac{\e^{-\I E_{\gamma}
t}g_1^{\gamma\gamma^\prime}}{\left(E_{\gamma}-E_{\gamma^\prime}\right)}+
(-\I G_\gamma^{(3)} t)\sum_{\gamma_1}\frac{\e^{-\I E_{\gamma}
t}g_1^{\gamma\gamma_1}g_1^{\gamma_1\gamma^\prime}
\eta^{\gamma\gamma^\prime}}{\left(E_{\gamma}-E_{\gamma_1}\right)\left(E_{\gamma}-E_{\gamma^\prime}\right)}
\nonumber\\
& &-(-\I G_\gamma^{(2)} t)\sum_{\gamma_1}\left[\frac{\e^{-\I
E_{\gamma} t} g_1^{\gamma\gamma_1}g_1^{\gamma_1\gamma}
g_1^{\gamma\gamma^\prime}}{\left(E_{\gamma}-E_{\gamma_1}\right)^2\left(E_{\gamma}-E_{\gamma^\prime}\right)}
+\frac{\e^{-\I E_{\gamma} t}
g_1^{\gamma\gamma_1}g_1^{\gamma_1\gamma}
g_1^{\gamma\gamma^\prime}}{\left(E_{\gamma}-E_{\gamma_1}\right)
\left(E_{\gamma}-E_{\gamma^\prime}\right)^2}\right]\nonumber\\
& & +(-\I G_\gamma^{(2)} t)\sum_{\gamma_1,\gamma_2}\frac{\e^{-\I
E_{\gamma} t}g_1^{\gamma\gamma_1}g_1^{\gamma_1\gamma_2}
g_1^{\gamma_2\gamma^\prime}\eta_{\gamma\gamma_2}\eta_{\gamma\gamma^\prime}}
{\left(E_{\gamma}-E_{\gamma_1}\right)\left(E_{\gamma}-E_{\gamma_2}\right)
\left(E_{\gamma}-E_{\gamma^\prime}\right)} \eeqa
\beqa A_5(t\e^{-\I E_{\gamma_1} t},{\rm N})&=&-(-\I
t)\sum_{\gamma_1}\frac{G_{\gamma_1}^{(3)}\e^{-\I
E_{\gamma_1}t}g_1^{\gamma\gamma_1}g_1^{\gamma_1\gamma^\prime}
\eta_{\gamma\gamma^\prime}}{\left(E_{\gamma}-E_{\gamma_1}\right)
\left(E_{\gamma_1}-E_{\gamma^\prime}\right)}
\nonumber\\
& &-(-\I t)\sum_{\gamma_1,\gamma_2}\frac{G_{\gamma_1}^{(2)}\e^{-\I
E_{\gamma_1}t}g_1^{\gamma\gamma_1}g_1^{\gamma_1\gamma_2}
g_1^{\gamma_2\gamma^\prime}\eta_{\gamma_1\gamma^\prime}\eta_{\gamma\gamma^\prime}}
{\left(E_{\gamma}-E_{\gamma_1}\right)\left(E_{\gamma_1}-E_{\gamma_2}\right)
\left(E_{\gamma_1}-E_{\gamma^\prime}\right)} \eeqa
\beqa A_5(t\e^{-\I E_{\gamma_2} t},{\rm N})&=&(-\I
t)\sum_{\gamma_1,\gamma_2}\frac{G_{\gamma_2}^{(2)}\e^{-\I
E_{\gamma_2}t}g_1^{\gamma\gamma_1}g_1^{\gamma_1\gamma_2}
g_1^{\gamma_2\gamma^\prime}\eta_{\gamma\gamma_2}\eta_{\gamma\gamma^\prime}}
{\left(E_{\gamma}-E_{\gamma_2}\right)\left(E_{\gamma_1}-E_{\gamma_2}\right)
\left(E_{\gamma_2}-E_{\gamma^\prime}\right)} \eeqa
\beqa A_5(t\e^{-\I E_{\gamma^\prime} t},{\rm N})&=&-(-\I
G_{\gamma^\prime}^{(4)} t)\frac{\e^{-\I E_{\gamma^\prime}
t}g_1^{\gamma\gamma^\prime}}{\left(E_{\gamma}-E_{\gamma^\prime}\right)}+
(-\I G_{\gamma^\prime}^{(3)} t)\sum_{\gamma_1}\frac{\e^{-\I
E_{\gamma^\prime} t}g_1^{\gamma\gamma_1}g_1^{\gamma_1\gamma^\prime}
\eta^{\gamma\gamma^\prime}}{\left(E_{\gamma}-E_{\gamma^\prime}\right)
\left(E_{\gamma_1}-E_{\gamma^\prime}\right)}
\nonumber\\
& &+(-\I G_{\gamma^\prime}^{(2)}
t)\sum_{\gamma_1}\left[\frac{\e^{-\I E_{\gamma^\prime} t}
g_1^{\gamma^\prime\gamma_1}g_1^{\gamma_1\gamma^\prime}
g_1^{\gamma\gamma^\prime}}{\left(E_{\gamma}-E_{\gamma^\prime}\right)^2
\left(E_{\gamma_1}-E_{\gamma^\prime}\right)} +\frac{\e^{-\I
E_{\gamma^\prime} t}
g_1^{\gamma^\prime\gamma_1}g_1^{\gamma_1\gamma^\prime}
g_1^{\gamma\gamma^\prime}}{\left(E_{\gamma}-E_{\gamma^\prime}\right)
\left(E_{\gamma_1}-E_{\gamma^\prime}\right)^2}\right]\nonumber\\
& & -(-\I G_{\gamma^\prime}^{(2)}
t)\sum_{\gamma_1,\gamma_2}\frac{\e^{-\I E_{\gamma^\prime}
t}g_1^{\gamma\gamma_1}g_1^{\gamma_1\gamma_2}
g_1^{\gamma_2\gamma^\prime}\eta_{\gamma_1\gamma^\prime}\eta_{\gamma\gamma^\prime}}
{\left(E_{\gamma}-E_{\gamma^\prime}\right)\left(E_{\gamma_1}-E_{\gamma^\prime}\right)
\left(E_{\gamma_2}-E_{\gamma^\prime}\right)} \eeqa

For the parts with $t^2\e$, we have \beqa A_5(t^2\e^{-\I E_{\gamma}
t},{\rm D})&=&\frac{(-\I t)^2}{2!} 2
G_{\gamma}^{(2)}G_{\gamma}^{(3)}\delta_{\gamma\gamma^\prime}\e^{-\I
E_{\gamma} t},\\
A_5(t^2\e^{-\I E_{\gamma_1} t},{\rm D})&=&A_5(t^2\e^{-\I
E_{\gamma_2} t},{\rm D})=A_5(t^2\e^{-\I E_{\gamma^\prime} t},{\rm
D})=0.\eeqa
\beqa A_5(t^2\e^{-\I E_{\gamma} t},{\rm N})&=&\frac{(-\I t)^2}{2!}
\left(G_{\gamma}^{(2)}\right)^2\frac{\e^{-\I
E_{\gamma} t}}{E_{\gamma}-E_{\gamma^\prime}}g_1^{\gamma\gamma^\prime},\\
A_5(t^2\e^{-\I E_{\gamma_2} t},{\rm N})&=&A_5(t^2\e^{-\I
E_{\gamma^\prime} t},{\rm N})=0,\\
A_5(t^2\e^{-\I E_{\gamma^\prime} t},{\rm N})&=&-\frac{(-\I t)^2}{2!}
\left(G_{\gamma^\prime}^{(2)}\right)^2\frac{\e^{-\I
E_{\gamma^\prime}
t}}{E_{\gamma}-E_{\gamma^\prime}}g_1^{\gamma\gamma^\prime}.\eeqa

It is clear that above diagonal and nondiagonal part about
$A_5^{\gamma\gamma^\prime}(t\e)$ and
$A_5^{\gamma\gamma^\prime}(t\e)$ indeed has the expected forms and
can be merged reasonably to the lower order terms in order to obtain
the improved forms of perturbed solutions.

\subsection{$l=6$ case}

Now let we consider the case of the sixth order approximation
($l=6$). From eq.(\ref{gpd}) it follows that the first
decompositions of $g$-product have $2^5=32$ terms. Like the $l=5$
case, they can be divided into 6 group \beq
A_6^{\gamma\gamma^\prime}= \sum_{i=0}^4
\mathcal{A}_6^{\gamma\gamma^\prime}(i;\eta),\eeq where $i$ indicates
the number of $\eta$ functions. Obviously \beq
\mathcal{A}_6^{\gamma\gamma^\prime}(0;\eta)={A}_6^{\gamma\gamma^\prime}(ccccc),\eeq
\beqa
\mathcal{A}_6^{\gamma\gamma^\prime}(1;\eta)&=&{A}_6^{\gamma\gamma^\prime}(ccccn)
+{A}_6^{\gamma\gamma^\prime}(cccnc)+{A}_6^{\gamma\gamma^\prime}(ccncc)\nonumber\\
& &
+{A}_6^{\gamma\gamma^\prime}(cnccc)+{A}_6^{\gamma\gamma^\prime}(ncccc),\eeqa
\beqa \mathcal{A}_6^{\gamma\gamma^\prime}(2;\eta)&=&
{A}_6^{\gamma\gamma^\prime}(cccn)+
{A}_6^{\gamma\gamma^\prime}(ccncn)+
{A}_6^{\gamma\gamma^\prime}(cnccn)+
{A}_6^{\gamma\gamma^\prime}(ncccn)\nonumber\\ & & +
{A}_6^{\gamma\gamma^\prime}(ccnnc) +
{A}_6^{\gamma\gamma^\prime}(cncnc) +
 {A}_6^{\gamma\gamma^\prime}(nccnc)+
 {A}_6^{\gamma\gamma^\prime}(cnncc)\nonumber\\ & &+
 {A}_6^{\gamma\gamma^\prime}(ncncc) +
 {A}_6^{\gamma\gamma^\prime}(nnccc),\eeqa \beqa
\mathcal{A}_6^{\gamma\gamma^\prime}(3;\eta)&=&
{A}_6^{\gamma\gamma^\prime}(ccnnn)
+{A}_6^{\gamma\gamma^\prime}(cncnn) +
{A}_6^{\gamma\gamma^\prime}(cnncn) +
{A}_6^{\gamma\gamma^\prime}(cnnnc)\nonumber\\ & &  +
{A}_6^{\gamma\gamma^\prime}(nccnn) +
{A}_6^{\gamma\gamma^\prime}(ncncn) +
{A}_6^{\gamma\gamma^\prime}(ncnnc) +
{A}_6^{\gamma\gamma^\prime}(nnccn)\nonumber\\ & &  +
{A}_6^{\gamma\gamma^\prime}(nncnc) +
{A}_6^{\gamma\gamma^\prime}(nnncc), \eeqa \beqa
\mathcal{A}_6^{\gamma\gamma^\prime}(4;\eta)&=&{A}_6^{\gamma\gamma^\prime}(cnnnn)
+ {A}_6^{\gamma\gamma^\prime}(ncnnn)
+{A}_6^{\gamma\gamma^\prime}(nncnn)\nonumber\\ & &  +
{A}_6^{\gamma\gamma^\prime}(nnncn)+
{A}_6^{\gamma\gamma^\prime}(nnnnc) \eeqa
 \beq
\mathcal{A}_6^{\gamma\gamma^\prime}(5;\eta)={A}_6^{\gamma\gamma^\prime}(nnnnn).\eeq

Furthermore considering the high order contraction or
anti-contraction, we have \beqa {A}_6^{\gamma\gamma^\prime}(cccnn)
&=& {A}_6^{\gamma\gamma^\prime}(cccnn, kkkc) +
{A}_6^{\gamma\gamma^\prime}(cccnn, kkkn),\\
{A}_6^{\gamma\gamma^\prime}(ccncn) &=&
{A}_6^{\gamma\gamma^\prime}(ccncn, kkc) +
{A}_6^{\gamma\gamma^\prime}(ccncn, kkn),\\
{A}_6^{\gamma\gamma^\prime}(ccnnc) &=&
{A}_6^{\gamma\gamma^\prime}(ccnnc, kkck) +
{A}_6^{\gamma\gamma^\prime}(ccnnc, kknk),\\
{A}_6^{\gamma\gamma^\prime}(cnccn) &=&
{A}_6^{\gamma\gamma^\prime}(cnccn, kc) +
{A}_6^{\gamma\gamma^\prime}(cnccn, kn),\\
{A}_6^{\gamma\gamma^\prime}(cncnc) &=&
{A}_6^{\gamma\gamma^\prime}(cncnc, kck) +
{A}_6^{\gamma\gamma^\prime}(cncnc, knk),\\
{A}_6^{\gamma\gamma^\prime}(cnncc) &=&
{A}_6^{\gamma\gamma^\prime}(cnncc, kckk) +
{A}_6^{\gamma\gamma^\prime}(cnncc, knkk),\\
{A}_6^{\gamma\gamma^\prime}(ncccn) &=&
{A}_6^{\gamma\gamma^\prime}(ncccn, c) +
{A}_6^{\gamma\gamma^\prime}(ncccn, n),\\
{A}_6^{\gamma\gamma^\prime}(nccnc) &=&
{A}_6^{\gamma\gamma^\prime}(nccnc, ck) +
{A}_6^{\gamma\gamma^\prime}(nccnc, nk),\\
{A}_6^{\gamma\gamma^\prime}(ncncc) &=&
{A}_6^{\gamma\gamma^\prime}(ncncc, ckk) +
{A}_6^{\gamma\gamma^\prime}(ncncc, nkk),\\
{A}_6^{\gamma\gamma^\prime}(nnccc) &=&
{A}_6^{\gamma\gamma^\prime}(nnccc, ckkk) +
{A}_6^{\gamma\gamma^\prime}(nnccc, nkkk) \eeqa
\beqa {A}_6^{\gamma\gamma^\prime}(ccnnn) &=&
 {A}_6^{\gamma\gamma^\prime}(ccnnn, kkcc)
 + {A}_6^{\gamma\gamma^\prime}(ccnnn, kkcn)
 + {A}_6^{\gamma\gamma^\prime}(ccnnn, kknc)\nonumber\\
 & & +
    {A}_6^{\gamma\gamma^\prime}(ccnnn, kknn, kkc) + {A}_6^{\gamma\gamma^\prime}(ccnnn, kknn,
    kkn),\\
{A}_6^{\gamma\gamma^\prime}(cncnn) &=&
  {A}_6^{\gamma\gamma^\prime}(cncnn, kkkc, kck)
  + {A}_6^{\gamma\gamma^\prime}(cncnn, kkkc, knk)
  + {A}_6^{\gamma\gamma^\prime}(cncnn, kkkn, kck)\nonumber\\ & &
  + {A}_6^{\gamma\gamma^\prime}(cncnn, kkkn, knk, kc)
    + {A}_6^{\gamma\gamma^\prime}(cncnn, kkkn, knk, kn),\\
{A}_6^{\gamma\gamma^\prime}(cnncn) &=&
  {A}_6^{\gamma\gamma^\prime}(cnncn, kckk, kkc)
  + {A}_6^{\gamma\gamma^\prime}(cnncn, kckk, kkn)
  + {A}_6^{\gamma\gamma^\prime}(cnncn, knkk, kkc)\nonumber\\
  & &+
    {A}_6^{\gamma\gamma^\prime}(cnncn, knkk, kkn, kc)
    + {A}_6^{\gamma\gamma^\prime}(cnncn, knkk, kkn, kn),\\
{A}_6^{\gamma\gamma^\prime}(cnnnc) &=&
  {A}_6^{\gamma\gamma^\prime}(cnnnc, kcck)
  + {A}_6^{\gamma\gamma^\prime}(cnnnc, kcnk)
  + {A}_6^{\gamma\gamma^\prime}(cnnnc, knck)\nonumber\\
    & &  +
    {A}_6^{\gamma\gamma^\prime}(cnnnc, knnk, kck)+ {A}_6^{\gamma\gamma^\prime}(cnnnc, knnk,
    knk),\\
{A}_6^{\gamma\gamma^\prime}(nccnn) &=&
  {A}_6^{\gamma\gamma^\prime}(nccnn, kkkc, ck)
  + {A}_6^{\gamma\gamma^\prime}(nccnn, kkkc, nk)
  + {A}_6^{\gamma\gamma^\prime}(nccnn, kkkn, ck)\nonumber\\
  & & +
    {A}_6^{\gamma\gamma^\prime}(nccnn, kkkn, nk, c) + {A}_6^{\gamma\gamma^\prime}(nccnn, kkkn, nk,
    n),\\
{A}_6^{\gamma\gamma^\prime}(ncncn) &=&
  {A}_6^{\gamma\gamma^\prime}(ncncn, ckc)
  + {A}_6^{\gamma\gamma^\prime}(ncncn, ckn)
  + {A}_6^{\gamma\gamma^\prime}(ncncn, nkc)\nonumber\\
  & & + {A}_6^{\gamma\gamma^\prime}(ncncn, nkn, c) +
    {A}_6^{\gamma\gamma^\prime}(ncncn, nkn, n),\\
{A}_6^{\gamma\gamma^\prime}(ncnnc) &=&
  {A}_6^{\gamma\gamma^\prime}(ncnnc, kkck, ckk)
  + {A}_6^{\gamma\gamma^\prime}(ncnnc, kkck, nkk)
  + {A}_6^{\gamma\gamma^\prime}(ncnnc, kknk, ckk)\nonumber\\
  & & +
    {A}_6^{\gamma\gamma^\prime}(ncnnc, kknk, nkk, ck)
    + {A}_6^{\gamma\gamma^\prime}(ncnnc, kknk, nkk, nk),\\
{A}_6^{\gamma\gamma^\prime}(nnccn) &=&
  {A}_6^{\gamma\gamma^\prime}(nnccn, cc)
  + {A}_6^{\gamma\gamma^\prime}(nnccn, cn)
  + {A}_6^{\gamma\gamma^\prime}(nnccn, nc)\nonumber\\
& &  + {A}_6^{\gamma\gamma^\prime}(nnccn, nn, c) +
    {A}_6^{\gamma\gamma^\prime}(nnccn, nn, n),\\
{A}_6^{\gamma\gamma^\prime}(nncnc) &=&
  {A}_6^{\gamma\gamma^\prime}(nncnc, ckkk, kck)
  + {A}_6^{\gamma\gamma^\prime}(nncnc, ckkk, knk)
  + {A}_6^{\gamma\gamma^\prime}(nncnc, nkkk, kck)\nonumber\\
  & & +
    {A}_6^{\gamma\gamma^\prime}(nncnc, nkkk, knk, ck)
    + {A}_6^{\gamma\gamma^\prime}(nncnc, nkkk, knk, nk),\\
{A}_6^{\gamma\gamma^\prime}(nnncc) &=&
  {A}_6^{\gamma\gamma^\prime}(nnncc, cckk)
  + {A}_6^{\gamma\gamma^\prime}(nnncc, cnkk)
  + {A}_6^{\gamma\gamma^\prime}(nnncc, nckk) \nonumber\\
  & & +
    {A}_6^{\gamma\gamma^\prime}(nnncc, nnkk, c) + {A}_6^{\gamma\gamma^\prime}(nnncc, nnkk, n)
\eeqa
\beqa {A}_6^{\gamma\gamma^\prime}(cnnnn) &=&
  {A}_6^{\gamma\gamma^\prime}(cnnnn, kccc)
  + {A}_6^{\gamma\gamma^\prime}(cnnnn, kccn)
  + {A}_6^{\gamma\gamma^\prime}(cnnnn, kcnc)\nonumber\\
  & &
  + {A}_6^{\gamma\gamma^\prime}(cnnnn, kncc)
  + {A}_6^{\gamma\gamma^\prime}(cnnnn, kcnn)
  + {A}_6^{\gamma\gamma^\prime}(cnnnn, kncn)\nonumber\\
  & & + {A}_6^{\gamma\gamma^\prime}(cnnnn, knnc)
  + {A}_6^{\gamma\gamma^\prime}(cnnnn, knnn),
\eeqa \beqa {A}_6^{\gamma\gamma^\prime}(cnnnn, kcnn) &=&
{A}_6^{\gamma\gamma^\prime}(cnnnn, kcnn, kkc) +
{A}_6^{\gamma\gamma^\prime}(cnnnn, kcnn, kkn),\\
{A}_6^{\gamma\gamma^\prime}(cnnnn, kncn) &=&
{A}_6^{\gamma\gamma^\prime}(cnnnn, kncn, kc) +
{A}_6^{\gamma\gamma^\prime}(cnnnn, kncn, kn),\\
{A}_6^{\gamma\gamma^\prime}(cnnnn, knnc) &=&
{A}_6^{\gamma\gamma^\prime}(cnnnn, knnc, kck) +
{A}_6^{\gamma\gamma^\prime}(cnnnn, knnc, knk), \eeqa \beqa
{A}_6^{\gamma\gamma^\prime}(cnnnn, knnn) &=&
  {A}_6^{\gamma\gamma^\prime}(cnnnn, knnn, kcc)
  + {A}_6^{\gamma\gamma^\prime}(cnnnn, knnn, kcn)\nonumber\\ & &
  + {A}_6^{\gamma\gamma^\prime}(cnnnn, knnn, knc)+
    {A}_6^{\gamma\gamma^\prime}(cnnnn, knnn, knn, kc)\nonumber\\ & &
    + {A}_6^{\gamma\gamma^\prime}(cnnnn, knnn, knn, kn).
\eeqa
\beqa {A}_6^{\gamma\gamma^\prime}(ncnnn) &=&
  {A}_6^{\gamma\gamma^\prime}(ncnnn, kkcc, ckk)
  + {A}_6^{\gamma\gamma^\prime}(ncnnn, kkcc, nkk)\nonumber\\ & &
  + {A}_6^{\gamma\gamma^\prime}(ncnnn, kkcn, ckk) +
    {A}_6^{\gamma\gamma^\prime}(ncnnn, kknc, ckk)\nonumber\\
  & &
    + {A}_6^{\gamma\gamma^\prime}(ncnnn, kkcn, nkk)
    + {A}_6^{\gamma\gamma^\prime}(ncnnn, kknc, nkk)\nonumber\\
& & +
    {A}_6^{\gamma\gamma^\prime}(ncnnn, kknn, ckk)
    + {A}_6^{\gamma\gamma^\prime}(ncnnn, kknn, nkk),
    \eeqa
\beqa {A}_6^{\gamma\gamma^\prime}(ncnnn, kkcn, nkk) &=&
{A}_6^{\gamma\gamma^\prime}(ncnnn, kkcn, nkk, c) +
{A}_6^{\gamma\gamma^\prime}(ncnnn, kkcn, nkk, n),\\
{A}_6^{\gamma\gamma^\prime}(ncnnn, kknc, nkk) &=&
  {A}_6^{\gamma\gamma^\prime}(ncnnn, kknc, nkk, ck)
  + {A}_6^{\gamma\gamma^\prime}(ncnnn, kknc, nkk, nk),\ \ \ \ \ \ \ \\
{A}_6^{\gamma\gamma^\prime}(ncnnn, kknn, ckk) &=&
{A}_6^{\gamma\gamma^\prime}(ncnnn, kknn, ckc) +
{A}_6^{\gamma\gamma^\prime}(ncnnn, kknn, ckn), \eeqa
 \beqa
{A}_6^{\gamma\gamma^\prime}(ncnnn, kknn, nkk) &=&
  {A}_6^{\gamma\gamma^\prime}(ncnnn, kknn, nkc, ck)
  + {A}_6^{\gamma\gamma^\prime}(ncnnn, kknn, nkc, nk)\nonumber\\
 & &
  +
    {A}_6^{\gamma\gamma^\prime}(ncnnn, kknn, nkn, ck)
    + {A}_6^{\gamma\gamma^\prime}(ncnnn, kknn, nkn, nk, c)\nonumber\\
 & & +
    {A}_6^{\gamma\gamma^\prime}(ncnnn, kknn, nkn, nk, n)
\eeqa
\beqa {A}_6^{\gamma\gamma^\prime}(nncnn) &=&
  {A}_6^{\gamma\gamma^\prime}(nncnn, ckkc, kck)
  + {A}_6^{\gamma\gamma^\prime}(nncnn, ckkc, knk)\nonumber\\
  & &
  + {A}_6^{\gamma\gamma^\prime}(nncnn, ckkn, kck)+
    {A}_6^{\gamma\gamma^\prime}(nncnn, nkkc, kck)\nonumber\\
  & &
    + {A}_6^{\gamma\gamma^\prime}(nncnn, ckkn, knk)
    + {A}_6^{\gamma\gamma^\prime}(nncnn, nkkc, knk)\nonumber\\
    & & +
    {A}_6^{\gamma\gamma^\prime}(nncnn, nkkn, kck)
    + {A}_6^{\gamma\gamma^\prime}(nncnn, nkkn, knk),
\eeqa \beqa {A}_6^{\gamma\gamma^\prime}(nncnn, ckkn, knk) &=&
  {A}_6^{\gamma\gamma^\prime}(nncnn, ckkn, knk, kc)
  + {A}_6^{\gamma\gamma^\prime}(nncnn, ckkn, knk, kn),\ \ \ \ \ \ \ \\
{A}_6^{\gamma\gamma^\prime}(nncnn, nkkc, knk) &=&
  {A}_6^{\gamma\gamma^\prime}(nncnn, nkkc, knk, ck)
  + {A}_6^{\gamma\gamma^\prime}(nncnn, nkkc, knk, nk),\\
{A}_6^{\gamma\gamma^\prime}(nncnn, nkkn, kck) &=&
{A}_6^{\gamma\gamma^\prime}(nncnn, nkkn, kck, c) +
{A}_6^{\gamma\gamma^\prime}(nncnn, nkkn, kck, n), \eeqa \beqa
{A}_6^{\gamma\gamma^\prime}(nncnn, nkkn, knk) &=&
  {A}_6^{\gamma\gamma^\prime}(nncnn, nkkn, knk, cc)
  + {A}_6^{\gamma\gamma^\prime}(nncnn, nkkn, knk, cn)\nonumber\\
  & &
  +
    {A}_6^{\gamma\gamma^\prime}(nncnn, nkkn, knk, nc)
    + {A}_6^{\gamma\gamma^\prime}(nncnn, nkkn, knk, nn, c)\nonumber\\
  & &  +
    {A}_6^{\gamma\gamma^\prime}(nncnn, nkkn, knk, nn, n)
\eeqa
\beqa {A}_6^{\gamma\gamma^\prime}(nnncn) &=&
  {A}_6^{\gamma\gamma^\prime}(nnncn, cckk, kkc)
  + {A}_6^{\gamma\gamma^\prime}(nnncn, cckk, kkn)\nonumber\\
  & &
  + {A}_6^{\gamma\gamma^\prime}(nnncn, cnkk, kkc)
  +
    {A}_6^{\gamma\gamma^\prime}(nnncn, nckk, kkc)\nonumber\\
  & &
    + {A}_6^{\gamma\gamma^\prime}(nnncn, cnkk, kkn)
    + {A}_6^{\gamma\gamma^\prime}(nnncn, nckk, kkn)\nonumber\\
    & & +
    {A}_6^{\gamma\gamma^\prime}(nnncn, nnkk, kkc)
    + {A}_6^{\gamma\gamma^\prime}(nnncn, nnkk, kkn),
\eeqa \beqa {A}_6^{\gamma\gamma^\prime}(nnncn, cnkk, kkn) &=&
  {A}_6^{\gamma\gamma^\prime}(nnncn, cnkk, kkn, kc)
  + {A}_6^{\gamma\gamma^\prime}(nnncn, cnkk, kkn, kn),\ \ \ \ \ \ \\
{A}_6^{\gamma\gamma^\prime}(nnncn, nckk, kkn) &=&
{A}_6^{\gamma\gamma^\prime}(nnncn, nckk, kkn, c) +
{A}_6^{\gamma\gamma^\prime}(nnncn, nckk, kkn, n),\\
{A}_6^{\gamma\gamma^\prime}(nnncn, nnkk, kkc) &=&
{A}_6^{\gamma\gamma^\prime}(nnncn, nnkk, ckc) +
{A}_6^{\gamma\gamma^\prime}(nnncn, nnkk, nkc), \eeqa \beqa
{A}_6^{\gamma\gamma^\prime}(nnncn, nnkk, kkn) &=&
  {A}_6^{\gamma\gamma^\prime}(nnncn, nnkk, ckn, kc)
  + {A}_6^{\gamma\gamma^\prime}(nnncn, nnkk, ckn, kn)\nonumber\\
  & &
  +
    {A}_6^{\gamma\gamma^\prime}(nnncn, nnkk, nkn, kc)
    + {A}_6^{\gamma\gamma^\prime}(nnncn, nnkk, nkn, kn, c)\nonumber\\
  & &  +
    {A}_6^{\gamma\gamma^\prime}(nnncn, nnkk, nkn, kn, n).
\eeqa
\beqa {A}_6^{\gamma\gamma^\prime}(nnnnc) &=&
  {A}_6^{\gamma\gamma^\prime}(nnnnc, ccck)
  + {A}_6^{\gamma\gamma^\prime}(nnnnc, ccnk)
  + {A}_6^{\gamma\gamma^\prime}(nnnnc, cnck)\nonumber\\
  & &
  + {A}_6^{\gamma\gamma^\prime}(nnnnc, ncck)+
    {A}_6^{\gamma\gamma^\prime}(nnnnc, cnnk)
    + {A}_6^{\gamma\gamma^\prime}(nnnnc, ncnk)\nonumber\\
    & &
    + {A}_6^{\gamma\gamma^\prime}(nnnnc, nnck)
    + {A}_6^{\gamma\gamma^\prime}(nnnnc, nnnk),
\eeqa \beqa {A}_6^{\gamma\gamma^\prime}(nnnnc, cnnk) &=&
{A}_6^{\gamma\gamma^\prime}(nnnnc, cnnk, kck) +
{A}_6^{\gamma\gamma^\prime}(nnnnc, cnnk, knk),\\
{A}_6^{\gamma\gamma^\prime}(nnnnc, ncnk) &=&
{A}_6^{\gamma\gamma^\prime}(nnnnc, ncnk, ck) +
{A}_6^{\gamma\gamma^\prime}(nnnnc, ncnk, nk),\\
{A}_6^{\gamma\gamma^\prime}(nnnnc, nnck) &=&
{A}_6^{\gamma\gamma^\prime}(nnnnc, nnck, c) +
{A}_6^{\gamma\gamma^\prime}(nnnnc, nnck, n),\eeqa \beqa
{A}_6^{\gamma\gamma^\prime}(nnnnc, nnnk) &=&
  {A}_6^{\gamma\gamma^\prime}(nnnnc, nnnk, cck)
  + {A}_6^{\gamma\gamma^\prime}(nnnnc, nnnk, cnk)\nonumber\\
  & &
  + {A}_6^{\gamma\gamma^\prime}(nnnnc, nnnk, nck) +
    {A}_6^{\gamma\gamma^\prime}(nnnnc, nnnk, nnk, ck)\nonumber\\
  & &
    + {A}_6^{\gamma\gamma^\prime}(nnnnc, nnnk, nnk, nk).
\eeqa
\beqa {A}_6^{\gamma\gamma^\prime}(nnnnn) &=&
  {A}_6^{\gamma\gamma^\prime}(nnnnn, cccc)
  + {A}_6^{\gamma\gamma^\prime}(nnnnn, cccn)
  + {A}_6^{\gamma\gamma^\prime}(nnnnn, ccnc)\nonumber\\ & &
  + {A}_6^{\gamma\gamma^\prime}(nnnnn, cncc)
  +
    {A}_6^{\gamma\gamma^\prime}(nnnnn, nccc)
    + {A}_6^{\gamma\gamma^\prime}(nnnnn, ccnn)\nonumber\\ & &
    + {A}_6^{\gamma\gamma^\prime}(nnnnn, cncn)
    +
    {A}_6^{\gamma\gamma^\prime}(nnnnn, cnnc)
    + {A}_6^{\gamma\gamma^\prime}(nnnnn, nccn)\nonumber\\ & &
    + {A}_6^{\gamma\gamma^\prime}(nnnnn, ncnc)
    +
    {A}_6^{\gamma\gamma^\prime}(nnnnn, nncc)
    + {A}_6^{\gamma\gamma^\prime}(nnnnn, cnnn)\nonumber\\
    & & + {A}_6^{\gamma\gamma^\prime}(nnnnn, ncnn)
    +
    {A}_6^{\gamma\gamma^\prime}(nnnnn, nncn)
    + {A}_6^{\gamma\gamma^\prime}(nnnnn, nnnc) \nonumber\\
  & &+ {A}_6^{\gamma\gamma^\prime}(nnnnn,
    nnnn),\eeqa
\beqa {A}_6^{\gamma\gamma^\prime}(nnnnn, ccnn) &=&
{A}_6^{\gamma\gamma^\prime}(nnnnn, ccnn, kkc) +
{A}_6^{\gamma\gamma^\prime}(nnnnn, ccnn, kkn),\\
{A}_6^{\gamma\gamma^\prime}(nnnnn, cncn) &=&
{A}_6^{\gamma\gamma^\prime}(nnnnn, cncn, kc) +
{A}_6^{\gamma\gamma^\prime}(nnnnn, cncn, kn),\\
{A}_6^{\gamma\gamma^\prime}(nnnnn, cnnc) &=&
{A}_6^{\gamma\gamma^\prime}(nnnnn, cnnc, kck) +
{A}_6^{\gamma\gamma^\prime}(nnnnn, cnnc, knk),\\
{A}_6^{\gamma\gamma^\prime}(nnnnn, nccn) &=&
{A}_6^{\gamma\gamma^\prime}(nnnnn, nccn, c) +
{A}_6^{\gamma\gamma^\prime}(nnnnn, nccn, n),\\
{A}_6^{\gamma\gamma^\prime}(nnnnn, ncnc) &=&
{A}_6^{\gamma\gamma^\prime}(nnnnn, ncnc, ck) +
{A}_6^{\gamma\gamma^\prime}(nnnnn, ncnc, nk),\\
{A}_6^{\gamma\gamma^\prime}(nnnnn, nncc) &=&
{A}_6^{\gamma\gamma^\prime}(nnnnn, nncc, ckk) +
{A}_6^{\gamma\gamma^\prime}(nnnnn, nncc, nkk), \eeqa \beqa
{A}_6^{\gamma\gamma^\prime}(nnnnn, cnnn) &=&
  {A}_6^{\gamma\gamma^\prime}(nnnnn, cnnn, kcc)
  + {A}_6^{\gamma\gamma^\prime}(nnnnn, cnnn, kcn)\nonumber\\& &
  + {A}_6^{\gamma\gamma^\prime}(nnnnn, cnnn, knc)
   +
    {A}_6^{\gamma\gamma^\prime}(nnnnn, cnnn, knn, kc)\nonumber\\& &
    + {A}_6^{\gamma\gamma^\prime}(nnnnn, cnnn, knn, kn),\\
{A}_6^{\gamma\gamma^\prime}(nnnnn, ncnn) &=&
  {A}_6^{\gamma\gamma^\prime}(nnnnn, ncnn, kkc, ck)
  + {A}_6^{\gamma\gamma^\prime}(nnnnn, ncnn, kkc, nk)\nonumber\\& &
  +
    {A}_6^{\gamma\gamma^\prime}(nnnnn, ncnn, kkn, ck)
    + {A}_6^{\gamma\gamma^\prime}(nnnnn, ncnn, kkn, nk, c) \nonumber\\& &+
    {A}_6^{\gamma\gamma^\prime}(nnnnn, ncnn, kkn, nk, n),\\
{A}_6^{\gamma\gamma^\prime}(nnnnn, nncn) &=&
  {A}_6^{\gamma\gamma^\prime}(nnnnn, nncn, ckk, kc)
  + {A}_6^{\gamma\gamma^\prime}(nnnnn, nncn, ckk, kn)\nonumber\\& &
  +
    {A}_6^{\gamma\gamma^\prime}(nnnnn, nncn, nkk, kc)
    + {A}_6^{\gamma\gamma^\prime}(nnnnn, nncn, nkk, kn, c)\nonumber\\& & +
    {A}_6^{\gamma\gamma^\prime}(nnnnn, nncn, nkk, kn, n),\\
{A}_6^{\gamma\gamma^\prime}(nnnnn, nnnc) &=&
  {A}_6^{\gamma\gamma^\prime}(nnnnn, nnnc, cck)
  + {A}_6^{\gamma\gamma^\prime}(nnnnn, nnnc, cnk)\nonumber\\& &
  + {A}_6^{\gamma\gamma^\prime}(nnnnn, nnnc, nck)+
    {A}_6^{\gamma\gamma^\prime}(nnnnn, nnnc, nnk, ck)\nonumber\\& &
    + {A}_6^{\gamma\gamma^\prime}(nnnnn, nnnc, nnk, nk),\\
{A}_6^{\gamma\gamma^\prime}(nnnnn, nnnn) &=&
  {A}_6^{\gamma\gamma^\prime}(nnnnn, nnnn, ccc)
  + {A}_6^{\gamma\gamma^\prime}(nnnnn, nnnn, ccn)\nonumber\\& &
  + {A}_6^{\gamma\gamma^\prime}(nnnnn, nnnn, cnc) +
    {A}_6^{\gamma\gamma^\prime}(nnnnn, nnnn, ncc)\nonumber\\& &
    + {A}_6^{\gamma\gamma^\prime}(nnnnn, nnnn, cnn)
    + {A}_6^{\gamma\gamma^\prime}(nnnnn, nnnn, ncn)\nonumber\\
    & & +
    {A}_6^{\gamma\gamma^\prime}(nnnnn, nnnn, nnc)
    + {A}_6^{\gamma\gamma^\prime}(nnnnn, nnnn, nnn),\eeqa
\beqa \!\!\!\!{A}_6^{\gamma\gamma^\prime}(nnnnn, nnnn, cnn)\!\!\!
&=&\!\!\!
  {A}_6^{\gamma\gamma^\prime}(nnnnn, nnnn, cnn, kc)
  + {A}_6^{\gamma\gamma^\prime}(nnnnn, nnnn, cnn, kn),\ \ \ \ \\
\!\!\!\!{A}_6^{\gamma\gamma^\prime}(nnnnn, nnnn, ncn)\!\!\!
&=&\!\!\! {A}_6^{\gamma\gamma^\prime}(nnnnn, nnnn, ncn, c) +
{A}_6^{\gamma\gamma^\prime}(nnnnn, nnnn, ncn, n),\ \ \ \ \ \\
\!\!\!\!{A}_6^{\gamma\gamma^\prime}(nnnnn, nnnn, nnc)\!\!\!
&=&\!\!\!
  {A}_6^{\gamma\gamma^\prime}(nnnnn, nnnn, nnc, ck)
  + {A}_6^{\gamma\gamma^\prime}(nnnnn, nnnn, nnc, nk),\eeqa
\beqa {A}_6^{\gamma\gamma^\prime}(nnnnn, nnnn, nnn) &=&
  {A}_6^{\gamma\gamma^\prime}(nnnnn, nnnn, nnn, cc)
  + {A}_6^{\gamma\gamma^\prime}(nnnnn, nnnn, nnn, cn)\nonumber\\& &
  +
    {A}_6^{\gamma\gamma^\prime}(nnnnn, nnnn, nnn, nc)
    + {A}_6^{\gamma\gamma^\prime}(nnnnn, nnnn, nnn, nn, c) \nonumber\\& &+
    {A}_6^{\gamma\gamma^\prime}(nnnnn, nnnn, nnn, nn, n).
\eeqa

Thus, we obtain the contribution from the six order approximation
made of $203$ terms after finding out all of contractions and
anti-contractions.

Just like we have done in the $l=4$ or 5 cases, we decompose
\beqa\label{A6dto1}
A_6^{\gamma\gamma^\prime}&=&A_6^{\gamma\gamma^\prime}(\e)+A_6^{\gamma\gamma^\prime}(t\e)
+A_6^{\gamma\gamma^\prime}(t^2\e)+A_6^{\gamma\gamma^\prime}(t^3\e)\\
\label{A6dto2}&=&A_6^{\gamma\gamma^\prime}(\e,t\e)+A_6^{\gamma\gamma^\prime}(t^2\e,t^3\e)
,\eeqa To our purpose, we only calculate the second term
$A_6^{\gamma\gamma^\prime}(t^2\e,t^3\e)$ in eq.(\ref{A6dto2}).
Without loss of generality, we decompose it into \beqa
A_6^{\gamma\gamma^\prime}(t^2\e,t^3\e)&=&A_6^{\gamma\gamma^\prime}(t^2\e^{-\I
E_{\gamma}t},t^3\e^{-\I
E_{\gamma}t})+A_6^{\gamma\gamma^\prime}(t^2\e^{-\I
E_{\gamma_1}t},t^3\e^{-\I
E_{\gamma_1}t})\nonumber\\
& &+A_6^{\gamma\gamma^\prime}(t^2\e^{-\I
E_{\gamma^\prime}t},t^3\e^{-\I E_{\gamma^\prime}t}).\eeqa While,
every term in above equations has its diagonal and non-diagonal
parts about $\gamma$ and $\gamma^\prime$, that is \beqa
A_6^{\gamma\gamma^\prime}(t^2\e^{-\I
E_{\gamma_i}t})&=&A_6^{\gamma\gamma^\prime}(t^2\e^{-\I
E_{\gamma_i}t};{\rm D})+A_6^{\gamma\gamma^\prime}(t^2\e^{-\I
E_{\gamma_i}t};{\rm N}),\\
A_6^{\gamma\gamma^\prime}(t^3\e^{-\I
E_{\gamma_i}t})&=&A_6^{\gamma\gamma^\prime}(t^3\e^{-\I
E_{\gamma_i}t};{\rm D})+A_6^{\gamma\gamma^\prime}(t^3\e^{-\I
E_{\gamma_i}t};{\rm N}).  \eeqa where $E_{\gamma_i}$ takes $
E_{\gamma}, E_{\gamma_1}$ and $E_{\gamma^\prime}$. In the following,
we respectively calculate them term by term, and we only write down
the non-zero expressions for saving space.

\beqa A_6(ccccc;t^2\e,t^3\e)&=&\sum_{\gamma_1}\left[\frac{(-\I
t)^3}{3!}\frac{\e^{-\I
E_{\gamma}t}}{\left(E_\gamma-E_{\gamma_1}\right)^3}-\frac{(-\I
t)^2}{2}\frac{3\e^{-\I E_{\gamma}t}}{\left(E_\gamma-E_{\gamma_1}\right)^4}\right.\nonumber\\
& &\left.+\frac{(-\I t)^2}{2}\frac{\e^{-\I
E_{\gamma_1}t}}{\left(E_\gamma-E_{\gamma_1}\right)^4}\right]
\left|g_1^{\gamma\gamma_1}\right|^6\delta_{\gamma\gamma^\prime}.\eeqa
\beqa A_6(ccccn;t^2\e,t^3\e)&=&\sum_{\gamma_1}\left[\frac{(-\I
t)^2}{2}\frac{\e^{-\I
E_{\gamma}t}}{\left(E_\gamma-E_{\gamma_1}\right)^3
\left(E_\gamma-E_{\gamma^\prime}\right)}\right.\nonumber\\
& &\left.-\frac{(-\I t)^2}{2}\frac{\e^{-\I
E_{\gamma_1}t}}{\left(E_\gamma-E_{\gamma_1}\right)^3
\left(E_{\gamma_1}-E_{\gamma^\prime}\right)}\right]
\left|g_1^{\gamma\gamma_1}\right|^4g_1^{\gamma\gamma_1}
g_1^{\gamma_1\gamma^\prime}\eta_{\gamma\gamma^\prime}.\eeqa
\beqa
A_6(cccnc;t^2\e,t^3\e)&=&\sum_{\gamma_1,\gamma_2}\left[\frac{(-\I
t)^3}{3!}\frac{\e^{-\I
E_{\gamma}t}}{\left(E_\gamma-E_{\gamma_1}\right)^2
\left(E_\gamma-E_{\gamma^\prime}\right)}-\frac{(-\I
t)^2}{2}\frac{\e^{-\I
E_{\gamma}t}}{\left(E_\gamma-E_{\gamma_1}\right)^2
\left(E_{\gamma}-E_{\gamma_2}\right)^2}\right.\nonumber\\
& &\left.-\frac{(-\I t)^2}{2}\frac{\e^{-\I
E_{\gamma}t}}{\left(E_\gamma-E_{\gamma_1}\right)^3
\left(E_{\gamma}-E_{\gamma_2}\right)}\right]
\left|g_1^{\gamma\gamma_1}\right|^4\left|g_1^{\gamma\gamma_2}\right|^2
\eta_{\gamma_1\gamma_2}\delta_{\gamma\gamma^\prime}.\eeqa
\beqa A_6(ccncc;t^2\e,t^3\e)&=&\sum_{\gamma_1}\frac{(-\I
t)^2}{2}\frac{\e^{-\I
E_{\gamma_1}t}}{\left(E_\gamma-E_{\gamma_1}\right)^2
\left(E_{\gamma_1}-E_{\gamma^\prime}\right)^2}
\left|g_1^{\gamma\gamma_1}\right|^2
\left|g_1^{\gamma_1\gamma^\prime}\right|^2g_1^{\gamma\gamma_1}
g_1^{\gamma_1\gamma^\prime}\eta_{\gamma\gamma^\prime}.\hskip
1.0cm\eeqa
\beqa
A_6(cnccc;t^2\e,t^3\e)&=&\sum_{\gamma_1,\gamma_2}\left[\frac{(-\I
t)^3}{3!}\frac{\e^{-\I
E_{\gamma}t}}{\left(E_\gamma-E_{\gamma_1}\right)
\left(E_\gamma-E_{\gamma_2}\right)^2}-\frac{(-\I
t)^2}{2}\frac{\e^{-\I
E_{\gamma}t}}{\left(E_\gamma-E_{\gamma_1}\right)
\left(E_{\gamma}-E_{\gamma_2}\right)^3}\right.\nonumber\\
& &\left.-\frac{(-\I t)^2}{2}\frac{\e^{-\I
E_{\gamma}t}}{\left(E_\gamma-E_{\gamma_1}\right)^2
\left(E_{\gamma}-E_{\gamma_2}\right)^2}\right]
\left|g_1^{\gamma\gamma_1}\right|^2\left|g_1^{\gamma\gamma_2}\right|^4
\eta_{\gamma_1\gamma_2}\delta_{\gamma\gamma^\prime}.\eeqa
\beqa A_6(ncccc;t^2\e,t^3\e)&=&\sum_{\gamma_1}\left[-\frac{(-\I
t)^2}{2}\frac{\e^{-\I
E_{\gamma_1}t}}{\left(E_\gamma-E_{\gamma_1}\right)
\left(E_{\gamma_1}-E_{\gamma^\prime}\right)^3}\right.\nonumber\\
& &\left.+\frac{(-\I t)^2}{2}\frac{\e^{-\I
E_{\gamma^\prime}t}}{\left(E_\gamma-E_{\gamma^\prime}\right)
\left(E_{\gamma_1}-E_{\gamma^\prime}\right)^3}\right]
\left|g_1^{\gamma_1\gamma^\prime}\right|^4g_1^{\gamma\gamma_1}
g_1^{\gamma_1\gamma^\prime}\eta_{\gamma\gamma^\prime}.\eeqa
\beqa A_6(cccnn,kkkc;t^2\e,t^3\e)&=&\sum_{\gamma_1}\left[\frac{(-\I
t)^2}{2}\frac{\e^{-\I
E_{\gamma}t}}{\left(E_\gamma-E_{\gamma_1}\right)
\left(E_{\gamma}-E_{\gamma^\prime}\right)^3}\right.\nonumber\\
& &\left.+\frac{(-\I t)^2}{2}\frac{\e^{-\I
E_{\gamma^\prime}t}}{\left(E_\gamma-E_{\gamma^\prime}\right)^3
\left(E_{\gamma_1}-E_{\gamma^\prime}\right)}\right]
\left|g_1^{\gamma\gamma^\prime}\right|^4g_1^{\gamma\gamma_1}
g_1^{\gamma_1\gamma^\prime}.\hskip 0.5cm\eeqa
\beqa
A_6(cccnn,kkkn;t^2\e,t^3\e)&=&\sum_{\gamma_1,\gamma_2}\frac{(-\I
t)^2}{2}\frac{\e^{-\I E_{\gamma}t}
\left|g_1^{\gamma\gamma_1}\right|^4g_1^{\gamma\gamma_2}
g_1^{\gamma_2\gamma^\prime}\eta_{\gamma\gamma^\prime}\eta_{\gamma_1\gamma_2}
\eta_{\gamma_1\gamma^\prime}}{\left(E_\gamma-E_{\gamma_1}\right)^2
\left(E_{\gamma}-E_{\gamma_2}\right)\left(E_{\gamma}-E_{\gamma^\prime}\right)}
.\hskip 0.5cm\eeqa
\beqa
A_6(ccncn,kkc;t^2\e,t^3\e)&=&\sum_{\gamma_1,\gamma_2}\left[\frac{(-\I
t)^2}{2}\frac{\e^{-\I
E_{\gamma}t}}{\left(E_\gamma-E_{\gamma_1}\right)^3
\left(E_{\gamma}-E_{\gamma^\prime}\right)}\right.\nonumber\\
& &\left.-\frac{(-\I t)^2}{2}\frac{\e^{-\I
E_{\gamma_1}t}}{\left(E_\gamma-E_{\gamma_1}\right)^3
\left(E_{\gamma_1}-E_{\gamma_2}\right)}\right]
\left|g_1^{\gamma\gamma_1}\right|^4\left|g_1^{\gamma_1\gamma_2}\right|^2
\eta_{\gamma\gamma_2}\delta_{\gamma\gamma^\prime}.\hskip 1.0cm\eeqa
\beqa
A_6(ccncn,kkn;t^2\e,t^3\e)&=&\sum_{\gamma_1,\gamma_2}\frac{(-\I
t)^2}{2}\frac{\e^{-\I E_{\gamma_1}t}
\left|g_1^{\gamma\gamma_1}\right|^2\left|g_1^{\gamma_1\gamma_2}\right|^2
g_1^{\gamma\gamma_1}g_1^{\gamma_1\gamma^\prime}\eta_{\gamma\gamma_2}\eta_{\gamma\gamma^\prime}
\eta_{\gamma_2\gamma^\prime}}{\left(E_\gamma-E_{\gamma_1}\right)^2
\left(E_{\gamma_1}-E_{\gamma_2}\right)\left(E_{\gamma_1}-E_{\gamma^\prime}\right)}
.\hskip 0.5cm\eeqa
\beqa A_6(ccnnc,kkck;t^2\e,t^3\e)&=&\sum_{\gamma_1}\frac{(-\I
t)^2}{2}\frac{\e^{-\I E_{\gamma}t}
\left|g_1^{\gamma\gamma_1}\right|^2\left|g_1^{\gamma\gamma^\prime}\right|^2g_1^{\gamma\gamma_1}
g_1^{\gamma_1\gamma^\prime}}{\left(E_\gamma-E_{\gamma_1}\right)^2
\left(E_{\gamma}-E_{\gamma^\prime}\right)^2} .\eeqa
\beqa A_6(cnccn,kc;t^2\e,t^3\e)&=&\sum_{\gamma_1}\frac{(-\I
t)^2}{2}\frac{\e^{-\I E_{\gamma}t}
\left|g_1^{\gamma\gamma_1}\right|^2\left|g_1^{\gamma\gamma^\prime}\right|^2g_1^{\gamma\gamma_1}
g_1^{\gamma_1\gamma^\prime}}{\left(E_\gamma-E_{\gamma_1}\right)^2
\left(E_{\gamma}-E_{\gamma^\prime}\right)^2} .\eeqa
\beqa A_6(cnccn,kn;t^2\e,t^3\e)&=&\sum_{\gamma_1,\gamma_2}\frac{(-\I
t)^2}{2}\frac{\e^{-\I E_{\gamma}t}
\left|g_1^{\gamma\gamma_1}\right|^2\left|g_1^{\gamma\gamma_2}\right|^2
g_1^{\gamma\gamma_2}g_1^{\gamma_2\gamma^\prime}\eta_{\gamma\gamma^\prime}\eta_{\gamma_1\gamma_2}
\eta_{\gamma_1\gamma^\prime}}{\left(E_\gamma-E_{\gamma_1}\right)
\left(E_{\gamma}-E_{\gamma_2}\right)^2\left(E_{\gamma}-E_{\gamma^\prime}\right)}
.\hskip 0.5cm\eeqa
\beqa
A_6(cncnc,kck;t^2\e,t^3\e)&=&\sum_{\gamma_1,\gamma_2}\left[\frac{(-\I
t)^3}{3!}\frac{\e^{-\I
E_{\gamma}t}}{\left(E_\gamma-E_{\gamma_1}\right)^2
\left(E_\gamma-E_{\gamma_2}\right)}\right.\nonumber\\
& &-\frac{(-\I t)^2}{2}\frac{\e^{-\I
E_{\gamma}t}}{\left(E_\gamma-E_{\gamma_1}\right)^2
\left(E_{\gamma}-E_{\gamma_2}\right)^2}\nonumber\\
& &\left.-\frac{(-\I t)^2}{2}\frac{\e^{-\I
E_{\gamma}t}}{\left(E_\gamma-E_{\gamma_1}\right)^3
\left(E_{\gamma}-E_{\gamma_2}\right)}\right]\nonumber\\
& &\times
\left|g_1^{\gamma\gamma_1}\right|^4\left|g_1^{\gamma\gamma_2}\right|^2
\eta_{\gamma_1\gamma_2}\delta_{\gamma\gamma^\prime}.\eeqa
\beqa
A_6(cncnc,knk;t^2\e,t^3\e)&=&\sum_{\gamma_1,\gamma_2,\gamma_3}\left[\frac{(-\I
t)^3}{3!}\frac{\e^{-\I
E_{\gamma}t}}{\left(E_\gamma-E_{\gamma_1}\right)
\left(E_\gamma-E_{\gamma_2}\right)\left(E_\gamma-E_{\gamma_3}\right)}\right.\nonumber\\
& &-\frac{(-\I t)^2}{2}\frac{\e^{-\I
E_{\gamma}t}}{\left(E_\gamma-E_{\gamma_1}\right)
\left(E_{\gamma}-E_{\gamma_2}\right)\left(E_\gamma-E_{\gamma_3}\right)^2}\nonumber\\
& & -\frac{(-\I t)^2}{2}\frac{\e^{-\I
E_{\gamma}t}}{\left(E_\gamma-E_{\gamma_1}\right)
\left(E_{\gamma}-E_{\gamma_2}\right)^2\left(E_\gamma-E_{\gamma_3}\right)}\nonumber\\
& &\left.-\frac{(-\I t)^2}{2}\frac{\e^{-\I
E_{\gamma}t}}{\left(E_\gamma-E_{\gamma_1}\right)^2
\left(E_{\gamma}-E_{\gamma_2}\right)\left(E_\gamma-E_{\gamma_3}\right)}\right]\nonumber\\
& &\times
\left|g_1^{\gamma\gamma_1}\right|^2\left|g_1^{\gamma\gamma_2}\right|^2
\left|g_1^{\gamma\gamma_3}\right|^2\eta_{\gamma_1\gamma_2}\eta_{\gamma_1\gamma_3}
\eta_{\gamma_2\gamma_3}\delta_{\gamma\gamma^\prime}.\eeqa
\beqa A_6(cnncc,kckk;t^2\e,t^3\e)&=&\sum_{\gamma_1}\frac{(-\I
t)^2}{2}\frac{\e^{-\I E_{\gamma^\prime}t}
\left|g_1^{\gamma_1\gamma^\prime}\right|^2\left|g_1^{\gamma\gamma^\prime}\right|^2g_1^{\gamma\gamma_1}
g_1^{\gamma_1\gamma^\prime}}{\left(E_\gamma-E_{\gamma^\prime}\right)^2
\left(E_{\gamma_1}-E_{\gamma^\prime}\right)^2} .\eeqa
\beqa A_6(ncccn,c;t^2\e,t^3\e)&=&\sum_{\gamma_1,\gamma_2}\frac{(-\I
t)^2}{2}\frac{\e^{-\I E_{\gamma_1}t}
\left|g_1^{\gamma\gamma_1}\right|^2\left|g_1^{\gamma_1\gamma_2}\right|^4
\eta_{\gamma\gamma_2}\delta_{\gamma\gamma^\prime}}
{\left(E_\gamma-E_{\gamma_1}\right)^2
\left(E_{\gamma_1}-E_{\gamma_2}\right)^2} .\eeqa
\beqa A_6(ncccn,n;t^2\e,t^3\e)&=&-\sum_{\gamma_1,\gamma_2}\frac{(-\I
t)^2}{2}\frac{\e^{-\I E_{\gamma_1}t}
\left|g_1^{\gamma_1\gamma_2}\right|^4
g_1^{\gamma\gamma_1}g_1^{\gamma_1\gamma^\prime}\eta_{\gamma\gamma^\prime}\eta_{\gamma\gamma_2}
\eta_{\gamma_2\gamma^\prime}}{\left(E_\gamma-E_{\gamma_1}\right)
\left(E_{\gamma_1}-E_{\gamma_2}\right)^2\left(E_{\gamma_1}-E_{\gamma^\prime}\right)}
.\hskip 0.5cm\eeqa
\beqa A_6(nccnc,ck;t^2\e,t^3\e)&=&\sum_{\gamma_1}\frac{(-\I
t)^2}{2}\frac{\e^{-\I E_{\gamma^\prime}t}
\left|g_1^{\gamma\gamma^\prime}\right|^2\left|g_1^{\gamma_1\gamma^\prime}\right|^2
g_1^{\gamma\gamma_1}g_1^{\gamma_1\gamma^\prime}}
{\left(E_\gamma-E_{\gamma^\prime}\right)^2
\left(E_{\gamma_1}-E_{\gamma^\prime}\right)^2}.\eeqa
\beqa A_6(nccnc,nk;t^2\e,t^3\e)&=&\sum_{\gamma_1,\gamma_2}\frac{(-\I
t)^2}{2}\frac{\e^{-\I E_{\gamma^\prime}t}
\left|g_1^{\gamma_1\gamma^\prime}\right|^2
\left|g_1^{\gamma_2\gamma^\prime}\right|^2
g_1^{\gamma\gamma_1}g_1^{\gamma_1\gamma^\prime}\eta_{\gamma\gamma^\prime}\eta_{\gamma\gamma_2}
\eta_{\gamma_1\gamma_2}}{\left(E_\gamma-E_{\gamma^\prime}\right)
\left(E_{\gamma_1}-E_{\gamma^\prime}\right)^2\left(E_{\gamma_2}-E_{\gamma^\prime}\right)}
.\hskip 0.8cm\eeqa
\beqa
A_6(ncncc,ckk;t^2\e,t^3\e)&=&\sum_{\gamma_1,\gamma_2}\left[\frac{(-\I
t)^2}{2}\frac{\e^{-\I
E_{\gamma}t}}{\left(E_\gamma-E_{\gamma_1}\right)^3
\left(E_\gamma-E_{\gamma_2}\right)}\right.\nonumber\\
& &\left.-\frac{(-\I t)^2}{2}\frac{\e^{-\I
E_{\gamma_1}t}}{\left(E_\gamma-E_{\gamma_1}\right)^3
\left(E_{\gamma_1}-E_{\gamma_2}\right)}\right]
\left|g_1^{\gamma\gamma_1}\right|^4\left|g_1^{\gamma_1\gamma_2}\right|^2
\eta_{\gamma\gamma_2}\delta_{\gamma\gamma^\prime}.\hskip 0.8cm\eeqa
\beqa
A_6(ncncc,nkk;t^2\e,t^3\e)&=&-\sum_{\gamma_1,\gamma_2}\frac{(-\I
t)^2}{2}\frac{\e^{-\I E_{\gamma_1}t}
\left|g_1^{\gamma_1\gamma_2}\right|^2
\left|g_1^{\gamma_1\gamma^\prime}\right|^2
g_1^{\gamma\gamma_1}g_1^{\gamma_1\gamma^\prime}\eta_{\gamma\gamma^\prime}\eta_{\gamma\gamma_2}
\eta_{\gamma_2\gamma^\prime}}{\left(E_\gamma-E_{\gamma_1}\right)
\left(E_{\gamma_1}-E_{\gamma_2}\right)\left(E_{\gamma_1}-E_{\gamma^\prime}\right)^2}
.\hskip 0.8cm\eeqa
\beqa A_6(nnccc,ckkk;t^2\e,t^3\e)&=&\sum_{\gamma_1}\left[\frac{(-\I
t)^2}{2}\frac{\e^{-\I
E_{\gamma}t}}{\left(E_\gamma-E_{\gamma_1}\right)
\left(E_\gamma-E_{\gamma^\prime}\right)^3}\right.\nonumber\\
& &\left.+\frac{(-\I t)^2}{2}\frac{\e^{-\I
E_{\gamma^\prime}t}}{\left(E_{\gamma}-E_{\gamma^\prime}\right)^3
\left(E_{\gamma_1}-E_{\gamma^\prime}\right)}\right]
\left|g_1^{\gamma\gamma^\prime}\right|^4g_1^{\gamma\gamma_1}g_1^{\gamma_1\gamma^\prime}
.\hskip 0.8cm\eeqa
\beqa
A_6(nnccc,nkkk;t^2\e,t^3\e)&=&\sum_{\gamma_1,\gamma_2}\frac{(-\I
t)^2}{2}\frac{\e^{-\I E_{\gamma^\prime}t}
\left|g_1^{\gamma_2\gamma^\prime}\right|^4
g_1^{\gamma\gamma_1}g_1^{\gamma_1\gamma^\prime}\eta_{\gamma\gamma^\prime}\eta_{\gamma\gamma_2}
\eta_{\gamma_1\gamma_2}}{\left(E_\gamma-E_{\gamma^\prime}\right)
\left(E_{\gamma_1}-E_{\gamma^\prime}\right)\left(E_{\gamma_2}-E_{\gamma^\prime}\right)^2}
.\hskip 0.8cm\eeqa
\beqa A_6(ccnnn,kkcc;t^2\e,t^3\e)&=&\sum_{\gamma_1}\left[\frac{(-\I
t)^2}{2}\frac{\e^{-\I
E_{\gamma}t}}{\left(E_\gamma-E_{\gamma_1}\right)
\left(E_\gamma-E_{\gamma^\prime}\right)^3}\right.\nonumber\\
& &\left.+\frac{(-\I t)^2}{2}\frac{\e^{-\I
E_{\gamma^\prime}t}}{\left(E_{\gamma}-E_{\gamma^\prime}\right)^3
\left(E_{\gamma_1}-E_{\gamma^\prime}\right)}\right]
\left|g_1^{\gamma\gamma^\prime}\right|^2g_1^{\gamma\gamma^\prime}g_1^{\gamma^\prime\gamma_1}
g_1^{\gamma_1\gamma}g_1^{\gamma\gamma^\prime} .\hskip 1.2cm\eeqa
\beqa
A_6(ccnnn,kkcn;t^2\e,t^3\e)&=&\sum_{\gamma_1,\gamma_2}\frac{(-\I
t)^2}{2}\frac{\e^{-\I E_{\gamma}t}
\left|g_1^{\gamma\gamma_1}\right|^2
g_1^{\gamma\gamma_1}g_1^{\gamma_1\gamma_2}g_1^{\gamma_2\gamma}g_1^{\gamma\gamma^\prime}
\eta_{\gamma_1\gamma^\prime}\eta_{\gamma_2\gamma^\prime}}
{\left(E_\gamma-E_{\gamma_1}\right)^2
\left(E_{\gamma}-E_{\gamma_2}\right)\left(E_{\gamma}-E_{\gamma^\prime}\right)}
.\hskip 0.8cm\eeqa
\beqa
A_6(ccnnn,kknc;t^2\e,t^3\e)&=&\sum_{\gamma_1,\gamma_2}\frac{(-\I
t)^2}{2}\frac{\e^{-\I E_{\gamma^\prime}t}
\left|g_1^{\gamma\gamma^\prime}\right|^2
g_1^{\gamma\gamma^\prime}g_1^{\gamma^\prime\gamma_1}g_1^{\gamma_1\gamma_2}g_1^{\gamma_2\gamma^\prime}
\eta_{\gamma\gamma_1}
\eta_{\gamma\gamma_2}}{\left(E_\gamma-E_{\gamma^\prime}\right)^2
\left(E_{\gamma_1}-E_{\gamma^\prime}\right)\left(E_{\gamma_2}-E_{\gamma^\prime}\right)}
.\hskip 0.8cm\eeqa
\beqa
A_6(ccnnn,kknn,kkc;t^2\e,t^3\e)&=&\sum_{\gamma_1,\gamma_2,\gamma_3}\frac{(-\I
t)^2}{2}\frac{\e^{-\I E_{\gamma}t}
\left|g_1^{\gamma\gamma_1}\right|^2
g_1^{\gamma\gamma_1}g_1^{\gamma_1\gamma_2}g_1^{\gamma_2\gamma_3}g_1^{\gamma_3\gamma}
\eta_{\gamma\gamma_2}
\eta_{\gamma_1\gamma_3}\delta_{\gamma\gamma^\prime}}{\left(E_\gamma-E_{\gamma_1}\right)^2
\left(E_{\gamma}-E_{\gamma_2}\right)\left(E_{\gamma}-E_{\gamma_3}\right)}
.\hskip 1.2cm\eeqa
\beqa A_6(cncnn,kkkc,kck;t^2\e,t^3\e)&=&\sum_{\gamma_1}\frac{(-\I
t)^2}{2}\frac{\e^{-\I E_{\gamma}t}
\left|g_1^{\gamma\gamma_1}\right|^2\left|g_1^{\gamma\gamma^\prime}\right|^2
g_1^{\gamma\gamma_1}g_1^{\gamma_1\gamma^\prime}}
{\left(E_\gamma-E_{\gamma_1}\right)^2
\left(E_{\gamma}-E_{\gamma^\prime}\right)^2}.\eeqa
\beqa
A_6(cncnn,kkkc,knk;t^2\e,t^3\e)&=&\sum_{\gamma_1,\gamma_2}\frac{(-\I
t)^2}{2}\frac{\e^{-\I E_{\gamma}t}
\left|g_1^{\gamma\gamma_1}\right|^2\left|g_1^{\gamma\gamma^\prime}\right|^2
g_1^{\gamma\gamma_2}g_1^{\gamma_2\gamma^\prime}
\eta_{\gamma_1\gamma_2}
\eta_{\gamma_1\gamma^\prime}}{\left(E_\gamma-E_{\gamma_1}\right)
\left(E_{\gamma}-E_{\gamma_2}\right)\left(E_{\gamma}-E_{\gamma^\prime}\right)^2}
.\hskip 1.2cm\eeqa
\beqa
A_6(cncnn,kkkn,kck;t^2\e,t^3\e)&=&\sum_{\gamma_1,\gamma_2}\frac{(-\I
t)^2}{2}\frac{\e^{-\I E_{\gamma}t}
\left|g_1^{\gamma\gamma_1}\right|^2\left|g_1^{\gamma\gamma_2}\right|^2
g_1^{\gamma\gamma_1}g_1^{\gamma_1\gamma^\prime}
\eta_{\gamma\gamma^\prime}\eta_{\gamma_1\gamma_2}
\eta_{\gamma_2\gamma^\prime}}{\left(E_\gamma-E_{\gamma_1}\right)^2
\left(E_{\gamma}-E_{\gamma_2}\right)\left(E_{\gamma}-E_{\gamma^\prime}\right)}
.\hskip 1.2cm\eeqa
\beqa
A_6(cncnn,kkkn,knk,kc;t^2\e,t^3\e)&=&\sum_{\gamma_1,\gamma_2}\frac{(-\I
t)^2}{2}\frac{\e^{-\I E_{\gamma}t}
\left|g_1^{\gamma\gamma_1}\right|^2\left|g_1^{\gamma\gamma^\prime}\right|^2
g_1^{\gamma\gamma_2}g_1^{\gamma_2\gamma^\prime}
\eta_{\gamma_1\gamma_2}
\eta_{\gamma_1\gamma^\prime}}{\left(E_\gamma-E_{\gamma_1}\right)
\left(E_{\gamma}-E_{\gamma_2}\right)\left(E_{\gamma}-E_{\gamma^\prime}\right)^2}
.\hskip 1.2cm\eeqa
\beqa & &
A_6(cncnn,kkkn,knk,kn;t^2\e,t^3\e)\nonumber\\
& & \quad =\sum_{\gamma_1,\gamma_2,\gamma_3}\frac{(-\I
t)^2}{2}\frac{\e^{-\I E_{\gamma}t}
\left|g_1^{\gamma\gamma_1}\right|^2\left|g_1^{\gamma\gamma_2}\right|^2
g_1^{\gamma\gamma_3}g_1^{\gamma_3\gamma^\prime}
\eta_{\gamma\gamma^\prime}
\eta_{\gamma_1\gamma_2}\eta_{\gamma_1\gamma_3}
\eta_{\gamma_1\gamma^\prime}\eta_{\gamma_2\gamma_3}
\eta_{\gamma_2\gamma^\prime}}{\left(E_\gamma-E_{\gamma_1}\right)
\left(E_{\gamma}-E_{\gamma_2}\right)\left(E_{\gamma}-E_{\gamma_3}\right)
\left(E_{\gamma}-E_{\gamma^\prime}\right)} .\hskip 1.2cm\eeqa
\beqa
A_6(cnncn,kckk,kkc;t^2\e,t^3\e)&=&\sum_{\gamma_1,\gamma_2}\frac{(-\I
t)^2}{2}\frac{\e^{-\I E_{\gamma}t}
\left|g_1^{\gamma\gamma_1}\right|^2\left|g_1^{\gamma_1\gamma_2}\right|^2
\left|g_1^{\gamma\gamma_2}\right|^2 \delta_{\gamma\gamma^\prime}
}{\left(E_\gamma-E_{\gamma_1}\right)^2
\left(E_{\gamma}-E_{\gamma_2}\right)^2} \eeqa
\beqa
A_6(cnncn,knkk,kkc;t^2\e,t^3\e)&=&\sum_{\gamma_1,\gamma_2,\gamma_3}\frac{(-\I
t)^2}{2}\frac{\e^{-\I E_{\gamma}t}
\left|g_1^{\gamma\gamma_1}\right|^2\left|g_1^{\gamma\gamma_2}\right|^2
\left|g_1^{\gamma_2\gamma_3}\right|^2
\eta_{\gamma\gamma_3}\eta_{\gamma_1\gamma_2}\eta_{\gamma_1\gamma_3}\delta_{\gamma\gamma^\prime}
}{\left(E_\gamma-E_{\gamma_1}\right)
\left(E_{\gamma}-E_{\gamma_2}\right)^2\left(E_{\gamma}-E_{\gamma_3}\right)}.\hskip
1.3cm \eeqa
\beqa A_6(cnnnc,kcck;t^2\e,t^3\e)&=&\sum_{\gamma_1}\left[\frac{(-\I
t)^2}{2}\frac{\e^{-\I
E_{\gamma}t}}{\left(E_\gamma-E_{\gamma_1}\right)
\left(E_{\gamma}-E_{\gamma^\prime}\right)^3}\right.\nonumber\\
& &\left.+\frac{(-\I t)^2}{2}\frac{\e^{-\I
E_{\gamma^\prime}t}}{\left(E_\gamma-E_{\gamma^\prime}\right)^3
\left(E_{\gamma_1}-E_{\gamma^\prime}\right)}\right]
\left|g_1^{\gamma\gamma^\prime}\right|^4g_1^{\gamma\gamma_1}
g_1^{\gamma_1\gamma^\prime}.\eeqa
\beqa
A_6(cnnnc,kcnk;t^2\e,t^3\e)&=&\sum_{\gamma_1,\gamma_2}\frac{(-\I
t)^2}{2}\frac{\e^{-\I E_{\gamma^\prime}t}
\left|g_1^{\gamma\gamma^\prime}\right|^2\left|g_1^{\gamma_2\gamma^\prime}\right|^2
g_1^{\gamma\gamma_1}g_1^{\gamma_1\gamma^\prime}
\eta_{\gamma\gamma_2}\eta_{\gamma_1\gamma_2}
}{\left(E_\gamma-E_{\gamma^\prime}\right)^2
\left(E_{\gamma_1}-E_{\gamma^\prime}\right)\left(E_{\gamma_2}-E_{\gamma^\prime}\right)}.\hskip
1.3cm \eeqa
\beqa
A_6(cnnnc,knck;t^2\e,t^3\e)&=&\sum_{\gamma_1,\gamma_2}\frac{(-\I
t)^2}{2}\frac{\e^{-\I E_{\gamma}t}
\left|g_1^{\gamma\gamma_1}\right|^2\left|g_1^{\gamma\gamma^\prime}\right|^2
g_1^{\gamma\gamma_2}g_1^{\gamma_2\gamma^\prime}
\eta_{\gamma_1\gamma_2}\eta_{\gamma_1\gamma^\prime}
}{\left(E_\gamma-E_{\gamma_1}\right)
\left(E_{\gamma}-E_{\gamma_2}\right)\left(E_{\gamma_2}-E_{\gamma^\prime}\right)^2}.\hskip
1.3cm \eeqa
\beqa A_6(nccnn,kkkc,ck;t^2\e,t^3\e)&=&\sum_{\gamma_1}\frac{(-\I
t)^2}{2}\frac{\e^{-\I E_{\gamma^\prime}t}
\left|g_1^{\gamma_1\gamma^\prime}\right|^2 g_1^{\gamma\gamma^\prime}
g_1^{\gamma^\prime\gamma_1}g_1^{\gamma_1\gamma}g_1^{\gamma\gamma^\prime}}
{\left(E_\gamma-E_{\gamma^\prime}\right)^2
\left(E_{\gamma_1}-E_{\gamma^\prime}\right)^2}.\eeqa
\beqa
A_6(nccnn,kkkc,nk;t^2\e,t^3\e)&=&\sum_{\gamma_1,\gamma_2}\frac{(-\I
t)^2}{2}\frac{\e^{-\I E_{\gamma^\prime}t}
\left|g_1^{\gamma_1\gamma^\prime}\right|^2
g_1^{\gamma\gamma^\prime}g_1^{\gamma^\prime\gamma_1}
g_1^{\gamma_1\gamma_2}g_1^{\gamma_2\gamma^\prime}
\eta_{\gamma\gamma_1}\eta_{\gamma\gamma_2}
}{\left(E_\gamma-E_{\gamma^\prime}\right)
\left(E_{\gamma_1}-E_{\gamma^\prime}\right)^2\left(E_{\gamma_2}-E_{\gamma^\prime}\right)}.\hskip
1.3cm \eeqa
\beqa A_6(ncncn,ckc;t^2\e,t^3\e)&=&\sum_{\gamma_1}\frac{(-\I
t)^2}{2}\frac{\e^{-\I E_{\gamma_1}t}
\left|g_1^{\gamma\gamma_1}\right|^2
\left|g_1^{\gamma_1\gamma^\prime}\right|^2
g_1^{\gamma\gamma_1}g_1^{\gamma_1\gamma^\prime}\eta_{\gamma\gamma^\prime}}
{\left(E_\gamma-E_{\gamma_1}\right)^2
\left(E_{\gamma_1}-E_{\gamma^\prime}\right)^2}.\eeqa
\beqa
A_6(ncncn,ckn;t^2\e,t^3\e)&=&\sum_{\gamma_1,\gamma_2}\frac{(-\I
t)^2}{2}\frac{\e^{-\I E_{\gamma_1}t}
\left|g_1^{\gamma\gamma_1}\right|^2
\left|g_1^{\gamma_1\gamma_2}\right|^2
g_1^{\gamma\gamma_1}g_1^{\gamma_1\gamma^\prime}\eta_{\gamma\gamma_2}
\eta_{\gamma\gamma^\prime}\eta_{\gamma_2\gamma^\prime}}
{\left(E_\gamma-E_{\gamma_1}\right)^2\left(E_{\gamma_1}-E_{\gamma_2}\right)
\left(E_{\gamma_1}-E_{\gamma^\prime}\right)}.\hskip 0.5cm\eeqa
\beqa
A_6(ncncn,nkc;t^2\e,t^3\e)&=&-\sum_{\gamma_1,\gamma_2}\frac{(-\I
t)^2}{2}\frac{\e^{-\I E_{\gamma_1}t}
\left|g_1^{\gamma_1\gamma^\prime}\right|^2
\left|g_1^{\gamma_1\gamma_2}\right|^2
g_1^{\gamma\gamma_1}g_1^{\gamma_1\gamma^\prime}
\eta_{\gamma\gamma^\prime}\eta_{\gamma\gamma_2}
\eta_{\gamma\gamma^\prime}\eta_{\gamma_2\gamma^\prime}}
{\left(E_\gamma-E_{\gamma_1}\right)\left(E_{\gamma_1}-E_{\gamma_2}\right)
\left(E_{\gamma_1}-E_{\gamma^\prime}\right)^2}.\hskip 1.2cm\eeqa
\beqa
A_6(ncncn,nkn,c;t^2\e,t^3\e)&=&\sum_{\gamma_1,\gamma_2,\gamma_3}\frac{(-\I
t)^2}{2}\frac{\e^{-\I E_{\gamma_1}t}
\left|g_1^{\gamma\gamma_1}\right|^2
\left|g_1^{\gamma_1\gamma_2}\right|^2
\left|g_1^{\gamma_1\gamma_3}\right|^2
\eta_{\gamma\gamma_2}\eta_{\gamma\gamma_3}
\eta_{\gamma_2\gamma_3}\delta_{\gamma\gamma^\prime}}
{\left(E_\gamma-E_{\gamma_1}\right)^2\left(E_{\gamma_1}-E_{\gamma_2}\right)
\left(E_{\gamma_1}-E_{\gamma_3}\right)}.\hskip 1.2cm\eeqa
\beqa & & A_6(ncncn,nkn,n;t^2\e,t^3\e) \nonumber\\ & &\quad
=-\sum_{\gamma_1,\gamma_2,\gamma_3}\frac{(-\I t)^2}{2}\frac{\e^{-\I
E_{\gamma_1}t} \left|g_1^{\gamma_1\gamma_2}\right|^2
\left|g_1^{\gamma_1\gamma_3}\right|^2
g_1^{\gamma\gamma_1}g_1^{\gamma_1\gamma^\prime}\eta_{\gamma\gamma_2}
\eta_{\gamma\gamma_3}
\eta_{\gamma\gamma^\prime}\eta_{\gamma_2\gamma_3}
\eta_{\gamma_2\gamma^\prime}\eta_{\gamma_3\gamma^\prime}}
{\left(E_\gamma-E_{\gamma_1}\right)\left(E_{\gamma_1}-E_{\gamma_2}\right)
\left(E_{\gamma_1}-E_{\gamma_3}\right)\left(E_{\gamma_1}-E_{\gamma^\prime}\right)}.\hskip
1.2cm\eeqa
\beqa
A_6(ncnnc,kkck,ckk;t^2\e,t^3\e)&=&\sum_{\gamma_1,\gamma_2}\frac{(-\I
t)^2}{2}\frac{\e^{-\I E_{\gamma}t}
\left|g_1^{\gamma\gamma_1}\right|^2\left|g_1^{\gamma\gamma_2}\right|^2
\left|g_1^{\gamma_1\gamma_2}\right|^2 \delta_{\gamma\gamma^\prime}}
{\left(E_\gamma-E_{\gamma_1}\right)^2
\left(E_{\gamma}-E_{\gamma_2}\right)^2}.\eeqa
\beqa
A_6(ncnnc,kknk,ckk;t^2\e,t^3\e)&=&\sum_{\gamma_1,\gamma_2,\gamma_3}\frac{(-\I
t)^2}{2}\frac{\e^{-\I E_{\gamma}t}
\left|g_1^{\gamma\gamma_1}\right|^2\left|g_1^{\gamma\gamma_3}\right|^2
\left|g_1^{\gamma_1\gamma_2}\right|^2\eta_{\gamma\gamma_2}
\eta_{\gamma_1\gamma_3}\eta_{\gamma_2\gamma_3}
\delta_{\gamma\gamma^\prime}} {\left(E_\gamma-E_{\gamma_1}\right)^2
\left(E_{\gamma}-E_{\gamma_2}\right)\left(E_{\gamma}-E_{\gamma_3}\right)}.\hskip
1.3cm\eeqa
\beqa A_6(nnccn,cc;t^2\e,t^3\e)&=&\sum_{\gamma_1}\frac{(-\I
t)^2}{2}\frac{\e^{-\I E_{\gamma}t}
\left|g_1^{\gamma\gamma_1}\right|^2g_1^{\gamma\gamma^\prime}g_1^{\gamma^\prime\gamma_1}
g_1^{\gamma_1\gamma}g_1^{\gamma\gamma^\prime}}
{\left(E_\gamma-E_{\gamma_1}\right)^2
\left(E_{\gamma}-E_{\gamma^\prime}\right)^2}.\eeqa
\beqa A_6(nnccn,cn;t^2\e,t^3\e)&=&\sum_{\gamma_1,\gamma_2}\frac{(-\I
t)^2}{2}\frac{\e^{-\I E_{\gamma}t}
\left|g_1^{\gamma\gamma_2}\right|^2g_1^{\gamma\gamma_1}g_1^{\gamma_1\gamma_2}
g_1^{\gamma_2\gamma}g_1^{\gamma\gamma^\prime}\eta_{\gamma_1\gamma^\prime}
\eta_{\gamma_2\gamma^\prime}}{\left(E_\gamma-E_{\gamma_1}\right)
\left(E_{\gamma}-E_{\gamma_2}\right)^2\left(E_{\gamma}-E_{\gamma^\prime}\right)}.
\hskip 1.3cm\eeqa
\beqa A_6(nncnc,ckkk,kck;t^2\e,t^3\e)&=&\sum_{\gamma_1}\frac{(-\I
t)^2}{2}\frac{\e^{-\I E_{\gamma^\prime}t}
\left|g_1^{\gamma\gamma^\prime}\right|^2
\left|g_1^{\gamma_1\gamma^\prime}\right|^2 g_1^{\gamma\gamma_1}
g_1^{\gamma_1\gamma^\prime}}
{\left(E_\gamma-E_{\gamma^\prime}\right)^2
\left(E_{\gamma_1}-E_{\gamma^\prime}\right)^2}.\eeqa
\beqa
A_6(nncnc,ckkk,knk;t^2\e,t^3\e)&=&\sum_{\gamma_1,\gamma_2}\frac{(-\I
t)^2}{2}\frac{\e^{-\I E_{\gamma^\prime}t}
\left|g_1^{\gamma\gamma^\prime}\right|^2
\left|g_1^{\gamma_2\gamma^\prime}\right|^2 g_1^{\gamma\gamma_1}
g_1^{\gamma_1\gamma^\prime}\eta_{\gamma\gamma_2}\eta_{\gamma_1\gamma_2}}
{\left(E_\gamma-E_{\gamma^\prime}\right)^2
\left(E_{\gamma_1}-E_{\gamma^\prime}\right)\left(E_{\gamma_2}-E_{\gamma^\prime}\right)}.\hskip
1.3cm\eeqa
\beqa
A_6(nncnc,nkkk,kck;t^2\e,t^3\e)&=&\sum_{\gamma_1,\gamma_2}\frac{(-\I
t)^2}{2}\frac{\e^{-\I E_{\gamma^\prime}t}
\left|g_1^{\gamma_1\gamma^\prime}\right|^2
\left|g_1^{\gamma_2\gamma^\prime}\right|^2 g_1^{\gamma\gamma_1}
g_1^{\gamma_1\gamma^\prime}\eta_{\gamma\gamma_2}\eta_{\gamma_1\gamma_2}\eta_{\gamma\gamma^\prime}}
{\left(E_\gamma-E_{\gamma^\prime}\right)
\left(E_{\gamma_1}-E_{\gamma^\prime}\right)^2\left(E_{\gamma_2}-E_{\gamma^\prime}\right)}.\hskip
1.3cm\eeqa
\beqa
A_6(nncnc,nkkk,knk,ck;t^2\e,t^3\e)&=&\sum_{\gamma_1,\gamma_2}\frac{(-\I
t)^2}{2}\frac{\e^{-\I E_{\gamma^\prime}t}
\left|g_1^{\gamma\gamma^\prime}\right|^2
\left|g_1^{\gamma_2\gamma^\prime}\right|^2 g_1^{\gamma\gamma_1}
g_1^{\gamma_1\gamma^\prime}\eta_{\gamma\gamma_2}\eta_{\gamma_1\gamma_2}}
{\left(E_\gamma-E_{\gamma^\prime}\right)^2
\left(E_{\gamma_1}-E_{\gamma^\prime}\right)\left(E_{\gamma_2}-E_{\gamma^\prime}\right)}.\hskip
1.3cm\eeqa
\beqa & &
A_6(nncnc,nkkk,knk,nk;t^2\e,t^3\e)\nonumber\\
& &\quad =\sum_{\gamma_1,\gamma_2,\gamma_3}\frac{(-\I
t)^2}{2}\frac{\e^{-\I E_{\gamma^\prime}t}
\left|g_1^{\gamma_2\gamma^\prime}\right|^2
\left|g_1^{\gamma_3\gamma^\prime}\right|^2 g_1^{\gamma\gamma_1}
g_1^{\gamma_1\gamma^\prime}\eta_{\gamma\gamma_2}\eta_{\gamma\gamma_3}
\eta_{\gamma\gamma^\prime}\eta_{\gamma_1\gamma_2}\eta_{\gamma_1\gamma_3}\eta_{\gamma_2\gamma_3}}
{\left(E_\gamma-E_{\gamma^\prime}\right)
\left(E_{\gamma_1}-E_{\gamma^\prime}\right)\left(E_{\gamma_2}-E_{\gamma^\prime}\right)
\left(E_{\gamma_3}-E_{\gamma^\prime}\right)}.\eeqa
\beqa A_6(nnncc,cckk;t^2\e,t^3\e)&=&\sum_{\gamma_1}\left[\frac{(-\I
t)^2}{2}\frac{\e^{-\I
E_{\gamma}t}}{\left(E_\gamma-E_{\gamma_1}\right)
\left(E_\gamma-E_{\gamma^\prime}\right)^3}\right.\nonumber\\
& &\left.+\frac{(-\I t)^2}{2}\frac{\e^{-\I
E_{\gamma^\prime}t}}{\left(E_{\gamma}-E_{\gamma^\prime}\right)^3
\left(E_{\gamma_1}-E_{\gamma^\prime}\right)}\right]
\left|g_1^{\gamma\gamma^\prime}\right|^2g_1^{\gamma\gamma^\prime}g_1^{\gamma^\prime\gamma_1}
g_1^{\gamma_1\gamma}g_1^{\gamma\gamma^\prime} .\hskip 1.2cm\eeqa
\beqa
A_6(nnncc,cnkk;t^2\e,t^3\e)&=&\sum_{\gamma_1,\gamma_2}\frac{(-\I
t)^2}{2}\frac{\e^{-\I E_{\gamma}t}
\left|g_1^{\gamma\gamma^\prime}\right|^2
g_1^{\gamma\gamma_1}g_1^{\gamma_1\gamma_2}
g_1^{\gamma_2\gamma}g_1^{\gamma\gamma^\prime}
\eta_{\gamma_1\gamma^\prime}\eta_{\gamma_2\gamma^\prime}}
{\left(E_\gamma-E_{\gamma_1}\right)
\left(E_{\gamma}-E_{\gamma_2}\right)\left(E_{\gamma}-E_{\gamma^\prime}\right)^2}.\hskip
1.3cm\eeqa
\beqa
A_6(nnncc,nckk;t^2\e,t^3\e)&=&\sum_{\gamma_1,\gamma_2}\frac{(-\I
t)^2}{2}\frac{\e^{-\I E_{\gamma^\prime}t}
\left|g_1^{\gamma_2\gamma^\prime}\right|^2
g_1^{\gamma\gamma^\prime}g_1^{\gamma^\prime\gamma_1}
g_1^{\gamma_1\gamma_2}g_1^{\gamma_2\gamma^\prime}
\eta_{\gamma\gamma_1}\eta_{\gamma\gamma_2}}
{\left(E_\gamma-E_{\gamma^\prime}\right)
\left(E_{\gamma_1}-E_{\gamma^\prime}\right)\left(E_{\gamma_2}-E_{\gamma^\prime}\right)^2}.\hskip
1.3cm\eeqa
\beqa
A_6(nnncc,nnkk,c;t^2\e,t^3\e)&=&\sum_{\gamma_1,\gamma_2,\gamma_3}\frac{(-\I
t)^2}{2}\frac{\e^{-\I E_{\gamma}t}
\left|g_1^{\gamma\gamma_3}\right|^2
g_1^{\gamma\gamma_1}g_1^{\gamma_1\gamma_2}
g_1^{\gamma_2\gamma_3}g_1^{\gamma_3\gamma}
\eta_{\gamma\gamma_2}\eta_{\gamma_1\gamma_3}\delta_{\gamma\gamma^\prime}}
{\left(E_\gamma-E_{\gamma_1}\right)
\left(E_{\gamma}-E_{\gamma_2}\right)\left(E_{\gamma}-E_{\gamma_3}\right)^2}.\hskip
1.3cm\eeqa
\beqa A_6(cnnnn,kccc;t^2\e,t^3\e)&=&\sum_{\gamma_1}\frac{(-\I
t)^2}{2}\frac{\e^{-\I E_{\gamma}t}
\left|g_1^{\gamma\gamma_1}\right|^2
g_1^{\gamma\gamma^\prime}g_1^{\gamma^\prime\gamma_1}
g_1^{\gamma_1\gamma}g_1^{\gamma\gamma^\prime}}
{\left(E_\gamma-E_{\gamma_1}\right)^2
\left(E_{\gamma}-E_{\gamma^\prime}\right)^2}.\hskip 1.3cm\eeqa
\beqa
A_6(cnnnn,kccn;t^2\e,t^3\e)&=&\sum_{\gamma_1,\gamma_2}\frac{(-\I
t)^2}{2}\frac{\e^{-\I E_{\gamma}t}
\left|g_1^{\gamma\gamma_1}\right|^2
g_1^{\gamma\gamma_2}g_1^{\gamma_2\gamma_1}
g_1^{\gamma_1\gamma}g_1^{\gamma\gamma^\prime}
\eta_{\gamma_1\gamma^\prime}\eta_{\gamma_2\gamma^\prime}}
{\left(E_\gamma-E_{\gamma_1}\right)^2
\left(E_{\gamma}-E_{\gamma_2}\right)\left(E_{\gamma}-E_{\gamma^\prime}\right)}.\hskip
1.3cm\eeqa
\beqa
A_6(cnnnn,kncc;t^2\e,t^3\e)&=&\sum_{\gamma_1,\gamma_2}\frac{(-\I
t)^2}{2}\frac{\e^{-\I E_{\gamma}t}
\left|g_1^{\gamma\gamma_1}\right|^2
g_1^{\gamma\gamma^\prime}g_1^{\gamma^\prime\gamma_2}
g_1^{\gamma_2\gamma}g_1^{\gamma\gamma^\prime}
\eta_{\gamma_1\gamma^\prime}\eta_{\gamma_1\gamma_2}}
{\left(E_\gamma-E_{\gamma_1}\right)
\left(E_{\gamma}-E_{\gamma_2}\right)\left(E_{\gamma}-E_{\gamma^\prime}\right)^2}.\hskip
1.3cm\eeqa
\beqa
A_6(cnnnn,kcnn,kkc;t^2\e,t^3\e)&=&\sum_{\gamma_1,\gamma_2,\gamma_3}\frac{(-\I
t)^2}{2}\frac{\e^{-\I E_{\gamma}t}
\left|g_1^{\gamma\gamma_1}\right|^2
g_1^{\gamma\gamma_2}g_1^{\gamma_2\gamma_1}
g_1^{\gamma_1\gamma_3}g_1^{\gamma_3\gamma}
\eta_{\gamma_2\gamma_3}\delta_{\gamma\gamma^\prime}}
{\left(E_\gamma-E_{\gamma_1}\right)^2
\left(E_{\gamma}-E_{\gamma_2}\right)\left(E_{\gamma}-E_{\gamma_3}\right)}.\hskip
1.3cm\eeqa
\beqa
A_6(cnnnn,kncn,kc;t^2\e,t^3\e)&=&\sum_{\gamma_1,\gamma_2}\frac{(-\I
t)^2}{2}\frac{\e^{-\I E_{\gamma}t}
\left|g_1^{\gamma\gamma^\prime}\right|^2
g_1^{\gamma\gamma_1}g_1^{\gamma_1\gamma_2}
g_1^{\gamma_2\gamma}g_1^{\gamma\gamma^\prime}
\eta_{\gamma_1\gamma^\prime}\eta_{\gamma_2\gamma^\prime}}
{\left(E_\gamma-E_{\gamma_1}\right)
\left(E_{\gamma}-E_{\gamma_2}\right)\left(E_{\gamma}-E_{\gamma^\prime}\right)^2}.\hskip
1.3cm\eeqa
\beqa & &
A_6(cnnnn,kncn,kn;t^2\e,t^3\e)\nonumber\\
& &\quad =\sum_{\gamma_1,\gamma_2,\gamma_3}\frac{(-\I
t)^2}{2}\frac{\e^{-\I E_{\gamma}t}
\left|g_1^{\gamma\gamma_1}\right|^2 g_1^{\gamma\gamma_2}
g_1^{\gamma_2\gamma_3}g_1^{\gamma_3\gamma}g_1^{\gamma\gamma^\prime}
\eta_{\gamma_1\gamma_2}\eta_{\gamma_1\gamma_3}
\eta_{\gamma_1\gamma^\prime}\eta_{\gamma_2\gamma^\prime}\eta_{\gamma_3\gamma^\prime}}
{\left(E_\gamma-E_{\gamma_1}\right)
\left(E_{\gamma}-E_{\gamma_2}\right)\left(E_{\gamma}-E_{\gamma_3}\right)
\left(E_{\gamma}-E_{\gamma^\prime}\right)}.\eeqa
\beqa A_6(cnnnn,knnn,kcc;t^2\e,t^3\e)
&=&\sum_{\gamma_1,\gamma_2,\gamma_3}\frac{(-\I t)^2}{2}\frac{\e^{-\I
E_{\gamma}t} \left|g_1^{\gamma\gamma_1}\right|^2
g_1^{\gamma\gamma_2}
g_1^{\gamma_2\gamma_3}g_1^{\gamma_3\gamma_1}g_1^{\gamma_1\gamma}
\eta_{\gamma\gamma_3}\eta_{\gamma_1\gamma_2}\delta_{\gamma\gamma^\prime}}
{\left(E_\gamma-E_{\gamma_1}\right)^2
\left(E_{\gamma}-E_{\gamma_2}\right)\left(E_{\gamma}-E_{\gamma_3}\right)}.
\hskip 1.3cm \eeqa
\beqa & &
A_6(cnnnn,knnn,knc;t^2\e,t^3\e)\nonumber\\
& &\quad =\sum_{\gamma_1,\gamma_2,\gamma_3,\gamma_4}\frac{(-\I
t)^2}{2}\frac{\e^{-\I E_{\gamma}t}
\left|g_1^{\gamma\gamma_1}\right|^2 g_1^{\gamma\gamma_2}
g_1^{\gamma_2\gamma_3}g_1^{\gamma_3\gamma_4}g_1^{\gamma_4\gamma}
\eta_{\gamma\gamma_3}\eta_{\gamma_1\gamma_2}\eta_{\gamma_1\gamma_3}
\eta_{\gamma_1\gamma_4}\eta_{\gamma_2\gamma_4}\delta_{\gamma\gamma^\prime}}
{\left(E_\gamma-E_{\gamma_1}\right)
\left(E_{\gamma}-E_{\gamma_2}\right)\left(E_{\gamma}-E_{\gamma_3}\right)
\left(E_{\gamma}-E_{\gamma_4}\right)}.\eeqa
\beqa A_6(ncnnn,kkcc,ckk;t^2\e,t^3\e)&=&\sum_{\gamma_1}\frac{(-\I
t)^2}{2}\frac{\e^{-\I E_{\gamma^\prime}t}
\left|g_1^{\gamma\gamma^\prime}\right|^2\left|g_1^{\gamma_1\gamma^\prime}\right|^2
g_1^{\gamma\gamma_1}g_1^{\gamma_1\gamma^\prime}}
{\left(E_\gamma-E_{\gamma^\prime}\right)^2
\left(E_{\gamma_1}-E_{\gamma^\prime}\right)^2}.\hskip 1.3cm\eeqa
\beqa A_6(ncnnn,kkcc,nkk;t^2\e,t^3\e)
&=&\sum_{\gamma_1,\gamma_2}\frac{(-\I t)^2}{2}\frac{\e^{-\I
E_{\gamma^\prime}t}\left|g_1^{\gamma_1\gamma^\prime}\right|^2
g_1^{\gamma\gamma^\prime}
g_1^{\gamma^\prime\gamma_2}g_1^{\gamma_2\gamma_1}g_1^{\gamma_1\gamma^\prime}
\eta_{\gamma\gamma_1}\eta_{\gamma\gamma_2}}
{\left(E_\gamma-E_{\gamma^\prime}\right)
\left(E_{\gamma_1}-E_{\gamma^\prime}\right)^2\left(E_{\gamma_2}-E_{\gamma^\prime}\right)}.
\hskip 1.3cm \eeqa
\beqa A_6(ncnnn,kknc,ckk;t^2\e,t^3\e)
&=&\sum_{\gamma_1,\gamma_2}\frac{(-\I t)^2}{2}\frac{\e^{-\I
E_{\gamma^\prime}t}\left|g_1^{\gamma\gamma^\prime}\right|^2
\left|g_1^{\gamma_1\gamma^\prime}\right|^2 g_1^{\gamma\gamma_2}
g_1^{\gamma_2\gamma^\prime}
\eta_{\gamma\gamma_1}\eta_{\gamma_1\gamma_2}}
{\left(E_\gamma-E_{\gamma^\prime}\right)^2
\left(E_{\gamma_1}-E_{\gamma^\prime}\right)\left(E_{\gamma_2}-E_{\gamma^\prime}\right)}.
\hskip 1.3cm \eeqa
\beqa A_6(ncnnn,kknc,nkk,ck;t^2\e,t^3\e)
&=&\sum_{\gamma_1,\gamma_2}\frac{(-\I t)^2}{2}\frac{\e^{-\I
E_{\gamma^\prime}t}\left|g_1^{\gamma_1\gamma^\prime}\right|^2
g_1^{\gamma\gamma^\prime} g_1^{\gamma^\prime\gamma_2}
g_1^{\gamma_2\gamma} g_1^{\gamma\gamma^\prime}
\eta_{\gamma\gamma_1}\eta_{\gamma_1\gamma_2}}
{\left(E_\gamma-E_{\gamma^\prime}\right)^2
\left(E_{\gamma_1}-E_{\gamma^\prime}\right)\left(E_{\gamma_2}-E_{\gamma^\prime}\right)}.
\hskip 1.3cm \eeqa
\beqa & &
A_6(ncnnn,kknc,nkk,nk;t^2\e,t^3\e)\nonumber\\
& &\quad =\sum_{\gamma_1,\gamma_2,\gamma_3}\frac{(-\I
t)^2}{2}\frac{\e^{-\I E_{\gamma^\prime}t}
\left|g_1^{\gamma_1\gamma^\prime}\right|^2 g_1^{\gamma\gamma^\prime}
g_1^{\gamma^\prime\gamma_2}g_1^{\gamma_2\gamma_3}g_1^{\gamma_3\gamma^\prime}
\eta_{\gamma\gamma_1}\eta_{\gamma\gamma_2}\eta_{\gamma\gamma_3}
\eta_{\gamma_1\gamma_2}\eta_{\gamma_1\gamma_3}}
{\left(E_\gamma-E_{\gamma^\prime}\right)
\left(E_{\gamma_1}-E_{\gamma^\prime}\right)\left(E_{\gamma_2}-E_{\gamma^\prime}\right)
\left(E_{\gamma_3}-E_{\gamma^\prime}\right)}.\eeqa
\beqa
A_6(nncnn,ckkc,kck;t^2\e,t^3\e)&=&\sum_{\gamma_1,\gamma_2}\frac{(-\I
t)^2}{2}\frac{\e^{-\I E_{\gamma}t}
\left|g_1^{\gamma\gamma_1}\right|^2\left|g_1^{\gamma\gamma_2}\right|^2
\left|g_1^{\gamma_1\gamma_2}\right|^2\delta_{\gamma\gamma^\prime}}
{\left(E_\gamma-E_{\gamma_1}\right)^2
\left(E_{\gamma}-E_{\gamma_2}\right)^2}.\eeqa
\beqa A_6(nncnn,ckkc,knk;t^2\e,t^3\e)
&=&\sum_{\gamma_1,\gamma_2,\gamma_3}\frac{(-\I t)^2}{2}\frac{\e^{-\I
E_{\gamma}t}\left|g_1^{\gamma\gamma_2}\right|^2 g_1^{\gamma\gamma_1}
g_1^{\gamma_1\gamma_2} g_1^{\gamma_2\gamma_3} g_1^{\gamma_3\gamma}
\eta_{\gamma_1\gamma_3}\delta_{\gamma\gamma^\prime}}
{\left(E_\gamma-E_{\gamma_1}\right)
\left(E_{\gamma}-E_{\gamma_2}\right)^2\left(E_{\gamma}-E_{\gamma_3}\right)}.
\hskip 1.3cm \eeqa
\beqa A_6(nnncn,cckk,kkc;t^2\e,t^3\e)&=&\sum_{\gamma_1}\frac{(-\I
t)^2}{2}\frac{\e^{-\I E_{\gamma}t}
\left|g_1^{\gamma\gamma_1}\right|^2\left|g_1^{\gamma\gamma^\prime}\right|^2
g_1^{\gamma\gamma_1}g_1^{\gamma_1\gamma^\prime}}
{\left(E_\gamma-E_{\gamma_1}\right)^2
\left(E_{\gamma}-E_{\gamma^\prime}\right)^2}.\eeqa
\beqa A_6(nnncn,cckk,kkn;t^2\e,t^3\e)
&=&\sum_{\gamma_1,\gamma_2}\frac{(-\I t)^2}{2}\frac{\e^{-\I
E_{\gamma}t}\left|g_1^{\gamma\gamma_1}\right|^2 g_1^{\gamma\gamma_1}
g_1^{\gamma_1\gamma_2} g_1^{\gamma_2\gamma}
g_1^{\gamma\gamma^\prime}
\eta_{\gamma_1\gamma^\prime}\eta_{\gamma_2\gamma^\prime}}
{\left(E_\gamma-E_{\gamma_1}\right)^2
\left(E_{\gamma}-E_{\gamma_2}\right)\left(E_{\gamma}-E_{\gamma^\prime}\right)}.
\hskip 1.3cm \eeqa
\beqa A_6(nnncn,cnkk,kkc;t^2\e,t^3\e)
&=&\sum_{\gamma_1,\gamma_2}\frac{(-\I t)^2}{2}\frac{\e^{-\I
E_{\gamma}t}\left|g_1^{\gamma\gamma_2}\right|^2
\left|g_1^{\gamma\gamma^\prime}\right|^2 g_1^{\gamma\gamma_1}
g_1^{\gamma_1\gamma^\prime}
\eta_{\gamma_1\gamma_2}\eta_{\gamma_2\gamma^\prime}}
{\left(E_\gamma-E_{\gamma_1}\right)
\left(E_{\gamma}-E_{\gamma_2}\right)\left(E_{\gamma}-E_{\gamma^\prime}\right)^2}.
\hskip 1.3cm \eeqa
\beqa A_6(nnncn,cnkk,kkn,kc;t^2\e,t^3\e)
&=&\sum_{\gamma_1,\gamma_2}\frac{(-\I t)^2}{2}\frac{\e^{-\I
E_{\gamma}t}\left|g_1^{\gamma\gamma_2}\right|^2
g_1^{\gamma\gamma^\prime} g_1^{\gamma^\prime\gamma_1}
g_1^{\gamma_1\gamma} g_1^{\gamma\gamma^\prime}
\eta_{\gamma_1\gamma_2}\eta_{\gamma_2\gamma^\prime}}
{\left(E_\gamma-E_{\gamma_1}\right)
\left(E_{\gamma}-E_{\gamma_2}\right)\left(E_{\gamma}-E_{\gamma^\prime}\right)^2}.
\hskip 1.3cm \eeqa
\beqa & &
A_6(nnncn,cnkk,kkn,kn;t^2\e,t^3\e)\nonumber\\
& &\quad =\sum_{\gamma_1,\gamma_2,\gamma_3}\frac{(-\I
t)^2}{2}\frac{\e^{-\I E_{\gamma}t}
\left|g_1^{\gamma\gamma_3}\right|^2 g_1^{\gamma\gamma_1}
g_1^{\gamma_1\gamma_2}g_1^{\gamma_2\gamma}g_1^{\gamma\gamma^\prime}
\eta_{\gamma_1\gamma_3}\eta_{\gamma_1\gamma^\prime}\eta_{\gamma_2\gamma_3}
\eta_{\gamma_2\gamma^\prime}\eta_{\gamma_3\gamma^\prime}}
{\left(E_\gamma-E_{\gamma_1}\right)
\left(E_{\gamma}-E_{\gamma_2}\right)\left(E_{\gamma}-E_{\gamma_3}\right)
\left(E_{\gamma}-E_{\gamma^\prime}\right)}.\eeqa
\beqa A_6(nnnnc,ccck;t^2\e,t^3\e)&=&\sum_{\gamma_1}\frac{(-\I
t)^2}{2}\frac{\e^{-\I E_{\gamma^\prime}t}
\left|g_1^{\gamma_1\gamma^\prime}\right|^2
g_1^{\gamma\gamma^\prime}g_1^{\gamma^\prime\gamma_1}g_1^{\gamma_1\gamma}g_1^{\gamma\gamma^\prime}}
{\left(E_\gamma-E_{\gamma^\prime}\right)^2
\left(E_{\gamma_1}-E_{\gamma^\prime}\right)^2}.\eeqa
\beqa
A_6(nnnnc,ccnk;t^2\e,t^3\e)&=&\sum_{\gamma_1,\gamma_2}\frac{(-\I
t)^2}{2}\frac{\e^{-\I E_{\gamma^\prime}t}
\left|g_1^{\gamma_2\gamma^\prime}\right|^2
g_1^{\gamma\gamma^\prime}g_1^{\gamma^\prime\gamma_1}
g_1^{\gamma_1\gamma}g_1^{\gamma\gamma^\prime}\eta_{\gamma\gamma_2}\eta_{\gamma_1\gamma_2}}
{\left(E_\gamma-E_{\gamma^\prime}\right)^2
\left(E_{\gamma_1}-E_{\gamma^\prime}\right)\left(E_{\gamma_2}-E_{\gamma^\prime}\right)}.
\hskip 1.0cm\eeqa
\beqa
A_6(nnnnc,ncck;t^2\e,t^3\e)&=&\sum_{\gamma_1,\gamma_2}\frac{(-\I
t)^2}{2}\frac{\e^{-\I E_{\gamma^\prime}t}
\left|g_1^{\gamma_1\gamma^\prime}\right|^2
g_1^{\gamma\gamma^\prime}g_1^{\gamma^\prime\gamma_1}
g_1^{\gamma_1\gamma_2}g_1^{\gamma_2\gamma^\prime}\eta_{\gamma\gamma_1}\eta_{\gamma\gamma_2}}
{\left(E_\gamma-E_{\gamma^\prime}\right)
\left(E_{\gamma_1}-E_{\gamma^\prime}\right)^2\left(E_{\gamma_2}-E_{\gamma^\prime}\right)}.
\hskip 1.0cm\eeqa
\beqa
A_6(nnnnc,ncnk,ck;t^2\e,t^3\e)&=&\sum_{\gamma_1,\gamma_2}\frac{(-\I
t)^2}{2}\frac{\e^{-\I E_{\gamma^\prime}t}
\left|g_1^{\gamma\gamma^\prime}\right|^2
g_1^{\gamma\gamma^\prime}g_1^{\gamma^\prime\gamma_1}
g_1^{\gamma_1\gamma_2}g_1^{\gamma_2\gamma^\prime}\eta_{\gamma\gamma_1}\eta_{\gamma\gamma_2}}
{\left(E_\gamma-E_{\gamma^\prime}\right)^2
\left(E_{\gamma_1}-E_{\gamma^\prime}\right)\left(E_{\gamma_2}-E_{\gamma^\prime}\right)}.
\hskip 1.0cm\eeqa
\beqa & &
A_6(nnnnc,ncnk,nk;t^2\e,t^3\e)\nonumber\\
& &\quad =\sum_{\gamma_1,\gamma_2,\gamma_3}\frac{(-\I
t)^2}{2}\frac{\e^{-\I E_{\gamma^\prime}t}
\left|g_1^{\gamma_3\gamma^\prime}\right|^2 g_1^{\gamma\gamma^\prime}
g_1^{\gamma^\prime\gamma_1}g_1^{\gamma_1\gamma_2}g_1^{\gamma_2\gamma^\prime}
\eta_{\gamma\gamma_1}\eta_{\gamma\gamma_2}\eta_{\gamma\gamma_3}
\eta_{\gamma_1\gamma_3}\eta_{\gamma_2\gamma_3}}
{\left(E_\gamma-E_{\gamma^\prime}\right)
\left(E_{\gamma_1}-E_{\gamma^\prime}\right)\left(E_{\gamma_2}-E_{\gamma^\prime}\right)
\left(E_{\gamma_3}-E_{\gamma^\prime}\right)}.\eeqa
\beqa
A_6(nnnnc,nnck,c;t^2\e,t^3\e)&=&\sum_{\gamma_1,\gamma_2,\gamma_3}\frac{(-\I
t)^2}{2}\frac{\e^{-\I E_{\gamma}t}
\left|g_1^{\gamma\gamma_2}\right|^2
g_1^{\gamma\gamma_1}g_1^{\gamma_1\gamma_2}
g_1^{\gamma_2\gamma_3}g_1^{\gamma_3\gamma}\eta_{\gamma_1\gamma_3}\delta_{\gamma\gamma^\prime}}
{\left(E_\gamma-E_{\gamma_1}\right)
\left(E_{\gamma}-E_{\gamma_2}\right)^2\left(E_{\gamma}-E_{\gamma_3}\right)}.
\hskip 1.0cm\eeqa
\beqa A_6(nnnnc,nnnk,cck;t^2\e,t^3\e)
&=&\sum_{\gamma_1,\gamma_2,\gamma_3}\frac{(-\I t)^2}{2}\frac{\e^{-\I
E_{\gamma}t} \left|g_1^{\gamma\gamma_1}\right|^2
g_1^{\gamma\gamma_1}
g_1^{\gamma_1\gamma_2}g_1^{\gamma_2\gamma_3}g_1^{\gamma_3\gamma}
\eta_{\gamma\gamma_2}\eta_{\gamma_1\gamma_3}
\delta_{\gamma\gamma^\prime}} {\left(E_\gamma-E_{\gamma_1}\right)^2
\left(E_{\gamma}-E_{\gamma_2}\right)\left(E_{\gamma}-E_{\gamma_3}\right)
}.\hskip 1.3cm\eeqa
\beqa & &
A_6(nnnnc,nnnk,cnk;t^2\e,t^3\e)\nonumber\\
& &\quad =\sum_{\gamma_1,\gamma_2,\gamma_3,\gamma_4}\frac{(-\I
t)^2}{2}\frac{\e^{-\I E_{\gamma}t}
\left|g_1^{\gamma\gamma_4}\right|^2 g_1^{\gamma\gamma_1}
g_1^{\gamma_1\gamma_2}g_1^{\gamma_2\gamma_3}g_1^{\gamma_3\gamma}
\eta_{\gamma\gamma_2}\eta_{\gamma\gamma_3}\eta_{\gamma_1\gamma_3}
\eta_{\gamma_1\gamma_4}\eta_{\gamma_2\gamma_4}\eta_{\gamma_3\gamma_4}\delta_{\gamma\gamma^\prime}}
{\left(E_\gamma-E_{\gamma_1}\right)
\left(E_{\gamma}-E_{\gamma_2}\right)\left(E_{\gamma}-E_{\gamma_3}\right)
\left(E_{\gamma}-E_{\gamma_4}\right)}.\hskip 1.3cm\eeqa
\beqa
A_6(nnnnn,cccc;t^2\e,t^3\e)&=&\sum_{\gamma_1,\gamma_2}\frac{(-\I
t)^2}{2}\frac{\e^{-\I E_{\gamma}t}\left(
g_1^{\gamma\gamma_1}g_1^{\gamma_1\gamma_2}
g_1^{\gamma_2\gamma}\right)^2\delta_{\gamma\gamma^\prime}}
{\left(E_\gamma-E_{\gamma_1}\right)^2
\left(E_{\gamma}-E_{\gamma_2}\right)^2}. \eeqa
\beqa
A_6(nnnnn,ccnc;t^2\e,t^3\e)&=&\sum_{\gamma_1,\gamma_2,\gamma_3}\frac{(-\I
t)^2}{2}\frac{\e^{-\I E_{\gamma}t}
g_1^{\gamma\gamma_1}g_1^{\gamma_1\gamma_2}
g_1^{\gamma_2\gamma}g_1^{\gamma\gamma_1}g_1^{\gamma_1\gamma_3}
g_1^{\gamma_3\gamma}\eta_{\gamma_2\gamma_3}\delta_{\gamma\gamma^\prime}}
{\left(E_\gamma-E_{\gamma_1}\right)^2
\left(E_{\gamma}-E_{\gamma_2}\right)\left(E_{\gamma}-E_{\gamma_3}\right)}.
\hskip 1.3cm\eeqa
\beqa
A_6(nnnnn,cncc;t^2\e,t^3\e)&=&\sum_{\gamma_1,\gamma_2,\gamma_3}\frac{(-\I
t)^2}{2}\frac{\e^{-\I E_{\gamma}t}
g_1^{\gamma\gamma_1}g_1^{\gamma_1\gamma_2}
g_1^{\gamma_2\gamma}g_1^{\gamma\gamma_3}g_1^{\gamma_3\gamma_2}
g_1^{\gamma_2\gamma}\eta_{\gamma_1\gamma_3}\delta_{\gamma\gamma^\prime}}
{\left(E_\gamma-E_{\gamma_1}\right)
\left(E_{\gamma}-E_{\gamma_2}\right)^2\left(E_{\gamma}-E_{\gamma_3}\right)}.
\hskip 1.3cm\eeqa
\beqa
A_6(nnnnn,cnnc,kck;t^2\e,t^3\e)&=&\sum_{\gamma_1,\gamma_2,\gamma_3}\frac{(-\I
t)^2}{2}\frac{\e^{-\I E_{\gamma}t}
g_1^{\gamma\gamma_1}g_1^{\gamma_1\gamma_2}
g_1^{\gamma_2\gamma}g_1^{\gamma\gamma_3}g_1^{\gamma_3\gamma_1}
g_1^{\gamma_1\gamma}\eta_{\gamma_2\gamma_3}\delta_{\gamma\gamma^\prime}}
{\left(E_\gamma-E_{\gamma_1}\right)^2
\left(E_{\gamma}-E_{\gamma_2}\right)\left(E_{\gamma}-E_{\gamma_3}\right)}.
\hskip 1.3cm\eeqa
\beqa & &
A_6(nnnnn,cnnc,knk;t^2\e,t^3\e)\nonumber\\
& & \quad =\sum_{\gamma_1,\gamma_2,\gamma_3,\gamma_4}\frac{(-\I
t)^2}{2}\frac{\e^{-\I E_{\gamma}t}
g_1^{\gamma\gamma_1}g_1^{\gamma_1\gamma_2}
g_1^{\gamma_2\gamma}g_1^{\gamma\gamma_3}g_1^{\gamma_3\gamma_4}
g_1^{\gamma_4\gamma}\eta_{\gamma_1\gamma_3}\eta_{\gamma_1\gamma_4}
\eta_{\gamma_2\gamma_3}\eta_{\gamma_2\gamma_4}\delta_{\gamma\gamma^\prime}}
{\left(E_\gamma-E_{\gamma_1}\right)
\left(E_{\gamma}-E_{\gamma_2}\right)\left(E_{\gamma}-E_{\gamma_3}\right)
\left(E_{\gamma}-E_{\gamma_4}\right)}. \eeqa
Based on above non-zero 91 terms and vanishing 112 terms with
$t^2\e, t^3\e$ factor parts in all of 203 contraction- and anti
contraction- expressions, we can, via the reorganization and
summation, obtain the following concise forms: \beqa A_6(t^2\e^{-\I
E_{\gamma}t},{\rm D})&=&\frac{(-\I
t)^2}{2!}\left(G_\gamma^{(3)}\right)^2\e^{-\I
E_{\gamma}t}\delta_{\gamma\gamma^\prime}+\frac{(-\I t)^2}{2!}2
G_\gamma^{(2)}G_\gamma^{(4)}\e^{-\I
E_{\gamma}t}\delta_{\gamma\gamma^\prime}\nonumber\\
& &-\frac{(-\I
t)^2}{2!}\sum_{\gamma_1}\frac{\left(G_\gamma^{(2)}\right)^2\e^{-\I
E_{\gamma}t}}{\left(E_\gamma-E_{\gamma_1}\right)^2}
g_1^{\gamma\gamma_1}g_1^{\gamma_1\gamma}\delta_{\gamma\gamma^\prime}.
\eeqa
\beqa A_6(t^2\e^{-\I E_{\gamma_1}t},{\rm D})&=&\frac{(-\I
t)^2}{2!}\sum_{\gamma_1}\frac{\left(G_\gamma^{(2)}\right)^2\e^{-\I
E_{\gamma_1}t}}{\left(E_\gamma-E_{\gamma_1}\right)^2}
g_1^{\gamma\gamma_1}g_1^{\gamma_1\gamma}\delta_{\gamma\gamma^\prime}.
\eeqa \beqa A_6(t^2\e^{-\I E_{\gamma^\prime}t},{\rm D})&=&0.\eeqa
\beqa A_6(t^2\e^{-\I E_{\gamma}t},{\rm N})&=&\frac{(-\I t)^2}{2!}2
G_\gamma^{(2)}G_\gamma^{(3)}\frac{\e^{-\I
E_{\gamma}t}}{\left(E_\gamma-E_{\gamma^\prime}\right)}g_1^{\gamma\gamma^\prime}\nonumber\\
& &+\frac{(-\I
t)^2}{2!}\sum_{\gamma_1}\frac{\left(G_\gamma^{(2)}\right)^2\e^{-\I
E_{\gamma}t}}{\left(E_\gamma-E_{\gamma_1}\right)\left(E_\gamma-E_{\gamma^\prime}\right)}
g_1^{\gamma\gamma_1}g_1^{\gamma_1\gamma^\prime}\eta_{\gamma\gamma^\prime}.
\eeqa
\beqa A_6(t^2\e^{-\I E_{\gamma_1}t},{\rm N})&=&-\frac{(-\I
t)^2}{2!}\sum_{\gamma_1}\frac{\left(G_{\gamma_1}^{(2)}\right)^2\e^{-\I
E_{\gamma_1}t}}{\left(E_\gamma-E_{\gamma_1}\right)\left(E_{\gamma_1}-E_{\gamma^\prime}\right)}
g_1^{\gamma\gamma_1}g_1^{\gamma_1\gamma^\prime}\eta_{\gamma\gamma^\prime}.
\eeqa
\beqa A_6(t^2\e^{-\I E_{\gamma^\prime}t},{\rm N})&=&-\frac{(-\I
t)^2}{2!}2
G_{\gamma^\prime}^{(2)}G_{\gamma^\prime}^{(3)}\frac{\e^{-\I
E_{\gamma^\prime}t}}{\left(E_\gamma-E_{\gamma^\prime}\right)}g_1^{\gamma\gamma^\prime}\nonumber\\
& &+\frac{(-\I
t)^2}{2!}\sum_{\gamma_1}\frac{\left(G_{\gamma^\prime}^{(2)}\right)^2\e^{-\I
E_{\gamma^\prime}t}}{\left(E_\gamma-E_{\gamma^\prime}\right)\left(E_{\gamma_1}-E_{\gamma^\prime}\right)}
g_1^{\gamma\gamma_1}g_1^{\gamma_1\gamma^\prime}\eta_{\gamma\gamma^\prime}.
\eeqa Their forms are indeed the same as expected and can be merged
reasonably to the lower order terms in order to obtain the improved
forms of perturbed solutions.

\end{appendix}




\begin{references}
\bibitem{seq}E. Schr\"{o}dinger, Ann. Phys. {\bf 79}, 489-527(1926)
\bibitem{diracpqm}P. A. M. Dirac, {\it The principles of Quantum
Mechanics}, 4th edn(revised), Oxford: Clarendon Press (1974)
\bibitem{vonneumann}J. von Neumann, {\it The Mathematical Foundations of Quantum
Mechanics}, Princeton: University Press (1955)
\bibitem{LSE}B. A. Lippmann and J. Schwinger, Phys. Rev. {\bf 79},
469(1950)
\bibitem{inout}L. Van Hove, Bulletin Acad. Roy. Belgique {\bf 37},
1055(1951); Physica {\bf 18}, 145(1952)
\bibitem{Fermi}E. Fermi, {\it Nuclear Physics}, Chicago: University of
Chicago Press (1950)
\bibitem{MyDD}An Min Wang, quant-ph/0601051(v3)
\end{references}
\end{document}